\documentclass[a4paper,12pt,oneside,openright]{book}

\usepackage{customisations}

\usepackage{booktabs} % booktabs provides professional formatting commands for tables
\usepackage{amsmath} % amsmath provides extra maths symbols
\usepackage{textcomp} % textcomp provides extra text symbols (like a degrees celsius symbol)
\usepackage{graphicx}
\usepackage{url}
\usepackage[usenames,dvipsnames]{xcolor}
\usepackage{amsmath, amsthm, mathrsfs, amsfonts, dsfont, amssymb}
\usepackage{tikz-feynman, contour}
\newcommand{\myeqref}[1]{(\ref{#1})}
\newcommand{\mysecref}[1]{\ref{#1}}

\def\Caravel{{\textsc{Caravel}}} 
\DeclareMathOperator\arctanh{arctanh}
\DeclareMathOperator\arcosh{arcosh}

\newcommand\tickg{{\color{ForestGreen}\checkmark}}
\newcommand\crossr{{\color{RedOrange}\text{\sffamily X}}}

\title{An Amplitudes Approach to Spinning Black Hole Scattering}
\author{Dogan Akpinar}
\date{July 30, 2026} % of submission

\begin{document}

% Thesis front matter - title page, abstract, acknowledgements, declaration and table of contents
% See customisations.sty to modify the title page or declaration
\singlespacing
\maketitlepage
\frontmatter
\eighteenptleading
\chapter{Lay Summary}

Black holes are among the most fascinating objects in our universe. Although their inner structure remains shrouded 
in mystery to this day, they nonetheless provide invaluable insights into gravity itself. Indeed, when black holes 
(and/or neutron stars) form binary systems---pairs of compact objects that spiral toward one another and eventually 
merge---they release powerful bursts of energy in the form of gravitational waves: ripples in the fabric of spacetime 
that carry detailed information about their masses, spins, and dynamics.

For over a century, gravitational waves remained an exciting theoretical prediction of Einstein’s theory of general 
relativity. But this changed dramatically in 2015, when the LIGO–Virgo–KAGRA collaboration reported the first direct 
detection of gravitational waves produced by a binary merger comprised of two black holes. Not only did this confirm 
that Einstein was indeed right, but it also ushered in a new era of gravitational-wave astronomy, opening an 
unprecedented observational window into the universe and enabling high-precision tests of gravity. Naturally, a major 
goal in advancing gravitational-wave astronomy boils down to improving the sensitivity of our current 
ground-based detectors, as well as developing future space-based interferometers. In fact, as these instruments 
become more sensitive, it becomes increasingly important to improve our theoretical understanding and modelling of 
compact binary systems in order to accurately interpret the signals they detect and extract the underlying physics. 
This translates to pushing the frontier of state-of-the-art computations as well as systematically incorporating 
physical effects such as spin, tidal deformations, absorption, etc.

The focus of this thesis is to advance precision computations involving spinning black holes. Our strategy is to 
compute classical observables that describe their scattering dynamics using quantum field theory, enabling an 
efficient workflow that draws on powerful theoretical methods developed for particle physics. We further shed new 
light on the symmetry and integrability properties of scattering processes involving Kerr black holes.

\chapter{Abstract}

The monumental detection of gravitational waves by the LIGO-Virgo-KAGRA collaboration has ushered in a new era for precision studies of gravity. Notably, it has provided unprecedented insight into the complex dynamics of binary systems composed of massive, compact astrophysical objects such as neutron stars and black holes. This observational leap has called for equally impressive advances in our theoretical understanding of these gravitational two-body systems. Towards this direction, scattering amplitudes have remarkably emerged as a robust tool for precisely this purpose, allowing us to leverage the rich machinery of high-energy physics. 

In this thesis, we  study the classical dynamics associated with the scattering of two black holes. Our overarching goal is to systematically include the effects of black hole spin---an important feature of real astrophysical systems---thereby describing the dynamics of Kerr black holes. To this end, we model Kerr black holes as massive point particles in finite-spin representations and compute their scattering dynamics perturbatively. In fact, this strategy introduces an ambiguity that mixes classical and quantum contributions, which we elegantly resolve using our \textit{spin interpolation} method. Using this method, we obtain a completely novel classical amplitude for Kerr black hole scattering accurate to the quartic order in spin and at the third post-Minkowskian order. From this amplitude we extract the gauge-invariant radial action, enabling the computation of classical non-aligned-spin observables---such as the linear and angular impulses---using covariant Dirac brackets.

Most excitingly, the classical amplitude we obtain exhibits a spin-shift symmetry: it remains invariant under a shift of the black hole spin by the momentum transfer in the scattering process. Motivated by this mysterious property, we examine the conserved quantities governing the scattering process and show that the dynamics of a probe black hole remains integrable at asymptotic infinity through the quartic order in probe spin. Finally, we show that the quartic-in-spin radial action, for a probe black hole in a Kerr background, at any order in perturbation theory is completely fixed by the aligned-spin dynamics, provided that it satisfies asymptotic integrability and spin-shift symmetry.

\singlespacing
% Uncomment this line if you need to declare published work which forms part of the thesis
%\declarationpublications{\cite{Akpinar:2024meg,Akpinar:2025bkt,Akpinar:2025huz}}
\makedeclaration

\chapter{Acknowledgements}

\noindent

\normalsize

First, I would like to thank my supervisor, Mao Zeng, for your constant support, guidance, and encouragement 
throughout my PhD. I am grateful for your patience when faced with my (infinitely many) naive and idiotic 
questions, alongside your willingness to teach me everything that you know. It is true that the majority of the 
knowledge and skills that I have acquired during my PhD are thanks to you. 

I am also grateful to my examiners, Donal O'Connell and Harald Ita, for reading my thesis and for a very enjoyable and pleasant viva experience! 

Similarly, I would like to thank my amazing collaborators (and friends): Manfred, Fernando, Riccardo, Graham, 
Michael, Vittorio, Irene, Matteo, Asaad, Pablo, Toshifumi, Katsuki, Andrea, Hyun, and Kaho. I am, in 
particular, grateful to Manfred, Fernando, Riccardo, Graham, Andrea, Toshifumi, Katsuki, Hyun, and Kaho for 
all of our invaluable and reassuring chats about the intricacies of both physics and life. Especially for all 
of the fun nights at numerous Izakaya (and in one instance karaoke) with Toshifumi, Katsuki, Andrea, Hyun, 
Kaho, and many many others. These are most definitely experiences that I will remember and cherish forever. I 
am extremely indebted to Katsuki and Andrea for inviting me many times to visit them at YITP, as well as 
Toshifumi for his hospitality and openness in allowing me to spend months at a time at the Komaba campus of the 
University of Tokyo with him, Hyun, Kaho, Qianhang, Masataka, Tomoya, and his many (and I mean \textit{many}) 
other students. I am grateful to Fernando for inviting me to FSU, to visit him, where I was, in fact, not 
attacked by a crocodile nor the infamous \textit{Florida man} (thankfully!). I am also very grateful to Matteo 
and Zvi for inviting me to UCLA, where I had a lot of delicious food, chatted a lot about physics, and 
witnessed strange scenarios, including, but not limited to, car bumpers in the middle of the pavement. 

I am grateful to my amazing office mates (some of whom were here temporarily) William, Mile, Athari, Idrus, 
Xiancong, Daniel, Asaad, João, and Giorgio (who was also my awesome flatmate during my time at Edinburgh) for 
our constant jokes, therapy sessions, chats about physics, the most random phases of spewing constant brain rot 
(mostly Italian), and karaoke. I am forever grateful for all of the people I have met and the friendships that 
I have formed, in Edinburgh and all around the world. An exhaustive, and yet ever more incomplete, list 
includes: Anton, Tej, Franz, Latham, Rafael, Giuseppe, Karthik, Lucile, Fabian, Giacomo, Panagiotis, Roy, Jaco, 
Jen, Anh, John, Matthew, Matthaios, Filippo, François, Anne-Maëlle, Amadeo, Andres, Kays, Jitze, Mathias, 
Gustav, Gus, George, Shao Zho, Copper, Briag, Kymani, Nathan, Paolo, Vats, Raikhik, Maegan, Maria, Laura, Kata, 
Giulio, Sid, Ryo, Harkirat, Jeremy, Alasdair, Conor, Henrique, Thanos, Estifa'a, Joseph, and many many many others. 

I thank my friends and family for all of their ongoing love and support. Especially to my parents who have 
always strived to provide the best for my brother and me. I am thankful to all of our adorable pets: 
Zara, Fenrir, Freya, Kratos, alongside our late Sasha and Eris whom I miss very much. It is also important that 
this list includes the distinguished gentleman (and indeed the most adorable maltipoo) Apollo Beethoven. 
Finally, above everything, I thank God for all of His guidance, as I would not be here if it were not for His 
mercy, love, and guidance. 

%Last, but in no way least, to the love of my life, Hayssa, for her endless encouragement, love, and support that have gotten me through the hardest parts of my PhD. I am thankful also to her family for their hospitality for all of my long stays in Japan, as well as the distinguished gentleman (and the most adorable maltipoo) Apollo Beethoven

\cleardoublepage
\phantomsection
\addcontentsline{toc}{chapter}{\contentsname}
\setcounter{tocdepth}{2}
\tableofcontents

\cleardoublepage
\phantomsection
\addcontentsline{toc}{chapter}{\listfigurename}
\listoffigures

\cleardoublepage
\phantomsection
\addcontentsline{toc}{chapter}{\listtablename}
\listoftables

% Include main matter here
\mainmatter
\eighteenptleading
%--------------------------------------------------------------------------------
\chapter{Introduction}
%--------------------------------------------------------------------------------

The theory of general relativity (GR), introduced by Albert Einstein in 1915, fundamentally transformed our 
understanding of gravity. Rather than describing gravity as a force acting at a distance, it is, instead,  
interpreted as a manifestation of the curvature of spacetime generated by mass and energy. In essence, massive 
objects deform the geometric structure of spacetime, and this curvature governs the motion of matter and 
radiation.

The intricate relationship between matter and geometry is encoded in the Einstein field equations (EFEs)
\begin{equation}
    R_{\mu \nu} - \frac{1}{2} g_{\mu \nu} R = 8\pi G T_{\mu \nu}\;.
    \label{eq:EinsteinFE}
\end{equation}
The left-hand side encapsulates the geometric properties of spacetime---through the Ricci tensor, the Ricci 
scalar, and the metric tensor---thereby describing how spacetime curves. The right-hand side contains the 
energy-momentum tensor $T_{\mu \nu}$, which characterises the distribution and flow of matter 
and energy. Although this geometric interpretation of gravity 
differs profoundly from the Newtonian notion of forces, it is nevertheless striking that 
Newton's universal gravitational constant $G$ still appears in the EFEs.

The central challenge in GR boils down to solving the 
EFEs for physically relevant systems. This standard task, however, is not 
straightforward in the slightest for GR, as it involves solving ten coupled, non-linear partial differential 
equations---a notoriously challenging problem. There is, indeed, no general, analytic solution; instead, 
solutions are obtained by exploiting physical assumptions and symmetry arguments. Among the most interesting 
predictions that emerge from the EFEs is the existence of gravitational waves (GWs); predicted by Einstein in 
1916~\cite{einstein1916, einstein1916_naeherungsweise}. In a nutshell, GWs are perturbations of spacetime itself 
that propagate at the speed of light, carrying energy, linear and angular momentum, and are 
detectable by ground-based (and future space-based) interferometers on Earth. Remarkably, this century-old 
prediction was not confirmed until 2015, with the first direct detection of GWs by the LIGO–Virgo–KAGRA 
collaboration~\cite{LIGOScientific:2016aoc, LIGOScientific:2016dsl, LIGOScientific:2016sjg, 
LIGOScientific:2017vwq, LIGOScientific:2017bnn}. This monumental achievement marked the 
beginning of an era of GW astronomy, opening an unprecedented observational window into the universe and 
enabling high-precision tests of gravity. 

The sources of GWs are binary mergers: systems in which celestial objects, such as black 
holes and/or neutron stars, inspiral and eventually coalesce. Following the first detection, GW science 
has progressed at an astonishing pace, with the expectation that over 200 merger events will be detected by the end 
of the current LIGO O4b run. Indeed, the highly anticipated next-generation detectors~\cite{Punturo:2010zz, 
LISA:2017pwj, Reitze:2019iox} will provide significantly higher sensitivity and 
therefore require a commensurate increase in the accuracy of theoretical waveform 
templates~\cite{Purrer:2019jcp}. Alongside these advances in precision, it is crucial to account for physical effects 
present in realistic binary mergers; a particularly important effect is the spin of the compact objects. This 
therefore motivates ongoing efforts to push the frontier of precision modelling for spinning binary dynamics.

\begin{figure}[h!]
	\centering
		\includegraphics[width=0.7\linewidth]{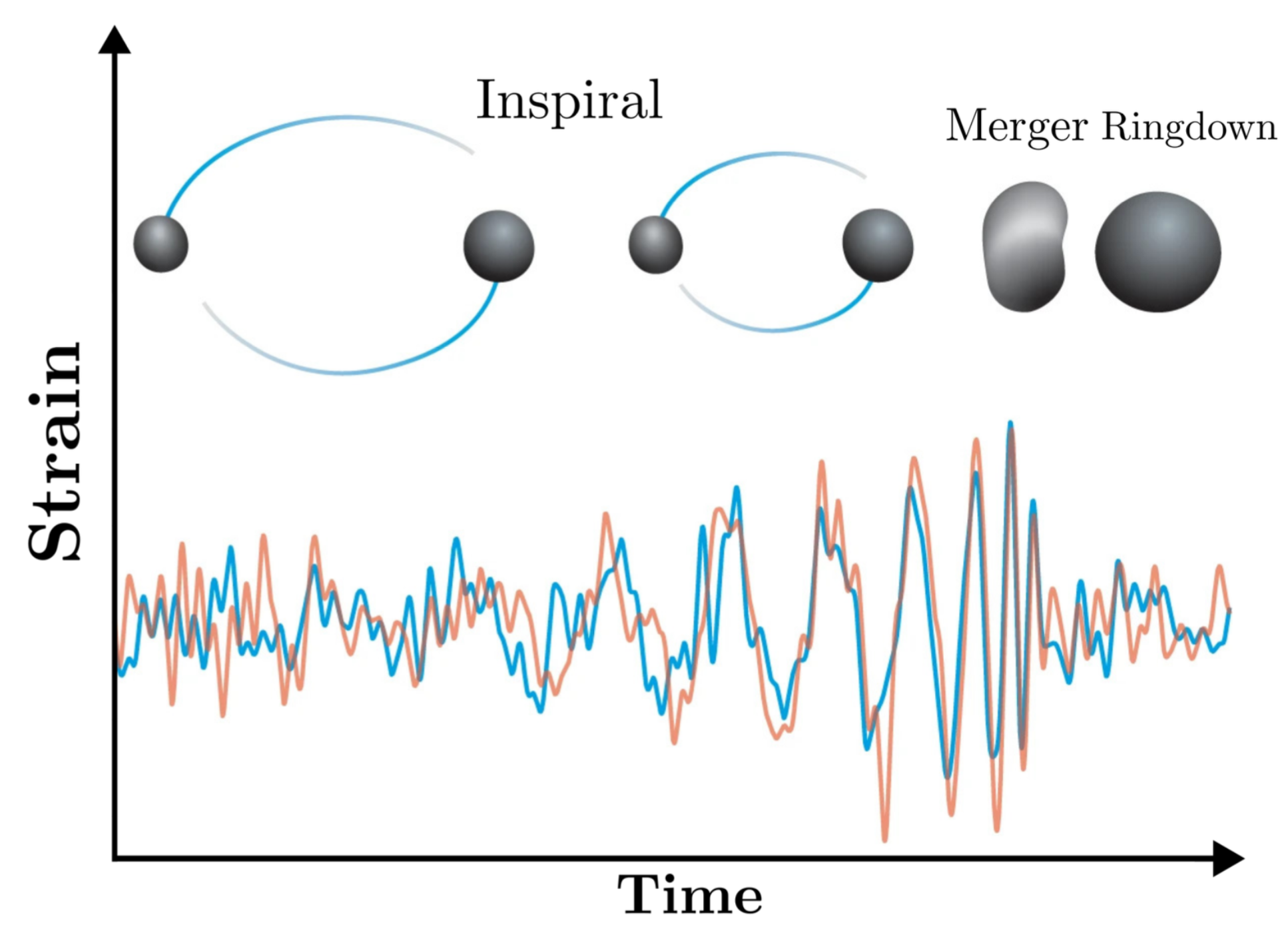}
	\caption{The stages of a generic binary merger~\cite{Bailes:2021tot}. The illustration (GWs are not depicted) shows the three different stages characterising the full motion of a binary merger event: inspiral, merger, and ringdown. Below this illustration is a very schematic strain-time signal detected on Earth throughout the whole motion of the merger event.}
	\label{fig:BinaryMerger}
\end{figure}

\begin{figure}[h!]
	\centering
		\includegraphics[width=0.7\linewidth]{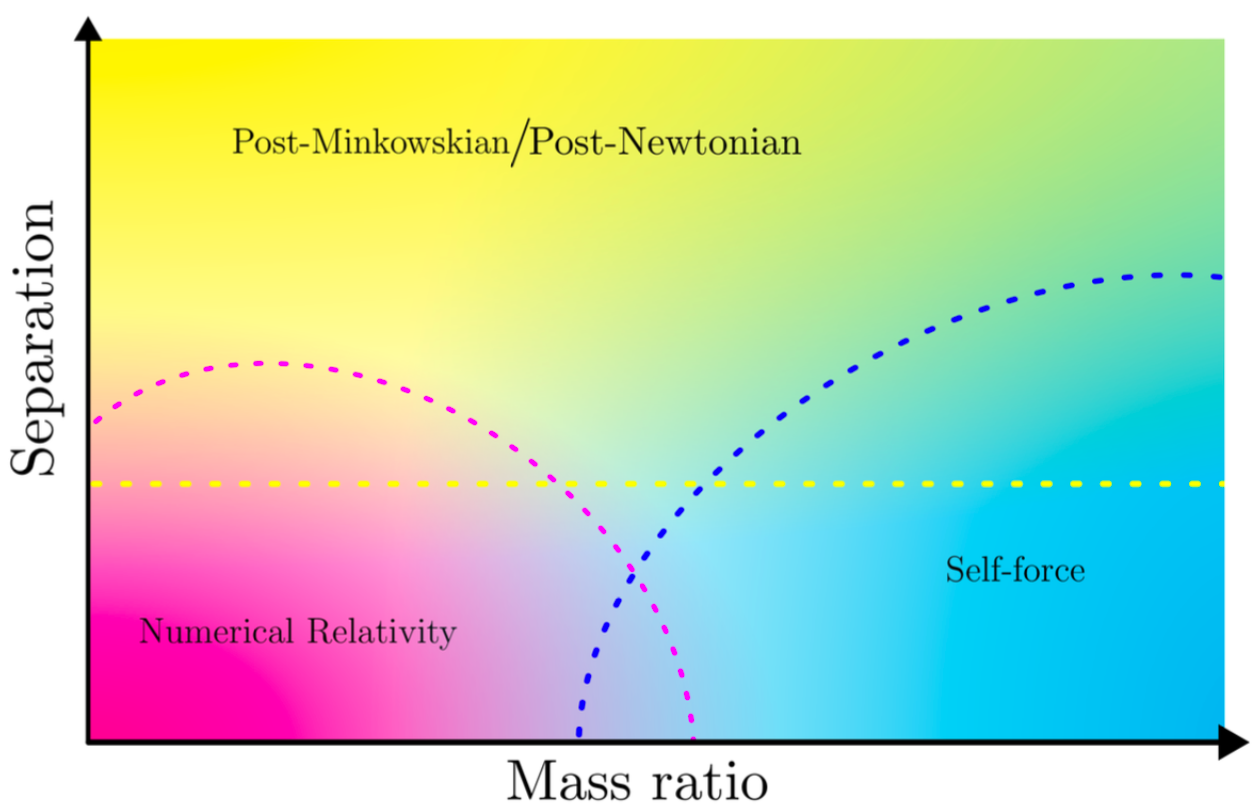}
	\caption{The parameter space of the two-body scattering problem in GR~\cite{Barack:2023oqp}. The coloured diagram schematically shows the domains of validity for different perturbative techniques. The regions of overlap provide points of comparison and exchange of information. In this thesis, we will explore the post-Minkowskian approximation, which is valid for large separation and arbitrary mass ratios.}
	\label{fig:ParameterSpace}
\end{figure}

Obtaining an exact, analytic solution for the full motion of the binary mergers is  
impossible. Nevertheless, the binary dynamics can be roughly organised into three different stages: the 
inspiral, the merger, and the ringdown (see Fig.~\ref{fig:BinaryMerger} for a schematic illustration). 
Each regime is governed by different physical scales and 
therefore admits very different theoretical treatments. A robust description of the full coalescence consequently 
requires combining complementary analytical and numerical approaches. If we are interested in analytic approaches 
(which we are in this thesis), perturbation theory provides a particularly powerful framework given that suitably 
small parameters are accessible to us. The gravitational two-body problem 
possesses several such expansion parameters, enabling systematic approximation schemes that are valid in different
regions of the parameter space (see Fig.~\ref{fig:ParameterSpace}). Depending on the regime of choice, and 
the stage of the evolution that we are interested in, different perturbative approaches can be employed. For the 
inspiral phase, the most convenient choices are the so-called post-Newtonian 
(PN)~\cite{Blanchet:2002av,Foffa:2013qca} and 
post-Minkowskian (PM)~\cite{Bern:2019crd,Buonanno:2022pgc,Bjerrum-Bohr:2022blt} expansions---we will discuss 
these below; the gravitational self-force (GSF) formalism describes extreme-mass-ratio systems and is valid in 
both the inspiral and merger phases~\cite{Barack:2018yvs,Pound:2021qin,Poisson:2011nh}; and, finally, black 
hole perturbation theory treats small perturbations of a single black hole background for the ringdown 
region~\cite{Teukolsky:1973ha, Kokkotas:1999bd}. If one is interested in numerical approaches, the 
standard strategy is numerical relativity~\cite{PhysRevLett.95.121101} which is applicable throughout the full 
motion; albeit it is costly to simulate and prone to numerical errors across many orbital cycles. Taken together, 
these methods provide powerful insights into binary merger dynamics across a wide range of mass ratios, velocities, 
and gravitational field strengths.

In this thesis, we are interested in employing perturbation theory for the inspiral phase of the dynamics. 
The natural choice for modelling this is the PN expansion. To understand what perturbative 
parameters are available to us, let us consider a binary system of two black holes with comparable masses 
$m_1 \sim m_2 \sim m$ and velocities $u_1 \sim u_2 \sim u$ separated by a distance $r$; working in units where 
$c = 1$. From these quantities and Newton's constant $G$, one can build the dimensionless parameter
\begin{equation}
	\frac{Gm}{r} \sim u^2\;,
\end{equation}
where the scaling follows from the virial theorem. When the separation satisfies $r \gg Gm$, corresponding 
to distances much larger than the Schwarzschild radius (of order $Gm$), both $Gm/r$ and $u^2$ are 
small and of comparable size. This identifies a natural expansion regime characterised by weak gravitational 
fields, $Gm/r \ll 1$, and small (non-relativistic) velocities, $u \ll 1$. This hierarchy constitutes the PN 
expansion, which has a long and fruitful history in providing theoretical predictions for the dynamics of binary black hole 
mergers~\cite{Damour:1985mt, Gilmore:2008gq, Damour:2001bu, 
Bjerrum-Bohr:2002gqz,Blanchet:2003gy, Itoh:2003fy,Foffa:2011ub,Jaranowski:2012eb, Damour:2014jta,Galley:2015kus,Damour:2015isa, 
Damour:2016abl, Bernard:2015njp, Bernard:2016wrg, Foffa:2012rn, 
Foffa:2016rgu,Porto:2017dgs,Porto:2017shd,Foffa:2019yfl,Blumlein:2020pog,Foffa:2019hrb,Blumlein:2019zku,Bini:2020wpo,
Blumlein:2020pyo,Blumlein:2020znm,Bini:2020uiq,Blumlein:2021txj, 
Porto:2005ac,Steinhoff:2011sya,Levi:2014gsa,Levi:2015msa,Maia:2017yok,Levi:2018nxp,Levi:2020uwu, Brunello:2025gpf}.  

Despite the PN expansion being a natural choice for describing the inspiral phase, here we are actually interested in 
obtaining results that are valid relativistically, i.e.~to all orders in the velocity expansion. To do this, we 
use the PM expansion. The system that we consider is again that of two black holes of 
comparable masses, but now in an \textit{unbound} or \textit{hyperbolic} orbit. This
essentially describes a scattering scenario, where the black holes enter from past infinity with velocities 
$u_1$ and $u_2$, and with an impact parameter (i.e.~separation) $b$, and then scatter to future infinity. The 
natural expansion parameter is then $Gm/b$, which is small provided $b \gg Gm$, i.e.,~for very far separated black 
holes. Because we now have an unbound orbit, the virial theorem is not applicable and we have access to 
relativistic dynamics. Thus, the PM expansion constitutes a weak field expansion, $Gm/b \ll 1$. Of course, one 
might protest that unbound orbits are unphysical, especially because they have not yet been observed by GW 
detectors~\cite{Morras:2021atg,Bini:2023gaj}. 
The advantage of this scenario, however, lies in the plethora of machinery available to us. More specifically, 
since our goal is to use quantum field theory (QFT) techniques to study the two-body problem, reformulating the 
scattering dynamics using the language of scattering amplitudes gives us access to the vast amount of machinery 
developed for particle physics\footnote{Here we treat gravity as a low-energy effective field theory, allowing 
us to make meaningful predictions when graviton energies are much smaller than the Planck mass.}. It is very crucial, 
however, to stress that, although recasting a classical problem as a quantum problem clearly seems counterintuitive, 
this approach has yielded remarkable progress in both amplitude- and worldline-based methods, both in the absence of 
spin~\cite{Cheung:2018wkq, Kosower:2018adc, Bern:2019nnu, Bern:2019crd, Cristofoli:2019neg, Bjerrum-Bohr:2019kec, 
Brandhuber:2021eyq, Bern:2021dqo, Bern:2021yeh, Damgaard:2023ttc, Kosmopoulos:2023bwc, Bern:2025zno, Kalin:2020mvi, Kalin:2020fhe, 
Dlapa:2021vgp, kalin:2022hph, Dlapa:2022lmu, Dlapa:2023hsl, Dlapa:2021npj, Mogull:2020sak, Jakobsen:2021smu, Jakobsen:2022psy, 
Jakobsen:2023oow, Driesse:2024xad, Driesse:2024feo, He:2025how, Cheung:2023lnj, Cheung:2024byb, Driesse:2026qiz, Goldberger:2004jt, 
Goldberger:2009qd, Goldberger:2022rqf, Goldberger:2005cd, Cristofoli:2021jas, Aoki:2026eos, Alessio:2025isu, Alessio:2026bdi, 
Gonzo:2026yha, Bjerrum-Bohr:2025bqg, Bjerrum-Bohr:2026fhs, Bautista:2026qse} and with spin~\cite{Maybee:2019jus, Porto:2005ac, 
Porto:2008jj, Porto:2016pyg, Bini:2017xzy, Bini:2018ywr, Vines:2017hyw, Vines:2018gqi, Guevara:2017csg, Guevara:2018wpp, 
Chung:2018kqs, Arkani-Hamed:2019ymq, Guevara:2019fsj, Chung:2019duq, Damgaard:2019lfh, Aoude:2020onz, Chung:2020rrz, 
Guevara:2020xjx, Bern:2020buy, Kosmopoulos:2021zoq, Chen:2021qkk, FebresCordero:2022jts, Bern:2022kto, Bern:2023ity, 
Menezes:2022tcs, Riva:2022fru, Damgaard:2022jem, Aoude:2022thd, Aoude:2022trd, Bautista:2022wjf, Gonzo:2023goe, Aoude:2023vdk, 
Lindwasser:2023zwo, Brandhuber:2023hhl, DeAngelis:2023lvf, Aoude:2023dui, Bohnenblust:2023qmy, Gatica:2024mur, Cristofoli:2021jas, 
Luna:2023uwd, Gatica:2023iws, Liu:2021zxr, Jakobsen:2021lvp, Jakobsen:2021zvh, Jakobsen:2022fcj, Jakobsen:2022zsx, 
Jakobsen:2023ndj, Jakobsen:2023hig, Heissenberg:2023uvo, Lindwasser:2023dcv, Bautista:2023sdf, Cangemi:2023ysz, 
Brandhuber:2024bnz, Chen:2024mmm, Correia:2024jgr, Bhattacharyya:2024kxj, Alaverdian:2024spu, Brandhuber:2024qdn, 
Brandhuber:2024lgl, Akpinar:2024meg, Akpinar:2025bkt, Bohnenblust:2024hkw, Haddad:2024ebn, Bonocore:2024uxk, 
Akpinar:2025huz, Bohnenblust:2025gir, Aoki:2024boe, Hoogeveen:2025tew, Akpinar:2025tct, Aoki:2025ihc, 
Aoude:2025xxq,DiVecchia:2021bdo,DiVecchia:2022nna, DiVecchia:2022piu, Bjerrum-Bohr:2026fhx}. This 
involves systematically extracting the classical dynamics from quantum scattering amplitudes (including 
loop level)~\cite{Kosower:2018adc, Bern:2019nnu}, and subsequently computing classically well-defined 
observables~\cite{Kosower:2018adc, Maybee:2019jus, Bern:2021dqo, Cheung:2018wkq, Amati:1990xe}; examples of which are 
the linear and angular impulses, scattering angle, and so on. Given the success of the scattering perspective, a 
large body of work has been dedicated to relating bound dynamics to unbound dynamics either via analytic 
continuations~\cite{Kalin:2019rwq, Kalin:2019inp, Adamo:2022ooq, Adamo:2024oxy} or by matching to an effective 
one-body theory~\cite{Buonanno:1998gg, Buonanno:2000ef, Damour:2000we, Damour:2001tu, Buonanno:2005xu}. This is 
something of an afterthought, as we will assume that the scattering dynamics will ultimately provide information about 
the bound dynamics.

Our goal in this thesis is to compute classical scattering observables that describe spinning binary systems---involving 
so-called Kerr black holes---within the PM expansion; specifically through the quartic order in spin and the third PM order 
(i.e.~3PM or two loops), which is based on Refs.~\cite{Akpinar:2024meg, Akpinar:2025bkt}. As a direct result of these 
computations, we observe that the classical limit of the scattering amplitudes for spinning point particles exhibit a mysterious 
symmetry---the so-called spin-shift symmetry---leading us to the study of conserved quantities and integrability from an 
on-shell and gauge-invariant perspective (based on Ref.~\cite{Akpinar:2025tct}). Taken together, this advances both the 
state-of-the-art and our theoretical understanding of Kerr black hole scattering.

This thesis is organised in the following way. To begin, in Chapter~\ref{sec:Chapter2} we introduce 
all of the key tools and techniques required for state-of-the-art computations. This is a lengthy 
discussion, but it summarises everything needed for our purposes.

In Chapter~\ref{sec:Chapter3} we work through the computation of Ref.~\cite{Akpinar:2024meg}, 
whereby we model Kerr black holes as point particles in (low) finite-spin-$s$ representations minimally 
coupled to gravity (as well as a toy model in electrodynamics), and we compute aligned-spin 
scattering observables through the quadratic order in spin and the 3PM order (and 3PL for electrodynamics). 
Practically speaking, the observables are obtained by first extracting the radial action---a 
gauge-invariant object that is well-defined classically---from non-aligned-spin classical scattering amplitudes. Because we 
model Kerr black holes as point particles in finite spin representations, the main challenge that 
we face is resolving the so-called \textit{spin-Casimir ambiguity}. To address this, we introduce 
the \textit{spin-interpolation} method and show that we can faithfully extract spin-Casimir terms 
from classical finite-spin amplitudes.

We then generalise the spin-interpolation method (in principle, to arbitrary orders in spin) in 
Chapter~\ref{sec:Chapter4} and use it to compute scattering dynamics to quartic order in spin, as 
done in Ref.~\cite{Akpinar:2025bkt}. In both of these chapters, we observe that the classical 
two-loop amplitudes exhibit spin-shift symmetry in the probe limit through 3PM order---an 
unexpected result. It is important to note that this property is also exhibited for 
electrodynamics. Furthermore, in Chapter~\ref{sec:Chapter4} we show that the radial action 
extracted from the classical scattering amplitudes actually encodes the full non-aligned-spin 
dynamics, which we then use to compute scattering observables via the covariant Dirac 
brackets~\cite{Gonzo:2024zxo}, specifically, the linear and angular impulses. In order to 
finalise our efforts in precision calculations, we compute all-orders-in-spin radiation-reaction 
corrections to observables using the results of Ref.~\cite{Alessio:2022kwv} for the real part of 
the eikonal phase. 

By leveraging the available data together with the theoretical machinery developed in the 
literature, and motivated by the observation of spin-shift symmetry at non-trivial orders in 
perturbation theory, we initiate an on-shell, gauge-invariant approach to studying conserved 
quantities and integrability in Chapter~\ref{sec:Chapter5}. Via this effort, we show that the 
known bulk conserved quantities for a quadrupolar probe particle moving in a Kerr background 
remarkably remain conserved---and, more specifically, that the dynamics remain integrable---at 
higher orders in the spin of the probe than previously known, albeit for scattering dynamics. By 
introducing a position-space differential operator for spin-shift symmetry, we further find that it 
persists to all orders in the PM expansion and seems to mysteriously accompany integrability, 
although it is not itself explained by it. A powerful consequence of this analysis is that conserved quantities 
and spin-shift symmetry provide strong constraints, thereby allowing one to bootstrap the radial action---for the motion of a 
spinning probe in a Kerr background to quartic order in spin, but at arbitrary PM order---solely in 
terms of aligned-spin data.

There are also a number of appendices. In Appendix~\ref{sec:YMFeynmanRules} we 
collect the Feynman rules for Yang-Mills theory. In Appendix~\ref{sec:ScatteringAngleAndImpulse} we discuss the 
relation between the scattering impulse and scattering angle when the motion is constrained to lie on a plane (valid 
in the absence of spin or for aligned-spin configurations). In Appendix~\ref{sec:ExpectationValuePauliLubanskiVector} we derive 
the spin vector from general QFT considerations following Ref.~\cite{Maybee:2019jus}, and present the spin vector in the 
spin-$1$ and $2$ representations. In Appendix~\ref{sec:CovariantBoostedpolarisations} we derive the 
all-orders-in-momentum-transfer Lorentz transformations that are used to homogenise the momentum dependence of polarisations. 
In Appendix~\ref{sec:EvaluatingRankedIntegrals} we evaluate all the necessary tensor integrals 
needed to Fourier transform from momentum space to position space. In Appendix~\ref{sec:CarterConstFromHJEquations} we derive 
the Carter constant for a (spinless) probe mass moving in a Kerr background by solving the Hamilton-Jacobi equations.

%--------------------------------------------------------------------------------
\subsection*{Packages}
%--------------------------------------------------------------------------------

In the completion of the publications that contribute to this thesis, the following {\tt Mathematica} packages were 
used: {\tt FeynCalc} \cite{Shtabovenko:2023idz} and {\tt FIRE6}~\cite{Smirnov:2019qkx, Smirnov:2023yhb}.

%--------------------------------------------------------------------------------
\section*{Conventions}
%--------------------------------------------------------------------------------

Before moving on to a technical overview, it is useful to summarise the various 
conventions used throughout this thesis. 

We work in the mostly minus signature $(+1,-1,-1,-1)$ together with 
$\epsilon_{0123} = +1$. Additionally, we work in natural units $c = \hbar = 1$ 
unless stated otherwise. 

Fourier transforms are defined by 
\begin{equation}
	f(x) = \int \hat{\mathrm{d}}^n k \ e^{-ik \cdot x} f(k)\;, \quad f(k) = \int \mathrm{d}^n x \ e^{i k \cdot x} f(x)\;,
\end{equation}
where we conveniently hide powers of $2\pi$ via
\begin{equation}
	\hat{\mathrm{d}}^n k := \frac{\mathrm{d}^n k}{(2\pi)^n}\;, \quad \hat{\delta}^{(n)}(k) := (2\pi)^n \delta^{(n)}(k)\;.
\end{equation}
Lorentz-invariant phase-space delta functions and measures are defined as 
\begin{equation}
	\delta_{\Phi}(p_1 - p_2) := 2 E_{p_1} \hat{\delta}^{(3)}(\boldsymbol{p}_1 - \boldsymbol{p}_2)\;, \quad \mathrm{d}\Phi(p) := \hat{\mathrm{d}}^4 p \ \hat{\delta}(p^2 - m^2) \Theta(p^0)\;,
\end{equation}
with $E_p^2 = m^2 + \boldsymbol{p}^2$. The states are defined as 
\begin{equation}
	|p\rangle := a^\dagger(p)|0\rangle\;, \quad [a(p_1), a^\dagger(p_2)] = \delta_{\Phi}(p_1 - p_2)\;,
\end{equation}
such that multi-particle states read
\begin{equation}
	|p_1, p_2, \ldots, p_n \rangle = |p_1\rangle \otimes |p_2\rangle \otimes \ldots \otimes |p_n\rangle\;,
\end{equation}
together with the multi-particle state measure
\begin{equation}
	\mathrm{d}(p_1, p_2, \ldots, p_n) := \mathrm{d}\Phi(p_1)\mathrm{d}\Phi(p_2)\ldots\mathrm{d}\Phi(p_n)\;.
\end{equation}

For a given product of tensors we define the averaged symmetrisation and antisymmetrisation as
\begin{align}
	& X^{(\mu_1} \cdots X^{\mu_n)} = \frac{1}{n!} ( X^{\mu_1} X^{\mu_2}\cdots X^{\mu_n} + X^{\mu_2} X^{\mu_1} \cdots X^{\mu_n}+ \cdots)\;, \\
	& X^{[\mu_1} \cdots X^{\mu_n]} = \frac{1}{n!} ( X^{\mu_1} X^{\mu_2}\cdots X^{\mu_n} - X^{\mu_2} X^{\mu_1} \cdots X^{\mu_n}+ \cdots)\;.
\end{align}

Scattering amplitudes are defined with the normalisation in which they are 
equal to $(-i)$ times the sum of Feynman diagrams
\begin{equation}
	M = i \mathcal{M} = \sum \text{Feynman diagrams}\;,
\end{equation}
with an all-outgoing convention for external legs. Amplitudes in electrodynamics and 
gravity are denoted as $A = i \mathcal{A}$ and $M = i\mathcal{M}$, respectively, and 
the latter is used for general amplitude discussions. 

The covariant derivative acting on a rank-$(n,m)$ tensor is 
\begin{equation}
	\nabla_{\alpha} T^{\mu_1 \ldots \mu_n}_{\nu_1 \ldots \nu_m} = \partial_\alpha T^{\mu_1 \ldots \mu_n}_{\nu_1 \ldots \nu_m} + \sum_{i = 1}^{n} \Gamma^{\mu_i}_{~\alpha \beta} T^{\mu_1 \ldots \beta \ldots \mu_n}_{\nu_1 \ldots \nu_m} -\sum_{i = 1}^{m} \Gamma^{\beta}_{~\alpha \nu_i}T^{\mu_1  \ldots \mu_n}_{\nu_1 \ldots \beta\ldots \nu_m}\;,
\end{equation}
together with the Christoffel symbol
\begin{equation}
	\Gamma^{\mu}_{~\rho \sigma} = \frac{1}{2}g^{\mu \alpha}\left(\partial_\rho g_{\alpha \sigma} + \partial_\sigma g_{\rho \alpha } - \partial_\alpha g_{\rho \sigma}\right)\;,
\end{equation}
associated with the metric tensor $g_{\mu \nu}$ of the spacetime. Finally, the Riemann 
tensor is defined as
\begin{equation}
R^{\mu}{}_{\nu\rho\sigma}=\partial_{\rho}\Gamma^{\mu}{}_{\sigma\nu}-\partial_{\sigma}\Gamma^{\mu}{}_{\rho\nu}+\Gamma^{\mu}{}_{\rho\lambda}\Gamma^{\lambda}{}_{\sigma\nu}-\Gamma^{\mu}{}_{\sigma\lambda}\Gamma^{\lambda}{}_{\rho\nu}\;.
\end{equation}

%--------------------------------------------------------------------------------
\chapter{A Field-Theoretic Framework for Classical Binary Systems}
\label{sec:Chapter2}
%--------------------------------------------------------------------------------
In this chapter, we assemble the key tools and techniques required for state-of-the-art 
calculations of classical binary systems using scattering amplitudes. To begin 
we put together a dictionary of the QFTs that are considered
throughout this thesis, which we use to model black holes. Then, we proceed by setting 
up the kinematics that describe the scattering of two black holes. Following this, we 
introduce a systematic framework for separating classical contributions from quantum 
corrections in quantum scattering amplitudes. Given this foundation, we then review the 
main methods that enable the computation of classical loop-level scattering amplitudes. 
Finally, we will end with a discussion of how these classical amplitudes can be used to 
extract physical observables.
%--------------------------------------------------------------------------------
\section{A Dictionary of Quantum Field Theories}
\label{sec:TheorySummary}
%--------------------------------------------------------------------------------

The ultimate objective of this thesis is to compute classical observables associated with the 
scattering of two black holes. The guiding principle throughout our work is the 
\textit{no-hair theorem}~\cite{Israel:1967za,Israel:1967wq,Carter:1971zc}, which asserts that 
black holes are completely characterised by just three parameters: mass, spin, and charge. Our 
focus will be on scenarios in which these two black holes remain widely separated compared to 
their Schwarzschild radii, thereby allowing us to approximate them as point particles endowed 
with some mass, spin, and charge. Therefore, we model the black holes as point-like objects whose 
dynamics are governed by QFTs describing point particles  
coupled to gravity. In everything that we do, we will be working with minimally coupled QFTs, 
which are the simplest theories that we can write down. One might reasonably argue that these 
theories are too simple, since we neglect information about physical effects such as 
tides~\cite{AccettulliHuber:2020dal,Henry:2019xhg,Marchand:2020fpt,Henry:2020ski}, absorption~\cite{Aoude:2023fdm,Aoude:2024jxd}, etc. In 
fact, these effects can be systematically included through the addition of higher-dimensional operators to the 
theory. However, for the purposes of this thesis, we will only focus on minimally coupled theories 
with the goal of describing Schwarzschild black holes in the spinless case and Kerr black holes 
in the spinning case. As a complementary simplification, we will also examine QFTs in which 
point particles are minimally coupled to the electromagnetic field, providing a useful toy 
model that captures some of the structural features of the gravitational case while being 
technically more tractable\footnote{In the case of electrodynamics we are not modelling Kerr black 
holes, but rather a disc of rotating charge.}.

Let us summarise all the QFTs that we consider throughout this thesis. In the case of 
electrodynamics, we will consider QFTs that describe:
\begin{enumerate}
	\item a massive charged scalar field minimally coupled to photons;
	\item a massive charged vector field (spin-$1$) minimally coupled to photons.
\end{enumerate}
While for gravity, we will consider QFTs that describe:
\begin{enumerate}
	\item a massive scalar field minimally coupled to gravity;
	\item a massive vector field (spin-$1$) minimally coupled to gravity;
	\item a massive tensor field (spin-$2$) minimally coupled to gravity.
\end{enumerate}

A natural question that immediately follows from our summary is \textit{why do quantum spin-$s$ 
fields capture the spin dynamics of Kerr black holes?} There is no rigorously proven answer to 
this question, but rather a large amount of quantitative evidence that seems to suggest this. The 
first thing to note is that Kerr black holes have the remarkable property that their multipole 
moments are completely determined by its mass $m$ and spin $S$~\cite{Hansen1974}, which is a 
manifestation of the no-hair theorem. Recent efforts seem to suggest that an on-shell 
realisation of this no-hair theorem is that black holes correspond to classical 
limits of quantum scattering amplitudes for massive spin-$s$ particles minimally coupled to 
gravity~\cite{Maybee:2019jus}. Indeed, this fact followed from a calculation showing that amplitudes describing  
the long-range gravitational scattering of spin-$1/2$ and $1$ particles give the universal 
spin-orbit couplings in the PN corrections to the gravitational 
potential~\cite{Ross:2007zza, Holstein:2008sx}. This was further extended to the quartic order 
in the Kerr black hole multipole expansion, i.e.~$\mathcal{O}(S^4)$, by considering spin-$2$ 
particles~\cite{Vaidya:2014kza}\footnote{It should be noted that this has also been observed in 
electromagnetism~\cite{Holstein:2008sw}.}. These astonishing results bore the expectation that 
the classical limit of the interactions of a massive spin-$s$ field can generate, at most, 
the first $2s$ classical spin multipoles. 

In fact, this expectation can be fleshed out a little more intuitively by investigating the 
relationship between polarisations---which encode the spin degrees of freedom of point 
particles---and classical spin vectors. This relationship will be derived in detail 
later (see \mysecref{sec:Classical_spin}), but the idea for this statement is roughly as follows. 
Massive spin-$1$ and spin-$2$ point particles carry a polarisation vector 
$\varepsilon^\mu$ and tensor $\varepsilon^{\mu \nu}$, respectively. These polarisations can 
be mapped onto classical spin vectors $S^\mu$ by identifying the classical spin with the 
expectation value of the Pauli–Lubanski vector~\cite{Maybee:2019jus}. In deriving the relation 
that translates polarisation structures into spin structures, we quickly find that polarisation vectors 
and tensors can generate, at most, quadratic- and quartic-in-spin contributions, respectively. 
This aligns with the general expectation that quantum spin-$s$ particles can probe the Kerr 
black hole multipoles only up to order 
$\mathcal{O}(S^{2s})$~\cite{Holstein:2008sw,Holstein:2008sx,Vaidya:2014kza}.

Given this expectation, the attentive reader might also rightfully ask \textit{why have we 
chosen to stop at a massive spin-$1$ particle for electromagnetism and a massive spin-$2$ 
particle for gravity?} The reasoning for this lies in the spin of the massless particles: the photon has 
spin-$1$ while the graviton has spin-$2$. The more general statement is that minimally coupling 
a massive spin-$s$ particle to a massless field of lower spin in flat spacetime develops inconsistencies 
due to the propagation of unphysical degrees of freedom~\cite{Deser:2001dt}. In order to 
cancel these unphysical degrees of freedom one can introduce auxiliary 
fields~\cite{Singh:1974qz}, 
or consider non-minimally coupled theories~\cite{Ochirov:2022nqz,PhysRevD.63.125027}. The inclusion of 
non-minimal couplings means that we no longer consider a fundamental point particle, but rather 
a composite particle whose finite size is encoded in the non-minimal coupling 
information. Along these lines, there have been several successful results in the literature 
that study the effects of non-minimal interactions for massive spin-$s$ particles with 
$s >2$~\cite{Bern:2020buy, Cangemi:2022bew, Cangemi:2023bpe, Cangemi:2023ysz}. However, the 
inclusion of non-minimal couplings no longer guarantees that we will faithfully probe 
Kerr black hole spin multipole structures. Although these theories elegantly compute the spin 
multipoles at the quintic-in-spin order and beyond, there is still tension regarding whether 
this is truly the dynamics of a Kerr black hole or instead the dynamics of some other compact 
spinning object. Indeed, this stems from a disagreement with black hole 
perturbation theory (BHPT)~\cite{Bautista:2021wfy, Bautista:2022wjf}. Furthermore, applying 
these higher-spin formalisms beyond the 2PM order remains technically challenging, an example 
of which are high-rank integrals that one must evaluate.

As our goal is to compute the corrections to black hole scattering dynamics that arise from 
these Kerr black hole spin multipoles to high precision, we will content ourselves with scattering 
at most massive spin-$2$ particles. In fact, we have chosen to model Kerr black holes as 
point particles in finite spin representations, which introduces its own flavour of complexity 
that must be tamed whilst simultaneously going beyond the 2PM order.

\subsection{Electrodynamics}
%--------------------------------------------------------------------------------

Before explicitly writing the QFT that minimally couples a massive charged scalar or vector field 
to photons, we will briefly overview the free theories for these fields. Then, we will present 
the minimally coupled theories.
%--------------------------------------------------------------------------------
\subsubsection{Free Massive Complex Scalar Field}
%--------------------------------------------------------------------------------
The theory encoding the free dynamics of a massive complex scalar field reads
\begin{equation}
	S_{\phi} = \int \mathrm{d}^D x \ \mathcal{L}_{\phi} = \int \mathrm{d}^D x \left(\partial_\mu \phi^\dagger \partial^\mu \phi - m^2_\phi \phi^\dagger \phi\right)\;,
	\label{eq:FreeComplexScalar}
\end{equation}
where the massive complex scalar field $\phi(x)$ has mass $m_\phi$. As there are no explicit 
interactions in the theory, all we can derive is the propagator associated to the complex 
scalar field. Explicitly, this can be obtained by solving the equation of motion (EoM)
\begin{equation}
	\frac{\delta S_\phi}{\delta \phi^\dagger} = (\Box + m^2_\phi)\phi = 0\;, \quad \Box = \partial_\mu \partial^\mu\;,
\end{equation}
using a Green's function $G(x-y)$ defined as\footnote{We insert a factor of $i$ on the right-hand side for 
later convenience.}
\begin{equation}
	\frac{\delta S_\phi}{\delta \phi^\dagger} = (\Box + m^2_\phi)G(x-y) = -i\delta^{(D)}(x-y)\;.
\end{equation}
Introducing the Fourier transform of $G(x-y)$
\begin{equation}
	G(x-y) = \int \hat{\mathrm{d}}^D k \ e^{-ik \cdot (x-y)} \tilde{G}(k)\;,
\end{equation}
we obtain
\begin{equation}
	(-k^2 + m^2_\phi)\tilde{G}(k) = -i,
\end{equation}
where we have used the fact that $\delta^{(D)}(x) = \int \hat{\mathrm{d}}^D k \ e^{-ik \cdot x}$. 
Then, solving for the Green's function yields the Feynman propagator
\begin{equation}
\begin{aligned}
     \raisebox{-0.25cm}{\includegraphics[scale=1.1]{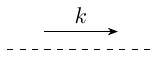}}
     = \tilde{G}(k) = \frac{i}{k^2 - m_\phi^2+ i0}\;,\\
\end{aligned}
\end{equation}
where $i0$ denotes the Feynman prescription, which has the effect of shifting the poles 
slightly off the real axis. This ensures that the propagator corresponds to the time 
ordered two-point function.
%--------------------------------------------------------------------------------
\subsubsection{Free Massive Complex Vector Field}
%--------------------------------------------------------------------------------
The theory describing the free dynamics of a massive complex vector field (spin-$1$) 
is encoded in the complex Proca action~\cite{Proca:1936fbw}
\begin{equation}
	S_{\mathrm{P}} = \int \mathrm{d}^D x \ \mathcal{L}_{\mathrm{P}} = \int \mathrm{d}^D x \left(-\frac{1}{2} V^\dagger_{\mu \nu}V^{\mu \nu} + m^2_{\tiny \mbox{V}} V_\mu^\dagger V^\mu\right)\;,
	\label{eq:FreeComplexProca}
\end{equation}
where the massive vector field $V_\mu(x)$ has mass $m_{\tiny \mbox{V}}$ and the vector field 
strength reads $V_{\mu \nu} = \partial_\mu V_\nu - \partial_\nu V_\mu$. Here the subscript P 
stands for \textit{Proca}, named after Alexandru Proca. Conveniently, the kinetic term may be simplified as
\begin{equation}
	V^\dagger_{\mu \nu}V^{\mu \nu} = V^\dagger_{\mu \nu}(\partial^\mu V^\nu - \partial^\nu V^\mu) = 2 V^\dagger_{\mu \nu} \partial^\mu V^\nu\;.
\end{equation}
The EoM is obtained by varying the action with respect to $V^\dagger_\nu$
\begin{equation}
	\frac{\delta S_{\mathrm{P}}}{\delta V^\dagger_\nu} = \partial_\mu V^{\mu \nu} + m_\mathrm{V}^2 V^\nu = (\eta^{\mu \nu}\Box- \partial^\mu \partial^\nu  + \eta^{\mu \nu}m_{\tiny \mbox{V}}^2)V_\mu = 0\;.
\end{equation}
If we contract the free index with $\partial_\nu$ we obtain
\begin{equation}
	\partial_\mu \partial_\nu V^{\mu \nu} +  m_{\tiny \mbox{V}}^2 \partial_\nu V^\nu = 0\;,
\end{equation}
which implies that $\partial_\nu V^\nu = 0$, provided that $ m_{\tiny \mbox{V}}^2 \neq 0$. 
Importantly, this is not a gauge choice imposed by hand, as one would do for a massless 
spin-$1$ field, since $V_\mu$ possesses no gauge freedom. Rather, it is a genuine 
constraint that must be satisfied by $V_\mu$.

As in the case of a scalar field, we obtain the propagator by solving the EoM with a 
Green's function
\begin{equation}
	(\eta^{\mu \nu}\Box- \partial^\mu \partial^\nu  + \eta^{\mu \nu}m_{\tiny \mbox{V}}^2)G_{\nu \sigma}(x-y) = i \delta^{\mu}_{~\sigma}\delta^{(D)}(x-y),
\end{equation}
which, after Fourier transforming, gives
\begin{equation}
	(-k^2 \eta^{\mu \nu} + k^\mu k^\nu + \eta^{\mu \nu}m_{\tiny \mbox{V}}^2)\tilde{G}_{\nu \sigma}(k) = i \delta^{\mu}_{~\sigma}\;.
\end{equation}
To obtain the Green's function we write an ansatz 
\begin{equation}
	\tilde{G}_{\nu \sigma}(k) = A(k)k^2 \eta_{\nu \sigma} + B(k) k_\nu k_\sigma\;,
\end{equation}
where $A,B$ are scalar functions of $k$. Plugging this in we find
\begin{equation}
	A(k)\delta^{\mu}_{~\sigma}k^2(-k^2 + m_{\tiny \mbox{V}}^2) + (A(k)k^2 + B(k) m_{\tiny \mbox{V}}^2)k^\mu k_\sigma = i \delta^{\mu}_{~\sigma}\;,
\end{equation}
such that 
\begin{equation}
	A(k) = \frac{-i}{k^2(k^2-m_{\tiny \mbox{V}}^2)}\;, \quad B(k) = \frac{i}{m_{\tiny \mbox{V}}^2(k^2-m_{\tiny \mbox{V}}^2)}\;.
\end{equation}
Thus, the Feynman propagator for the massive complex vector field is
\begin{equation}
    \raisebox{-0.25cm}{\includegraphics[scale=1.1]{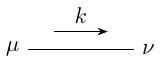}}
    = \tilde{G}^{\mu \nu}(k) = \frac{i}{k^2-m_{\mbox{\tiny V}}^2+ i0}\left(-\eta^{\mu \nu} + \frac{k^\mu k^\nu}{m_{\mbox{\tiny V}}^2}\right)\;.
\end{equation}
We further define the massive spin-1 transverse projector, which follows from the numerator of the 
propagator
\begin{equation}
    P_{(1)}^{\mu \nu }(k,m_{\mbox{\tiny V}}) = \sum_{\lambda_{\mbox{\tiny V}}} \varepsilon^{\star \mu}_{\lambda_{\mbox{\tiny V}}}(k)\varepsilon_{\lambda_{\mbox{\tiny V}}}^{\nu}(k) = -\eta^{\mu \nu} + \frac{k^\mu k^\nu}{m_{\mbox{\tiny V}}^2}\;,
    \label{eq:spin1-transverse-proj}
\end{equation}
where the subscript $(s)$ denotes the spin of the field, and 
$\varepsilon_{\lambda_{\mbox{\tiny V}}}^\mu$ are the polarisations of the massive vector field 
with physical states $\lambda_{\mbox{\tiny V}}$. 
%--------------------------------------------------------------------------------
\subsubsection{Minimal Coupling to Photons}
%--------------------------------------------------------------------------------
Given that we have understood the free theories for the massive complex scalar and vector, 
let us now minimally couple these fields to the photon. To start off, the dynamics of the 
photon are described by the Maxwell action
\begin{equation}
    S_{\mathrm{Maxwell}} = \int \mathrm{d}^4 x \ \mathcal{L}_{\mathrm{Maxwell}}
    = \int \mathrm{d}^4 x \left(-\frac{1}{4} F_{\mu \nu} F^{\mu \nu} - \frac{1}{2\xi}(\partial_\mu A^\mu)^2\right),
    \label{eq:FreeMaxwell}
\end{equation}
where $A_\mu(x)$ is the photon field, $F_{\mu \nu} = \partial_\mu A_\nu - \partial_\nu A_\mu$ is the  
abelian field strength tensor, and $\xi$ is a Lagrange multiplier. The second term in 
$\mathcal{L}_{\mathrm{Maxwell}}$ is a gauge-fixing term, introduced to handle the gauge freedom of 
the photon field: $A_\mu(x) \rightarrow A_\mu(x) + \partial_\mu \Lambda(x)$. This freedom implies 
that the kinetic operator has a null space spanned by pure gauge fields, 
$A_\mu = \partial_\mu \Lambda$, and is therefore not invertible. The gauge-fixing term removes this 
redundancy and renders the operator invertible, allowing us to define the propagator. Note that only 
the transverse components of this propagator correspond to the physical photon polarisations.

From this action, the Feynman propagator for the photon can be obtained:
\begin{equation}
    \raisebox{-0.25cm}{\includegraphics[scale=1.1]{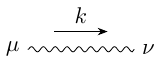}}
    = \frac{i}{k^2 + i0}\left(-\eta^{\mu \nu} + (1-\xi)\frac{k^\mu k^\nu}{k^2}\right),
\end{equation}
where we choose $\xi = 1$, corresponding to Feynman gauge.

Equipped with all of this knowledge, we can go ahead and minimally couple the massive 
complex scalar and vector fields to the photon. For the former, we acquire 
\begin{equation}
    \mathcal{L}_{\text{scalar}} = (D_\mu \phi)^{\dagger}(D^\mu \phi) - m_{\phi}^2 \phi^{\dagger}\phi\;, 
    \label{eq:ComplexScalarWPhoton}
\end{equation} 
where $D_\mu = \partial_\mu + i q_\phi e A_\mu$ is the usual covariant derivative and 
$q_\phi e$ is the charge of the scalar field. For the latter, however, if we 
simply take \myeqref{eq:FreeComplexProca} and minimally couple it to the photon, we 
will miss an additional three-point operator of the form 
$-i e V^{\dagger \mu}F_{\mu \nu}V^{\nu}$~\cite{Bautista:2022ewu}, 
which is needed to obtain the expected root-Kerr three-point and Compton 
amplitudes~\cite{Chiodaroli:2021eug,Cangemi:2023ysz}. Instead, if we consider the 
spontaneous symmetry breaking (SSB) of an $\mathrm{SO}(3)$ gauge theory, which leads, 
essentially, to a theory of $W^{\pm}$ bosons, then we will land on the correct theory 
with this additional operator. In order to illustrate this SSB, let us consider an $\mathrm{SO}(3)$ 
gauge theory with gauge fields $A_\mu^a(x)$ (with $a = 1,2,3$) and a real scalar triplet 
$\Phi^a$ transforming under the adjoint representation of $\mathrm{SO}(3)$. This theory is 
described by
\begin{equation}
	\mathcal{L}_{\mathrm{SO(3)}} = -\frac{1}{4}G_{\mu \nu}^a G^{a\mu \nu} + \frac{1}{2}(D_\mu \Phi)^a(D^\mu \Phi)^a -V(\Phi)\;, 
	\label{eq:SO3Theory}
\end{equation}
where the non-Abelian field strength and the covariant derivative read
\begin{align}
	& G_{\mu \nu}^a = \partial_\mu A_\nu^a - \partial_\nu A_\mu^a + g \epsilon^{abc}A^b_\mu A^c_\nu\;, \quad (D_\mu \Phi)^a = \partial_\mu \Phi^a + g \epsilon^{abc} A_\mu^b \Phi^c\;.
\end{align}
The scalar potential is the usual Mexican hat
\begin{equation}
	V(\Phi) = - \frac{1}{2}\mu^2 \Phi^a \Phi^a + \frac{\lambda}{4} (\Phi^a \Phi^a)^2\;, \quad \lambda >0\;,
\end{equation}
such that if $\mu^2 > 0$, the minimum is not at $\Phi^a = 0$ and SSB occurs. Importantly, 
the potential is minimised when
\begin{equation}
	\Phi^a \Phi^a = \frac{\mu^2}{\lambda} \equiv v^2\;,
\end{equation}
and exploiting rotational invariance allows us to choose the vacuum expectation value (VEV) as 
$\langle \Phi^a \rangle = (0,0,v)^T$. Evaluating the Lagrangian on this VEV, the gauge 
boson terms become
\begin{align}
	\mathcal{L}^{\mathrm{SSB}}_{\mathrm{SO(3)}} & = -\frac{1}{4}G_{\mu \nu}^a G^{a\mu \nu} + \frac{1}{2}g^2v^2 (\epsilon^{abc}\delta^{c3}A_\mu^b) (\epsilon^{ade}\delta^{e3}A^{d\mu}) \\
	& = -\frac{1}{4}G_{\mu \nu}^a G^{a\mu \nu} + \frac{1}{2}g^2v^2(\delta^{bd}-\delta^{b3}\delta^{d3})A_\mu^bA^{d\mu} \\
	& = -\frac{1}{4}G_{\mu \nu}^a G^{a\mu \nu} + \frac{1}{2}g^2v^2(A^1_\mu A^{1\mu} + A^2_\mu A^{2\mu})\;,
\end{align}
so that $A^1_\mu, A^2_\mu$ acquire a mass $m_{\tiny \mbox{V}} = g v$, while 
$A^3_\mu$ stays massless. This corresponds to the usual Higgs mechanism, in which one 
considers perturbations around the VEV, $\Phi^a = \langle \Phi^a \rangle + \varphi^a$. 
In this picture, $\varphi^3$ corresponds to the Higgs field, while the scalar 
fluctuations $\varphi^1$ and $\varphi^2$ are ``eaten'' by the gauge fields to provide 
their longitudinal components. Equivalently, one may work in the unitary gauge, where the Goldstone bosons are gauged away.
Here we focus only on the mass spectrum of the gauge bosons and do not consider the dynamics of the scalar 
fluctuations $\varphi^a$. In particular, we will not discuss the Higgs sector and restrict attention to the resulting massive 
and massless spin-$1$ fields.

It is convenient to work in the charged field basis
\begin{align}
	& V_\mu = \frac{1}{\sqrt{2}}(A_\mu^1 + i A_\mu^2)\;, \quad V^\dagger_\mu = \frac{1}{\sqrt{2}}(A_\mu^1 - i A_\mu^2)\;, \\
	& A^1_\mu = \frac{1}{\sqrt{2}}(V_\mu + V^\dagger_\mu)\;, \quad A^2_\mu = -\frac{i}{\sqrt{2}}(V_\mu -V^\dagger_\mu)\;,
\end{align}
so that $V_\mu$ manifestly corresponds to the charged vector boson. Focusing on the terms quadratic 
in the gauge field, we find 
\begin{align}
	\mathcal{L}^{\mathrm{SSB}}_{\mathrm{SO(3)}} \rvert_{\mathrm{A^2}}& = -\frac{1}{4}F_{\mu \nu} F^{\mu \nu} - \frac{1}{2}V^\dagger_{\mu \nu}V^{\mu \nu} + m_{\tiny \mbox{V}}^2 V^\dagger_\mu V^\mu\;,
\end{align}
where we relabel $A^3_\mu \rightarrow A_\mu$ such that $F_{\mu \nu}$ is the usual 
abelian field strength tensor and $V_{\mu \nu}$ is the vector field strength tensor 
defined below \myeqref{eq:FreeComplexProca}. Furthermore, the cubic and quartic 
interaction terms in the charged field basis read
\begin{align}
	& \mathcal{L}^{\mathrm{SSB}}_{\mathrm{SO(3)}} \rvert_{\mathrm{A^3}} = - ig \left(V^\dagger_{\mu \nu} A^\nu V^\mu - V_{\mu \nu}A^\nu V^{\dagger \mu} + V^{\dagger \mu}F_{\mu \nu}V^\nu\right)\;, \\
	& \mathcal{L}^{\mathrm{SSB}}_{\mathrm{SO(3)}} \rvert_{\mathrm{A^4}} = g^2 \left(A_{\mu}A_{\nu} V^{\mu}V^{\dagger \nu} - A_\nu A^\nu V^\dagger_\mu V^\mu \right)\;.
\end{align}
Putting all of this information together, we obtain 
\begin{align}
	\mathcal{L}_{\text{vector}} = - \frac{1}{2}V^\dagger _{\mu \nu}V^{\mu \nu} & + m_{\mbox{\tiny V}}^2 V^\dagger _{\mu} V^{\mu} + i q_{\mbox{\tiny V}} e V^{\dagger \mu} V_{\mu \nu}A^\nu 
        - i q_{\mbox{\tiny V}} e V^{\dagger }_{\mu \nu}A^\nu V^{\mu} \\ 
        & - i q_{\mbox{\tiny V}} e V^{\dagger \mu}F_{\mu \nu}V^{\nu} + q_{\mbox{\tiny V}}^2 e^2 A_\mu A_\nu V^\mu V^{\dagger \nu} - q_{\mbox{\tiny V}}^2e^2A^2 V^\mu V^\dagger_\mu\;, \nonumber 
\end{align}
where we identify $g = q_{\mbox{\tiny V}} e$ as the charge of the massive complex vector 
field.

To summarise, the dynamics of charged scalars and vectors, minimally coupled to photons, 
can be described by 
\begin{equation}
  \mathcal{L}_{\mathrm{EM}} = \mathcal{L}_{\text{Maxwell}} + \mathcal{L}_{\text{scalar}} + \mathcal{L}_{\text{vector}}\;.
  \label{eq:EMLagrangian}
\end{equation}
From this we may derive the Feynman rules that describe interactions. The vertex rules for scalar-photon
interactions read 
{\allowdisplaybreaks
\begin{align}
    & \raisebox{-1.25cm}{\includegraphics{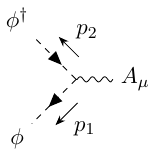}} \ = \ i q_\phi e(p_1 - p_2)_\mu\;, \\
    & \raisebox{-1.25cm}{\includegraphics{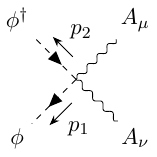}} \ = \ 2i q_\phi^2 e^2 \eta_{\mu \nu}\;.
\end{align}
}
and the vertex rules for vector-photon interactions read 
{\allowdisplaybreaks
\begin{align}
    & \raisebox{-1.25cm}{\includegraphics{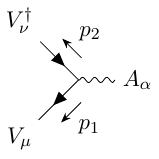}} \ = \ -iq_{\mbox{\tiny V}} e\left[\eta_{\mu \nu}(p_1 - p_2)_\alpha - \eta_{\mu \alpha}(2p_1 + p_2)_\nu + \eta_{\nu \alpha}(2p_2 + p_1)_\mu\right]\;, \\
    & \raisebox{-1.25cm}{\includegraphics{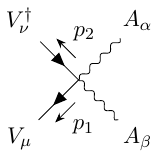}} \ = \ iq_{\mbox{\tiny V}}^2 e^2 \left[\eta_{\mu \alpha}\eta_{\nu \beta} + \eta_{\mu \beta}\eta_{\nu \alpha} - 2\eta_{\mu \nu}\eta_{\alpha \beta}\right]\;.
\end{align}
}

%--------------------------------------------------------------------------------
\subsection{Gravity}
%--------------------------------------------------------------------------------
Now we turn our attention to QFTs that describe massive real scalars, vectors, and 
tensors minimally coupled to gravity. As before, we start with a discussion of the 
free theories and then show how to minimally couple them. Since we have already 
discussed the scalar and vector, we focus on the massive tensor.
%--------------------------------------------------------------------------------
\subsubsection{Free Massive Tensor Field}
%--------------------------------------------------------------------------------
The free dynamics of a massive \textit{symmetric} tensor (spin-$2$) field, $H_{\mu \nu}$, of mass 
$m_{\tiny \mbox{H}}$ is encoded in the Fierz-Pauli action~\cite{Fierz:1939ix}
\begin{align}
    S_{\text{FP}} = \int \mathrm{d}^D x \ \mathcal{L}_{\text{FP}} = \int \mathrm{d}^D x \ \Bigg[ & \frac{1}{2}\partial_\rho H_{\mu \nu}\partial^\rho H^{\mu \nu} - \partial_\mu H_{\nu \rho} \partial^\nu H^{\mu \rho}\label{eq:FierzPauli_Action}\\
    &  \hspace{-1cm} + \partial_\mu H^{\mu \nu} \partial_\nu H - \frac{1}{2}\partial_\mu H \partial^\mu H - \frac{m_{\tiny \mbox{H}}^2}{2}(H_{\mu \nu}H^{\mu \nu} - H^2) \Bigg]\nonumber \;,
\end{align}
where $H = \eta^{\mu \nu}H_{\mu \nu}$. Note that the kinetic terms differ from other 
references by an overall minus sign because we work with a negative-signature metric 
(see for example Ref.~\cite{Vaidya:2014kza}). The EoM follows from varying the action 
with respect to $H^{\mu \nu}$
\begin{align}
    \partial^2 H_{\mu \nu} - \partial_\rho \partial_\mu H^{\rho}_{~\nu} - \partial_\rho \partial_\nu H^{\rho}_{~\mu} + \partial_\mu \partial_\nu H & + \eta_{\mu \nu} \partial_\rho \partial_\sigma H^{\rho \sigma} \label{eq:Spin2_EOM}\\
    & \hspace{-1cm} - \eta_{\mu \nu} \partial^2 H + m_{\tiny \mbox{H}}^2(H_{\mu\nu} - \eta_{\mu \nu}H) = 0 \nonumber \;.
\end{align}
Now we contract \myeqref{eq:Spin2_EOM} with $\partial^\mu$ to find
\begin{equation}
    \partial^\mu H_{\mu\nu} - \partial_\nu H = 0\;,
    \label{eq:Spin2_EOM_Constraint1}
\end{equation}
which is a constraint on the dynamics. Substituting this into \myeqref{eq:Spin2_EOM} 
gives
\begin{equation}
    \partial^2 H_{\mu \nu} - \partial_\mu \partial_\nu H + m_{\tiny \mbox{H}}^2(H_{\mu \nu} - \eta_{\mu \nu} H) = 0\;.
    \label{eq:Spin2_EOM_Simplified1}
\end{equation}
Then, contracting \myeqref{eq:Spin2_EOM_Simplified1} with $\eta^{\mu \nu}$ yields
\begin{equation}
    m_{\tiny \mbox{H}}^2(D-1)H = 0~~\forall~~m_{\tiny \mbox{H}}^2(D-1)\;,
\end{equation}
which shows that the field must be \textit{traceless}
\begin{equation}
    H = 0\;.
    \label{eq:Spin2_Traceless}
\end{equation}
Consequently, the constraint \myeqref{eq:Spin2_EOM_Constraint1} simplifies to 
\begin{equation}
    \partial^\mu H_{\mu\nu} = 0\;,
    \label{eq:Spin2_Transversality}
\end{equation}
which implies \textit{transversality}. In this way, we see that the EoM gives 
the following set of equations
\begin{equation}
    (\partial^2 + m_{\tiny \mbox{H}}^2)H_{\mu \nu} = 0\;, \quad  \partial^\mu H_{\mu\nu} = 0\;, \quad H = 0\;.
    \label{eq:Spin2_EOM_Set}
\end{equation}
The first equation is the Klein-Gordon equation for $H_{\mu \nu}$, which has $D(D+1)/2$ degrees 
of freedom (d.o.f.). Then, transversality and tracelessness provide $(D+1)$ constraints, thereby 
reducing the total number of d.o.f. to $(D-1)(D+2)/2$. More specifically, in $D = 4$ dimensions the 
$10$ d.o.f. of $H_{\mu \nu}$ reduces to $5$ physical d.o.f. This is consistent with a spin-$s$ 
particle having $2s+1$ independent components in $D=4$.

To determine the propagator, we recast the action as
\begin{equation}
    S_{\text{FP}} = \int \mathrm{d}^D x \ \left(\frac{1}{2} H_{\mu \nu} \Delta^{\mu \nu \rho \sigma}H_{\rho \sigma}\right)\;,
    \label{eq:FierzPauli_KernelAction}
\end{equation}
where the kernel reads
\begin{equation}
    \Delta^{\mu \nu}_{~~\rho \sigma} = -(\delta^{(\mu}_{~~\rho} \delta^{\nu)}_{~~\sigma} - \eta^{\mu \nu}\eta_{\rho \sigma})(\partial^2 + m_{\tiny \mbox{H}}^2) + 2 \partial^{(\mu}\partial_{(\rho}\delta^{\nu)}_{~~\sigma)} -  \partial^{\mu} \partial^{\nu}\eta_{\rho \sigma}-\partial_{\rho} \partial_{\sigma}\eta^{\mu \nu}\;.
\end{equation}
The goal is to invert this kernel, which we will do via an ansatz. Importantly, we expect the 
general form of the inverse to read
\begin{equation}
    G^{\mu \nu \rho \sigma}(x) = \int \hat{\mathrm{d}}^D k \ e^{-ik\cdot x}\tilde{G}^{\mu \nu \rho \sigma}(k) \;,
\end{equation}
with
\begin{equation}
    \tilde G^{\mu \nu \rho \sigma}(k) = \frac{A(k)}{2}\Big(P^{\mu \rho}_{(1)} P^{\nu \sigma}_{(1)} + P^{\mu \sigma}_{(1)} P^{\nu \rho}_{(1)}\Big) + B(k)P^{\mu \nu}_{(1)}P^{\rho \sigma}_{(1)}\;,
\end{equation}
where $P^{\mu \nu}_{(1)}(k,m_{\mbox{\tiny H}})$ is the transverse projector for a 
massive spin-$1$ defined in \myeqref{eq:spin1-transverse-proj}. Then, inverting 
follows from solving
\begin{equation}
    \Delta^{\mu \nu \alpha \beta}\tilde{G}_{\alpha \beta \rho \sigma} = \frac{i}{2}(\delta^{\mu}_{~\rho} \delta^{\nu}_{~\sigma} + \delta^{\mu}_{~\sigma} \delta^{\nu}_{~\rho})\delta^{(D)}(x)\;,
\end{equation}
for $A(k), B(k)$. Explicitly, we find
\begin{equation}
    A(k) = \frac{i}{k^2-m_{\tiny \mbox{H}}^2}\;, \qquad B(k) = -\frac{i}{k^2-m_{\tiny \mbox{H}}^2}~\frac{1}{D-1}\;,
\end{equation}
such that the massive spin-$2$ propagator reads
\begin{equation}
     \tilde{G}^{\mu \nu \rho \sigma} = \frac{i}{k^2 - m_{\tiny \mbox{H}}^2 + i0} ~ \left( P^{\mu (\rho}_{(1)} P^{\sigma)\nu }_{(1)} - \frac{1}{D-1}P^{\mu \nu}_{(1)}P^{\rho \sigma}_{(1)}\right),
\end{equation}
which we diagrammatically represent as 
\begin{equation}
	\raisebox{-0.25cm}{\includegraphics[scale=1.1]{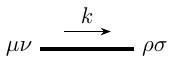}}
    = \tilde{G}^{\mu \nu \rho \sigma}\;.
\end{equation}
We further define the massive spin-2 transverse projector, which follows from the numerator of the 
propagator
\begin{equation}
    P_{(2)}^{\mu \nu \rho \sigma}(k) = \sum_{\lambda_{\mbox{\tiny H}}} \varepsilon_{\lambda_{\mbox{\tiny H}}}^{\star \mu \nu}(k)\varepsilon_{\lambda_{\mbox{\tiny H}}}^{\rho \sigma}(k) = P^{\mu (\rho}_{(1)} P^{\sigma)\nu }_{(1)} - \frac{1}{D-1}P^{\mu \nu}_{(1)}P^{\rho \sigma}_{(1)}\;,
    \label{eq:spin2-transverse-proj}
\end{equation}
where we have suppressed the arguments of $P^{\mu \nu}_{(1)}$, and 
$\varepsilon_{\lambda_{\mbox{\tiny H}}}^{\mu \nu}$ are the polarisations of the massive tensor 
field with physical states $\lambda_{\mbox{\tiny H}}$. 

%--------------------------------------------------------------------------------
\subsubsection{Minimal Coupling to Gravity}
%--------------------------------------------------------------------------------
The dynamics of gravitons in Einstein gravity can be described by the Einstein-Hilbert action 
in the harmonic gauge
\begin{equation}
    \begin{split}
    S_{\mathrm{GR}} & = \int \mathrm{d}^D x  \left(\mathcal{L}_{\mathrm{EH}}  + \mathcal{L}_{\mathrm{GF}}\right) =  \int \mathrm{d}^D x  \sqrt{-g} \left( - \frac{2}{\kappa^2} R  + \frac{1}{2\xi} g_{\mu \nu} \Gamma^\mu \Gamma^\nu\right),
    \end{split}
    \label{eq:GravityAction}
\end{equation}
where $\kappa^2 = 32 \pi G$, $g_{\mu \nu}$ is the metric with determinant $g$, 
$\Gamma^\mu = \Gamma^{\mu}_{~\rho \sigma} g^{\rho \sigma}$, and $\xi$ is a free paremeter. 
From the path integral gauge fixing procedure there would be a ghost term, which we ignore since it is 
not relevant in the classical limit. To obtain the dynamics of the graviton we perform a weak field 
expansion around flat spacetime
\begin{equation}
	g_{\mu \nu} = \eta_{\mu \nu} + \kappa h_{\mu \nu}\;, \quad \eta_{\mu \nu} = \text{diag}(1,-1,\cdots,-1)\;,
\end{equation} 
with $h_{\mu \nu}$ interpreted as the graviton field. Expanding \myeqref{eq:GravityAction} to quadratic 
order in the graviton field and inverting the EoM, yields the graviton propagator
\begin{equation}
	\raisebox{-0.25cm}{\includegraphics[scale=1.1]{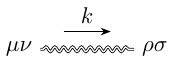}}
    = \frac{i}{k^2 + i0}  \left((\mathcal{P}^{-1})^{\mu\nu\alpha\beta} - 2(1 - \xi) \mathcal{I}^{\mu \nu \rho \kappa} \frac{k_\rho k_\sigma}{k^2} \mathcal{I}^{\kappa \sigma \alpha \beta} \right)\;,
\end{equation}
where 
\begin{equation}
	(\mathcal{P}^{-1})^{\mu\nu\alpha\beta} =  \mathcal{I}^{\mu\nu \alpha\beta} - \frac{1}{D-2}\eta^{\mu \nu}\eta^{\alpha \beta} \;, \quad \mathcal{I}^{\mu\nu \alpha\beta} = \eta^{\mu (\alpha} \eta^{\beta)\nu}\;.
\end{equation}
As for photons we choose the Feynman gauge $\xi = 1$.

Now we proceed with minimally coupling the massive scalar, vector, and tensor 
to gravity. Explicitly, we consider the dynamics of
\begin{equation}
	\frac{\mathcal{L}_\textrm{GR}}{\sqrt{-g}} = \mathcal{L}_\textrm{EH} + \mathcal{L}_\textrm{GF} + \mathcal{L}_\textrm{scalar} + \mathcal{L}_\textrm{vector} + \mathcal{L}_\textrm{tensor} + \ldots\;,
\end{equation}
with matter Lagrangians
\allowdisplaybreaks{
\begin{align}
	&  \mathcal{L}_\textrm{scalar} =  \frac{1}{2}g^{\mu\nu}\partial_\mu\phi \partial_\nu\phi - \frac{1}{2}m_\phi^2\phi^2\;, \label{eq:L_scalar}\\
	&  \mathcal{L}_\textrm{vector} = -\frac{1}{4}g^{\mu\alpha}g^{\nu\beta}V_{\alpha\beta}V_{\mu\nu} +\frac{1}{2}m_{\mbox{\tiny V}}^2 g^{\mu\nu} V_\mu V_\nu\;, \label{eq:L_vector}\\
	&  \mathcal{L}_\textrm{tensor} =  g^{\mu\nu}g^{\alpha\beta}g^{\omega\sigma}\Big\{\frac{1}{2}(\nabla_\omega H_{\mu\alpha})(\nabla_\sigma H_{\nu\beta}) - (\nabla_\mu H_{\alpha\omega})(\nabla_\beta H_{\nu\sigma}) \nonumber \label{eq:L_tensor}\\
	& \hspace{3cm} + (\nabla_\mu H_{\nu\alpha})(\nabla_\beta H_{\omega\sigma}) - \frac{1}{2}(\nabla_\mu H_{\alpha\beta})(\nabla_\nu H_{\omega\sigma}) \Big\}  \\
	& \hspace{3cm} - \frac{1}{2}m_{\tiny \mbox{H}}^2 g^{\mu\nu}g^{\alpha\beta}\big( H_{\mu\alpha}H_{\nu\beta} - H_{\mu\nu}H_{\alpha\beta}\big)\;,\nonumber
\end{align}
}
where $\nabla_\mu$ is the covariant derivative defined with a Christoffel
connection and $\ldots$ denote higher-dimensional operators~\cite{Donoghue:1994dn} 
that we ignore.

At this stage, one might be tempted to crank the wheel and derive Feynman rules 
for the graviton-graviton and graviton-matter interaction vertices. However, this 
quickly becomes cumbersome. To circumvent this tedious approach, we will use a 
more practical strategy for obtaining graviton tree-level amplitudes, which will 
be discussed in more detail at appropriate points in this thesis. More 
specifically, our starting point will be pure gluon amplitudes, for which we 
convert gluon legs into massive scalar or vector legs through a dimensional 
reduction of a massless $(D+2)$-dimensional theory into a $D$-dimensional theory 
with massive particles~\cite{Bern:2019crd} (see \mysecref{sec:Tree_amplitudes_via_dimensional_reduction}). Following this, 
we employ the double-copy construction (see \mysecref{sec:DoubleCopy}) to obtain 
graviton tree-level amplitudes directly from the corresponding gluon amplitudes. 
It should be stressed that the dimensional reduction can be either done at the 
level of pure gluon amplitudes or pure graviton amplitudes, as it commutes with 
the double copy procedure. For the latter one may convert graviton legs 
into a massive scalar, vector, or tensor.

%--------------------------------------------------------------------------------
\section{Kinematic Setup}
\label{sec:KinematicSetUp}
\begin{figure}[h!]
 \centering
 \includegraphics{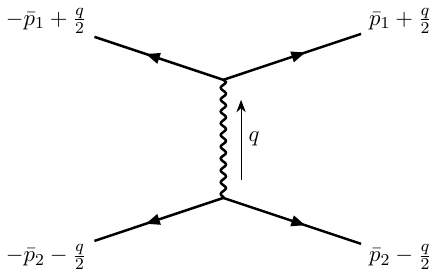}
 \caption{Momentum assignments for the particles in the scattering process. The lower external lines represent massive scalar particles, the upper external lines represent massive spin-$s$ particles, and wavy lines represent exchanged massless particles. External momenta are taken to be outgoing.}
 \label{fig:kinematics}
\end{figure}

%--------------------------------------------------------------------------------
In this thesis, we are interested in the scattering dynamics of spinless and spinning 
black holes. To model such a system, we consider the $2 \rightarrow 2$ scattering 
processes involving a massive spin-$s$ point particle $X$ of mass $m_1$ and a massive 
scalar $\phi$ of mass $m_2$. In both cases, the particles satisfy the 
\textit{on-shell} condition
\begin{equation}
	p_X^2 = m_1^2\;, \quad p_\phi^2 = m_2^2\;.
\end{equation}
Since our goal is to capture the effects of Kerr black hole multipoles to the 
quartic-in-spin order using point particles in finite spin representations, we must 
consider processes where the spin of $X$ is $0,1$ and $2$. In other words, we 
must compute three different scattering processes. The exact reasoning for this will 
be explained later, but it follows from an ambiguity associated to quantum point 
particles in finite spin representations. Thus, for our purposes, the 
spin-$s$ particle is restricted to have $s \leq 2$ and therefore carries a 
polarisation vector $\varepsilon^{\mu}(p)$ ($s = 1$) or tensor $\varepsilon^{\mu \nu}
(p)$ ($s=2$), which satisfy
\begin{enumerate}
	\item transversality: $p_{\mu} \varepsilon^{\mu}(p) = 0$ and $p_{\mu} \varepsilon^{\mu \nu}(p)= p_{\nu} \varepsilon^{\mu \nu}(p) = 0$;
	\item tracelessness: $\varepsilon^{\mu}_{~\mu}(p) = 0$.
\end{enumerate}
These particles interact at a distance by exchanging massless particles: gravitons in 
Einstein gravity and photons in electrodynamics. Specifically, we are interested 
in the following processes
\begin{equation}
  X(-p_1, \varepsilon_1) + \phi(-p_2) \rightarrow \phi(p_3) + X(p_4, \varepsilon^\star_4)\;, \quad X \in \{\Phi, V^\mu, H^{\mu \nu}\}\;,
  \label{eq:phi_phi_V_V}
\end{equation}
where the signs follow our all-outgoing convention for external momenta and 
the fields $\Phi$, $V^\mu$, and $H^{\mu \nu}$ are the massive scalar, vector, 
and tensor respectively. In order to facilitate the extraction of classical physics, it 
is convenient to choose a useful parametrisation of external variables. We choose 
the external momenta, as shown in Fig.~\ref{fig:kinematics}, 
which are parametrised by
\begin{equation}
   p_1 = -\bar{p}_1 + \frac{q}{2}\;, \qquad p_2 = -\bar{p}_2 - \frac{q}{2}\;, \qquad p_3 = \bar{p}_2 - \frac{q}{2}\;, 
    \qquad p_4 = \bar{p}_1 + \frac{q}{2}\;. 
    \label{eq:KinematicSetUp}
\end{equation}
The barred momenta $\bar p_i$ are considered to be much larger than $q$, as 
explained in \mysecref{sec:ClassicalLim}, and are interpreted to be the
average momentum on top of which there are $\mathcal O(q)$
fluctuations. The barred momenta are also orthogonal to $q$
\begin{equation}
  p_1^2 - p_4^2 = 2 (\bar{p}_1 \cdot q) = 0\;, \qquad p_2^2 - p_3^2 = -2 (\bar{p}_2 \cdot q) = 0\;,
\end{equation}
following from on-shell conditions: $p_1^2 = p_4^2 = m_1^2, \  p_2^2 = p_3^2 = m_2^2$. 
Furthermore, we define velocity vectors,
\begin{equation}
 u_1^{\mu} = \frac{\bar{p}_1^{\mu}}{\bar{m}_1}\;, \qquad  u_2^{\mu} = \frac{\bar{p}_2^{\mu}}{\bar{m}_2}\;,
\end{equation}
with 
\begin{equation}
  \bar{m}_i^2 = \bar{p}_i^2 = m_i^2 - \frac{q^2}{4}\;,
  \label{eq:mBarDef}
\end{equation}
which, by virtue of normalisation, satisfy
\begin{equation}
  u_1^2 =  u_2^2 = 1\;, \qquad  u_1 \cdot q =  u_2 \cdot q = 0\;.
  \label{eq:velvectorsproperties}
\end{equation}
Finally, we define the kinematic variable
\begin{equation}
   y = u_1 \cdot u_2 = \frac{\bar{p}_1 \cdot \bar{p}_2} {\bar{m}_1 \bar{m}_2} \;,
   \label{eq:yVarDef}
 \end{equation}
which is a measure of the relative velocity or the
rapidity difference between the two massive particles.

%-------------------------------------------------------------------------------
\section{The Classical Limit of Scattering Amplitudes}
\label{sec:ClassicalLim}
%--------------------------------------------------------------------------------

In reformulating the scattering dynamics of classical black holes in the 
language of QFT, we have necessarily recast a classical problem into a quantum 
problem. It is, therefore, crucial that we understand how to systematically extract 
classical physics from quantum objects, like scattering 
amplitudes. Since we are only truly interested in classical dynamics, it is 
appropriate to retain the minimal amount of information necessary when describing
scattering processes. To illustrate how this is achieved in practice, let us first 
restrict our discussion to scalar particles in gravity. The distinction between 
classical and quantum information is then dictated by an appropriate 
expansion---known as a \textit{soft} expansion---which follows from the method of 
regions (see \mysecref{sec:MethodOfRegions})~\cite{Beneke:1997zp,Smirnov:2012gma}. More 
specifically, this involves identifying the main scales in the process as
\begin{equation}
\begin{split}
    & \text{Compton length:} \ \ell_c \sim \frac{\hbar}{m}\;, \\
    & \text{Schwarzschild radius:} \ r_s \sim Gm\;, \\
    & \text{Impact parameter:} \ |b| \sim \frac{\hbar}{|q|}\;,
\end{split}
\end{equation}
where $m \sim m_1 + m_2$ is some common mass scale, $q$ is the momentum 
transfer, and $b$ is the relative transverse distance (Fourier conjugate to $q$) 
of the system. Having identified these scales, we should now understand their 
relative sizes. Firstly, we are describing celestial objects, which, on average, 
have extremely large masses. This means that $\ell_c \ll r_s$. Secondly, we imagine 
the scattering process of two black holes that are far separated at past and future 
infinity, such that their interactions are dictated by long-range forces. 
Therefore, $b$ is by far the largest scale in the game. Together, the relative 
sizes of these scales define a hierarchy~\cite{Herrmann:2021tct}
\begin{equation}
  \ell_c \ll r_s \ll |b| \quad \Longleftrightarrow \quad \frac{\hbar}{m} \ll Gm \ll \frac{\hbar}{|q|}\;,
\end{equation}
where we keep powers of $\hbar$ for clarity. From this, it is 
straightforward to see
\begin{equation}
    \frac{\ell_c}{|b|} \ll \frac{r_s}{|b|} \ll 1  \quad \Longleftrightarrow \quad \frac{|q|}{m} \ll \frac{Gm|q|}{\hbar} \ll 1\;.
    \label{eq:HierarchyOfScales}
\end{equation}
This equation tells us everything that we need to know about the classical and 
quantum pieces of scattering amplitudes. The smallest quantity $|q|/m$ is 
identified as the scaling of $\mathcal O(\hbar)$ quantum corrections. Henceforth,
we explicitly set $\hbar = 1$, relying on the fact that
a small-$q$ expansion is a convenient proxy for the small-$\hbar$
expansion when the mass $m$ is treated as an $\mathcal O(1)$ quantity,
to isolate classical information from quantum corrections.

At the same time, the second largest quantity tells us about the classical 
scaling, controlled by the dimensionless perturbative expansion parameter 
$r_s/|b| \sim Gm|q|$. An expansion in the combination $r_s/|b|$ defines the 
PM expansion, which allows us to systematically 
study the perturbative (weak field) dynamics of the scattering process. As this 
parameter contains a factor of $G$ always accompanied by a factor of $|q|$, every 
additional power of $G$ requires that the amplitude is expanded to one order higher in 
$|q|$. More generally, the rule is that at a given $\mathcal{O}(G^n)$ we expand 
the scattering amplitude up to $\mathcal{O}(|q|^{n-3})$, in order to extract 
the classical information. This is reliant on the fact that the leading order 
contribution in the PM expansion arises from a single $t$-channel exchange, which scales 
as $\mathcal{O}(1/|q|^2)$.
An important point is that at one loop and beyond the 
classical order is not simply the leading piece in the expansion. In fact, there 
are superclassical/hyperclassical orders---classically divergent terms---which 
start at $\mathcal{O}(1/|q|^2)$ and, therefore, precede the classical 
order. These superclassical terms cancel or vanish in observables that have a 
meaningful classical limit, and therefore play no physical role.

When considering spinning external particles, this narrative needs to be extended.
This follows from the classical spin vector associated to the spinning particle $S$ 
which scales as $\mathcal{O}(1/|q|)$. This scaling is quickly understood by realising 
that the classical limit follows from the spin quantum number $s$ becoming large as 
$\mathcal{O}(1/|q|)$ in the correspondence limit. The consequence is that, to extract the 
full classical information at a given $\mathcal{O}(G^n)$, we must expand to higher 
orders in $|q|$ relative to the leading superclassical order. The order needed 
is completely determined by the expected order in the Kerr black hole multipole 
expansion we wish to probe. Thus, in order to extract the $\mathcal{O}(S^{2s})$ 
corrections we must expand our amplitudes to $\mathcal{O}(|q|^{n -3 + 2s})$.

An identical discussion applies to electrodynamics. In this case, the Schwarzschild 
radius becomes the size of a classical particle, $r_s \sim \alpha/m$, where 
$\alpha = e^2/(4\pi)$ is the electromagnetic coupling which is used as the expansion parameter.
This is often referred to as the \textit{post-Lorenzian} (PL) expansion~\cite{Bern:2021xze}.

%--------------------------------------------------------------------------------
\section{The Generalised Unitarity Framework }
\label{sec:generalised_unitarity}
%--------------------------------------------------------------------------------

The first step in computing higher-order corrections to classical observables is to  
obtain the loop-level amplitudes that capture all relevant classical contributions. 
Towards this direction, generalised 
unitarity~\cite{Bern:1994zx,Bern:1994cg,Bern:1997sc,Britto:2004nc,Bern:2007ct} 
conveniently allows for the construction of loop-level integrands as products of tree-level 
amplitudes. This follows from integrands being rational functions, and thereby having  
poles associated to on-shell propagators. Propagators are put on-shell by replacing the inverse
propagators with a delta function whose argument enforces the on-shell
conditions. Then, we consider the residues of these poles---or
\textit{generalised cuts}---to construct the integrand as a product of tree-level
amplitudes summed over all physical states that can cross the 
cut
\begin{equation}
 M^{(L)}_{\text{residue}} \equiv \sum_{\text{states}} M^{(0)}_{1}
 M^{(0)}_{2}\ldots M^{(0)}_{n-1}\;M^{(0)}_{n}\;.
 \label{eq:factorization}
\end{equation}
The action of multiplying amplitudes together and summing over the exchanged on-shell 
states is known as \textit{sewing}. Of course, since generalised unitarity relies on the 
poles of integrands, it can never capture the effects of contributions that have no 
propagators at all. This is an important feature that does not affect our 
discussion, as such diagrams correspond to quantum interactions.

The summation over physical states necessarily introduces transverse projectors for 
the propagating field. For both massless and massive spin-$1$ fields we need the transverse 
projector for the photon 
\begin{equation}
	\begin{split}
		& P_{\gamma}^{\mu\nu}(k)\equiv \sum_{\lambda_\gamma} \epsilon_{\lambda_\gamma}^{\star\mu}(k) 
	 	\epsilon_{\lambda_\gamma}^\nu(k) = -\eta^{\mu \nu} + \frac{k^\mu r^\nu + k^\nu r^\mu}{k\cdot r}\;, \\
	\end{split}
\label{eq:qed_state_projector}
\end{equation}
as well as for the massive vector boson \myeqref{eq:spin1-transverse-proj}. Here $\lambda_i$ 
labels physical states, $k$ is the momentum of the particle that crosses the cut, and $r$ is a 
massless reference momentum. Notice that we use $\epsilon_i^{\mu_1 \ldots \mu_n}$ and 
$\varepsilon_i^{\mu_1 \ldots \mu_n}$ to denote the polarisations of massless and massive 
spinning particles, respectively. For cuts involving gravitons and the massive spin-$2$ field we 
require the $D_s$-dimensional transverse projector of the graviton\footnote{Here $D_s$ denotes the 
dimensionality of the spin space in which the polarisation tensors and physical-state sums are defined, 
whereas $D$ denotes the spacetime dimension in which momenta live. Keeping $D_s$ and $D$ distinct 
is often useful as it makes the number of internal physical states, and hence the dimensional-regularisation 
scheme, explicit.}
\begin{equation}
    P_{h}^{\mu\nu \rho \sigma}(k)\equiv \sum_{\lambda_{h}} \epsilon_{\lambda_{h}}^{* \mu \nu}(k)\epsilon_{\lambda_{h}}^{\rho \sigma}(k) = \frac{1}{2}\left(P_{\gamma}^{\mu \rho}P_{\gamma}^{\nu \sigma} + P_{\gamma}^{\mu \sigma}P_{\gamma}^{\nu \rho}\right) - \frac{1}{D_s - 2}P_{\gamma}^{\mu \nu} P_{\gamma}^{\rho \sigma}\;,
    \label{eq:gravity_state_projector}
\end{equation}
alongside the projector of the massive tensor \myeqref{eq:spin2-transverse-proj}. As 
before, here $\epsilon_{\lambda_{h}}^{\mu \nu}(k)$ are the polarisation tensors for the 
gravitons that cross the cut. Keep in mind that we have shown the generic $D_s$ dependence 
in the above equation, but we specialise to $D_s=D=4-2\epsilon$, i.e.\ conventional 
dimensional regularisation, in our calculations.

We then express the loop integrand as a linear combination of Feynman diagrams, each 
with a set of propagators and a numerator. The complete quantum integrand can explicitly 
be obtained by a spanning set of cuts, namely, the set of cuts that allows one to fix 
every term in the linear combination. Since we are interested in only the classical 
information in four-point scattering amplitudes, the number of diagrams that we 
need to consider is drastically reduced~\cite{Bern:2019crd, Cheung:2018wkq, 
Neill:2013wsa, Bjerrum-Bohr:2018xdl}. In fact, we never 
expect a contribution from diagrams that have pinched photon/graviton lines, since these 
correspond to zero-point quantum interactions with no classical consequence. 
So the number of spanning cuts needed to construct the classically relevant 
part of the loop integrand is reduced, and constitutes a subset of the full spanning set. 
%

%--------------------------------------------------------------------------------
\subsection{A Brief Word on Integration-By-Parts Reduction}
%--------------------------------------------------------------------------------
After successfully constructing the relevant integrands (discussed below), 
we are faced with the task of reducing the, often large, number of diagrams to a minimal basis 
of \textit{master integrals} (MIs). This can be done using an \textit{integration-by-
parts reduction} (IBP) and is conveniently performed using e.g.~{\tt FIRE6}
~\cite{Smirnov:2019qkx, Smirnov:2023yhb} or {\tt FIRE7}~\cite{Smirnov:2025prc}. The result is 
that $L$-loop amplitudes are recast as
\begin{equation}
 M^{(L)} = \sum_n c_n \mathcal{I}_n\;,
\end{equation}
where the $c_n$ are rational functions of kinematic variables and the MIs $\mathcal{I}_n$, 
which are scalar integrals, have the general form
\begin{equation}
    G_{[\alpha_1, \alpha_2, \ldots,  \alpha_{m-1}, \alpha_m]} = \int \prod_{i = 1}^{L} \left(\mu^{2\epsilon} \hat{\mathrm{d}}^D \ell_i\right) \frac{1}{\rho_1^{\alpha_1} \rho_2^{\alpha_2} \ldots \rho_{m-1}^{\alpha_{m-1}} \rho_m^{\alpha_m}}\;,
    \label{eq:GeneralIBPIntegral}
\end{equation}
with loop momenta $\ell_i$, propagators denoted by $\rho_i$, and integer indices 
$\alpha_i \in \mathbb{Z}$. Loop integrals are regularised using dimensional 
regularisation, $D = 4 - 2\epsilon$, and involves a factor of 
the regularisation scale $\mu^{2\epsilon}$ per loop. Importantly, the most general form of the (quadratic) 
propagators is $\rho_i = q_i^2 - m^2_i$, where $q_i$ is the momentum of the propagator, which is generically a 
function of loop-momenta, external momenta, and mass. The explicit values of these MIs can be 
determined using standard techniques found within the literature, for example differential 
equations~\cite{Kotikov:1990kg,Bern:1993kr,Remiddi:1997ny,Gehrmann:1999as,Goncharov:2010jf,Henn:2013pwa, 
Herrmann:2021tct,Parra-Martinez:2020dzs}. It is also appropriate to mention 
that in the literature the convention is to write
\begin{equation}
	\mu^{2\epsilon} \hat{\mathrm{d}}^D \ell_i = \mu^{2\epsilon}\frac{\mathrm{d}^D \ell_i}{(2\pi)^D} = \frac{i\bar{\mu}^{2\epsilon}}{(4\pi)^2}\frac{\mathrm{d}^D \ell_i}{i\pi^{D/2}}e^{\epsilon \gamma_E}\;,
\end{equation}
where we strip off the overall factor of $i\bar{\mu}^{2\epsilon}/(4\pi)^2$. Here 
$\bar{\mu}^2 = (4\pi e^{-\gamma_E}\mu^2)^\epsilon$ defines the modified subtraction scheme, 
$\gamma_E$ is Euler's constant, and $\epsilon$ is the dimensional regulator. The MIs 
are then evaluated with this different integral measure, and, at the very end, the removed 
factor is reintroduced to obtain the correct expression. Since the focus of this thesis is not  
integration techniques, but rather the physics of black hole scattering, we will not 
discuss this and simply cite the literature. 

Given this general workflow, the objective becomes to determine the numerators that 
appear in this linear combination of Feynman diagrams, which can be done in a few 
different ways. In this effort, we will outline two methods with a particular 
emphasis on analytic generalised unitarity. It should be noted, however, that some of 
the results of this thesis were obtained with numerical generalised unitarity, see for 
example Refs.~\cite{Ita:2015tya,Abreu:2017idw, Abreu:2017xsl,Abreu:2017hqn, 
Akpinar:2024meg,Akpinar:2025bkt}.

%--------------------------------------------------------------------------------
\subsection{A Top-Down Construction}
%--------------------------------------------------------------------------------

Within the context of generalised unitarity one often resorts to an ansatz 
to fix the aforementioned numerators at the multi-loop order~\cite{Ita:2015tya, 
Abreu:2017xsl, Bern:2019crd}. Now, although this approach seems conceptually straightforward, 
there are technical issues that arise beyond one-loop. These include, but are not limited to,
the large size of the ansatz, the presence of many redundant coefficients, and the 
challenges involved in solving the linear systems required to determine these coefficients. There exists alternative strategies that forego an ansatz for the numerator altogether, see 
for example Refs.~\cite{Bourjaily:2017wjl,Bourjaily:2023apy}. As a matter of fact, the methods that we have in mind 
will provide a direct way of obtaining the classically relevant contributions by simply sewing together tree-level 
amplitudes. The first of these strategies will be referred to in this thesis as a \textit{top-down} construction. 
Here we will illustrate, at least schematically, how this procedure works at the one-loop order, keeping in mind 
that we wish to fix the \textit{integrand}. To be as general as possible, we will do this in a 
somewhat theory-agnostic way---albeit focusing on theories like gauge theory and gravity---recalling that we 
will target diagrams that are relevant for classical physics in the scattering of two massive particles at 
one loop.

At this order in perturbation theory, it is well understood that the superclassical terms (scaling as 
$1/q^2$) are given by the box and crossed box diagrams~\cite{Kosower:2018adc}\footnote{In explicit diagrams 
we understand that all legs are outgoing.}
%This means that we suppress the minus signs on the momenta $-p_1$ and $-p_2$.}
\begin{equation}
  \vcenter{
    \hbox{
      \scalebox{0.8}{\begin{tikzpicture}[scale=1,every node/.style={transform shape}]
        \begin{feynman}[every blob={/tikz/fill=gray!30,/tikz/inner sep=2pt}]
          \vertex[draw, circle, minimum size=2pt, 
            inner sep=2pt, pattern=north east lines, 
            pattern color=black, fill=white, line width=0.6pt] (b1) at (-0.75,1) {};
          \vertex[draw, circle, minimum size=2pt, 
            inner sep=2pt, pattern=north east lines, 
            pattern color=black, fill=white, line width=0.6pt] (b2) at (-0.75,-1) {};
           \vertex[draw, circle, minimum size=2pt, 
            inner sep=2pt, pattern=north east lines, 
            pattern color=black, fill=white, line width=0.6pt] (b3) at (0.75,1) {};
         \vertex[draw, circle, minimum size=2pt, 
            inner sep=2pt, pattern=north east lines, 
            pattern color=black, fill=white, line width=0.6pt] (b4) at (0.75,-1) {};
          \vertex (l1) at (-2,1) {$p_1$};
          \vertex (l2) at (-2,-1) {$p_2$};
          \vertex (r1) at (2,1) {$p_4$};
          \vertex (r2) at (2,-1) {$p_3$};
          \diagram*{
            (b1) -- [draw=red, very thick] (l1),
      		(b3) -- [draw=red, very thick] (r1),
      		(b1) -- [draw=red, very thick] (b3),
      		(b2) -- [draw=blue, very thick] (l2),
      		(b4) -- [draw=blue, very thick] (r2),
      		(b4) -- [draw=blue, very thick] (b2),
      		(b2) -- [boson] (b1),
      		(b3) -- [boson] (b4),
          };
        \end{feynman}
      \end{tikzpicture}
      }
    }
  }
  \mathrel{\!\!\!\!\!\!\scalebox{0.8}{$+$}\!\!}
  \vcenter{
    \hbox{
      \scalebox{0.8}{\begin{tikzpicture}[scale=1,every node/.style={transform shape}]
        \begin{feynman}[every blob={/tikz/fill=gray!30,/tikz/inner sep=2pt}]
          \vertex[draw, circle, minimum size=2pt, 
            inner sep=2pt, pattern=north east lines, 
            pattern color=black, fill=white, line width=0.6pt] (b1) at (-0.75,1) {};
          \vertex[draw, circle, minimum size=2pt, 
            inner sep=2pt, pattern=north east lines, 
            pattern color=black, fill=white, line width=0.6pt] (b2) at (-0.75,-1) {};
           \vertex[draw, circle, minimum size=2pt, 
            inner sep=2pt, pattern=north east lines, 
            pattern color=black, fill=white, line width=0.6pt] (b3) at (0.75,1) {};
         \vertex[draw, circle, minimum size=2pt, 
            inner sep=2pt, pattern=north east lines, 
            pattern color=black, fill=white, line width=0.6pt] (b4) at (0.75,-1) {};
          \vertex (l1) at (-2,1) {$p_1$};
          \vertex (l2) at (-2.35,-1) {$p_2$};
          \vertex (r1) at (2,1) {$p_4$};
          \vertex (r2) at (2,-1) {$p_3$};
          \diagram*{
            (b1) -- [draw=red, very thick] (l1),
      		(b3) -- [draw=red, very thick] (r1),
      		(b1) -- [draw=red, very thick] (b3),
      		(b2) -- [draw=blue, very thick] (l2),
      		(b4) -- [draw=blue, very thick] (r2),
      		(b4) -- [draw=blue, very thick] (b2),
      		(b1) -- [boson] (b4),
      		(b2) -- [boson] (b3),
          };
        \end{feynman}
      \end{tikzpicture}
      }
    }
  }\;,
\end{equation}
while the classical terms (scaling as $1/q$) are given by the triangle and inverted triangle diagrams
\begin{equation}
  \vcenter{
    \hbox{
      \scalebox{0.8}{\begin{tikzpicture}[scale=1,every node/.style={transform shape}]
        \begin{feynman}[every blob={/tikz/fill=gray!30,/tikz/inner sep=2pt}]
          \vertex[draw, circle, minimum size=2pt, 
            inner sep=2pt, pattern=north east lines, 
            pattern color=black, fill=white, line width=0.6pt] (b1) at (-0.75,-1) {};
          \vertex[draw, circle, minimum size=2pt, 
            inner sep=2pt, pattern=north east lines, 
            pattern color=black, fill=white, line width=0.6pt] (b2) at (0,1) {};
           \vertex[draw, circle, minimum size=2pt, 
            inner sep=2pt, pattern=north east lines, 
            pattern color=black, fill=white, line width=0.6pt] (b3) at (0.75,-1) {};
          \vertex (l1) at (-2,1) {$p_1$};
          \vertex (l2) at (-2,-1) {$p_2$};
          \vertex (r1) at (2,1) {$p_4$};
          \vertex (r2) at (2,-1) {$p_3$};
          \diagram*{
            (b2) -- [draw=red, very thick] (l1),
      		(b2) -- [draw=red, very thick] (r1),
      		(b1) -- [draw=blue, very thick] (l2),
      		(b3) -- [draw=blue, very thick] (r2),
      		(b3) -- [draw=blue, very thick] (b1),
      		(b1) -- [boson] (b2),
      		(b3) -- [boson] (b2),
          };
        \end{feynman}
      \end{tikzpicture}
      }
    }
  }
  \mathrel{\!\!\!\!\!\!\scalebox{0.8}{$+$}\!\!}
  \vcenter{
    \hbox{
      \scalebox{0.8}{\begin{tikzpicture}[scale=1,every node/.style={transform shape}]
        \begin{feynman}[every blob={/tikz/fill=gray!30,/tikz/inner sep=2pt}]
          \vertex[draw, circle, minimum size=2pt, 
            inner sep=2pt, pattern=north east lines, 
            pattern color=black, fill=white, line width=0.6pt] (b1) at (-0.75,1) {};
          \vertex[draw, circle, minimum size=2pt, 
            inner sep=2pt, pattern=north east lines, 
            pattern color=black, fill=white, line width=0.6pt] (b2) at (0,-1) {};
           \vertex[draw, circle, minimum size=2pt, 
            inner sep=2pt, pattern=north east lines, 
            pattern color=black, fill=white, line width=0.6pt] (b3) at (0.75,1) {};
          \vertex (l1) at (-2,1) {$p_1$};
          \vertex (l2) at (-2,-1) {$p_2$};
          \vertex (r1) at (2,1) {$p_4$};
          \vertex (r2) at (2,-1) {$p_3$};
          \diagram*{
            (b1) -- [draw=red, very thick] (l1),
      		(b3) -- [draw=red, very thick] (r1),
      		(b1) -- [draw=red, very thick] (b3),
      		(b2) -- [draw=blue, very thick] (l2),
      		(b2) -- [draw=blue, very thick] (r2),
      		(b4) -- [draw=blue, very thick] (b2),
      		(b1) -- [boson] (b2),
      		(b3) -- [boson] (b2),
          };
        \end{feynman}
      \end{tikzpicture}
      }
    }
  }
 \;.
 \label{eq:TriDiagrams}
\end{equation}
In gauge theory and gravity, one also has triangle diagrams with a triple-particle 
vertex that may be embedded into \myeqref{eq:TriDiagrams}. All other contributions 
are quantum suppressed and are therefore discarded. The exact way that 
we determine what is superclassical, classical, or quantum suppressed will be explained 
in \mysecref{sec:MethodOfRegions}~\cite{Bern:2019crd}. Thus, the linear combination of 
Feynman diagrams that we consider is 
\begin{equation}
\hspace{-0.4cm}
  \vcenter{
    \hbox{
      \scalebox{0.8}{\begin{tikzpicture}[scale=1,every node/.style={transform shape}]
        \begin{feynman}[every blob={/tikz/fill=gray!80, /tikz/inner sep=6pt}]
           \vertex[draw, circle, minimum size=1cm, 
            fill=gray, line width=1pt] (b1) at (0,0) {};
          % Inner white circle (donut hole)
    	  \node[circle, minimum size=0.5cm, draw=black, thick, fill=white] at (0,0) {};
          \vertex (l1) at (-1,1) {$p_1$};
          \vertex (l2) at (-1,-1) {$p_2$};
          \vertex (r1) at (1,1) {$p_4$};
          \vertex (r2) at (1,-1) {$p_3$};
          \diagram*{
            (b1) -- [draw=red, very thick] (l1),
      		(b1) -- [draw=red, very thick] (r1),
      		(b1) -- [draw=blue, very thick] (l2),
      		(b1) -- [draw=blue, very thick] (r2),
          };
        \end{feynman}
      \end{tikzpicture}
      }
    }
  }
  \mathrel{\!\!\!\!\!\!\scalebox{0.8}{$=$}}
  \vcenter{
    \hbox{
      \scalebox{0.8}{\begin{tikzpicture}[scale=1,every node/.style={transform shape}]
        \begin{feynman}[every blob={/tikz/fill=gray!30,/tikz/inner sep=2pt}]
          \vertex[draw, circle, minimum size=2pt, 
            inner sep=2pt, pattern=north east lines, 
            pattern color=black, fill=white, line width=0.6pt] (b1) at (-0.45,0.6) {};
          \vertex[draw, circle, minimum size=2pt, 
            inner sep=2pt, pattern=north east lines, 
            pattern color=black, fill=white, line width=0.6pt] (b2) at (-0.45,-0.6) {};
           \vertex[draw, circle, minimum size=2pt, 
            inner sep=2pt, pattern=north east lines, 
            pattern color=black, fill=white, line width=0.6pt] (b3) at (0.45,0.6) {};
         \vertex[draw, circle, minimum size=2pt, 
            inner sep=2pt, pattern=north east lines, 
            pattern color=black, fill=white, line width=0.6pt] (b4) at (0.45,-0.6) {};
          \vertex (l1) at (-1.35,0.6) {$p_1$};
          \vertex (l2) at (-1.35,-0.6) {$p_2$};
          \vertex (r1) at (1.35,0.6) {$p_4$};
          \vertex (r2) at (1.35,-0.6) {$p_3$};
          \diagram*{
            (b1) -- [draw=red, very thick] (l1),
      		(b3) -- [draw=red, very thick] (r1),
      		(b1) -- [draw=red, very thick] (b3),
      		(b2) -- [draw=blue, very thick] (l2),
      		(b4) -- [draw=blue, very thick] (r2),
      		(b4) -- [draw=blue, very thick] (b2),
      		(b2) -- [boson] (b1),
      		(b3) -- [boson] (b4),
          };
        \end{feynman}
      \end{tikzpicture}
      }
    }
  }
  \mathrel{\!\!\!\!\scalebox{0.8}{$+$}\!}
  \vcenter{
    \hbox{
      \scalebox{0.8}{\begin{tikzpicture}[scale=1,every node/.style={transform shape}]
        \begin{feynman}[every blob={/tikz/fill=gray!30,/tikz/inner sep=2pt}]
          \vertex[draw, circle, minimum size=2pt, 
            inner sep=2pt, pattern=north east lines, 
            pattern color=black, fill=white, line width=0.6pt] (b1) at (-0.45,0.6) {};
          \vertex[draw, circle, minimum size=2pt, 
            inner sep=2pt, pattern=north east lines, 
            pattern color=black, fill=white, line width=0.6pt] (b2) at (-0.45,-0.6) {};
           \vertex[draw, circle, minimum size=2pt, 
            inner sep=2pt, pattern=north east lines, 
            pattern color=black, fill=white, line width=0.6pt] (b3) at (0.45,0.6) {};
         \vertex[draw, circle, minimum size=2pt, 
            inner sep=2pt, pattern=north east lines, 
            pattern color=black, fill=white, line width=0.6pt] (b4) at (0.45,-0.6) {};
          \vertex (l1) at (-1.35,0.6) {$p_1$};
          \vertex (l2) at (-1.35,-0.6) {$p_2$};
          \vertex (r1) at (1.35,0.6) {$p_4$};
          \vertex (r2) at (1.35,-0.6) {$p_3$};
          \diagram*{
            (b1) -- [draw=red, very thick] (l1),
      		(b3) -- [draw=red, very thick] (r1),
      		(b1) -- [draw=red, very thick] (b3),
      		(b2) -- [draw=blue, very thick] (l2),
      		(b4) -- [draw=blue, very thick] (r2),
      		(b4) -- [draw=blue, very thick] (b2),
      		(b1) -- [boson] (b4),
      		(b2) -- [boson] (b3),
          };
        \end{feynman}
      \end{tikzpicture}
      }
    }
  }
  \mathrel{\!\!\!\!\scalebox{0.8}{$+$}\!}
  \vcenter{
    \hbox{
      \scalebox{0.8}{\begin{tikzpicture}[scale=1,every node/.style={transform shape}]
        \begin{feynman}[every blob={/tikz/fill=gray!30,/tikz/inner sep=2pt}]
          \vertex[draw, circle, minimum size=2pt, 
            inner sep=2pt, pattern=north east lines, 
            pattern color=black, fill=white, line width=0.6pt] (b1) at (-0.45,-0.6) {};
          \vertex[draw, circle, minimum size=2pt, 
            inner sep=2pt, pattern=north east lines, 
            pattern color=black, fill=white, line width=0.6pt] (b2) at (0,0.6) {};
           \vertex[draw, circle, minimum size=2pt, 
            inner sep=2pt, pattern=north east lines, 
            pattern color=black, fill=white, line width=0.6pt] (b3) at (0.45,-0.6) {};
          \vertex (l1) at (-1.35,0.6) {$p_1$};
          \vertex (l2) at (-1.35,-0.6) {$p_2$};
          \vertex (r1) at (1.35,0.6) {$p_4$};
          \vertex (r2) at (1.35,-0.6) {$p_3$};
          \diagram*{
            (b2) -- [draw=red, very thick] (l1),
      		(b2) -- [draw=red, very thick] (r1),
      		(b1) -- [draw=blue, very thick] (l2),
      		(b3) -- [draw=blue, very thick] (r2),
      		(b3) -- [draw=blue, very thick] (b1),
      		(b1) -- [boson] (b2),
      		(b3) -- [boson] (b2),
          };
        \end{feynman}
      \end{tikzpicture}
      }
    }
  }
  \mathrel{\!\!\!\!\scalebox{0.8}{$+$}\!}
  \vcenter{
    \hbox{
      \scalebox{0.8}{\begin{tikzpicture}[scale=1,every node/.style={transform shape}]
        \begin{feynman}[every blob={/tikz/fill=gray!30,/tikz/inner sep=2pt}]
          \vertex[draw, circle, minimum size=2pt, 
            inner sep=2pt, pattern=north east lines, 
            pattern color=black, fill=white, line width=0.6pt] (b1) at (-0.45,0.6) {};
          \vertex[draw, circle, minimum size=2pt, 
            inner sep=2pt, pattern=north east lines, 
            pattern color=black, fill=white, line width=0.6pt] (b2) at (0,-0.6) {};
           \vertex[draw, circle, minimum size=2pt, 
            inner sep=2pt, pattern=north east lines, 
            pattern color=black, fill=white, line width=0.6pt] (b3) at (0.45,0.6) {};
          \vertex (l1) at (-1.35,0.6) {$p_1$};
          \vertex (l2) at (-1.35,-0.6) {$p_2$};
          \vertex (r1) at (1.35,0.6) {$p_4$};
          \vertex (r2) at (1.35,-0.6) {$p_3$};
          \diagram*{
            (b1) -- [draw=red, very thick] (l1),
      		(b3) -- [draw=red, very thick] (r1),
      		(b1) -- [draw=red, very thick] (b3),
      		(b2) -- [draw=blue, very thick] (l2),
      		(b2) -- [draw=blue, very thick] (r2),
      		(b1) -- [boson] (b2),
      		(b3) -- [boson] (b2),
          };
        \end{feynman}
      \end{tikzpicture}
      }
    }
  }
  \;,
\end{equation}
where the red and blue solid lines represent particles of different mass, while the wavy lines 
represent the massless mediators. As explained before, we know which propagators appear in each 
diagram while the numerators are unknown. Here we wish to fix these numerators by simply 
performing generalised cuts without ever referring to an ansatz. To outline this top-down 
construction explicitly, we should choose a momentum routing for the internal legs: 
\begin{equation}
	\begin{tikzpicture}[baseline=(current bounding box.center)]
        \begin{feynman}[every blob={/tikz/fill=gray!30,/tikz/inner sep=2pt}]
          \vertex[draw, circle, minimum size=2pt, 
            inner sep=2pt, pattern=north east lines, 
            pattern color=black, fill=white, line width=0.6pt] (b1) at (-0.75,1) {};
          \vertex[draw, circle, minimum size=2pt, 
            inner sep=2pt, pattern=north east lines, 
            pattern color=black, fill=white, line width=0.6pt] (b2) at (-0.75,-1) {};
           \vertex[draw, circle, minimum size=2pt, 
            inner sep=2pt, pattern=north east lines, 
            pattern color=black, fill=white, line width=0.6pt] (b3) at (0.75,1) {};
         \vertex[draw, circle, minimum size=2pt, 
            inner sep=2pt, pattern=north east lines, 
            pattern color=black, fill=white, line width=0.6pt] (b4) at (0.75,-1) {};
          \vertex (l1) at (-2,1) {$p_1$};
          \vertex (l2) at (-2,-1) {$p_2$};
          \vertex (r1) at (2,1) {$p_4$};
          \vertex (r2) at (2,-1) {$p_3$};
          \diagram*{
            (b1) -- [draw=red, very thick] (l1),
      		(b3) -- [draw=red, very thick] (r1),
      		(b1) -- [draw=red, very thick, edge node={node[midway, above] {$\rho_1$}}] (b3),
      		(b2) -- [draw=blue, very thick] (l2),
      		(b4) -- [draw=blue, very thick] (r2),
      		(b4) -- [draw=blue, very thick, edge node={node[midway, below] {$\rho_2$}}] (b2),
      		(b2) -- [boson, momentum={[arrow shorten=0.35] $\ell_1$}, edge node={node[midway, right] {$\rho_3$}}] (b1),
      		(b3) -- [boson, momentum={[arrow shorten=0.35] $\ell_2$}, edge node={node[midway, left] {$\rho_4$}}] (b4),
          };
        \end{feynman}
      \end{tikzpicture} 
      \ = \ \int \hat{\mathrm{d}}^D \ell_1 \frac{N_{\scriptscriptstyle\Box}}{\rho_1 \rho_2 \rho_3 \rho_4},
\end{equation}
where, choosing the red and blue lines to have mass $m_1$ and $m_2$ respectively, the inverse 
propagators are 
\begin{equation}
    \rho_1 = (\ell_1 - p_1)^2 - m_1^2 , \ \rho_2 = (\ell_1 + p_2)^2 - m_2^2, \ \rho_3 = \ell_1^2, \ \rho_4 = \ell_2^2 = (\ell_1 - q)^2, \\
    \label{eq:boxPropagators}
\end{equation}
together with $q = p_1 + p_4 = - p_2 - p_3$. The choice of internal momentum routing is the 
same for the crossed box, triangle, and inverted triangle diagrams, and we denote their 
numerators by $N_{\scriptscriptstyle\boxtimes}$, $N_{\scriptscriptstyle\bigtriangleup}$, and 
$N_{\scriptscriptstyle\bigtriangledown}$, respectively. Here, the use of \textit{crossed} refers to 
the crossing from the s-channel to the u-channel, e.g.\ by exchanging $p_1$ and $p_4$ or $p_2$ 
and $p_3$, meaning that the propagators of the crossed box are related to that of the 
box by precisely this exchange. 

\begin{figure}[h!]
	\centering
		\includegraphics[width=1\linewidth]{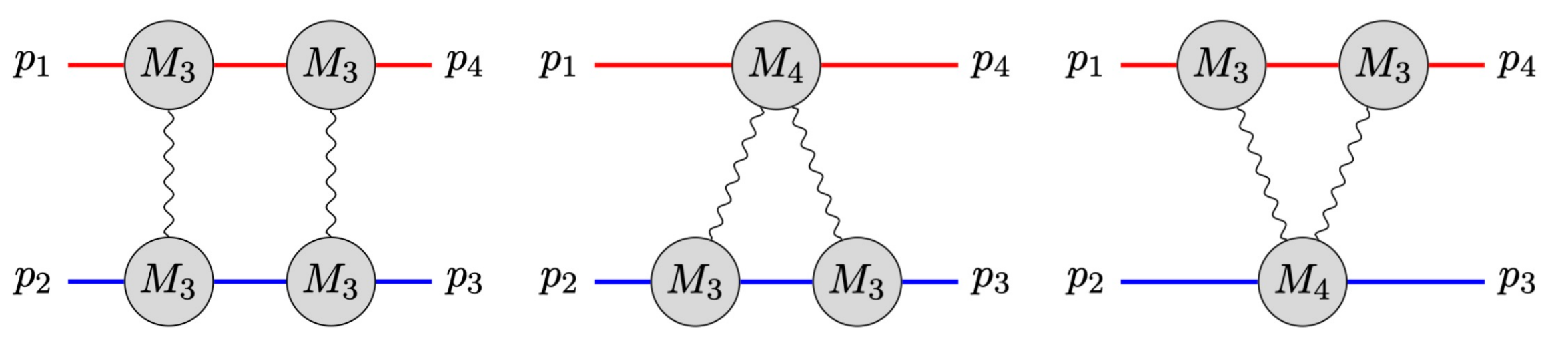}
	\caption{The cuts that allow one to fix both the superclassical and classical 
	contributions at the one-loop order. Left to right: maximal cut, the 
	triangle cut, and the inverted triangle cut. The red and blue solid lines represent 
	massive particles of different mass, whereas the wavy line denotes a massless 
	mediator. Furthermore, the blobs are on-shell tree-level amplitudes and all exposed 
	internal propagators are understood to be cut. }
	\label{fig: oneLoop_SpanningCuts}
\end{figure}

The superclassical numerators are fixed by considering the 
\textit{maximal cut}~\cite{Bern:2019crd,Britto:2004nc}, which follows from setting all internal 
propagators on-shell. More specifically, taking the residue as all the propagators in the box diagram 
go on-shell completely kills the crossed box, triangle, and inverted triangle diagrams. This  
yields
\begin{equation}
	\hspace{-0.3cm}
	N_{\scriptscriptstyle\Box} = \sum_{\mathrm{states}} M^{(0)}_{3}(p_1, \tilde{p}_1, -\ell_1) M^{(0)}_{3} (p_2, -\tilde{p}_2, \ell_1) M^{(0)}_{3}(\tilde{p}_2, p_3, -\ell_2) M^{(0)}_{3}(-\tilde{p}_1,p_4,\ell_2),
\end{equation}
where $M^{(0)}_{3}(p_1, \tilde{p}_1, \ell_1)$ is the tree-level amplitude involving 
two outgoing massive legs of momenta $p_1$ and $\tilde{p}_1 = -p_1 - \ell_1$, and one outgoing 
massless leg of momentum $\ell_1$. To obtain the numerator of the crossed box we can either 
consider the maximal cut with $p_2$ and $p_3$ exchanged or simply by crossing the box 
numerator. Now that the numerators of the superclassical terms have been fixed, we may move 
onto the classical terms. For concreteness, let us begin with  
$N_{\scriptscriptstyle\bigtriangleup}$. To do this, we consider the triple cut involving the 
sewing of two three-point amplitudes on the blue line and a Compton amplitude on the red line 
(see Fig.~\ref{fig: oneLoop_SpanningCuts}); this will be referred to as the \textit{triangle 
cut}. This selects the triangle cut of the box and crossed box, as well as the triangle 
numerator\footnote{The triangle numerator has an explicit factor of $i$ because generalised cuts 
trade propagators for delta functions, which are accompanied by factors of the imaginary unit.}:
\begin{align}
\hspace{-0.6cm}
  \sum_{\mathrm{states}} & M^{(0)}_{4}(p_1, p_4, -\ell_1, \ell_2) M^{(0)}_{3} (p_2, -\tilde{p}_2, \ell_1) M^{(0)}_{3}(\tilde{p}_2, p_3, -\ell_2) \\
  & \hspace{3.5cm} \mathrel{\scalebox{0.8}{$=$}\!\!\!}
  \vcenter{
    \hbox{
      \scalebox{0.8}{\begin{tikzpicture}[scale=1,every node/.style={transform shape}]
        \begin{feynman}[every blob={/tikz/fill=gray!30,/tikz/inner sep=2pt}]
          \vertex[draw, circle, minimum size=2pt, 
            inner sep=2pt, pattern=north east lines, 
            pattern color=black, fill=white, line width=0.6pt] (b1) at (-0.45,0.6) {};
          \vertex[draw, circle, minimum size=2pt, 
            inner sep=2pt, pattern=north east lines, 
            pattern color=black, fill=white, line width=0.6pt] (b2) at (-0.45,-0.6) {};
           \vertex[draw, circle, minimum size=2pt, 
            inner sep=2pt, pattern=north east lines, 
            pattern color=black, fill=white, line width=0.6pt] (b3) at (0.45,0.6) {};
         \vertex[draw, circle, minimum size=2pt, 
            inner sep=2pt, pattern=north east lines, 
            pattern color=black, fill=white, line width=0.6pt] (b4) at (0.45,-0.6) {};
          \vertex (l1) at (-1.35,0.6) {$p_1$};
          \vertex (l2) at (-1.35,-0.6) {$p_2$};
          \vertex (r1) at (1.35,0.6) {$p_4$};
          \vertex (r2) at (1.35,-0.6) {$p_3$};
          \diagram*{
            (b1) -- [draw=red, very thick] (l1),
      		(b3) -- [draw=red, very thick] (r1),
      		(b1) -- [draw=red, very thick] (b3),
      		(b2) -- [draw=blue, very thick] (l2),
      		(b4) -- [draw=blue, very thick] (r2),
      		(b4) -- [draw=blue, very thick] (b2),
      		(b2) -- [boson] (b1),
      		(b3) -- [boson] (b4),
          };
         % Draw dashed line
         \draw [plain, color = black] ($(0,-0.8)$) -- ($(0,-0.4)$);
         \draw [plain, color = black] ($(-0.65,0)$) -- ($(-0.25,0)$);
         \draw [plain, color = black] ($(0.65,0)$) -- ($(0.25,0)$);
        \end{feynman}
      \end{tikzpicture}
      }
    }
  }
  \mathrel{\!\!\!\!\scalebox{0.8}{$+$}\!}
  \vcenter{
    \hbox{
      \scalebox{0.8}{\begin{tikzpicture}[scale=1,every node/.style={transform shape}]
        \begin{feynman}[every blob={/tikz/fill=gray!30,/tikz/inner sep=2pt}]
          \vertex[draw, circle, minimum size=2pt, 
            inner sep=2pt, pattern=north east lines, 
            pattern color=black, fill=white, line width=0.6pt] (b1) at (-0.45,0.6) {};
          \vertex[draw, circle, minimum size=2pt, 
            inner sep=2pt, pattern=north east lines, 
            pattern color=black, fill=white, line width=0.6pt] (b2) at (-0.45,-0.6) {};
           \vertex[draw, circle, minimum size=2pt, 
            inner sep=2pt, pattern=north east lines, 
            pattern color=black, fill=white, line width=0.6pt] (b3) at (0.45,0.6) {};
         \vertex[draw, circle, minimum size=2pt, 
            inner sep=2pt, pattern=north east lines, 
            pattern color=black, fill=white, line width=0.6pt] (b4) at (0.45,-0.6) {};
          \vertex (l1) at (-1.35,0.6) {$p_1$};
          \vertex (l2) at (-1.35,-0.6) {$p_3$};
          \vertex (r1) at (1.35,0.6) {$p_4$};
          \vertex (r2) at (1.35,-0.6) {$p_2$};
          \diagram*{
            (b1) -- [draw=red, very thick] (l1),
      		(b3) -- [draw=red, very thick] (r1),
      		(b1) -- [draw=red, very thick] (b3),
      		(b2) -- [draw=blue, very thick] (l2),
      		(b4) -- [draw=blue, very thick] (r2),
      		(b4) -- [draw=blue, very thick] (b2),
      		(b2) -- [boson] (b1),
      		(b3) -- [boson] (b4),
          };
         % Draw dashed line
         \draw [plain, color = black] ($(0,-0.8)$) -- ($(0,-0.4)$);
         \draw [plain, color = black] ($(-0.65,0)$) -- ($(-0.25,0)$);
         \draw [plain, color = black] ($(0.65,0)$) -- ($(0.25,0)$);
        \end{feynman}
      \end{tikzpicture}
      }
    }
  }  \mathrel{\!\!\!\!\scalebox{0.8}{$-$}}
  iN_{\scriptscriptstyle\bigtriangleup} \nonumber \;,
\end{align}
where $M^{(0)}_{4}(p_1, p_4, \ell_1, \ell_2)$ is the tree-level four-point Compton 
amplitude involving two outgoing massive legs of momenta $p_1$ and $p_4 = -p_1 - \ell_1 - \ell_2$, 
and two outgoing massless legs of momenta $\ell_1$ and $\ell_2$. Furthermore, the solid black 
line on internal legs indicate a generalised cut, i.e.~they are on-shell, and we 
have ``unwrapped" the crossed box by writing it as a box with $p_2$ and $p_3$ swapped. 
One may proceed with solving this equation for $N_{\scriptscriptstyle\bigtriangleup}$ by 
subtracting the triangle cut of the box and crossed box from the sewing of tree-level 
amplitudes. In fact, we know that this has been done correctly once all massive propagators 
cancel and we are left with a rational function. The inverted triangle numerator is determined 
using exactly the same procedure, with the difference being that the red propagator is cut, as 
opposed to the blue propagator; we call this the \textit{inverted triangle cut}. 

%--------------------------------------------------------------------------------
\subsection{A Direct Method for Fixing Numerators}
\label{eq:GeneralisedUni_DirectWay}
%--------------------------------------------------------------------------------

Although this top-down method is particularly convenient at one-loop since there are 
only a few diagrams to fix, going beyond this order becomes more involved. In particular, at 
two-loops and beyond, the number of diagrams to fix grows rapidly, and so this approach 
becomes tedious. One can therefore ask if there is a way to obtain all the relevant information 
without writing an ansatz for the numerator and also without this systematic subtraction of 
diagrams. 

The strategy we will use relies on the fact that after fixing the 
relevant numerators we will nevertheless perform a soft expansion, as explained in 
\mysecref{sec:MethodOfRegions}, in order to really access the classical physics. By 
virtue of this soft expansion, the quadratic propagators $\rho_i$ that we have seen thus far are 
expanded into linear propagators, thereby requiring a subsequent IBP reduction into a minimal 
basis of MIs with linear propagators~\cite{Parra-Martinez:2020dzs, Herrmann:2021tct}. With this information 
in hand, we will use the following strategy to mesh all the steps together into one pipeline. 
First, we construct the spanning cuts by sewing together the appropriate tree-level amplitudes. 
Secondly, treating each spanning cut as a loop-level numerator, we perform the soft expansion to 
the required order needed to access the classical physics. Finally, we perform the IBP reduction with 
linearised propagators. The outcome is a sum of MIs for each spanning cut which must be 
assembled into the classical loop-level integrand, while excluding any duplicate copies of 
integrand terms that are fixed by more than one spanning cut. There are two important points to 
keep in mind regarding this workflow. First, because a given integrand term can be determined by 
more than one spanning cut, this redundancy provides a non-trivial consistency check of our 
construction. Secondly, after performing the IBP reduction with linearised propagators, we find MIs 
that have pinches (i.e.~contact vertices) on propagators that were put on-shell in the spanning 
cut. These contributions cannot be fixed correctly because the generalised unitarity framework relies 
on the existence of a simple pole, and thus we simply discard them. Indeed one might 
rightfully worry that in doing this we discard essential terms in the construction. In 
practice, such terms are either determined by a different spanning cut or make no contribution to 
the classical physics.

Let us be more specific about how this is done to construct loop-level integrands 
for electromagnetism and gravity. At one-loop we will not consider the maximal cut and 
instead just construct the triangle and inverted triangle cuts. These cuts contain all the information
regarding the box, crossed box, and triangles. Treating these cuts as the numerator of our 
integrand, we perform the soft expansion and subsequently IBP reduce. The result is 
\begin{equation}
    M^{\text{(1)}} = c_{\scriptscriptstyle\Box}\mathcal{I}_{\scriptscriptstyle\Box} + c_{\scriptscriptstyle\boxtimes}\mathcal{I}_{\scriptscriptstyle\boxtimes} + c_{\scriptscriptstyle\bigtriangledown}\mathcal{I}_{\scriptscriptstyle\bigtriangledown} + c_{\scriptscriptstyle\bigtriangleup}\mathcal{I}_{\scriptscriptstyle\bigtriangleup} + \ldots,
\end{equation}
where $c_i$ are the master coefficients, which are functions of external kinematics and 
spacetime dimensions, and the MIs are written in terms of linearised propagators. Here 
the $\ldots$ are quantum suppressed contributions, such as the bubble, which we discard. Now 
about discarding terms that cannot be fixed by a cut. For example, after IBP reducing the 
expanded triangle cut, we obtain a non-zero contribution to the inverted triangle master. Since 
this involves a pinch in a propagator which was put on-shell in the spanning cut, we simply 
discard it as it cannot be correct using this cut. 

\begin{figure}[h!]
	\centering
		\includegraphics[width=1\linewidth]{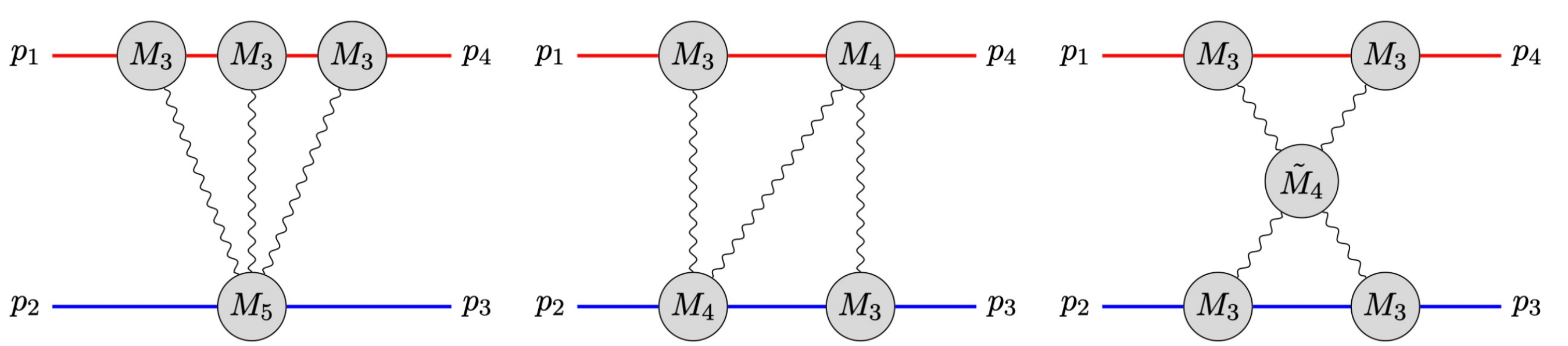}
	\caption{The independent generalised cuts needed to fix the superclassical and 
	classical contributions at the two-loop order. Left to right: inverted double-triangle, the 
	N cut, and the H cut. Note that we have not included the cuts which follow by 
	flips or crossing. The red and blue solid lines represent 
	massive particles of different mass, whereas the wavy line denotes a massless 
	mediator. Furthermore, the blobs are on-shell tree-level amplitudes and all exposed 
	internal propagators are understood to be cut. In the case of 
	the H cut, the pure 	graviton tree-level amplitude is denoted with a tilde.}
	\label{fig: twoLoop_SpanningCuts}
\end{figure}

The two-loop integrand, on the other hand, is more involved as there are many more
diagrams in the linear combination (see for example Ref.~\cite{Bern:2019crd} for details). 
At this loop level, the relevant cuts for classical physics in electromagnetism are the 
double-triangle cut, the inverted double-triangle cut, the N cut, the horizontally flipped 
N cut, and the crossed N cut (see Fig.~\ref{fig: twoLoop_SpanningCuts}), using a naming 
convention that describes what the massless part of the diagram looks like. In the case of 
gravity, we further have the H cut because of graviton self interactions. It is crucial to note 
that these spanning cuts allow us to determine only the conservative dynamics 
(see \mysecref{sec:MethodOfRegions}). If we wish to describe radiative dynamics---which includes
both non-dissipative\footnote{In this thesis, we will refer to the non-dissipative part of 
the radiative dynamics as \textit{radiation-reaction}, which follows from the real (or 
time-reversal symmetric) part.} and dissipative effects---then we must 
modify the spanning cuts (see Ref.~\cite{Herrmann:2021tct}). 

%If we wish to describe dissipative dynamics, i.e where 
%energy is lost from the system, then there will be more spanning cuts (see 
%Ref.~\cite{Herrmann:2021tct}). 
%

As the two-loop computation is complex, alongside the fact that the focus of this thesis 
is the dynamics of black hole scattering, we will omit specific details and will therefore 
only present the final result. 

%--------------------------------------------------------------------------------
\subsection{A Note on Sewing Amplitudes}
\label{sec:GWI}
%--------------------------------------------------------------------------------

The essential step in the generalised unitarity framework is the sewing of tree-level 
amplitudes. This procedure generically involves massless mediators such as photons or 
gravitons, which, after summing over the states that cross the cut, yield factors of the 
transverse projectors, i.e.~\myeqref{eq:qed_state_projector} or 
\myeqref{eq:gravity_state_projector}. These projectors contain a spurious light-cone 
singularity---given by the presence of the reference momentum---that we 
drag along with the expectation that it cancels at the very end. In fact, we 
can completely omit this spurious singularity by simply replacing 
$P_{\gamma}^{\mu \nu}$ by $-\eta^{\mu \nu}$, which is guaranteed if the tree-level amplitudes 
that we sew together satisfy the so-called \textit{generalised Ward identities} 
(GWIs)~\cite{Kosmopoulos:2020pcd}.

\begin{figure}[h!]
    \centering
    \begin{tikzpicture}[scale=1,every node/.style={transform shape}]
      \begin{feynman}[every blob={/tikz/fill=gray!30,/tikz/inner sep=2pt}]
        % Define vertices, their positions, and their labels
        \vertex[blob] (b1) at (0,1) {$ A_4$};
        \vertex[blob] (b2) at (0,-1) {$ A_4$};
        \vertex (v1) at (-0.75,0);
        \vertex (v1) at (-0.75,0);
        \vertex (l1) at (-1.5,1) {$p_1$};
        \vertex (l2) at (-1.5,-1) {$p_2$};
        \vertex (r1) at (1.5,1) {$p_4$};
        \vertex (r2) at (1.5,-1) {$p_3$};
        % Draw Feynman diagram
        \diagram*{
        (b1) -- [draw=red, very thick] (l1),
        (b1) -- [draw=red, very thick] (r1),
        (b2) -- [draw=blue, very thick] (l2),
        (b2) -- [draw=blue, very thick] (r2),
        (b2) -- [boson, quarter right, momentum'= $k_3$] (b1),
        (b1) -- [boson, quarter right,momentum'= $k_2$] (b2),
        };
      \end{feynman}
    \end{tikzpicture}
    \caption{The bubble cut: an example cut that involves the sewing of two tree-level Compton amplitudes.}
    \label{fig: bubbleCut}
\end{figure}
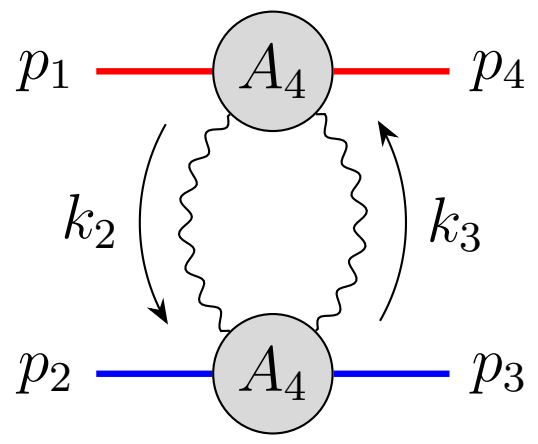

To understand this in detail, let us provide an example following the discussion of 
Ref.~\cite{Kosmopoulos:2020pcd}. For concreteness we will consider computing the bubble cut, 
shown in Fig.~\ref{fig: bubbleCut}, involving massive scalar fields interacting via photon 
exchanges. This cut requires the following two tree-level Compton amplitudes
\begin{equation}
    \begin{split}
        & A_{4,\mathrm{u}}^{(0)} = 2ie^2 q_{\phi_1}^2 \left(\epsilon_2 \cdot 
        \epsilon_{3} + \frac{(p_1 \cdot \epsilon_2)(p_4 \cdot \epsilon_{3})}{k_2\cdot (k_3 - p_4)} + 
        \frac{(p_1 \cdot \epsilon_{3})(p_4 \cdot\epsilon_{2})}{k_2\cdot (k_3-p_1)} \right)\;, \\
        & A_{4,\mathrm{d}}^{(0)} = 2ie^2 q_{\phi_2}^2\left(\epsilon_{2} \cdot 
        \epsilon_{3} - \frac{(p_2\cdot \epsilon_{2})(p_3\cdot \epsilon_3)}{k_2\cdot (k_3 + p_3)} -  
        \frac{(p_2\cdot \epsilon_{3})(p_3\cdot \epsilon_{2})}{k_2\cdot (k_3+p_2)}\right)\;,
    \end{split}
    \label{eq:ComptonAmplitudes}
\end{equation}
where the subscripts $u,d$ refer to the ``upper" and ``lower" amplitudes shown in Fig.~\ref{fig: bubbleCut}. 
We use this labelling system to suppress the arguments of the amplitude: 
$A_{4,\mathrm{u}}^{(0)} = A_{4}^{(0)}(p_1, p_4, k_2, -k_3)$ and 
$A_{4,\mathrm{d}}^{(0)} = A_{4}^{(0)}(p_2, p_3, -k_2, k_3)$. 

Since the amplitude is a gauge-invariant object, it must satisfy the Ward identity (WI) for each gauge 
particle. This statement is realised by the vanishing of the amplitude when the polarisation vector of a 
single gauge particle, say $\epsilon$, is replaced with the particle's momentum, say $p$:
\begin{equation}
	 A_{n}\rvert_{\epsilon_\mu \rightarrow p_\mu} = 0\;.
\end{equation}
Crucially, this property only holds once we have imposed the on-shell conditions, which include the 
mass-shell condition $p_i^2 = m_i^2$, total momentum conservation, alongside the null condition and 
transversality of the polarisation vectors: $\epsilon^2_i = 0$ and $p_i \cdot \epsilon_i = 0$. On the other 
hand, the GWIs are a stronger version of the WIs. It states that
the amplitude vanishes when replacing polarisations with momenta without ever invoking null condition and 
transversality of the polarisation vectors. This has far reaching consequences when sewing together 
massless legs, which we will highlight below.

Let us consider just $ A_{4,\mathrm{u}}^{(0)}$ for simplicity. Replacing all the $\epsilon_i$ with 
$\omega_i$, where $\omega_i$ is assumed to have no special properties, and setting the coupling to unity 
$e^2 q_{\phi_1}^2 \rightarrow 1$ yields
\begin{equation}
	\bar{A}_{4,\mathrm{u}}^{(0)} = 2i \left(\omega_2 \cdot 
        \omega_{3} + \frac{(p_1 \cdot \omega_2)(p_4 \cdot \omega_{3})}{k_2\cdot (k_3 - p_4)} + 
        \frac{(p_1 \cdot \omega_{3})(p_4 \cdot\omega_{2})}{k_2\cdot (k_3-p_1)} \right)\;,
\end{equation} 
which we will call an off-shell amplitude~\cite{Kosmopoulos:2020pcd}. Then, setting 
$\omega_2 \rightarrow k_2$ gives
\begin{align}
	\bar{A}_{4,\mathrm{u}}^{(0)}\rvert_{\omega_2 \rightarrow k_2} & = 2i ( p_1 + p_4 + k_2) \cdot \omega_3 = 2i(k_3 \cdot \omega_3) \neq 0\;,
\end{align}
which clearly does not vanish unless $k_3 \cdot \omega_3 = 0$. Thus, this amplitude does not 
satisfy the GWI but it does satisfy the WI. Notice that the replacement 
$\omega_2 \rightarrow k_2$ introduces a $k_2 \cdot \omega_{3}$ term that is not present in the 
off-shell amplitude. To ensure that GWI is satisfied, we need to find a basis that contains all 
products that might arise from $\omega_2 \rightarrow k_2$. This basis can be readily determined 
by simply eliminating the momentum of one of the massive legs in the on-shell amplitude 
\myeqref{eq:ComptonAmplitudes}, say, using $p_4 = k_3 - k_2 - p_1$\footnote{We could 
equivalently solve momentum conservation as $p_1 = k_3 - k_2 -p_4$.}. Using the mass-shell 
conditions and special properties of the polarisation vectors to simplify the amplitude, and, 
finally, replacing $\epsilon_i \rightarrow \omega_i$  we find 
\begin{align}
    \bar{A}_{4,\mathrm{u}}^{\prime (0)}= 2i\left(\omega_2 \cdot \omega_{3} - 
        \frac{(p_1 \cdot \omega_2)(p_1 + k_2) \cdot \omega_{3} }{k_2 \cdot k_3} + 
        \frac{(p_1 \cdot \omega_{3})(k_3 - p_1) \cdot \omega_{2}}{k_2\cdot (k_3 - p_1)} \right)\;.
    \label{eq:ComptonAmplitudeForGWI}
\end{align}
Here we have used a prime to distinguish between off-shell amplitudes, but we can loosely think of 
them as off-shell amplitudes calculated in different gauges~\cite{Kosmopoulos:2020pcd}. It is 
then straightforward to see that the GWI is now manifestly satisfied
\begin{equation}
	\bar{A}_{4,\mathrm{u}}^{\prime (0)}\rvert_{\omega_2 \rightarrow k_2} = 0\;.
\end{equation}
Therefore, using momentum conservation on the tree-level Compton amplitudes in 
\myeqref{eq:ComptonAmplitudes} puts them in a form where the GWIs are manifest:
\begin{equation}
    \begin{split}
        & A_{4,\mathrm{u}}^{(0)} = 2ie^2 q_{\phi_1}^2 \left(\omega_2 \cdot \omega_{3} - 
        \frac{(p_1 \cdot \omega_2)(p_1 + k_2) \cdot \omega_{3}}{k_2\cdot p_1} + 
        \frac{(p_1 \cdot \omega_{3})(k_3 - p_1) \cdot\omega_{2}}{k_2\cdot (k_3 - p_1)} \right)\;, \\
        & A_{4,\mathrm{d}}^{(0)} = 2ie^2 q_{\phi_2}^2\left(\omega_{2} \cdot \omega_{3} - 
        \frac{(p_2\cdot \omega_{2})(k_2 - p_2)\cdot \omega_3}{k_2\cdot p_2} -  
        \frac{(p_2\cdot \omega_{3})(k_3 + p_2)\cdot \omega_{2}}{k_2\cdot (k_3 + p_2)}\right)\;.
    \end{split}
    \label{eq:ComptonAmplitudesPhoton1GWI}
\end{equation}
Given this nice property, we may ask: \textit{how does this simplify the sewing procedure}? To 
see the simplifications that follow let us consider sewing the photon labelled by momentum 
$k_2$. We start by replacing $\omega_2 \rightarrow \bar{\omega}_2$ in 
$\mathcal{A}_{4,\mathrm{d}}^{(0)}$ (to keep track of which polarisation comes from which 
amplitude), such that sewing this photon gives
\begin{align}
	 \sum_{\lambda_\gamma} A_{4,\mathrm{u}}^{(0)} A_{4,\mathrm{d}}^{(0)} = \mathbb{A}_{\mu \nu} \sum_{\lambda_\gamma} \bar{\omega}^\mu_{2,\lambda_\gamma} \omega^\nu_{2,\lambda_\gamma} = \mathbb{A}_{\mu \nu} P^{\mu \nu}_{\gamma}\;.
\end{align}
Now we analyse what happens to the spurious singularity in this contraction. To see this more 
clearly, we explicitly write 
\begin{align}
	\left(\sum_{\lambda_\gamma} A_{4,\mathrm{u}}^{(0)} A_{4,\mathrm{d}}^{(0)}\right)_{\mathrm{sing.}} & = \frac{1}{k_2 \cdot r}\Big(A_{4,\mathrm{u}}^{(0)}\rvert_{\omega_2 \rightarrow k_2}\ A_{4,\mathrm{d}}^{(0)}\rvert_{\bar{\omega}_{2} \rightarrow r} + A_{4,\mathrm{u}}^{(0)}\rvert_{\omega_2 \rightarrow r} \ A_{4,\mathrm{d}}^{(0)}\rvert_{\bar{\omega}_{2} \rightarrow k_2}\Big) \nonumber \\
	& = 0 \;,
\end{align}
which clearly vanishes because the tree-level amplitudes satisfy the GWI
\begin{equation}
	A_{4,\mathrm{u}}^{(0)}\rvert_{\omega_2 \rightarrow k_2} = 
A_{4,\mathrm{d}}^{(0)}\rvert_{\bar{\omega}_{2} \rightarrow k_2} = 0\;.
\end{equation} 
Therefore, this has the net effect of simply performing the sewing with
\begin{equation}
	P_{\gamma}^{\mu \nu} = - \eta^{\mu \nu}\;,
	\label{eq:qed_state_projector_simplified}
\end{equation}
i.e.~we drop the second term of \myeqref{eq:qed_state_projector}.

The discussion for gravity follows identically. In this case the polarisation vector is replaced 
with a polarisation tensor, however since we are interested in Einstein gravity we always 
decompose them into a product of polarisation vectors (see 
\mysecref{sec:DoubleCopy} for more details). The consequence of 
$P_{\gamma}^{\mu \nu} \rightarrow - \eta^{\mu \nu}$ is that the graviton transverse projector 
becomes:
\begin{equation}
    P_{h}^{\mu\nu \rho \sigma} = \frac{1}{2}\left(\eta^{\mu \rho}\eta^{\nu \sigma} + \eta^{\mu \sigma}\eta^{\nu \rho}\right) - \frac{1}{D_s - 2}\eta^{\mu \nu} \eta^{\rho \sigma}\;.
    \label{eq:gravity_state_projector_simplified}
\end{equation}
Throughout this thesis, all tree-level amplitudes that we feed into generalised cuts will exhibit 
GWIs explicitly, such that we always use these simplified transverse projectors. Furthermore, in 
addition to the massless legs, we will also have to sew together massive legs. In fact, sewing 
massive scalar legs follows from a simple multiplication, as there are no spin degrees of freedom 
to sum over. However, this is not the case when sewing massive vector or massive tensor legs. 
Unfortunately, the transverse propagators associated to these massive spinning particles do not 
enjoy the same simplifications as the massless spinning legs, even if the amplitude satisfies 
GWIs. 

%--------------------------------------------------------------------------------
\section{The Method of Regions}
\label{sec:MethodOfRegions}
%--------------------------------------------------------------------------------
Having understood how loop-level integrands can be obtained by sewing together 
tree-level amplitudes, we now wish to understand how the discussion of 
\mysecref{sec:ClassicalLim} allows us to systematically extract the classical dynamics at the loop level. 
In this section we will make explicit reference to gravitons, but the discussion is exactly the 
same for photons.

Given an internal graviton line with four-momentum $\ell = (\omega, \boldsymbol{\ell})$, we can identify four 
ways in which its four-momentum components scale~\cite{Bern:2019crd, Parra-Martinez:2020dzs}:
\begin{equation}
	\begin{split}
		& \text{hard:} \ (\omega,  \boldsymbol{\ell} ) \sim m(1,1), \\
		& \text{soft:} \  (\omega,\boldsymbol{\ell} ) \sim ( |\boldsymbol{q}| , |\boldsymbol{q}| ) \sim J^{-1}(m |\boldsymbol{v}|, m |\boldsymbol{v}|), \\
		& \text{potential:} \  (\omega,\boldsymbol{\ell} ) \sim ( |\boldsymbol{q}||\boldsymbol{v}| , |\boldsymbol{q}| ) \sim J^{-1}(m |\boldsymbol{v}|^2, m |\boldsymbol{v}| ), \\
		& \text{radiation:} \  (\omega,\boldsymbol{\ell} ) \sim ( |\boldsymbol{q}||\boldsymbol{v}| , |\boldsymbol{q}||\boldsymbol{v}| ) \sim J^{-1}(m |\boldsymbol{v}|^2, m |\boldsymbol{v}|^2 ),
	\end{split}
\end{equation}
where $m \sim m_1 + m_2$, $|\boldsymbol{v}| = q^0/|\boldsymbol{q}|$ is the velocity, and each scale 
defines a \textit{region}. Here $J = m |\boldsymbol{v}||\boldsymbol{q}|^{-1}$ is the angular momentum 
of the system, and the $q$-expansion can be expressed as an expansion in $J^{-1}$ (since $J \gg 1$).

Importantly, these regions involve two small parameters, $|\boldsymbol{q}|$ and $|\boldsymbol{v}|$, 
which define the classical and non-relativistic limits respectively. The method
of regions instructs us to expand the integrands using the scaling corresponding to a given region, 
and then integrate over the whole space of loop momenta in dimensional regularisation. In actual 
practice we simply take the loop momenta to scale as $\ell \sim q$, thereby allowing us to perform the 
soft expansion of loop-level integrands which we truncate at the classical order following the 
considerations of \mysecref{sec:ClassicalLim}. Then, using the discussion of 
Ref.~\cite{Parra-Martinez:2020dzs} regarding the evaluation of integrals, we choose to evaluate all loop 
integrals in the \textit{potential region}, which isolates the \textit{conservative} dynamics, at least 
through the two-loop order. Crucially, this describes part of the non-dissipative dynamics, i.e.~the 
scenario where energy is \textit{not} lost from the system. If one wishes to account for radiative 
effects, then, alongside modifying the spanning cuts (as mentioned in \ref{sec:generalised_unitarity}), 
the integrals must be evaluated in the soft region~\cite{Herrmann:2021tct}. Since in this thesis we are 
only interested in the conservative dynamics, we restrict the components of the momentum transfer to obey
\begin{equation}
	 |\boldsymbol{q}| \gg q^0,
	 \label{eq:MomentumTransfer}
\end{equation}
so that the graviton mediates instantaneous long-range interactions~\cite{Parra-Martinez:2020dzs}. 
%--------------------------------------------------------------------------------
\subsection{Soft Expanding Propagators}
%--------------------------------------------------------------------------------
Recalling the nice parametrisation we chose for the external momentum in \ref{sec:KinematicSetUp}, we 
want to now discuss how to soft expand our loop integrands. This expansion is 
characterised by
\begin{equation}
	|\ell| \sim |q| = \sqrt{t} \ll |\bar{p}_i|, \ \bar{m}_i, \ \sqrt{s}\;,
\end{equation}
where $s = (p_1 + p_2)^2 = (p_3 + p_4)^2$ and $t = (p_1 + p_4)^2 = (p_2 + p_3)^2$ are the usual 
Mandelstam variables and $\ell$ stands for any combination of graviton momenta.

The massless graviton propagators take the form 
\begin{equation}
	\frac{1}{\ell^2}, \ \frac{1}{(\ell - q)^2}\;,
\end{equation}
such that they automatically have power counting $\mathcal{O}(1/|q|^2)$. Meanwhile, massive 
propagators are a sum of external momenta and internal momenta, so we need to expand them in the 
soft region~\cite{Parra-Martinez:2020dzs}
\begin{equation}
	\begin{split}
		\frac{1}{(\ell + \bar{p}_i \pm q/2)^2 - m_i^2} & = \frac{1}{\ell^2 + (\bar{p}_i \pm q/2)^2 + 2\bar{p}_i \cdot \ell \pm \ell \cdot q - m_i^2} \\
		& =  \frac{1}{2\bar{p}_i \cdot \ell} - \frac{\ell^2 \pm \ell \cdot q}{(2\bar{p}_i \cdot \ell)^2} +\cdots \;, 
	\end{split}
	\label{eq:MatterSoftExpansion}
\end{equation}
and thus all matter propagators are replaced by linearised propagators. Choosing to write 
this in terms of the velocity factors out the scale associated to $\bar{p}_i$ from the 
propagators
\begin{equation}
	\begin{split}
	 	\frac{1}{2\bar{p}_i \cdot \ell} & = \frac{1}{2\ell \cdot (\bar{m}_i u_i)} \\
	 		& = \frac{1}{2u_i \cdot \ell  \sqrt{m_i^2 - q^2/4}} \\
	 		& = \frac{1}{2 u_i \cdot \ell} \left( \frac{1}{m_i} +  \frac{q^2}{8m_i^3} +  \frac{3q^4}{128m_i^5} + \cdots \right)\;.
	 \end{split}
\end{equation}
This choice has the advantage that each order is homogeneous in $|q|$, by virtue of the absence of 
Lorentz-products between external and graviton momenta in the numerators. 

To summarise, in the soft region all the graviton propagators remain unexpanded, whereas the matter 
propagators have the form $1/(2 \bar{p}_i \cdot \ell)$ raised to higher powers beyond the leading 
order in the soft expansion. Therefore, at any loop order the power counting rules in the soft 
region is:
\begin{equation}
	\begin{split}
		& \text{graviton propagators} \sim 1/|q|^2, \\ 
		& \text{matter propagators} \sim 1/|q|, \\ 
		& \text{integration measure per loop} \sim |q|^4. \\ 
	\end{split}
\end{equation}
%

%--------------------------------------------------------------------------------
\subsection{Determining the $q$ Scaling of Diagrams}
%--------------------------------------------------------------------------------

These rules equip us with the knowledge to predict the $|q|$ scaling of 
the expected MIs before even constructing loop-level integrands via our generalised unitarity 
pipeline. To illustrate this, let us consider the tree-level, one-loop, and two-loop cases and 
analyse their explicit $|q|$ scalings, which we will do diagram-by-diagram.  

\textbf{Tree-level:} Since \myeqref{eq:EMLagrangian} and \myeqref{eq:L_scalar} do not contain 
interaction vertices directly coupling matter fields of different masses, the tree-level amplitude 
only receives a contribution from the $t$-channel exchange diagram. Indeed this is by construction, since 
an $s$- or $u$-channel diagram would constitute quantum corrections. As the relevant diagram goes 
as $1/t$, the $|q|$-scaling is trivially
\begin{equation}
	\scalebox{0.8}{\begin{tikzpicture}[baseline=(current bounding box.center)]
        \begin{feynman}[every blob={/tikz/fill=gray!30,/tikz/inner sep=2pt}]
          \vertex[draw, circle, minimum size=2pt, 
            inner sep=2pt, pattern=north east lines, 
            pattern color=black, fill=white, line width=0.6pt] (b1) at (0,1) {};
          \vertex[draw, circle, minimum size=2pt,
            inner sep=2pt, pattern=north east lines, 
            pattern color=black, fill=white, line width=0.6pt] (b2) at (0,-1) {};
          \vertex (l1) at (-2,1) {$p_1$};
          \vertex (l2) at (-2,-1) {$p_2$};
          \vertex (r1) at (2,1) {$p_4$};
          \vertex (r2) at (2,-1) {$p_3$};
          \diagram*{
            (b1) -- [draw=red, very thick] (l1),
      		(b1) -- [draw=red, very thick] (r1),
      		(b2) -- [draw=blue, very thick] (l2),
      		(b2) -- [draw=blue, very thick] (r2),
            (b1) -- [boson] (b2)
          };
        \end{feynman}
      \end{tikzpicture}
      }
      \sim \ 1/|q|^2\;.
\end{equation}

\textbf{One-loop:} At this order things become more interesting. By performing the soft expansion of 
\myeqref{eq:boxPropagators}, we find following scalings 
\begin{equation}
	\vcenter{
    \hbox{
      \scalebox{0.8}{\begin{tikzpicture}[baseline=(current bounding box.center)]
        \begin{feynman}[every blob={/tikz/fill=gray!30,/tikz/inner sep=2pt}]
          \vertex[draw, circle, minimum size=2pt, 
            inner sep=2pt, pattern=north east lines, 
            pattern color=black, fill=white, line width=0.6pt] (b1) at (-0.75,1) {};
          \vertex[draw, circle, minimum size=2pt, 
            inner sep=2pt, pattern=north east lines, 
            pattern color=black, fill=white, line width=0.6pt] (b2) at (-0.75,-1) {};
           \vertex[draw, circle, minimum size=2pt, 
            inner sep=2pt, pattern=north east lines, 
            pattern color=black, fill=white, line width=0.6pt] (b3) at (0.75,1) {};
         \vertex[draw, circle, minimum size=2pt, 
            inner sep=2pt, pattern=north east lines, 
            pattern color=black, fill=white, line width=0.6pt] (b4) at (0.75,-1) {};
          \vertex (l1) at (-2,1) {$p_1$};
          \vertex (l2) at (-2,-1) {$p_2$};
          \vertex (r1) at (2,1) {$p_4$};
          \vertex (r2) at (2,-1) {$p_3$};
          \diagram*{
            (b1) -- [draw=red, very thick] (l1),
      		(b3) -- [draw=red, very thick] (r1),
      		(b1) -- [draw=red, very thick] (b3),
      		(b2) -- [draw=blue, very thick] (l2),
      		(b4) -- [draw=blue, very thick] (r2),
      		(b4) -- [draw=blue, very thick] (b2),
      		(b2) -- [boson] (b1),
      		(b3) -- [boson] (b4),
          };
        \end{feynman}
      \end{tikzpicture}
      }
    }
  }
  ,
  \vcenter{
    \hbox{
      \scalebox{0.8}{\begin{tikzpicture}[baseline=(current bounding box.center)]
        \begin{feynman}[every blob={/tikz/fill=gray!30,/tikz/inner sep=2pt}]
          \vertex[draw, circle, minimum size=2pt, 
            inner sep=2pt, pattern=north east lines, 
            pattern color=black, fill=white, line width=0.6pt] (b1) at (-0.75,1) {};
          \vertex[draw, circle, minimum size=2pt, 
            inner sep=2pt, pattern=north east lines, 
            pattern color=black, fill=white, line width=0.6pt] (b2) at (-0.75,-1) {};
           \vertex[draw, circle, minimum size=2pt, 
            inner sep=2pt, pattern=north east lines, 
            pattern color=black, fill=white, line width=0.6pt] (b3) at (0.75,1) {};
         \vertex[draw, circle, minimum size=2pt, 
            inner sep=2pt, pattern=north east lines, 
            pattern color=black, fill=white, line width=0.6pt] (b4) at (0.75,-1) {};
          \vertex (l1) at (-2,1) {$p_1$};
          \vertex (l2) at (-2.35,-1) {$p_2$};
          \vertex (r1) at (2,1) {$p_4$};
          \vertex (r2) at (2,-1) {$p_3$};
          \diagram*{
            (b1) -- [draw=red, very thick] (l1),
      		(b3) -- [draw=red, very thick] (r1),
      		(b1) -- [draw=red, very thick] (b3),
      		(b2) -- [draw=blue, very thick] (l2),
      		(b4) -- [draw=blue, very thick] (r2),
      		(b4) -- [draw=blue, very thick] (b2),
      		(b1) -- [boson] (b4),
      		(b2) -- [boson] (b3),
          };
        \end{feynman}
      \end{tikzpicture}
      }
    }
  }
  \sim \ 1/|q|^2\;,
\end{equation}
and 

\begin{equation}
	\hspace{-0.2cm}
	\vcenter{
    \hbox{
      \scalebox{0.8}{\begin{tikzpicture}[baseline=(current bounding box.center)]
        \begin{feynman}[every blob={/tikz/fill=gray!30,/tikz/inner sep=2pt}]
          \vertex[draw, circle, minimum size=2pt, 
            inner sep=2pt, pattern=north east lines, 
            pattern color=black, fill=white, line width=0.6pt] (b1) at (-0.75,-1) {};
          \vertex[draw, circle, minimum size=2pt, 
            inner sep=2pt, pattern=north east lines, 
            pattern color=black, fill=white, line width=0.6pt] (b2) at (0,1) {};
           \vertex[draw, circle, minimum size=2pt, 
            inner sep=2pt, pattern=north east lines, 
            pattern color=black, fill=white, line width=0.6pt] (b3) at (0.75,-1) {};
          \vertex (l1) at (-2,1) {$p_1$};
          \vertex (l2) at (-2,-1) {$p_2$};
          \vertex (r1) at (2,1) {$p_4$};
          \vertex (r2) at (2,-1) {$p_3$};
          \diagram*{
            (b2) -- [draw=red, very thick] (l1),
      		(b2) -- [draw=red, very thick] (r1),
      		(b1) -- [draw=blue, very thick] (l2),
      		(b3) -- [draw=blue, very thick] (r2),
      		(b3) -- [draw=blue, very thick] (b1),
      		(b1) -- [boson] (b2),
      		(b3) -- [boson] (b2),
          };
        \end{feynman}
      \end{tikzpicture}
      }
    }
  }
  ,
  \vcenter{
    \hbox{
      \scalebox{0.8}{\begin{tikzpicture}[baseline=(current bounding box.center)]
        \begin{feynman}[every blob={/tikz/fill=gray!30,/tikz/inner sep=2pt}]
          \vertex[draw, circle, minimum size=2pt, 
            inner sep=2pt, pattern=north east lines, 
            pattern color=black, fill=white, line width=0.6pt] (b1) at (-0.75,1) {};
          \vertex[draw, circle, minimum size=2pt, 
            inner sep=2pt, pattern=north east lines, 
            pattern color=black, fill=white, line width=0.6pt] (b2) at (0,-1) {};
           \vertex[draw, circle, minimum size=2pt, 
            inner sep=2pt, pattern=north east lines, 
            pattern color=black, fill=white, line width=0.6pt] (b3) at (0.75,1) {};
          \vertex (l1) at (-2,1) {$p_1$};
          \vertex (l2) at (-2,-1) {$p_2$};
          \vertex (r1) at (2,1) {$p_4$};
          \vertex (r2) at (2,-1) {$p_3$};
          \diagram*{
            (b1) -- [draw=red, very thick] (l1),
      		(b3) -- [draw=red, very thick] (r1),
      		(b1) -- [draw=red, very thick] (b3),
      		(b2) -- [draw=blue, very thick] (l2),
      		(b2) -- [draw=blue, very thick] (r2),
      		(b4) -- [draw=blue, very thick] (b2),
      		(b1) -- [boson] (b2),
      		(b3) -- [boson] (b2),
          };
        \end{feynman}
      \end{tikzpicture}
      }
    }
  }
  \sim \ 1/|q|\;.
\end{equation}

Following the discussion of \mysecref{sec:ClassicalLim}, the triangle and inverted triangle 
scale classically, while the box and crossed box are superclassical since their 
scaling precedes the classical order in the soft expansion. 

\textbf{Two loops:} At this loop order the argument is exactly the same as before. Thus, we will not 
describe all diagrams explicitly, but we will provide three examples. The first of these 
will be the double-box diagram
\begin{equation}
	\vcenter{
    \hbox{
      \scalebox{0.8}{\begin{tikzpicture}[baseline=(current bounding box.center)]
        \begin{feynman}[every blob={/tikz/fill=gray!30,/tikz/inner sep=2pt}]
          \vertex[draw, circle, minimum size=2pt, 
            inner sep=2pt, pattern=north east lines, 
            pattern color=black, fill=white, line width=0.6pt] (b1) at (-1.5,1) {};
          \vertex[draw, circle, minimum size=2pt, 
            inner sep=2pt, pattern=north east lines, 
            pattern color=black, fill=white, line width=0.6pt] (b2) at (-1.5,-1) {};
          \vertex[draw, circle, minimum size=2pt, 
            inner sep=2pt, pattern=north east lines, 
            pattern color=black, fill=white, line width=0.6pt] (b3) at (0,1) {};
         \vertex[draw, circle, minimum size=2pt, 
            inner sep=2pt, pattern=north east lines, 
            pattern color=black, fill=white, line width=0.6pt] (b4) at (0,-1) {};
         \vertex[draw, circle, minimum size=2pt, 
            inner sep=2pt, pattern=north east lines, 
            pattern color=black, fill=white, line width=0.6pt] (b5) at (1.5,1) {};
         \vertex[draw, circle, minimum size=2pt, 
            inner sep=2pt, pattern=north east lines, 
            pattern color=black, fill=white, line width=0.6pt] (b6) at (1.5,-1) {};
          \vertex (l1) at (-3,1) {$p_1$};
          \vertex (l2) at (-3,-1) {$p_2$};
          \vertex (r1) at (3,1) {$p_4$};
          \vertex (r2) at (3,-1) {$p_3$};
          \diagram*{
            (b1) -- [draw=red, very thick] (l1),
            (b1) -- [draw=red, very thick] (b3),
            (b3) -- [draw=red, very thick] (b5),
      		(b5) -- [draw=red, very thick] (r1),
      		(b2) -- [draw=blue, very thick] (l2),
      		(b4) -- [draw=blue, very thick] (b2),
      		(b4) -- [draw=blue, very thick] (b6),
      		(b6) -- [draw=blue, very thick] (r2),
      		(b2) -- [boson] (b1),
      		(b3) -- [boson] (b4),
      		(b5) -- [boson] (b6),
          };
        \end{feynman}
      \end{tikzpicture}
      }
    }
  }
  \sim \ 1/|q|^2\;,
\end{equation}
the second is the IV diagram
\begin{equation}
	\vcenter{
    \hbox{
      \scalebox{0.8}{\begin{tikzpicture}[baseline=(current bounding box.center)]
        \begin{feynman}[every blob={/tikz/fill=gray!30,/tikz/inner sep=2pt}]
          \vertex[draw, circle, minimum size=2pt, 
            inner sep=2pt, pattern=north east lines, 
            pattern color=black, fill=white, line width=0.6pt] (b1) at (-1.5,1) {};
          \vertex[draw, circle, minimum size=2pt, 
            inner sep=2pt, pattern=north east lines, 
            pattern color=black, fill=white, line width=0.6pt] (b2) at (-1.5,-1) {};
          \vertex[draw, circle, minimum size=2pt, 
            inner sep=2pt, pattern=north east lines, 
            pattern color=black, fill=white, line width=0.6pt] (b3) at (0,1) {};
         \vertex[draw, circle, minimum size=2pt, 
            inner sep=2pt, pattern=north east lines, 
            pattern color=black, fill=white, line width=0.6pt] (b4) at (0.75,-1) {};
         \vertex[draw, circle, minimum size=2pt, 
            inner sep=2pt, pattern=north east lines, 
            pattern color=black, fill=white, line width=0.6pt] (b5) at (1.5,1) {};
          \vertex (l1) at (-3,1) {$p_1$};
          \vertex (l2) at (-3,-1) {$p_2$};
          \vertex (r1) at (3,1) {$p_4$};
          \vertex (r2) at (3,-1) {$p_3$};
          \diagram*{
            (b1) -- [draw=red, very thick] (l1),
            (b1) -- [draw=red, very thick] (b3),
            (b3) -- [draw=red, very thick] (b5),
      		(b5) -- [draw=red, very thick] (r1),
      		(b2) -- [draw=blue, very thick] (l2),
      		(b4) -- [draw=blue, very thick] (b2),
      		(b4) -- [draw=blue, very thick] (r2),
      		(b2) -- [boson] (b1),
      		(b3) -- [boson] (b4),
      		(b5) -- [boson] (b4),
          };
        \end{feynman}
      \end{tikzpicture}
      }
    }
  }
  \sim \ 1/|q|\;,
\end{equation}
and, finally, the third is the inverted double-triangle diagram
\begin{equation}
	\vcenter{
    \hbox{
      \scalebox{0.8}{\begin{tikzpicture}[baseline=(current bounding box.center)]
        \begin{feynman}[every blob={/tikz/fill=gray!30,/tikz/inner sep=2pt}]
          \vertex[draw, circle, minimum size=2pt, 
            inner sep=2pt, pattern=north east lines, 
            pattern color=black, fill=white, line width=0.6pt] (b1) at (-1.5,1) {};
          \vertex[draw, circle, minimum size=2pt, 
            inner sep=2pt, pattern=north east lines, 
            pattern color=black, fill=white, line width=0.6pt] (b2) at (0,-1) {};
          \vertex[draw, circle, minimum size=2pt, 
            inner sep=2pt, pattern=north east lines, 
            pattern color=black, fill=white, line width=0.6pt] (b3) at (0,1) {};
         \vertex[draw, circle, minimum size=2pt, 
            inner sep=2pt, pattern=north east lines, 
            pattern color=black, fill=white, line width=0.6pt] (b5) at (1.5,1) {};
          \vertex (l1) at (-3,1) {$p_1$};
          \vertex (l2) at (-3,-1) {$p_2$};
          \vertex (r1) at (3,1) {$p_4$};
          \vertex (r2) at (3,-1) {$p_3$};
          \diagram*{
            (b1) -- [draw=red, very thick] (l1),
            (b1) -- [draw=red, very thick] (b3),
            (b3) -- [draw=red, very thick] (b5),
      		(b5) -- [draw=red, very thick] (r1),
      		(b2) -- [draw=blue, very thick] (l2),
      		(b2) -- [draw=blue, very thick] (r2),
      		(b1) -- [boson] (b2),
      		(b3) -- [boson] (b2),
      		(b5) -- [boson] (b2),
          };
        \end{feynman}
      \end{tikzpicture}
      }
    }
  }
  \sim \ |q|^0\;. 
\end{equation}
From the two-loop order and beyond, the organisation of superclassical diagrams is  
not as simple as compared to one loop. Since at two loops the classical order is given by 
$\mathcal{O}(|q|^0)$, there are two orders that precede this: 
$\mathcal{O}(1/|q|^2)$ and $\mathcal{O}(1/|q|)$. They are both superclassical, but are organised 
in terms of even and odd powers of $|q|$~\cite{Parra-Martinez:2020dzs}. Thus, the double-box 
(even in $|q|$) and IV (odd in $|q|$) diagrams are superclassical, while the inverted 
double-triangle is classical. 	

%--------------------------------------------------------------------------------
\section{The Double Copy}
\label{sec:DoubleCopy}
%--------------------------------------------------------------------------------

In all of our discussions thus far, we have seen that tree-level amplitudes are the 
key ingredients in constructing loop-level amplitudes. Indeed, in gauge theory one can 
straightforwardly obtain tree-level amplitudes via Feynman rules. Whereas in gravity one 
would not call this endeavour ``simple", but rather complicated and 
tedious\footnote{This is because the number of interaction vertices 
that must be derived grow very rapidly as the multiplicity increases, alongside the fact that the 
expressions are generally more complicated because the graviton is a rank-$2$ field.}. Remarkably, 
there exists a correspondence between gauge theory and 
gravity---called the \textit{double copy}---that allows one to obtain amplitudes 
in gravity by somewhat ``squaring" amplitudes in gauge 
theory~\cite{Bern:2008qj,Bern:2010ue}
\begin{equation}
    M_{\text{gravity}} \sim A_{\text{gauge}} \times A_{\text{gauge}}\;.
\end{equation}
The first incarnation of this was in the context of tree-level string 
amplitudes~\cite{Kawai:1985xq}. The power of this correspondence lies in the fact that one can 
bypass ever deriving Feynman rules for gravitational theories and simply obtain gravity 
amplitudes from gauge theory amplitudes. This relation has been crucial for high-loop 
calculations of classical black hole scattering, underpinning recent 5PM results in
$\mathcal{N} = 8$ supergravity~\cite{Bern:2024adl,Bern:2025zno} and Einstein 
gravity~\cite{Bern:2025wyd}. 

There are a few different manifestations of the double copy. In this thesis we are interested 
in two strains of the double copy: the BCJ double copy and the KLT relations. In fact, the advantage of 
the double-copy will be two-fold. The first is that it will be particularly powerful in 
constructing tree-level amplitudes involving massive spinning legs up to spin $ s = 2$. The second is that 
it meshes extremely well with the generalised unitarity framework since it provides the necessary 
tree-level ingredients in a GWI form, reliant on the gauge theory amplitudes already being in this form. 

%The former is particularly useful at four-points, while the latter shines when interested in higher-point amplitudes.

%--------------------------------------------------------------------------------
\subsection{The BCJ Double Copy}
\label{sec:BCJDoubleCopy}
%--------------------------------------------------------------------------------

Let us consider the \textit{BCJ double copy}, which was first developed by Bern, Carrasco, and 
Johansson~\cite{Bern:2008qj}. To begin we will discuss the non-uniqueness of the 
kinematic numerators in an example followed by the general strategy for 
applying the BCJ double copy; the example is the tree-level four-gluon 
amplitude. Importantly, the gravity theory we obtain from this procedure includes the graviton 
alongside two additional fields, the B-field and the dilaton, which are removed because we 
are interested in Einstein gravity.

%--------------------------------------------------------------------------------
\subsubsection{Colour-Kinematics Duality: An Example}
\label{sec:NonUniquess4GluonAmplitude}
%--------------------------------------------------------------------------------
%
Let us consider the four-gluon tree amplitude. The general form of 
the amplitude is written as a sum over cubic graphs (with the coupling $g$ absorbed into 
$n_i$)
\begin{equation}
    A^{\mathrm{(0)}}_4(1,2,3,4) = \sum_{i \in \text{cubic}} \frac{c_i n_i}{\rho_i} = \frac{c_s n_s}{s}+ \frac{c_t n_t}{t} + \frac{c_u n_u}{u}\;,
    \label{eq:4GluonAmplitudeGen}
\end{equation}
%which is com of the diagrams shown in Fig.~\ref{fig:4GluonTreeAmplitudeContributions}, which
% 
where $\rho_i$ are inverse propagators, which can be written in terms of the 
Mandelstam variables $s,t,u$; $n_i$ are the kinematic numerators, which are functions of 
kinematic variables (i.e. momentum and polarisation); and $c_i$ are the colour factors, which 
are given by the structure constants
\begin{equation}
    c_s = f^{abc} f^{cde}\;, \ c_t = f^{a e c}f^{c b d}\;, \ c_u = f^{d a c}f^{c b e}\;,
\end{equation}
with $a,b,d,e$ being the colour indices associated to legs $1,2,3,4$. These structure 
constants satisfy the Jacobi identity: $c_s + c_t + c_u = 0$. By virtue of momentum conservation
$s+t+u = 0$, the following shifts in the colour factors~\cite{Brown:2016mrh}
\begin{equation}
    c_s\rightarrow c_s + \beta s\;, \ c_t\rightarrow c_t + \beta t\;, \ c_u\rightarrow c_u + \beta u\;,
    \label{eq:ColourShift}
\end{equation}
preserves the Jacobi identity for some arbitrary $\beta$. The statement that 
\myeqref{eq:4GluonAmplitudeGen} is invariant under this shift implies the kinematic Jacobi 
relation~\cite{Bern:2008qj, Brown:2016mrh}
\begin{equation}
    n_s + n_t + n_u = 0\;.
    \label{eq:KinJacobi}
\end{equation}
The existence of kinematic numerators that satisfy this identity is referred to as the 
\textit{colour-kinematic duality}. In fact, these kinematic numerators are not uniquely defined 
because the amplitude is invariant under transformations of the form
\begin{equation}
    n_s\rightarrow n_s + \beta s\;, \ n_t\rightarrow n_t + \beta t\;, \ n_u\rightarrow n_u + \beta u\;,
    \label{eq:GenGaugeTrans}
\end{equation}
again for some arbitrary $\beta$. To see this invariance explicitly, we 
substitute \myeqref{eq:GenGaugeTrans} into \myeqref{eq:4GluonAmplitudeGen} to find that 
\begin{equation}
    \delta A^{\mathrm{(0)}}_4(1,2,3,4) = \beta(c_s +  c_u + c_t) = 0\;,
    \label{eq:GaugeCondition}
\end{equation}
which vanishes due to the Jacobi identities. In the case of the four-gluon tree-level 
amplitude, one automatically finds a set of kinematic numerators that satisfies 
\myeqref{eq:KinJacobi}, after naively absorbing the contact term into the cubic numerators 
(their explicit forms will be revisited later). The consequence of the 
gauge dependence \myeqref{eq:GenGaugeTrans} is that the embedding of the contact term 
into the cubic numerators may be reshuffled~\cite{Bern:2008qj}. Thus, the BCJ procedure 
(see details below) can be readily applied to obtain the four-graviton tree amplitude. For 
$n$-gluon tree- and loop-level amplitudes, however, this is not always so trivial since one must 
look at each contributing diagram and decompose them into a sum of cubic diagrams. This is done 
by absorbing all four-point vertices into the cubic numerators, which one can do by multiplying 
and dividing by appropriate factors of inverse propagators~\cite{Brown:2016mrh,Bern:2019crd}. 
As discussed in Ref.~\cite{Bern:2008qj}, not all choices will give a set of numerators 
that satisfy \myeqref{eq:KinJacobi} such that the first problem faced by double copiers is 
finding numerators that satisfy the colour-kinematics duality. 

%--------------------------------------------------------------------------------
\subsubsection{A General Procedure}
\label{sec:DCProcedureOutline}
%--------------------------------------------------------------------------------
The general, but perturbative, statement of the BCJ double copy begins with an $n$-point and 
$L$-loop YM amplitude
\begin{equation}
    A_{n}^{(L)} = i^{L-1} g^{n-2+2L}\sum_{i\in \text{cubic}} \int \prod_{a = 1}^{L} \hat{\mathrm{d}}^4 \ell_a \frac{c_i n_i}{\prod_{\alpha_i}\rho_{\alpha_i}}\;,
\end{equation}
where $g$ is the YM coupling, $\rho_{\alpha_i}$ are the propagators, and all 
other variables are as defined before. One may identify triplets of colour factors that satisfy 
the Jacobi identity 
\begin{equation}
    c_i \pm c_j \pm c_k = 0\;,
    \label{eq:GenJacobiIdentity}
\end{equation}
where sign differences arise from permuting the colour indices on a given graph. Exploiting 
gauge redundancies one can always find a representation of kinematic numerators, which multiply the 
colour factors of the above triplets, that satisfy the colour-kinematics duality
\begin{equation}
    n_i \pm n_j \pm n_k = 0\;.
    \label{eq:GenKinematicJacobiIdentity}
\end{equation}
Triplets of kinematic numerators that satisfy this duality are called \textit{colour duals}, 
and amplitudes that have such kinematic numerators are said 
to be in the \textit{BCJ representation}. Needless to say, once one has successfully found 
the colour duals, the gravity amplitude is given by a simple replacement rule:
\begin{equation}
\begin{split}
    M_{n}^{(L)} & = A_{n}^{(L)} \rvert_{c_i \rightarrow n_i, \ g\rightarrow \frac{\kappa}{2}}\\
    & = i^{L-1} g^{n-2+2L}\sum_{i\in \text{cubic}} \int \prod_{a = 1}^{L} \hat{\mathrm{d}}^4 \ell_a \frac{c_i n_i}{\prod_{\alpha_i}\rho_{\alpha_i}} \biggr\rvert_{c_i \rightarrow n_i, \ g\rightarrow \frac{\kappa}{2}}\\
    & = i^{L-1} \left(\frac{\kappa}{2}\right)^{n-2+2L} \sum_{i\in \text{cubic}} \int \prod_{a = 1}^{L} \hat{\mathrm{d}}^4 \ell_a  \frac{n_i n_i}{\prod_{\alpha_i}\rho_{\alpha_i}}\;,
\end{split}
\end{equation}
where $\kappa = \sqrt{32\pi G}$ is the gravitational coupling and $G$ is Newton's gravitational 
constant. There are three important points to note regarding our discussion thus far. The 
first is that the double copy has been proven at only tree-level~\cite{Bern:2010yg}, and is 
conjectured to hold at the loop level (see e.g.~\cite{Bern:2012uf, Bern:2010ue}). The second is 
that, when replacing the colour factors with kinematic numerators, we may do so with a distinct set 
of numerators that satisfy the colour-kinematics duality. In other words, for a given combination 
$c_i n_i$ we can replace $c_i \rightarrow \tilde{n}_i$ where $\tilde{n}_i$ satisfies the 
colour-kinematics duality but can come from a different theory. The consequence is that the gravity 
theory that comes out depends on which numerators that we use, thereby leading to a whole zoo of 
interesting theories~\cite{Adamo:2022dcm}. Finally, the third is that we do not necessarily land 
on a gravity theory that only has the graviton, meaning that, if we are only interested in Einstein 
gravity, we should remove all other fields. 

This final point is better understood with an example. Let us consider the case in which 
both numerators come from pure non-supersymmetric YM. The outcome is GR coupled to a 
$2$-form, known as the B-field, and a scalar called the dilaton. More specifically, these modes are 
better seen by decomposing the tensor product into irreducible representations (irreps):
\begin{equation}
	A_\mu \otimes A_\nu = h_{(\mu \nu)} \oplus B_{[\mu \nu]} \oplus \phi\;,
	\label{eq:irreps_decomp}
\end{equation}
with the symmetric traceless, anti-symmetric, and scalar irreps understood to be the graviton, 
B-field, and dilaton respectively. In fact, these are the modes that one finds in the low 
energy limit of string theory. This is somewhat unsurprising as the tree-level double copy 
realised through the KLT relations~\cite{Kawai:1985xq} is understood to be a relation between 
open and closed strings. It is therefore crucial that we ensure that only graviton modes are 
kept in our amplitudes, as we are interested in Einstein gravity. 

Our application of the (BCJ or KLT) double copy will be at the level of tree-level amplitudes, 
which directly feeds into our generalised unitarity framework. Since we are interested in 
Einstein gravity, we must ensure that all modes other than the graviton are projected out in 
our generalised cuts. As the double copy involves the product of kinematic numerators, which 
are functions of the 
polarisation vectors $\epsilon^\mu$ and $\tilde{\epsilon}^\nu$, the polarisation tensor 
in the corresponding gravity theory appears as the product of these polarisation vectors: 
$\epsilon^{\mu \nu} = \epsilon^\mu \tilde{\epsilon}^\nu$. This general structure of the 
polarisation tensor contains the irrep modes described in \mysecref{eq:irreps_decomp}. If 
we wish to keep only the graviton modes then we simply need to ensure that the polarisation 
tensor is symmetric and traceless. This can be done via the replacement
\begin{equation}
	 \epsilon^\mu \tilde{\epsilon}^\nu \rightarrow P_h^{\mu \nu \alpha \beta}\epsilon_\alpha \tilde{\epsilon}_\beta\;.
	 \label{eq:eps_to_graviton_eps}
\end{equation} 
Since these amplitudes feed into generalised cuts, using the transverse projector 
\myeqref{eq:gravity_state_projector} (or \myeqref{eq:gravity_state_projector_simplified} if GWI is 
satisfied) when sewing amplitudes directly extracts the symmetric traceless modes; this 
is precisely the strategy used in state-of-the-art 
computations~\cite{Bern:2024adl,Bern:2025zno,Bern:2021dqo,Bern:2021yeh}. 

As an aside, if one is interested in a computation with external graviton states, then a 
trivial shortcut is to take $\tilde{\epsilon} = \epsilon$ whilst imposing the null 
condition of polarisation vectors. Of course, this can also be done for the internal graviton 
states appearing in generalised cuts, but one must nonetheless use 
$P_h^{\mu \nu \alpha \beta}$.

%--------------------------------------------------------------------------------
\subsection{The KLT Relations}
%--------------------------------------------------------------------------------

As previously mentioned, the KLT relations were the first realisation of the double copy, born 
from the observation that closed-string tree amplitudes factorise into products of 
open-string tree amplitudes~\cite{Kawai:1985xq}. Importantly, the convenience of the KLT relations is 
that it relies not on the existence of colour-kinematics duality, making it particularly 
powerful at higher multiplicity. As we have already gone into the technical details regarding the 
outcome of the double copy procedure, here we will simply note the relations for three-point, 
four-point, and five-point amplitudes\footnote{Keep in mind that the KLT relations produce the same 
result as the BCJ double copy.}. For our purposes, we require the KLT relations through five points
\begin{align}
	M_3^{\mathrm{(0)}}(1,2,3) & = i \tilde{A}_3^{\mathrm{(0)}}(1,2,3) \tilde{A}_3^{\mathrm{(0)}}(1,2,3)\;, \nonumber \\
	M_4^{\mathrm{(0)}}(1,2,3,4) & = - i s_{12} \tilde{A}_4^{\mathrm{(0)}}(1,2,3,4) \tilde{A}_4^{\mathrm{(0)}}(1,2,4,3)\;, \nonumber \\
	M_5^{\mathrm{(0)}}(1,2,3,4,5) & = i s_{12} s_{34} \tilde{A}_5^{\mathrm{(0)}}(1,2,3,4,5)\tilde{A}_5^{\mathrm{(0)}}(2,1,4,3,5) \\
	& \hspace{1cm} + i s_{13} s_{24} \tilde{A}_5^{\mathrm{(0)}}(1,3,2,4,5) \tilde{A}_5^{\mathrm{(0)}}(3,1,4,2,5)\;,\nonumber
\end{align}
with $\tilde{A}_n^{\mathrm{(0)}}$ being partial amplitudes whose colour factors and coupling 
constant are stripped off. Here we define $s_{ij} = (p_i + p_j)^2$ and we suppress the 
gravitational coupling $(\kappa/2)^{n-2}$ at $n$-points. These relations hold in any spacetime 
dimension and are formulated in terms of massless amplitudes. Despite this, we will 
find that these relations also hold for amplitudes involving massive spinning legs, following from 
dimensional reduction~\cite{Bern:2019crd} (see \mysecref{sec:Tree_amplitudes_via_dimensional_reduction}). For 
completeness, an extension of the KLT relations to any number of external legs can be found in 
Ref.~\cite{Bern:1998sv}.

%--------------------------------------------------------------------------------
\section{From Amplitudes to Classical Observables}
%--------------------------------------------------------------------------------

The final step in studying the dynamics of black hole scattering is the extraction of observables 
from the loop-level amplitudes we have constructed. A particularly convenient object that 
allows one to achieve this is the so-called radial action~\cite{Bern:2021dqo}. Its explicit relation to 
scattering amplitudes should be understood as conjectural, supported by
non-trivial consistency checks and explicit calculations rather than by a
general proof. Importantly, the 
radial action only encodes information about the conservative dynamics. It turns out 
that we can compute observables, such as the impulse and spin kick, for general 
non-aligned spin configurations from this radial action through the covariant Dirac bracket 
formalism proposed by Refs.~\cite{Gonzo:2024zxo, Alessio:2025flu} (see also 
Refs.~\cite{Kim:2024grz, Kim:2024svw, Kim:2025hpn} for a related Poisson bracket approach). This 
formalism anticipates that the radial action is explicitly written in terms of classical spin 
vectors. For now we will assume that the amplitude is already expressed in terms of classical spin 
vectors, but we will discuss in detail how to convert polarisation structures to classical spin 
structures in Chapter~\mysecref{sec:Chapter3}.

%--------------------------------------------------------------------------------
\subsection{The Radial Action}
\label{sec:RadialAction}
%--------------------------------------------------------------------------------

The essential task in the scattering-amplitude-based program for classical
physics is the extraction of the desired classical observables from 
scattering amplitudes. This is highly non-trivial since the amplitude itself is
not a classical object, as seen by the fact that even the tree-level amplitude
is divergent in the classical limit, i.e.~as $\hbar\to 0$. Conceptually it is most 
straightforward to directly target classical observables using the Kosower-Maybee-O'Connell 
(KMOC) formalism~\cite{Kosower:2018adc}. The reasoning for this is 
simple, yet powerful:~suitably defined quantum observables must have a smooth classical limit. 
The most important observable for conservative dynamics is arguably the \textit{impulse}, which defines the 
change of momentum during a scattering event between the asymptotic past and future. Then, two
important observables are obtained from projections of the impulse: the
centre-of-mass scattering angle\footnote{In general, multiple angles are needed to 
define the scattering event if the scattered particles are not confined 
to a scattering plane, e.g. when spin is involved.} and the total radiated energy, or, 
equivalently, the missing energy in the two-body system of the heavy point particles. Similarly, 
one can compute the total radiated angular momentum and displacement of the centre-of-mass during 
the scattering event. In a subsequent paper~\cite{Maybee:2019jus} another important
observable in the context of spinning bodies was pointed out. The
\textit{spin kick} is the change of the orientation (and perhaps 
magnitude~\cite{Bern:2023ity, Alaverdian:2025jtw}) 
of the spin of the individual particles. Despite the fact that the KMOC
formalism is conceptually straightforward, in practice it can become cumbersome
because of the different types of contributions that have to be assembled and
computed. 

For conservative spinless observables, a powerful method to directly extract
the classical observables from the amplitude, known as the
\textit{amplitude-action relation}, was put forward in
Ref.~\cite{Bern:2021dqo}. Most crucially, this has proven extremely efficient in
the extraction of the scattering angle for a variety of different problems up to the fourth
order in perturbation 
theory~\cite{Bern:2021dqo,Bern:2021yeh,Barack:2023oqp,Bern:2023ccb,Bern:2024adl}.
These relations were also found to remarkably hold when leading-order radiative
effects are included~\cite{Bern:2021xze}. Furthermore, in this thesis we will also find that it 
holds in the presence of spin. 

Let us review the amplitude-action relation for spinless particles \cite{Bern:2021dqo}; 
the relation is unaltered in the presence of spin. The exact statement reads
\begin{equation}
  \label{eq:aarelation}
           i \mathcal{M}(E,\boldsymbol{q})= \int_J e^{i \boldsymbol b
             \cdot \boldsymbol q} (e^{i I_r(E, J(\boldsymbol b))} - 1) \;,\quad \int_J := 4E |\boldsymbol p| \int d^{D-2} \boldsymbol b\;,
\end{equation}
with 
\begin{equation}
	\boldsymbol b^2 = -b^2 \equiv |b|^2\;, \quad \boldsymbol q^2 = -q^2 \equiv |q|^2\;, \quad J = |\boldsymbol{b}||\boldsymbol{p}|\;,
\end{equation}
where $\boldsymbol{b}$ and $\boldsymbol{q}$ denote the transverse spatial impact-parameter and 
momentum-transfer respectively, while $J(\boldsymbol{b})$ is the orbital angular momentum at past infinity. In addition, 
\begin{equation}
	|\boldsymbol{p}| = \frac{m_1 m_2 \sqrt{y^2-1}}{E}\;, \quad E = \sqrt{s} = \sqrt{m_1^2 + m_2^2 + 2m_1 m_2 y}\;,
\end{equation}
denote the three-momentum and total energy at past infinity. Here $I_r(E,J)$ 
is the radial action~\cite{Landau:1975pou}, which is defined via the integral of the radial 
momentum, $p_r$, along the scattering trajectory
\begin{equation}
  I_r(E,J)=\int p_r(E,J)\; \mathrm{d}r\;.
\end{equation}
Importantly, the radial action conveniently serves as a generating functional for scattering 
observables, such as the scattering angle
\begin{equation}
  \chi(E,J)=-\frac{\partial}{\partial J}I_r(E,J)\;.
  \label{eq:angleradaction}
\end{equation}
Expanding the exponential in \myeqref{eq:aarelation} perturbatively, we find iteration terms of the 
form (for $n \geq 2$)~\cite{Bern:2024adl}
\begin{equation}
	\int_J \frac{(i I_r(E,J))^n}{n!} = i \int_{\boldsymbol{\ell}}\frac{\tilde I_r (\boldsymbol{\ell}_1) \ldots \tilde I_r (\boldsymbol{\ell}_n)}{Z_1 \ldots Z_{n-1}}\;, 
\end{equation}
together with
\begin{equation}
	\tilde{I}_r = \int_J e^{i \boldsymbol b
             \cdot \boldsymbol q} I_r\;,
 \quad Z_i = -4E|\boldsymbol{p}| ((\boldsymbol{\ell}_1 + \boldsymbol{\ell}_2 + \ldots + \boldsymbol{\ell}_i) \cdot \hat{\boldsymbol{z}} + i0)\;,
\end{equation}
where $\hat{\boldsymbol{z}} = \boldsymbol{p}/|\boldsymbol{p}|$ denotes the unit vector in the 
direction of the three-momentum in the centre-of-mass frame. Furthermore, the integration measure
reads
\begin{equation}
	\int_{\boldsymbol{\ell}} := \mu^{2(n-1)\epsilon} \int \left(\prod_{i=1}^{n} \hat{\mathrm{d}}^{D-1} \boldsymbol{\ell}_i\right) \hat{\delta}^{(D-1)}\left(\sum_{j = 1}^{n} \boldsymbol{\ell}_j - \boldsymbol{q}\right)\;.
\end{equation}
Crucially, the dimensional regularisation scale $\mu$ is included to 
regulate all loop integrals because of divergences that appear at the three-loop level and beyond; see 
Refs.~\cite{Bern:2021dqo, Bern:2021yeh, Bern:2024adl, Bern:2025zno, Bern:2025wyd} for more details. For 
our purposes, $\mu$ will play no important role as we limit our discussion to the two-loop order.

In the context of perturbation theory, it is useful to perform this matching order by order, starting 
from abstract expansions in powers of the coupling constant
\begin{equation}
  \label{eq:pexpansions}
  \mathcal{M}=\sum_{k} \mathcal{M}_k\;,\hskip 1 cm  {\tilde{I}}_r=\sum_{k}{\tilde{I}}_{r,k}\;,
\end{equation}
where $\mathcal{M}_k$ and $\tilde{I}_{r,k}$ are of $k$-th order in the coupling constant (in the case
at hand either the fine-structure constant $\alpha$ or Newton's constant $G$).  
The first few orders of \eqref{eq:aarelation} read
\begin{align}
  \mathcal{M}_{1}={}&{\tilde{I}}_{r,1}\;,\\
  \label{eq:aarelationPert2}
  \mathcal{M}_{2}={}&{\tilde{I}}_{r,2}+\int_{\boldsymbol{\ell}}\frac{{\tilde{I}}_{r,1}{\tilde{I}}_{r,1}}{Z_1}\;,\\
  \label{eq:aarelationPert3}
  \mathcal{M}_{3}={}&{\tilde{I}}_{r,3}+\int_{\boldsymbol{\ell}}\frac{{\tilde{I}}_{r,1}{\tilde{I}}_{r,2}}{Z_1}+\int_{\boldsymbol{\ell}}\frac{{\tilde{I}}_{r,1}{\tilde{I}}_{r,1}{\tilde{I}}_{r,1}}{Z_1Z_2}\;,
\end{align}
where the sum over permutations is understood to be implicit; for example:
\begin{equation}
	{\tilde{I}}_{r,1}{\tilde{I}}_{r,2} := {\tilde{I}}_{r,1}(\boldsymbol{\ell}_1){\tilde{I}}_{r,2}(\boldsymbol{\ell}_2) + {\tilde{I}}_{r,1}(\boldsymbol{\ell}_2){\tilde{I}}_{r,2}(\boldsymbol{\ell}_1)\;.
\end{equation}
The exponentiation of the radial action is manifest via iteration integrals
involving products of lower-order terms alongside
linearised three-dimensional propagators $Z_i$. Structurally, these relations are very similar to the 
relations one would find, for example, when matching to an effective field theory
\cite{Neill:2013wsa,Cheung:2018wkq}. The amplitude is equal to some remainder
$\tilde{I}_r$ plus terms originating from iterations of lower-order objects.
Importantly, the main difference here is that, in terms of the amplitude-action relation, we have a
direct way to identify the part of the amplitude that must cancel with
the iteration terms on the right-hand side of \eqref{eq:aarelationPert2}
and \eqref{eq:aarelationPert3}.

More specifically, when evaluating the necessary Feynman integrals to compute the
amplitude $\mathcal{M}_n$ within perturbation theory, we will generally find
integrals in which, after evaluating all energy integrals, we are left with
integrals that have $Z$-poles. The extremely crucial point is 
that the lower-order iterations of the radial action precisely correspond to these superclassical 
divergences, which precede the classical order, such that the radial action is simply the finite part 
of loop-level amplitudes. In practice, after soft expansion, we isolate the radial action through two 
loops by simply dropping the double- and triple-iteration ladder-like master integrals and keeping the 
real part of the amplitude. This is how the radial action was extracted in the spinless case in 
Ref.~\cite{Bern:2021xze} as well as in the spinning case in 
Refs.~\cite{Akpinar:2024meg, Akpinar:2025bkt}. 

%--------------------------------------------------------------------------------
\subsection{Covariant Dirac Brackets}
\label{sec:CovariantDiracBrackets}
%--------------------------------------------------------------------------------

As mentioned, one can obtain observables like the scattering angle via a simple derivative of the 
radial action with respect to the total angular momentum $J$. For the conservative scattering of 
spinless point particles, the impulse and scattering angle are directly related; see 
Appendix~\ref{sec:ScatteringAngleAndImpulse} for more details. In the case of spinning particles, 
this is no longer the case because the scattering process is not confined to a single plane; meaning
that several angles parametrise the scattering. For certain aligned-spin configurations 
we find the same relation as in the spinless case, since the scattering is 
confined to the equatorial plane. This, however, does not encode the general non-aligned spin dynamics, 
which is crucial if we wish to obtain the full spinning dynamics. 

A remarkably simple, and yet powerful, method for computing observables, such as the impulse and spin 
kick, was proposed in Ref.~\cite{Gonzo:2024zxo}. This approach makes use of covariant Dirac brackets. 
In this section we will derive these brackets which, together with the radial action, are used to 
compute non-aligned spin observables. Although in this thesis we will consider the dynamics of systems 
with only one spinning point particle, we derive the brackets for the cases 
of two spinning point particles following Ref.~\cite{Gonzo:2024zxo, Hanson:1974qy}. 

To this end, let us consider two spinning point particles in the asymptotic past, where they follow straight 
line trajectories 
\begin{equation}
    x_i^\mu(\tau_i) = u_i^\mu \tau_i + b_i^\mu\;, \quad i = 1,2\;,
\end{equation}
where $\tau_i$ is the proper time along the trajectory and $b_i$ is the initial trajectory. Here $u_i$ 
denotes the particle four-velocity and the spin dynamics are parametrised by the spin tensor
$\mathcal{S}_i^{\mu \nu}$. It is important to note that, since we have extracted the classical 
amplitude to obtain the radial action, here we can safely ignore quantum corrections, for 
example, in the difference between $\bar{m}_i$ and $m_i$. Therefore, when extracting observables, 
we always choose to work with $m_i$. The \textit{kinematic phase space} for the 
classical gravitational scattering of these two spinning bodies is given by\footnote{Here $2\mathrm{B}$ 
stands for two-body.}
\begin{equation}
	\mathcal{H}_{\mathrm{2B}} = \{ u_1^\mu, u_2^\mu, a_1^{\mu}, a_2^{\mu}, b^\mu\}\;,
	\label{eq:FullPhaseSpaceVars}
\end{equation}
where the normalised spin vectors $a_i^\mu = S_i^\mu/m_i$ are related to the spin tensor via
\begin{equation}
	\mathcal{S}_i^{\mu \nu} = m_i \epsilon^{\mu \nu \rho \sigma} u_{i, \rho} a_{i, \sigma}\;,
\end{equation}
and $b^\mu = b_2^\mu - b_1^\mu$ defines the initial impact parameter. The non-vanishing Poisson 
brackets are\footnote{The sign on $\{ b^\mu, u_i^\nu\}_{\text{P.B.}}$ is tied to the metric 
convention, i.e.~the negative signature.}
\begin{equation}
\begin{split}
	& \hspace{2cm} \{ b_i^\mu, u_i^\nu\}_{\text{P.B.}} = -\eta^{\mu \nu}/m_i\;, \\
	& \{\mathcal{S}_i^{\mu \nu}, \mathcal{S}_i^{\alpha \beta} \}_{\text{P.B.}} = \mathcal{S}_i^{\mu \alpha} \eta^{\nu \beta} - \mathcal{S}_i^{\mu \beta} \eta^{\nu \alpha} - (\mu \leftrightarrow \nu) \;.
\end{split}
\end{equation}
Since the classical scattering process is subject to constraints, the dynamics live on a 
subspace of the full kinematic phase space, known as the \textit{physical phase 
space}. These Poisson brackets are defined on the full kinematic phase space, such that we wish to 
define a set of brackets that acts solely on this physical phase space; these are the so-called 
Dirac brackets. Therefore, by imposing all constraints on the system, we will promote these Poisson 
brackets to Dirac brackets. In total, there are four constraints that must be imposed. There are 
two spin-related constraints, namely the spin-supplementary condition (SSC) and a proper-frame 
gauge condition,
\begin{align}
    (1)\quad & V^\mu = u_{i\nu}\,\mathcal{S}_i^{\mu\nu} = 0\;, \\
    (2)\quad & \chi^\mu = \Lambda_{i,0}^{~\mu} - u_i^\mu = 0\;,
    \label{eq:SpinConstraints}
\end{align}
where $\Lambda_{i,0}^{~\mu}$ denotes the timelike component of the body-fixed frame (see 
Ref.~\cite{Hanson:1974qy} for details). The first constraint fixes the representative centre-of-mass 
worldline of the spinning body. Since the antisymmetric spin tensor $\mathcal{S}^{\mu\nu}$ has six 
independent components while physical spin only carries three rotational degrees of freedom, this 
condition removes the unphysical components associated with a mass dipole moment relative to the 
chosen worldline. The second constraint fixes the timelike 
part of the body-fixed tetrad by aligning it with the particle's four-velocity. In other words, 
this removes the boost freedom of the local Lorentz frame and leaves only spatial rotations as the 
remaining physical rotational d.o.f. describing the orientation of the body. 

We define the second class constraints associated with these constraints as 
\begin{equation}
    \phi^\alpha = \{ V^0, V^1, V^2, V^3, \chi^0, \chi^1, \chi^2, \chi^3\}\;,
    \label{eq:SpinConstraints_Components}
\end{equation}
and implement them at the level of brackets to define the first iteration of Dirac 
brackets\footnote{This is not the final set of Dirac brackets yet as we have not imposed all 
constraints.}
\begin{equation}
    \{a,b\}' = \{a,b\}_{\text{P.B.}} - \{a,\phi^\alpha\}_{\text{P.B.}} C_{\alpha \beta}^{-1} \{\phi^\beta,b\}_{\text{P.B.}}\;,
\end{equation}
where $C^{\alpha \beta} = \{\phi^\alpha, \phi^\beta\}_{\text{P.B.}}$. The non-zero Poisson bracket 
for $\Lambda^{\mu \nu}$ reads
\begin{equation}
	\{\Lambda^{\mu \nu}, \mathcal{S}_i^{\alpha \beta}\}_{\text{P.B.}} = \Lambda^{\mu \alpha}\mathcal{S}_i^{\nu \beta} - \Lambda^{\mu \beta} \mathcal{S}_i^{\alpha \nu}\;.
\end{equation}
Then, given the non-zero brackets between 
\begin{align}
    \{V^\mu, V^\nu\}_{\text{P.B.}} & = \mathcal{S}^{\mu \nu}_i\;, \\
    \{V^\mu, \chi^\nu\}_{\text{P.B.}} & = - \eta^{\mu \nu}\;, \\
    \{\chi^\mu, \chi^\nu\}_{\text{P.B.}} & = 0\;,
\end{align}
we find $C^{\alpha \beta}$ and $C_{\alpha \beta}^{-1}$ to be
\begin{equation}
    C^{\alpha \beta} = \begin{pmatrix}
        \mathcal{S}^{\mu \nu}_i & \eta^{\mu \nu} \\
        - \eta^{\mu \nu}  & 0 
    \end{pmatrix}\;, \qquad 
    C^{-1}_{\alpha \beta} = \begin{pmatrix}
         0 & -\eta^{\mu \nu} \\
        \eta^{\mu \nu}  & \mathcal{S}^{\mu \nu}_i 
    \end{pmatrix}\;.
\end{equation}
By determining the non-zero Poisson brackets involving \myeqref{eq:FullPhaseSpaceVars} and 
\myeqref{eq:SpinConstraints_Components}
\begin{align}
    \{b_i^\mu, V^\nu\}_{\text{P.B.}} & = -\mathcal{S}^{\mu \nu}_i/m_i\;, \\
    \{b_i^\mu, \chi^\nu\}_{\text{P.B.}} & = \eta^{\mu \nu}/m_i\;, \\
    \{\mathcal{S}_i^{\mu \nu}, V^\beta\}_{\text{P.B.}} & = u_i^\mu \mathcal{S}_i^{\nu \beta} - u_i^\nu \mathcal{S}_i^{\mu \beta}\;, \\
    \{\mathcal{S}_i^{\mu \nu}, \chi^\beta\}_{\text{P.B.}} & = u_i^\nu \eta^{\mu \beta} - u_i^\mu \eta^{\nu \beta}\;,
\end{align}
it follows that the non-zero first iteration of Dirac brackets are
\begin{align}
    \{b^\mu, u_i^\nu \}' & = -\operatorname{sgn}_i \eta^{\mu \nu}/m_i\;, \\
    \{b^\mu, b^\nu \}' & = \mathcal{S}^{\mu \nu}_1/m_1^2 + \mathcal{S}^{\mu \nu}_2/m_2^2\;, \\
    \{b^\mu, \mathcal{S}_i^{\nu \rho} \}' & =  \operatorname{sgn}_i (u_i^\rho \mathcal{S}_i^{\mu \nu}-u_i^\nu \mathcal{S}_i^{\mu \rho} )/m_i\;, \\
    \{\mathcal{S}_i^{\mu \nu}, \mathcal{S}_i^{\rho \sigma} \}' & = \mathcal{S}^{\mu \rho}_i (\eta^{\nu \sigma} + u_i^\nu u_i^\sigma) - \mathcal{S}^{\mu \sigma}_i (\eta^{\nu \rho} + u_i^\nu u_i^\rho) \\
    & \hspace{2cm} - \mathcal{S}^{\nu \rho}_i (\eta^{\mu \sigma} + u_i^\mu u_i^\sigma) + \mathcal{S}^{\nu \sigma}_i (\eta^{\mu \rho} + u_i^\mu u_i^\rho) \nonumber\;,
\end{align}
together with $b^\mu = b^\mu_2 - b^\mu_1$ and $\operatorname{sgn}_i \in \{ -1,1 \}$ 
for $i \in \{1,2\}$. There are four more constraints that enforce the on-shell 
and transversality conditions
\begin{equation}
    \begin{split}
        & \phi_1 = b \cdot u_1 = 0\;, \qquad \phi_2 = u_1^2 = 1\;, \\
        & \phi_3 = b \cdot u_2 = 0\;, \qquad \phi_4 = u_2^2 = 1\;,
    \end{split}
\end{equation}
which allows us to define the Dirac brackets that live on the physical subspace
\begin{equation}
    \{a,b\}_{\text{D.B.}}  = \{a,b\}'- \{a,\phi_i\}' K_{i j}^{-1} \{\phi_j,b\}'\;,
\end{equation}
where $K^{ij} = \{\phi_i, \phi_j\}'$. Following the same steps as before, the matrix is
\begin{equation}
    K^{ij} = \begin{pmatrix}
    0 & 2/m_1 & 0 & -2y/m_2 \\
    -2/m_1 & 0 & -2y/m_1 & 0 \\
    0 & 2y/m_1 & 0 & -2/m_2 \\ 
    2y/m_2 & 0 & 2/m_2 & 0 
    \end{pmatrix}\;, 
\end{equation}
and its inverse reads
\begin{equation}
    (K^{-1})_{ij} = \frac{1}{2(y^2-1)}\begin{pmatrix}
        0 & m_1 & 0 & ym_2 \\
        -m_1 & 0 & ym_1 & 0 \\
        0 & -ym_1 & 0 & -m_2 \\ 
        -ym_2 & 0 & m_2 & 0 
        \end{pmatrix}\;,
\end{equation}
Let us start with the products involving $b^\mu$
\begin{align}
    & \{b^\mu, \phi_1\}' = \mathcal{S}^{\mu \nu}_2 u_{1\nu}/m_2^2 + b^\mu/m_1\;, \\
    & \{b^\mu, \phi_2\}' = 2u_1^\mu/m_1\;, \\
    & \{b^\mu, \phi_3\}' = \mathcal{S}^{\mu \nu}_1 u_{2\nu}/m_1^2 - b^\mu/m_2\;, \\
    & \{b^\mu, \phi_4\}' = -2u_2^\mu/m_2\;.
\end{align}
Then we look at the products involving $u_1$
\begin{align}
    & \{u_1^\mu, \phi_1\}' = -u_1^\mu/m_1\;, \\
    & \{u_1^\mu, \phi_2\}' = 0\;, \\
    & \{u_1^\mu, \phi_3\}' = -u_2^\mu/m_1\;, \\
    & \{u_1^\mu, \phi_4\}' = 0\;,
\end{align}
and $u_2$
\begin{align}
    & \{u_2^\mu, \phi_1\}' = u_1^\mu/m_2\;, \\
    & \{u_2^\mu, \phi_2\}' = 0\;, \\
    & \{u_2^\mu, \phi_3\}' = u_2^\mu/m_2\;, \\
    & \{u_2^\mu, \phi_4\}' = 0\;.
\end{align}
Finally, we also need the brackets involving the spin tensors
\begin{align}
    & \{\mathcal{S}_i^{\mu \nu}, \phi_1\}' =  \operatorname{sgn}_i (u_i^\nu \mathcal{S}^{\rho \mu}_i- u_i^\mu \mathcal{S}^{\rho \nu}_i ) u_{1 \rho}/m_i\;, \\
    & \{\mathcal{S}_i^{\mu \nu}, \phi_2\}' = 0\;, \\
    & \{\mathcal{S}_i^{\mu \nu}, \phi_3\}' =  \operatorname{sgn}_i (u_i^\nu \mathcal{S}^{\rho \mu}_i-u_i^\mu \mathcal{S}^{\rho \nu}_i ) u_{2 \rho}/m_i\;, \\
    & \{\mathcal{S}_i^{\mu \nu}, \phi_4\}' = 0\;.
\end{align}
In this way, the non-vanishing Dirac brackets are 
\begin{eqnarray}
&&\hspace{-0.4cm}\{b^\mu, u_i^\nu\}_{\text{D.B.}} = -\operatorname{sgn}_i \frac{(y^2-1)b^\mu b^\nu + l^\mu l^\nu}{m_i b^2 (y^2-1)}\;, \\
&&\hspace{-0.4cm}\{b^\mu, a_i^\nu\}_{\text{D.B.}} = \operatorname{sgn}_i \frac{u_i^\nu ((y^2-1)b^\mu(b\cdot a_i) + l^\mu(l \cdot a_i))}{m_i b^2 (y^2-1)}\;, \\
&&\hspace{-0.4cm}\{b^\mu,b^\nu\}_{\text{D.B.}} = \frac{(b^2(y u_2^\mu - u_1^\mu) - l^\mu(a_1 \cdot u_2))b^\nu - (\mu \leftrightarrow \nu)}{m_1b^2(y^2-1)}\\
&& \hspace{6cm} + (1\leftrightarrow 2, b \rightarrow -b)\;,  \nonumber\\
&&\hspace{-0.4cm}\{a^\mu_i, a^\nu_i\}_{\text{D.B.}} = \frac{\epsilon^{\mu \nu}_{~~\rho \sigma} u_i^\rho a_i^\sigma}{m_i}\;,
\label{eq:FundamentalDBs}
\end{eqnarray}
where we define
\begin{eqnarray}
	\quad l^\mu = \epsilon^{\mu}_{~ \nu \rho \sigma} b^\nu u_1^\rho u_2^\sigma\;.
\end{eqnarray} 
Note that these brackets have sign differences as compared to Ref.~\cite{Gonzo:2024zxo}, which is a 
direct consequence of using the opposite metric signature alongside a different sign in defining the 
spin vector. One should marvel at how simple these brackets are. 

The real power of these brackets becomes apparent when computing observables. Given a radial action, 
$I_r$, we may compute the change in an observable via~\cite{Gonzo:2024zxo}
\begin{equation}
    \Delta \lambda^\mu = \sum_{n=0}^{\infty} \frac{1}{n!}\{\underbrace{I_r,\{I_r,\ldots,\{I_r}_{n \ \text{times}},\lambda^\mu\}_{\text{D.B.}}\ldots\}_{\text{D.B.}}\}_{\text{D.B.}}\;,
    \label{eq:obsFormula}
\end{equation}
where  $\lambda^\mu \in \{u_1^\mu, u_2^\mu, a_1^\mu, a_2^\mu\}$. Then by expanding the radial action in 
the coupling constant, the observable can be computed order-by-order in perturbation theory. The iterated brackets involving lower-order pieces of the radial action play precisely the same 
role as the contributions in the KMOC formalism that come from the amplitude with a two-particle cut. 

In general, some observable can be projected into a basis formed by 
$\{u_1^\mu, u_2^\mu, b^\mu, l^\mu\}$ such that
\begin{equation}
	\Delta \lambda^\mu = (\Delta \lambda^\mu)_b \frac{b^\mu}{|b|} + (\Delta \lambda^\mu)_{u_1} u_1^\mu + (\Delta \lambda^\mu)_{u_2} u_2^\mu + (\Delta \lambda^\mu)_\ell \frac{l^\mu}{|l|}\;,
	\label{eq:ObservableBasis}
\end{equation}
where $|l| = |b|\sqrt{y^2-1}$. Here $b, l$ and $u_1, u_2$ define the transverse and 
longitudinal directions relative to the momentum of each particle, respectively. Furthermore, since we 
are focused on conservative dynamics, there are certain constraints that the observables must obey. 
Specifically, conservation of momentum
\begin{equation}
	(u_i + \Delta u_i)^2 - u_i^2 = 2 u_i \cdot \Delta u_i + (\Delta u_i)^2 = 0\;,
\end{equation}
yields the relation
\begin{equation}
	2 u_i \cdot \Delta u_i = (\Delta u_i)^2\;.
	\label{eq:consv_mon_constraint}
\end{equation}
Then, expanding $\Delta u_i$ in the coupling as
\begin{equation}
	\Delta u_i^\mu = \sum_{n} g^{n+1}\Delta u_{i}^{\mu, (n)}\;, \quad g \in \{\alpha, G\}\;,
\end{equation}
shows that the longitudinal parts of $\Delta u_i$ are completely fixed by contributions from lower 
orders in the coupling. For example:
\begin{align}
	& u_i \cdot \Delta u_{i}^{(0)} = 0\;, \\
	& 2 u_i \cdot \Delta u_{i}^{(1)} = - (\Delta u_{i}^{(0)})^2\;, \\
	& u_i \cdot \Delta u_{i}^{(2)} = - \Delta u_{i}^{(0)} \cdot \Delta u_{i}^{(1)}\;,
\end{align}
and so on. Another useful constraint comes from the SSC condition
\begin{equation}
	(u_i + \Delta u_i) \cdot (a_i + \Delta a_i) - u_i \cdot a_i = 0\;,
\end{equation}
which yields
\begin{equation}
	u_i \cdot \Delta a_i + \Delta u_i \cdot a_i + \Delta u_i \cdot \Delta a_i = 0\;.
	\label{eq:u_dot_a_constraint}
\end{equation}

%--------------------------------------------------------------------------------
\subsection{From Commutators to Dirac Brackets}
\label{sec:CommutatorsToCovariantDiracBrackets}
%--------------------------------------------------------------------------------

Consider the variation of an operator $\mathcal{Q}(u_i,a_i,b)$ within the KMOC 
formalism~\cite{Kosower:2018adc}
\begin{align}
 \Delta \hat{\mathcal{Q}} &= \hat{S}^{\dagger} \hat{\mathcal{Q}} \hat{S} - \hat{\mathcal{Q}} = \hat{S}^{\dagger}[\hat{\mathcal{Q}}, \hat{S}]\;, 
 \label{eq:operator_var}
\end{align}
where $S$ is the usual S-matrix that evolves quantum states. Using the 
exponential representation of the S-matrix~\cite{Damgaard:2021ipf,Damgaard:2023ttc}
\begin{equation}
    \hat{S} = \exp\left(i \hat{N}/\hbar\right)\,,
\end{equation}
\myeqref{eq:operator_var} takes the form
\begin{align}
 \Delta \hat{\mathcal{Q}} = \sum_{n = 1}^{\infty} \frac{(-i)^n}{\hbar^n n!} \underbrace{[\hat{N},[\hat{N}, \ldots,[\hat{N}}_{n \text{ times}}, \hat{\mathcal{Q}}]]]\,.
 \label{eq:operator_var_exp}
\end{align}
Within the KMOC formalism, the initial two-particle state are prepared as well-
separated localised wavepackets 
\begin{equation}
	|\psi \rangle_{\mathrm{in}} := \int \mathrm{d}\Phi(p_1,p_2) e^{i (p_1 \cdot b_1 + p_2 \cdot b_2)} \phi(p_1)\phi(p_2)|p_1, p_2 \rangle\;,
\end{equation}
where $\phi(p_i)$ are appropriately normalised wavefunctions 
\begin{equation}
	\int \mathrm{d}\Phi(p_i)|\phi(p_i)| = 1\;,
\end{equation}
peaked around the classical momentum. Restricting ourselves to the conservative 
scattering process, one defines the radial action via the Fourier transform of the elastic four-point 
matrix element of the $\hat{N}$-operator in the classical limit
\begin{align}
 \hspace{-15pt}I_r(\{u_i,a_i,b\}) =  \int  \hat{\mathrm{d}}^{D} q \,\hat{\delta}\left(2 \bar{p}_1 \cdot q\right) \hat{\delta}\left(2 \bar{p}_2 \cdot q\right) \mathrm{e}^{i q \cdot b / \hbar} \langle p_4 p_3 |\hat{N}|p_1 p_2\rangle \Big|_{\hbar \to 0}\;,
 \label{eq:radial_act_def}
\end{align}
as proven rigorously in the spinless case~\cite{Damgaard:2023ttc}. By retracing the connection between 
the radial action and the amplitude defined from the usual expansion $\hat{S} = 1 + i \hat{T}$, we 
arrive at the amplitude-action relation defined in \myeqref{eq:aarelation}. Importantly, the covariant 
Dirac bracket formalism~\cite{Gonzo:2024zxo} that we derived provides the 
classical limit of commutators in the physical phase space $\mathcal{H}_{\rm 2B}$~\cite{Akpinar:2025tct}
\begin{align}
    [f(\cdot),g(\cdot)] \stackrel{\hbar \to 0}{\to} i \hbar \{f(\cdot),g(\cdot)\}_{\text{D.B.}}\;,
\end{align}
so that the total variation of the expectation value $\langle \Delta \hat{\mathcal{Q}} \rangle$ 
\eqref{eq:operator_var} in the scattering process becomes, after inserting the two-body phase space 
completeness relation and Fourier transforming to impact parameter space,
\begin{align}
 \langle \Delta \hat{\mathcal{Q}} \rangle \Big|_{\mathrm{cons}} &= \sum_{n = 1}^{\infty} \frac{1}{ n!} \underbrace{\{I_r,\{I_r, \ldots,\{I_r}_{n \text{ times}}, \mathcal{Q}\}_{\text{D.B.}}\}_{\text{D.B.}}\}_{\text{D.B.}}\,.
 \label{eq:operator_var_final}
\end{align}
This is precisely the same relation as \myeqref{eq:obsFormula}.
%--------------------------------------------------------------------------------
\chapter{Spinning Black Hole Scattering at Quadratic Order in Spin}
\label{sec:Chapter3}
%--------------------------------------------------------------------------------

Having reviewed the extensive machinery required to perform the computations presented in this 
thesis, we now (finally!) turn to the study of black hole scattering dynamics. This 
chapter will focus on the results of Ref.~\cite{Akpinar:2024meg}. To begin, we will show 
how one can obtain tree-level amplitudes involving massive point particles using a dimensional 
reduction of massless legs. Following this, we discuss a form factor decomposition that 
encodes the spin degrees of freedom in tensor structures that obey parity 
invariance and crossing symmetry. Then we explain how to obtain classical spin vectors from 
polarisation vectors through an expectation value of the Pauli-Lubanski vector. Naively translating 
polarisation structures into spin vector structures will yield incorrect results because of an 
apparent ambiguity associated to finite-spin representations. To resolve this ambiguity, and 
thereby faithfully extract the spinning dynamics, we utilise the \textit{spin interpolation} method 
to obtain the finite part of the classical amplitude through quadratic order in spin and 3PM. 
Remarkably, we observe that the two-loop amplitude exhibits a spin-shift symmetry---previously 
observed at one-loop---for the probe structures. Finally, from the finite part of the classical 
amplitude, we obtain the radial action and compute the aligned-spin scattering angle.  

Importantly, we will show that the spin interpolation amounts to computing the 
difference---an interpolation---between the spin-diagonal parts of the scalar-scalar and 
scalar-vector scattering amplitudes, by including their quantum corrections. As we go to higher 
orders in spin, the spin interpolation no longer constitutes a simple difference between 
spin-diagonal terms because off-diagonal contributions play an essential role. Therefore, we will 
develop a more sophisticated strategy for performing the spin interpolation, making it 
applicable to higher orders in the spin expansion (see Chapter~\mysecref{sec:Chapter4} for details). 

%--------------------------------------------------------------------------------
\section{Tree-Level Amplitudes via Dimensional Reduction}
\label{sec:Tree_amplitudes_via_dimensional_reduction}
%--------------------------------------------------------------------------------

To begin, let us discuss a systematic way of obtaining tree-level amplitudes that involve massive 
scalar or vector legs from tree-level gluon (or gravity) amplitudes. For simplicity, we 
will consider starting with gluon amplitudes, which we will use to obtain four-point and five-point 
amplitudes involving massive scalar or massive vector legs via a \textit{dimensional 
reduction}~\cite{Bern:2019crd}. 

%--------------------------------------------------------------------------------
\subsection{Gluon Amplitudes}
\label{sec:Gluon_amplitudes}
%--------------------------------------------------------------------------------

As the dimensional reduction is exactly the same at $n$-points, we will (appropriately) 
restrict explicit discussions to the four-point case. 

%--------------------------------------------------------------------------------
\subsubsection*{Four-Point Gluon Amplitude}
%--------------------------------------------------------------------------------

The four-gluon tree-level amplitude is a standard result that can be found in most standard 
textbooks (see for example Ref.~\cite{Schwartz:2014sze}). The diagrams that contribute to this  
amplitude can be found in Fig.~\ref{fig:4GluonTreeAmplitudeContributions}. The total amplitude is 
given as\footnote{See Appendix~\ref{sec:YMFeynmanRules} for the YM Feynman rules.}
\begin{equation}
    A^{\mathrm{(0)}}_4(1,2,3,4) = g^2\left(\frac{c_s n_s}{s} + \frac{c_u n_u}{u} + \frac{c_t n_t}{t}\right)\;,
    \label{eq:4GluonTreeAmplitude}
\end{equation}
where we have absorbed the four-point contribution to find kinematic numerators
\begin{equation}
\begin{split}
    & n_s = N_s - \frac{i}{2}s\left[(\epsilon_1 \cdot \epsilon_4)(\epsilon_2 \cdot \epsilon_3) - (\epsilon_1 \cdot \epsilon_3)(\epsilon_2 \cdot \epsilon_4)\right]\;,\\
    & n_u = N_u - \frac{i}{2}u\left[(\epsilon_1 \cdot \epsilon_4)(\epsilon_2 \cdot \epsilon_3) - (\epsilon_1 \cdot \epsilon_2)(\epsilon_3 \cdot \epsilon_4)\right]\;,\\
    & n_t = N_t - \frac{i}{2}t\left[(\epsilon_1 \cdot \epsilon_2)(\epsilon_3 \cdot \epsilon_4) - (\epsilon_1 \cdot \epsilon_3)(\epsilon_2 \cdot \epsilon_4)\right]\;,
\end{split}
\label{eq:4GluonKinematicNumerators}
\end{equation}
and the explicit forms of $N_s, N_u, N_t$ are omitted here. In particular, given $N_s$ one can determine 
$N_u$ and $N_t$ via crossing symmetry. In deriving these expressions, we have used the transversality 
property of polarisation vectors, the explicit form of the Mandelstam variables
\begin{equation}
\begin{split}
    & s = (k_1 + k_2)^2 = (k_3 + k_4)^2\;, \\
    & u = (k_1 + k_3)^2 = (k_2 + k_4)^2\;, \\
    & t = (k_1 + k_4)^2 = (k_2 + k_3)^2\;, 
\end{split}
\end{equation}
and defined the colour factors as
\begin{equation}
    c_s = \tilde{f}^{a_1 a_2 b}\tilde{f}^{b a_3 a_4}, \ c_t = \tilde{f}^{a_4 a_1 b}\tilde{f}^{b a_2 a_3}, \ c_u = \tilde{f}^{a_3 a_1 b}\tilde{f}^{b a_4 a_2},
    \label{eq:ColourStructure}
\end{equation}
which satisfy the Jacobi identity $c_s - c_u - c_t = 0$.
\begin{figure}[h!]
\centering
\hspace{-0.7cm}
%-------------------- (s) --------------------
\begin{minipage}{0.22\textwidth}
\centering
\begin{tikzpicture}[baseline=(current bounding box.center)]
  \begin{feynman}
    \vertex (v1) at (-0.5,0.5);
    \vertex (v2) at (0.5,0.5);
    \vertex (l1) at (-1.5,0.5) {$k_1$};
    \vertex (l2) at (-0.5,-1) {$k_2$};
    \vertex (r1) at (1.5,0.5) {$k_4$};
    \vertex (r2) at (0.5,-1) {$k_3$};

    \diagram*{
      (v1) -- [gluon] (l1),
      (l2) -- [gluon] (v1),
      (r1) -- [gluon] (v2),
      (r2) -- [gluon] (v2),
      (v2) -- [gluon] (v1),
    };
  \end{feynman}
\end{tikzpicture}

\vspace{0.5em}
{\small (s)}
{\small\itshape s-channel}
\end{minipage}
\hfill
%-------------------- (u) --------------------
\begin{minipage}{0.22\textwidth}
\centering
\begin{tikzpicture}[baseline=(current bounding box.center)]
  \begin{feynman}
    \vertex (v1) at (-0.5,0.5);
    \vertex (v2) at (0.5,0.5);
    \vertex (l1) at (-1.5,0.5) {$k_1$};
    \vertex (l2) at (-0.5,-1) {$k_3$};
    \vertex (r1) at (1.5,0.5) {$k_4$};
    \vertex (r2) at (0.5,-1) {$k_2$};

    \diagram*{
      (v1) -- [gluon] (l1),
      (l2) -- [gluon] (v1),
      (r1) -- [gluon] (v2),
      (r2) -- [gluon] (v2),
      (v2) -- [gluon] (v1),
    };
  \end{feynman}
\end{tikzpicture}

\vspace{0.5em}
{\small (u)}
{\small\itshape u-channel}
\end{minipage}
\hfill
%-------------------- (t) --------------------
\begin{minipage}{0.22\textwidth}
\centering
\begin{tikzpicture}[baseline=(current bounding box.center)]
  \begin{feynman}
    \vertex (v1) at (0,0.5);
    \vertex (v2) at (0,-1);
    \vertex (l1) at (-1.5,0.5) {$k_1$};
    \vertex (l2) at (-1.5,-1) {$k_2$};
    \vertex (r1) at (1.5,0.5) {$k_4$};
    \vertex (r2) at (1.5,-1) {$k_3$};

    \diagram*{
      (v1) -- [gluon] (l1),
      (l2) -- [gluon] (v2),
      (r1) -- [gluon] (v1),
      (v2) -- [gluon] (r2),
      (v2) -- [gluon] (v1),
    };
  \end{feynman}
\end{tikzpicture}

\vspace{0.5em}
{\small (t)}
{\small\itshape t-channel}
\end{minipage}
\hfill
%-------------------- (f) --------------------
\begin{minipage}{0.22\textwidth}
\centering
\begin{tikzpicture}[baseline=(current bounding box.center)]
  \begin{feynman}
    \vertex (v1) at (0,0.5);
    \vertex (l1) at (-1.5,0.5) {$k_1$};
    \vertex (l2) at (-1,-1) {$k_2$};
    \vertex (r1) at (1.5,0.5) {$k_4$};
    \vertex (r2) at (1,-1) {$k_3$};

    \diagram*{
      (v1) -- [gluon] (l1),
      (l2) -- [gluon] (v1),
      (r1) -- [gluon] (v1),
      (v1) -- [gluon] (r2),
    };
  \end{feynman}
\end{tikzpicture}

\vspace{0.5em}
{\small (f)}
{\small\itshape contact}
\end{minipage}

\caption{The explicit Feynman diagrams that contribute to the tree-level four-gluon amplitude shown in \myeqref{eq:4GluonTreeAmplitude}. Here each diagram is labelled by its corresponding Mandelstam variable, whereas $\mathrm{(f)}$ labels the four-point contact vertex. Importantly, writing the amplitude in terms of cubic vertices involves embedding $\mathrm{(f)}$ into $\mathrm{(s)}$, $\mathrm{(u)}$, and $\mathrm{(t)}$. These diagrams were obtained with the Feynman rules in Appendix~\mysecref{sec:YMFeynmanRules}.}
\label{fig:4GluonTreeAmplitudeContributions}
\end{figure}

%--------------------------------------------------------------------------------
\subsubsection*{Five-Point Gluon Amplitude (and Beyond)}
%--------------------------------------------------------------------------------

Since the five-point (and more generally the $n$-point) tree-level gluon amplitude is much 
more involved to derive using Feynman rules, we will take a nice shortcut. The massless YM 
amplitudes required are conveniently taken from the software {\tt IncreasingTrees} developed by Edison 
and Teng~\cite{Edison:2020ehu}, which spits out tree-level gluon and
graviton amplitudes to high multiplicity. Importantly, the output from {\tt
IncreasingTrees} is organised in terms of the distinct diagrams that contribute
to the five-point amplitude, making it particularly useful for our purposes. 

%--------------------------------------------------------------------------------
\subsection{Massive Scalars from Massless States}
\label{sec:Dimensional_reduction_to_scalar_legs}
%--------------------------------------------------------------------------------

To obtain the amplitude involving two massive scalar legs and two gluon legs, we perform a dimensional 
reduction of a $(D+2)$-dimensional YM theory to find a $D$-dimensional theory with a 
massive scalar~\cite{Bern:2019crd}. Indeed, the dimensional reduction essentially compactifies the extra 
dimensions where masses and polarisations can be hidden. In order to illustrate this, let us 
consider a pure YM theory living in $(D+2)$-dimensions and let us convert gluon legs $1$ 
and $4$ to massive scalars. Since we wish to ``reduce" these legs to scalar legs, we should ``hide" the 
polarisations in the compactified dimensions by choosing the polarisation vectors as 
\begin{equation}
    \epsilon_1^A = \epsilon_4^A = (0,0,\cdots,0,1)\;. 
\end{equation}
Note that we use capital Latin letters to denote indices in $(D+2)$-dimensional space, while Greek letters 
denote indices in the $D$-dimensional subspace. At the same time, we define their momenta to 
be\footnote{The metric signature here is $\text{diag}(+,-,\cdots,-)$ such that the sign choice for the 
mass entry depends on momentum conservation.}
\begin{equation}
    k_1^A = (p_1^\mu, m, 0)\;, \quad k_4^A = (p_4^\mu, -m, 0)\;.
\end{equation}
From the perspective of the $D$-dimensional subspace legs these are massive scalars 
with mass $m$, whereas from the perspective of the $(D+2)$-dimensional space they are massless 
\begin{equation}
    k_1 \bullet k_1 = p_1^2 - m^2 = 0\;, \quad k_4 \bullet k_4 = p_4^2 - m^2 = 0\;.
\end{equation}
Here $\bullet$ and $\cdot$ denote dot products in the $(D+2)$-dimensional and $D$-dimensional subspace 
respectively\footnote{Example: $k_1 \bullet k_1 = k_1^A k_{1A}$ and $p_1 \cdot p_1 = p_1^2 = p_1^\mu p_{1\mu}$.}. 
The polarisation vectors and momenta of the remaining legs live in the $D$-dimensional subspace 
defined by the first $D$ entries, so
\begin{equation}
    \begin{split}
        & k_2^A = (k_2^\mu,0,0) , \ k_3^A = (k_3^\mu,0,0), \\
        & \epsilon_2^A = (\epsilon_2^\mu,0,0) , \ \epsilon_3^A = (\epsilon_3^\mu,0,0)\;.
    \end{split}
\end{equation}
Then, with these choices, we find the following dot products
\begin{equation}
\begin{split}
    & \epsilon_2 \bullet \epsilon_{3} = \epsilon_2\cdot \epsilon_3, \ \epsilon_1 \bullet \epsilon_{4} = -1\;, \\ 
    & \epsilon_1 \bullet \epsilon_{i} = \epsilon_4 \bullet \epsilon_{i} = 0, \ \text{for} \ i = 2,3 \;, \\
    & k_{1}\bullet \epsilon_1 = k_{4}\bullet \epsilon_1 = k_{1}\bullet \epsilon_4 = k_{4}\bullet \epsilon_4 =0\;, \\
    & k_{2}\bullet \epsilon_1 = k_{3}\bullet \epsilon_1 = k_{2}\bullet \epsilon_4 = k_{3}\bullet \epsilon_4 = 0\;.
    \label{eq:DimensionalReductionIdentities}
\end{split}
\end{equation}
Now we may implement dimensional reduction to find the tree-level amplitude for the scattering of two 
scalars with two gluons. Since we have crossing symmetry, we only consider the $s$-channel 
contribution which will generate the contribution from the other channels. Concretely, taking $n_s$ in 
\eqref{eq:4GluonKinematicNumerators} and applying the identities  
\eqref{eq:DimensionalReductionIdentities} we find
\begin{equation}
\begin{split}
    n_s & = n(1^\text{s},2,3,4^\text{s})\\
    & = \frac{i}{2}\Big\{ (s-m^2) (\epsilon_2 \cdot \epsilon_3) - 2(p_1\cdot \epsilon_2)(p_{12} \cdot \epsilon_3) - 2(p_{34}\cdot \epsilon_2)(p_4 \cdot \epsilon_3)\Big\}\;,
\end{split}
\end{equation}
where $p_{ij} = p_i + p_j$\footnote{Note that the superscript s denotes massive \textit{scalar} particles.}. 
In order to make the crossing symmetry between the $s$- and $u$-channels manifest, we have split the term associated 
to the cubic diagram into two and rewritten it using momentum conservation. We have also used 
\begin{equation}
\begin{split}
    & s' = (k_1 + k_2)^2 = 2p_1\cdot k_2\;, \\
    & s = (p_1 + k_2)^2 = m^2 + 2p_1\cdot k_2\;,
\end{split}
\end{equation}
to get $s' = s-m^2$. The $u$-channel kinematic numerator follows by crossing labels $2$ and $3$ 
\begin{equation}
\begin{split}
    n_u & = n(1^\text{s},3,2,4^\text{s}) \\
    & = \frac{i}{2}\Big\{ (u-m^2) (\epsilon_2 \cdot \epsilon_3) - 2(p_1\cdot \epsilon_3)(p_{13} \cdot \epsilon_2) - 2(p_{24}\cdot \epsilon_3)(p_4 \cdot \epsilon_2)\Big\}\;,
\end{split}
\end{equation}
and, finally, the $t$-channel kinematic numerator follows from the kinematic Jacobi 
identity\footnote{This has been 
confirmed by starting from the kinematic numerators in \eqref{eq:4GluonKinematicNumerators} and applying 
the identities from \eqref{eq:DimensionalReductionIdentities}.}
\begin{equation}
    n_t = n_s - n_u\;.
\end{equation}
Thus, the tree amplitude for the scattering of two massive scalars with two gluons is
\begin{equation}
    A^{\mathrm{(0)}}_{4}(1^\text{s},2,3,4^\text{s}) = g^2\left(\frac{c_s n_s}{s-m^2}+ \frac{c_t n_t}{t} + \frac{c_u n_u}{u-m^2} \right)\;,
    \label{eq:2Gluon2scalarAmplitude}
\end{equation}
which is conveniently in the BCJ representation, since the kinematic numerators satisfy the 
colour-kinematic duality. 

%--------------------------------------------------------------------------------
\subsection{Massive Vectors from Massless States}
\label{sec:Dimensional_reduction_to_vector_legs}
%--------------------------------------------------------------------------------

To dimensionally reduce two gluon legs to two massive vector legs, the discussion is almost identical. In 
this case, we choose the polarisation vectors of the legs we wish to make massive to be
\begin{equation}
    \epsilon_1^A = (\varepsilon_1^\mu,0,0), \ \epsilon_4^A = (\varepsilon_4^{\mu},0,0)\;,
\end{equation}
with all other momenta and polarisations chosen as before. Then, implementing the dimensional reduction 
boils down to imposing
\begin{equation}
    s' = s - m^2, \ k_1 \bullet k_4 = p_1 \cdot p_4 + m^2\;.
\end{equation}
In this way, the tree amplitude for the scattering of two massive vectors with two gluons is 
\begin{equation}
    A^{(0)}_{4}(1^\text{v},2,3,4^\text{v}) = g^2\left(\frac{c_s n_s}{s-m^2}+ \frac{c_t n_t}{t} + \frac{c_u n_u}{u-m^2} \right),
    \label{eq:2Gluon2vectorAmplitude}
\end{equation}
where we omit the explicit form of the kinematic numerators. The superscript v denotes 
massive \textit{vector} particles.

%--------------------------------------------------------------------------------
\subsection{YM to Electrodynamics and Gravity}
%--------------------------------------------------------------------------------

Having obtained the four-point and five-point tree-level amplitudes involving massive scalars or 
massive vectors in gauge theory, we may acquire the corresponding amplitudes in 
electrodynamics and gravity as follows. 

\textbf{Electrodynamics.} Take the Abelian limit by choosing the colour indices to take 
specific values, such that all colour factors $f^{abc}$ are $-1,0$ or $1$. More specifically:
\begin{itemize}
	\item \textbf{Four-point:} Legs $1$ and $4$ are massive such that we choose $a_1 = 1$, $a_4 = 2$ and $a_2 = a_3 = 3$.
	\item \textbf{Five-point:} Legs $1$ and $5$ are massive such that we choose $a_1 = 1$, $a_5 = 2$ and $a_2 = a_3 = a_4 = 3$.
\end{itemize}
The resulting amplitude has two legs corresponding to massive charged (scalar or vector) particles while 
all remaining legs are photons.

\textbf{Gravity.} Use the double copy (BCJ or KLT). For gravity amplitudes involving massive scalars, both gauge theory copies in the procedure should involve massive scalars. For gravity amplitudes involving massive vectors, one gauge copy should involve a massive scalar while the other should involve a massive vector.

It is important to note that this procedure also works at the level of pure graviton amplitudes. In this 
case, we should first perform the double copy of the pure gluon amplitude and then perform a dimensional 
reduction. The upside of this is that graviton legs can directly be converted into massive scalar, 
massive vector, or even massive tensor legs. 

%--------------------------------------------------------------------------------
\section{A Form Factor Decomposition for Spin-1}
\label{sec:Form_factor_decomp}
%--------------------------------------------------------------------------------

The presence of polarisation vectors in the $2\to 2$ scattering amplitude for a pair of massive spin-$0$ 
and spin-$s$ particles introduces additional complexity, especially at loop level. This is evident in 
any attempt to perform IBP reduction, as one must define additional propagators---called irreducible 
scalar products (ISPs)---to handle the reduction of Lorentz products between polarisation vectors and loop 
momentum. As a matter of fact, difficulties posed by spinning d.o.f. can be 
easily circumvented by decomposing the amplitude into a basis of linearly independent tensor 
structures~\cite{FebresCordero:2022jts}. 

In this chapter we focus on the $2\to 2$ scattering amplitude for a pair of massive 
spin-$0$ and spin-$1$ particles\footnote{We also compute the scattering amplitude for a pair of massive 
spin-$0$ particles, but a form factor decomposition is not needed in this case.}
\begin{equation}
	V(-p_1, \varepsilon_1) + \phi(-p_2) \rightarrow \phi(p_3) + V(p_4, \varepsilon^\star_4)\;,
\end{equation}
so let us perform the form factor decomposition in this case. We start by 
stripping off the polarisation vectors\footnote{We use subscripts to denote what scattering amplitude we 
are considering. In this case it is the $2\to 2$ scattering amplitude between a massive vector and massive 
scalar.}
\begin{equation}
	M_{\phi \tiny{\mbox{V}}} = \varepsilon^\star_{4 \mu} T^{\mu \nu}(\bar{p}_1, \bar{p}_2, q)\varepsilon_{1\nu}\;,
\end{equation}
where in general $T^{\mu \nu}$ is decomposed as
\begin{equation}
	T^{\mu \nu}(\bar{p}_1, \bar{p}_2, q) = A \eta^{\mu \nu} + \sum_{i,j = 1}^{3} B_{ij} k^\mu_i k^\nu_j\;, \quad k_i \in \{\bar{p}_1, \bar{p}_2, q\}\;,
\end{equation}
which has $10$ structures in total. If we impose transversality of the polarisations,
$p_1 \cdot \varepsilon_1 = p_4 \cdot \varepsilon_4^\star = 0$, we can eliminate all $\bar{p}_1$
using our symmetric parameterisation \eqref{eq:KinematicSetUp}, such that $k_i$ can only be $\bar{p}_2$ 
and $q$. This reduces the total number of possible structures to $5$, from which we construct the 
linearly independent structures~\cite{FebresCordero:2022jts, Akpinar:2024meg}	
\begin{equation}
\Big\{T_1^{\mu \nu},\ldots,T_5^{\mu \nu}\Big\} = \Big\{ \eta^{\mu \nu},\; q^\mu q^\nu,\;
 q^2 \bar{p}_2^\mu \bar{p}_2^\nu,\; 2\,q_{\phantom{2}}^{[\mu} \bar{p}_2^{\nu]},\;2\,q_{\phantom{2}}^{(\mu} \bar{p}_2^{\nu)}\Big\}\;,
 \label{eq:Spin1TensorStructures}
\end{equation}
thereby forming a basis set for the amplitude
\begin{equation}
    M_{\phi \tiny{\mbox{V}}} = \sum_{n=1}^5 \mathbb{M}_n T_n\;, \qquad  T_n = \varepsilon^\star_{4\mu} T_n^{\mu \nu} \varepsilon_{1\nu}\;.
    \label{eq:tensorBasis}
  \end{equation}
In fact, the convenience of this decomposition is that the entire spin dependence is
encapsulated in the structures $T_n$ for any theory, while the coefficients $\mathbb{M}_n$---the 
\textit{form factors}---are theory-dependent quantities. The former point is what makes 
this decomposition particularly useful, as we will relate polarisation structures to spin vector structures.
Note that we are working with parity-invariant theories such as Einstein gravity and electrodynamics, and 
therefore do not include Levi-Civita tensors. Moreover, the coefficient of $T_5$, which exists in our loop 
integrand, vanishes upon explicit integration, which is expected because $T_5$ does not obey the crossing symmetry of 
the scattering amplitude.

In the case of analytic computations, we obtain the form factors according to the following 
procedure\footnote{In Ref.~\cite{Akpinar:2024meg} amplitudes were also constructed numerically, for which 
the strategy for obtaining the form factors is slightly different.}. Suppressing the subscript of 
$M_{\phi \tiny{\mbox{V}} }$, we multiply the matrix element with a conjugated tensor and sum over all 
polarisations
\begin{equation}
  \sum_{\text{spin}} M T_m^\dagger = \sum_{n=1}^5 \mathbb{M}_n \sum_{\text{spin}}T_n T^\dagger_m \equiv \sum_{n=1}^5 \mathbb{M}_n \mathcal{P}_{nm}\;,
\end{equation}
where we have defined the projection matrix
\begin{equation}
  \mathcal{P}_{nm} = \sum_{\text{spin}}\left(\varepsilon_{4\mu}^\star T_n^{\mu \nu} \varepsilon_{1\nu}\right) 
  \left(\varepsilon^\star_{1\alpha}T_m^{\dagger\alpha\beta}\varepsilon_{4\beta} \right) 
  = P_{4\mu\beta} T_n^{\mu \nu}P_{1\nu\alpha}T_m^{\dagger \alpha \beta} 
  = \textrm{Tr}\left[P_4 T_n P_1 T_m^\dagger\right]\;,
\end{equation}
with $P_i^{\mu\nu}$ being the polarisation sum \eqref{eq:spin1-transverse-proj} for the vector particle 
legs $p_1$ and $p_4$ both with mass $m_1$. Then,  
\begin{equation}
    \sum_{n=1}^5 \mathbb{M}_n \mathcal{P}_{nm} =  \sum_{n=1}^5 \mathbb{M}_n P_{4\mu\beta} T_n^{\mu \nu}P_{1\nu\alpha}T_m^{\dagger \alpha \beta} 
    = P_{4\mu\beta} M^{\mu \nu}P_{1\nu\alpha}T_m^{\dagger \alpha \beta} 
    = \textrm{Tr}\left[P_4 M P_1 T_m^\dagger\right]\;,
\end{equation}
where the right-hand side defines a vector that corresponds to the form factor
coefficients once acted on with the inverse projection matrix
\begin{equation}
   \mathbb{M}_k = \sum_{n,m=1}^5 \mathbb{M}_n\mathcal{P}_{nm}\left(\mathcal{P}^{-1}\right)_{mk} = \sum_{m=1}^5 \textrm{Tr}\left[P_4 M P_1 T_m^\dagger\right]\left(\mathcal{P}^{-1}\right)_{mk}\;.
\end{equation}
In explicit computations, we have always used a $5 \times 5$ projection matrix 
instead of a $4 \times 4$ projection matrix to explicitly check the vanishing of the coefficient of $T_5$
after integration, rather than making the assumption a priori. 

%--------------------------------------------------------------------------------
\section{Classical Spin from Quantum Field Theory}
\label{sec:Classical_spin}
%--------------------------------------------------------------------------------

The $ 2 \rightarrow 2$ amplitudes that we compute include polarisation 
vectors corresponding to each external spin-$1$ massive particle. 
Here we discuss how one determines spin-dependent quantities
from these polarised amplitudes. 

%--------------------------------------------------------------------------------
\subsection{Extracting Spin Dependence}
\label{sec:ExtractingSpin}
%--------------------------------------------------------------------------------
The initial step towards making the spin-dependence manifest in amplitudes is to
understand what spin means from a quantum mechanical point of view. The proposed
quantum mechanical interpretation of the classical spin vector is simply the
expectation value of the Pauli-Lubanski operator~\cite{Maybee:2019jus}
\begin{equation}
   \mathds{W}^\mu = \frac{1}{2}\epsilon^{\mu \nu \rho \sigma} \mathds{P}_\nu \mathds{J}_{\rho \sigma}\;,
\end{equation}
where $\mathds{P}^\nu$, $\mathds{J}^{\rho \sigma}$ are the translation and
Lorentz generators, respectively. Explicitly, the spin operator is defined as
\begin{equation}
     \mathds{S}^\mu \equiv \frac{\mathds{W}^\mu}{m}\;,
\end{equation}
where $m$ is the mass of the spinning particle. The generators fulfil the Lorentz algebra
\begin{equation}
\begin{split}
  & [\mathds{J}^{\mu \nu},\mathds{P}^\rho] = i (\eta^{\mu \rho}\mathds{P}^\nu - \eta^{\nu \rho}\mathds{P}^\mu)\;, \\
  & [\mathds{J}^{\mu \nu},\mathds{J}^{\rho \sigma}] = i(\eta^{\nu \rho}\mathds{J}^{\mu \sigma} - \eta^{\mu \rho}\mathds{J}^{\nu \sigma} - \eta^{\nu \sigma}\mathds{J}^{\mu \rho} + \eta^{\mu \sigma}\mathds{J}^{\nu \rho})\;,
\end{split}
\end{equation}
and one finds that 
\begin{equation}
 [\mathds{P}^\mu, \mathds{W}^\nu] = 0\;, \qquad 
 [\mathds{J}^{\mu \nu},\mathds{W}^\rho] = i (\eta^{\mu \rho}\mathds{W}^\nu - \eta^{\nu \rho}\mathds{W}^\mu)\;,
\end{equation}
where the latter is expected because $\mathds{W}^\mu$ is a vector operator. 
One can further determine that
\begin{equation}
    [\mathds{W}^\mu, \mathds{W}^\nu] = -i \epsilon^{\mu \nu \rho \sigma}\mathds{W}_\rho \mathds{P}_\sigma\;.
\end{equation}
In the rest frame of the massive particle we find $\mathds{W}^0 = 0$, which
implies that the remaining generators satisfy 
\begin{equation}
    [\mathds{W}^i, \mathds{W}^j] = -i \epsilon^{i j k }\mathds{W}_k\;.
\end{equation}
The important observation is that these operators are precisely the 
generators of the little group, which serves to justify the chosen interpretation.

Now the goal is to define a spin vector in terms of polarisation
vectors. In order to do this, we consider a massive spin-$1$ particle with
momentum $\bar p_1^{\,\prime \mu} = \bar p_1^{\,\mu}(m_1/\bar m_1)$, which
will be taken as the central value of the momentum in the wave packet
we consider within the KMOC formalism. For readers familiar with
heavy-particle effective theory \cite{Aoude:2020onz, Aoude:2022trd},
the central momentum is analogous to the large label momentum in 
such formalism. The spin vector, as an operator in spin states $i, j$ of the 
little-group representation, takes the form~\cite{Maybee:2019jus,Chung:2019duq}\footnote{See 
Appendix~\mysecref{sec:ExpectationValuePauliLubanskiVector} (and
Appendix~B of Ref.~\cite{Maybee:2019jus}) for more details on the derivation of 
\myeqref{eq:SpinVectorOperator}.} 
\begin{equation}
    \mathds S_{ij}^\mu =  -\frac{i}{m_1} \epsilon^{\mu \nu \alpha \beta} \bar p^{\,\prime}_{1\nu} \bar \varepsilon^{\,\star}_{i \alpha}(\bar{p}_1^{\,\prime}) \bar \varepsilon_{j \beta}(\bar{p}_1^{\,\prime})\;,
    \label{eq:SpinVectorOperator}
\end{equation}
where $\bar \varepsilon_{j \beta}$ denotes a polarisation vector with little-group 
index $j$ and Lorentz index $\beta$. Consider a massive particle with polarisation 
vector
\begin{equation}
  \bar \varepsilon^{\,\mu} = \sum_j w_j \bar \varepsilon_j^{\,\mu}\ ,
  \label{eq:PolVecWeightedSum}
\end{equation}
that is a complex weighted sum of polarisation vectors of different spin
states, then the expectation value of the spin vector is
\begin{equation}
  S^\mu = -\frac{i}{m_1} \epsilon^{\mu \nu \alpha \beta} \bar p_{1\nu}^{\,\prime} \, w_i^{\,\star} \bar \varepsilon^{\,\star}_{i \alpha}(\bar p_1^{\,\prime}) \, w_j \bar \varepsilon_{j \beta}(\bar p_1^{\,\prime})
  = -\frac{i}{m_1} \epsilon^{\mu \nu \alpha \beta} \bar p^{\,\prime}_{1\nu} \bar \varepsilon^{\,\star}_\alpha(\bar p_1^{\,\prime}) \bar \varepsilon_\beta(\bar p_1^{\,\prime})\;.
    \label{eq:SpinVectorPolVecs}
\end{equation}
Given that the wave packet for an incoming massive particle within the KMOC formalism has a spread in 
momentum, the polarisation vectors $\varepsilon_1(-p_1)$ and $\varepsilon^\star_4(p_4)$ in the wave 
packet can be defined by Lorentz transformations of 
$\bar \varepsilon(\bar p_1^{\,\prime})$; we refer to these as boosted/barred polarisations. The 
Lorentz transformations take the central value of the momentum in the wave packet, $\bar p_1^{\,\prime}$, 
to $-p_1 = \bar p_1 - q/2$ and $p_4 = \bar p_1 + q/2$, respectively, and trivially preserve the 
transversality condition of the polarisation vectors\footnote{Note that this is different 
from Ref.~\cite{Maybee:2019jus} where they boost $p_4$ to $p_1$, since we treat the two momenta in a 
symmetric manner.}. The explicit derivation of the transformation is shown in 
Appendix~\mysecref{sec:CovariantBoostedpolarisations}. More specifically, the transformed 
polarisation vectors take the form
\begin{equation}
\begin{split}
  & \varepsilon^\mu_{1} =\Lambda(\psi)^{\mu}_{~\nu}\bar{\varepsilon}^\nu = \bar{\varepsilon}^\mu + \Lambda_1 (q\cdot\bar{\varepsilon}) \bar{p}_1^\mu + \Lambda_2 (q\cdot\bar{\varepsilon}) q^\mu\;, \\
  & \varepsilon^{\star\mu}_{4} =\Lambda(-\psi)^{\mu}_{~\nu}\bar{\varepsilon}^{\,\star\nu} = \bar{\varepsilon}^{\,\star\mu} - \Lambda_1 (q\cdot\bar{\varepsilon}^{\,\star}) \bar{p}_1^\mu + \Lambda_2 (q\cdot\bar{\varepsilon}^{\,\star}) q^\mu\;,
\end{split}
\label{eq:polarisationBoost}
\end{equation}
where the boost coefficients are 
\begin{equation}
\Lambda_1 = \frac{1}{\sqrt{-q^2\bar{m}_1^2}}\sinh(\psi)\;, \qquad\qquad  \Lambda_2 = \frac{1}{q^2}\left(\cosh(\psi) - 1\right)\;, 
	\label{eq:BoostCoeffDefns}
\end{equation}
with
\begin{equation}
 \psi = \arctanh\left(\frac{\sqrt{-q^2}}{{2\bar{m}_1}}\right)\;.
 \label{eq:BoostAngleDefns}
\end{equation}
From the spin vector as an operator in spin states of a definite momentum \eqref{eq:SpinVectorOperator} 
the squared spin operator follows as 
\begin{equation}
	\begin{split}
    (\mathds S^\mu \mathds S^\nu)_{i j} & = -\frac{1}{m_1^2} \epsilon^{\mu \alpha \beta \gamma} \epsilon^{\nu \lambda \rho \sigma} \bar{p}^{\,\prime}_{1\alpha}\bar{p}^{\,\prime}_{1\lambda}\bar{\varepsilon}^{\,\star}_{i \beta}\bar{\varepsilon}_{j \sigma}\sum_{k} \bar{\varepsilon}^{\,\star}_{k \gamma}\bar{\varepsilon}_{k \rho} \\
    & = -\bar{\varepsilon}^{\,\star\mu}_{i}\bar{\varepsilon}^\nu_{j} - \delta_{ij}\left(\eta^{\mu \nu} - \frac{\bar{p}_1^{\,\prime\mu} \bar{p}_1^{\,\prime\nu}}{m_1^2} \right)\;.
    \end{split}
\end{equation}
This equation can be used to define the linear-in-spin and quadratic-in-spin terms for a general weighted sum of individual spin states \eqref{eq:PolVecWeightedSum} following the same manipulations of \eqref{eq:SpinVectorPolVecs} and using the normalisation $\sum_i w_i^\star w_i = 1$,
\begin{equation}
\begin{split}
   & \bar{\varepsilon}^{\,\star\mu} (\bar p_1^{\,\prime}) \bar{\varepsilon}^\nu (\bar p_1^{\,\prime}) - \bar{\varepsilon}^{\,\star\nu} (\bar p_1^{\,\prime}) \bar{\varepsilon}^\mu (\bar p_1^{\,\prime})  = \frac{i}{m_1}\epsilon^{\mu \nu \alpha \beta} \bar{p}_{1 \alpha}^{\,\prime} S_{\beta}\;, \\
   & \bar{\varepsilon}^{\,\star\mu} (\bar p_1^{\,\prime}) \bar{\varepsilon}^\nu (\bar p_1^{\,\prime}) + \bar{\varepsilon}^{\,\star\nu} (\bar p_1^{\,\prime}) \bar{\varepsilon}^\mu (\bar p_1^{\,\prime}) = -(S^\mu S^\nu + S^\nu S^\mu)  - 2 \left(\eta^{\mu \nu} - \frac{\bar{p}_1^{\,\prime\mu} \bar{p}_1^{\,\prime\nu}}{m_1^2}\right)\;.
\end{split}
\end{equation}
If we average over the two expressions we find~\cite{Cangemi:2022abk}\footnote{Note that $S^\mu S^\nu$ should 
be interpreted as an expectation value $\langle S^\mu S^\nu \rangle = \sum_{i,k,j}\omega_i^\star 
\mathds{S}^{(\mu}_{ik}\mathds{S}^{\nu)}_{kj}\omega_j$. See \myeqref{eq:prodOfSpin} 
and Chapter~\mysecref{sec:Chapter4} for more details.}
\begin{equation}
    \begin{split}
        \bar{\varepsilon}^{\,\star\mu} \bar{\varepsilon}^\nu & = - \left(\eta^{\mu \nu} - \frac{\bar{p}_1^{\,\prime\mu} \bar{p}_1^{\,\prime\nu}} {m_1^2}\right) + \frac{i}{2m_1}\epsilon^{\mu \nu \alpha \beta} \bar{p}_{1 \alpha}^{\,\prime} S_{\beta} - \frac{1}{2}(S^\mu S^\nu + S^\nu S^\mu)\;,
    \end{split}
    \label{eq:Spin1Decomposition}
\end{equation}
where the first term in \myeqref{eq:Spin1Decomposition} contains information on both the 
spin-independent and Casimir structures, through the identification 
$\eta_{\mu \nu}(S^{\mu}S^{\nu}) = -2$ for the spin-1 representation~\cite{Chung:2019duq}. This is a 
crucial point, since this allows us to extract the full Casimir dependence.

For completeness, let us discuss how to re-express the tensor structures of 
\mysecref{sec:Form_factor_decomp} in terms of  spin vectors. As explained, the first step is performing 
the appropriate Lorentz transformations, where only the leading 
contributions in $q$ are needed
\begin{equation}
 \Lambda_1 = \frac{1}{2\bar{m}_1^2} + \mathcal{O}(q)\;, \qquad \Lambda_2 = -\frac{1}{8\bar{m}_1^2} + \mathcal{O}(q)\;.
\end{equation}
Then \myeqref{eq:Spin1Decomposition} recasts the tensor structures of 
\myeqref{eq:tensorBasis} as (up to $\mathcal{O}(q^2)$)
\begin{equation}
	\begin{split}
		& T_1 = -1 + \frac{q^2}{2\bar{m}_1^2} + \frac{1}{2\bar{m}_1^2}(q \cdot S)^2\;, \\
		& T_2 = -q^2 - (q \cdot S)^2\;, \\[8pt]
		& T_3 = \bar{m}_2^2(y^2-1)q^2 - \bar{m}_2^2q^2(u_2 \cdot S)^2\;, \\[4pt]
		& T_4 = -i \bar{m}_2 \epsilon^{\mu \nu \rho \sigma} q_\mu u_{1\nu}u_{2\rho} S_\sigma + \frac{y\bar{m}_2}{\bar{m}_1}q^2 + \frac{y\bar{m}_2}{\bar{m}_1}(q \cdot S)^2\;,
	\end{split}
	\label{eq:Spin1TensorBoostToSpin}
\end{equation}
which we have evaluated in $D=4$ dimensions and used $S^2 = -2$ explicitly. 
Once again, we emphasise that the terms with no spin dependence
contain information about both spin-independent and spin Casimir structures.
Another important point is that the coefficient of $T_1 =
\varepsilon_4^\star \cdot \varepsilon_1$ in the leading small-$|q|$ limit is 
expected to be the negative of the spin-$0$ amplitude, because it corresponds to the 
spin-independent piece of the spin-$1$ amplitude but picks up a minus
sign from the mostly-minus metric signature.

%--------------------------------------------------------------------------------
\subsection{Spin Interpolation}
\label{sec:SpinInterpolation}
%--------------------------------------------------------------------------------

When extracting the spin dependence of classical dynamics, fixed-spin
theories have an apparent ambiguity that stems from the so-called spin Casimir, 
i.e.~the square of the spin vector $S^2= -s(s+1)$. The consequence is
that any explicit spin Casimir terms are simply numbers, like $S^2 = -2$ for
$s = 1$, and so it seems that we are missing the necessary information
to faithfully capture the spin dependence. In fact, the combination
$q^2 S^2$ is what actually appears, as this combination is of
$\mathcal O(\hbar^0)$ in the classical limit where the spin quantum
number becomes large. This becomes proportional to $q^2$ for a fixed
low spin and is seemingly indistinguishable from the quantum
corrections of the spin-independent part of the amplitude.
This is one motivation to turn to generic-spin theories
or worldline EFTs, which make the spin structures manifest from the very
start, thereby bypassing the problem altogether.

Here we propose a new method that allows one to extract the full spin
dependence starting from fixed-spin theories. There are two main observations
that allow for this. The first is that, from the ambiguity itself, spin 
Casimir terms can, naively, be hidden as quantum suppressed terms in the 
spin-independent structure of the amplitude. The second is that there exists a 
form of \textit{spin universality} which makes lower-spin amplitudes reappear in 
certain terms of higher-spin amplitudes. In fact, a notion of spin universality 
is known to be exhibited at the level of classical coefficients of the 
amplitudes~\cite{Holstein:2008sx, Holstein:2008sw, Siemonsen:2019dsu}, but our 
method relies on the fact that even quantum corrections to these coefficients 
are spin universal. To explain how this method works we will 
focus on the scattering between a spinless black hole and a spinning black hole, 
up to quadratic order in spin.

At tree level the $q^2 S^2$ term cancels the $1/q^2$ pole of the tree
amplitude and leaves no non-analytic dependence on $q$ required for classical 
interactions. Thus, the spin Casimir ambiguity only plays a role at one-loop order 
and beyond. Schematically, after suitable subtraction of superclassical
divergences (the details of which do not matter), the finite remainder of the 
four-point amplitudes for the scattering of a scalar particle and a spinning 
particle is captured by the following ansatz which holds for \emph{any} spin 
representation,
\begin{equation}
M  = \sum_{n, i} c^{(n,i)} \mathcal{J}^{(n,i)}\;, 
 \label{eq:GeneralAnsatz_sSquared}
\end{equation}
where $n$ is the spin power in a given structure, and $i$ labels all spin 
structures with a given $n$. The coefficients $c^{(n,i)}$ are generally series 
expansions in the coupling constant and $q$. Expanding \eqref{eq:GeneralAnsatz_sSquared} 
in $|q|$, each contribution reads
\begin{align}
\displaybreak[0]
  & c^{(0,1)}\mathcal{J}^{(0,1)} = 1 + c_{\text{quantum}} q^2 + \mathcal O(q^4)\; , \nonumber \\
  & c^{(1,1)}\mathcal{J}^{(1,1)} = i(n \cdot a) \, [ 1 + \mathcal O(q^2)] \;, \nonumber \\
  & c^{(2,1)}\mathcal{J}^{(2,1)} = (q \cdot a)^2 \, [ 1 + \mathcal O(q^2)] \;, \label{eq: SpinStructures}\\ 
  & c^{(2,2)}\mathcal{J}^{(2,2)} = q^2 (u_2 \cdot a)^2 \, [ 1 + \mathcal O(q^2)] \;, \nonumber \\ 
  & c^{(2,3)}\mathcal{J}^{(2,3)} = q^2 a^2 \, [ 1 + \mathcal O(q^2)] \;, \nonumber 
\end{align}
where we define 
\begin{equation}
	n^\mu = -\epsilon^{\mu \nu \rho \sigma} q_\nu u_{1\rho}u_{2\sigma}\;.
\end{equation}
Furthermore, we have used the normalised spin vector, $a^\mu = S^\mu/\bar{m}_1$, 
so that the mass dependence is uniform across all the spin structures. 
The spin-diagonal part of the amplitude, again expanded in $|q|$, is
\begin{equation}
  \label{eq:ampSpinDiagonal}
  \mathcal M_{\text{spin-diagonal}} = c^{(0,1)} (1 +
  c_{\text{quantum}} q^2 ) + c^{(2,3)}
  q^2 a^2 + \mathcal{O}(q^4)\;,
\end{equation}
which contains the spinless term, a quantum correction to the spinless term, and 
a spin Casimir term. The spin Casimir term can be isolated from the quantum
correction term at $\mathcal O(|q|^2)$ by an interpolation between the spin-1 
amplitude and the spin-0 amplitude\footnote{The mass of the 
spinning particle is $m_1$, while the scalar particle always has mass $m_2$.}. 
More explicitly, using $S^2 = -s(s+1)$, we have
\begin{equation}
  \mathcal
  M_{\text{spin-diagonal}} \Big |_{s=1}  - \mathcal M_{\text{spin-diagonal}}
  \Big |_{s=0} = -2  c^{(2,3)} q^2 / \bar m_1^2\; ,
  \label{eq:spin1MinusSpin0}
\end{equation}
where this identity is understood up to $\mathcal{O}(q^4)$.
In this way, we unambiguously identify the spin Casimir coefficient
$c^{(2,3)}$ from the naively quantum-suppressed $\mathcal O(|q|^2)$ part
of the amplitude, as desired. Here we have relied on spin universality,
whereby the spinless term and the quantum correction term
are identical for the $s=1$ and $s=0$ representations, and therefore 
cancel out in the difference on the left-hand side of \eqref{eq:spin1MinusSpin0}. 

As we will see, using this procedure we faithfully determine the spin
Casimir up to the two-loop level---that is up to the third order in the coupling
constant---for both gravity and electrodynamics. This method 
will subsequently be extended to higher spins, allowing us to obtain novel 
results through the quartic order in spin. In this case, the ansatz contains many more terms 
and the spin interpolation will require the three amplitudes mentioned in 
\mysecref{sec:KinematicSetUp}: scalar-scalar, scalar-vector, and scalar-tensor. Indeed, a simple 
subtraction will no longer fix all the coefficients in the ansatz as there are 
spin Casimir terms associated to different spin structures. To this end, we will propose a more 
convenient strategy involving solving linear systems of equations. This provides a systematic method 
for determining the full spin dependence of the classical two-body dynamics from scattering amplitudes 
of fixed-spin theories.

%--------------------------------------------------------------------------------
\section{Classical Amplitudes in Electrodynamics and Gravity}
\label{sec:ScatteringAmplitudes}
%-------------------------------------------------------------------------------

In this section we provide the results for the $2\to 2$ amplitudes in electrodynamics 
and Einstein gravity. As the results presented here are the classical 
amplitudes, we can safely ignore quantum corrections, for example, in the difference between $\bar{m}_i$ 
and $m_i$.

Before presenting the explicit results, a quick word on the amplitude constructions is in order. This was 
done using two different approaches in Ref.~\cite{Akpinar:2025bkt}. The first is an  
analytic construction using the techniques and objects described in Chapter~\mysecref{sec:Chapter2} and 
\mysecref{sec:Tree_amplitudes_via_dimensional_reduction}. The second 
approach is a numerical construction of all electrodynamics and gravity amplitudes using 
the \Caravel{} framework~\cite{Abreu:2020xvt}; we found exact agreement with the analytic 
amplitudes in overlapping cases\footnote{The numerical computations were completed by my amazing 
collaborators: Manfred Kraus and Fernando Febres Cordero.}. Furthermore, as another strong validation of the 
numerical implementation, we have found that all three-, four-, and five-point tree-level amplitudes involved in 
generalised unitarity agree numerically with analytic constructions. 

This strategy gives the amplitude as an expansion in the dimensional regularisation 
parameter $\epsilon$, from the $1/\epsilon^2$ order to the $1/\epsilon$ order, as well as an expansion over 
small $|q|$, from order $1/|q|^2$ to order $|q|^2$, at numerical values of the remaining kinematic variables 
$(\bar m_1, \bar m_2, y)$. Moreover, the only nontrivial functions involved, other than rational functions, 
are $\operatorname{arcosh}(y)$. After the subtraction of iteration terms, the infrared divergence 
$\sim 1/\epsilon^2$ and superclassical divergence $\sim 1/|q|^2, 1/|q|$ disappear. This leaves us 
with an amplitude whose $\epsilon$ expansion starts at $\mathcal O(1/\epsilon)$, which combined with the 
Taylor expansion of the overall factor $(-q^2)^{-2\epsilon}$ generates a non-analytic $\log(-q^2)$ term 
responsible for classical interactions at this order. The $|q|$ expansion now includes the classical 
$\mathcal O(|q|^0)$ and the (naively) quantum order $\mathcal O(|q|^2)$.

The final analytic answer was subsequently obtained by performing an analytic reconstruction from multiple 
numerical evaluations of the amplitude in different kinematic configurations (see 
Ref.~\cite{Akpinar:2024meg} for details).

%--------------------------------------------------------------------------------
\subsection{The Electromagnetic Case}
\label{sec:ScatteringAmplitudesQED}
%-------------------------------------------------------------------------------

The resulting finite remainder, written in a form for massive particles of any spin, reads
\begin{equation}
 \mathcal{A}_{\text{finite}}^{(L)} = \frac{\alpha_{\textrm{eff}}^{1+L}}{(-q^2)^{1-L/2+L\epsilon}}\sum_{n,i}^{} c^{(L,n,i)} \mathcal{J}^{(n,i)}\;,
 \label{eq:QED_finite_sSquared}
\end{equation}
with $\alpha_{\textrm{eff}}= q_\phi q_{\mbox{\tiny V}} \alpha$ and $\alpha = e^2/(4\pi)$. Here $L$ denotes 
the loop order and, truncating to the quadratic order in spin, the structures $\mathcal{J}^{(n,i)}$ are defined in 
\myeqref{eq:GeneralAnsatz_sSquared}. Then, the coefficients are as follows. At tree level we have
\begin{equation}
\begin{split}
 &c^{(0,0,1)} = -16y \pi m_1 m_2\;, \qquad c^{(0,1,1)} = -16 \pi m_1 m_2\;, \\
 & c^{(0,2,1)} = - 8 \pi m_1 m_2\;, \qquad ~~~ c^{(0,2,2)} = c^{(0,2,3)} = 0\;.
 \end{split}
\end{equation}
At one loop we find
\begin{equation}
\begin{split}
 & c^{(1,0,1)} = 4\pi^2(m_1 + m_2)\;,  \\
 & c^{(1,1,1)} = 4\pi^2y\frac{2m_1 + m_2}{(y^2-1)}\;,  \\
 & c^{(1,2,1)} = - c^{(1,2,3)} = \pi^2\frac{(5y^2-3)m_1+2y^2 m_2}{(y^2-1)}\;, \\
 &  c^{(1,2,2)} = - \pi^2\frac{4(2y^2-1)m_1+2(y^2+1)m_2}{(y^2-1)^2}\;,
\end{split}
\end{equation}
which is the sum of two types of mass structures: $m_1$ which dominates in the 
spinless probe limit, $m_1 \gg m_2$; and $m_2$ which dominates in the spinning probe 
limit, $m_2 \gg m_1$. In other words, the sum of the two opposite probe limits 
completely fixes the one-loop amplitude. Practically, one can extract each probe limit 
from the amplitude by simply taking the limit as the heavy mass goes to infinity.
The two-loop result, on the other hand, contains terms that dominate in either probe 
limit, as well an additional mass structure corresponding to the 
first self-force (1SF) correction. Organising the results into probe structures and 
1SF structures according to
\begin{equation}
c^{(2,n,i)} = c^{(2,n,i)}_{\text{probe}} + c^{(2,n,i)}_{\text{1SF}}\;,
\end{equation}
truncated to leading order in the dimensional regulator $\epsilon$, yields 
\begin{itemize}
\item Probe structures: 
{\allowdisplaybreaks
\begin{align}
  c^{(2,0,1)}_{\text{probe}} & = -2\pi y\frac{(2y^2-3)(m_1^2+m_2^2)}{3\epsilon(y^2-1)^2m_1m_2}\;, \nonumber \\
  c^{(2,1,1)}_{\text{probe}} & = -2\pi\frac{(2y^2-1)(3m_1^2+m_2^2)}{3\epsilon(y^2-1)^2m_1m_2}\;,  \\
  c^{(2,2,1)}_{\text{probe}} & = - c^{(2,2,3)}_{\text{probe}} = -\pi y\frac{(10y^2-9)m_1^2+(2y^2-1)m_2^2}{3\epsilon(y^2-1)^2m_1m_2}\;, \nonumber \\
  c^{(2,2,2)}_{\text{probe}} & = 2\pi y\frac{(7 y^2-6)m_1^2 +y^2m_2^2}{3 \epsilon (y^2-1)^3m_1 m_2}\;, \nonumber 
\end{align}
}

\item 1SF structures:
{\allowdisplaybreaks
\begin{align}
  c^{(2,0,1)}_{\text{1SF}} & = \frac{4\pi(y^4 - 3y^2 + 3)}{3\epsilon(y^2-1)^2}\;, \nonumber \\
  c^{(2,1,1)}_{\text{1SF}} & = \frac{4\pi y(y^2-3)}{3\epsilon(y^2-1)^2}-\frac{4\pi}{\epsilon(y^2-1)^{3/2}}\arcosh(y)\;, \nonumber \\
  c^{(2,2,1)}_{\text{1SF}} & = \frac{2\pi(y^4+1)}{3\epsilon(y^2-1)^2}- 
  \frac{2\pi y(2y^2-1)}{\epsilon(y^2-1)^{5/2}}\arcosh(y)\;, \\
  c^{(2,2,2)}_{\text{1SF}} & = -\frac{\pi(2y^4-11y^2-10)}{3\epsilon(y^2-1)^3} + \frac{\pi y(4y^2-9)}{\epsilon(y^2-1)^{7/2}}\arcosh(y)\;, \nonumber \\
  c^{(2,2,3)}_{\text{1SF}} & = -\frac{\pi(2y^4-3y^2-4)}{3\epsilon(y^2-1)^2} + \frac{\pi y(4y^2-5)}{\epsilon(y^2-1)^{5/2}}\arcosh(y)\; \nonumber.
\end{align}
}
\end{itemize}
%c^{(2,2,1)}_{\text{1SF}} & = -\frac{2\pi(y^4-12y^2-3)}{9\epsilon(y^2-1)^2}- 
%  \frac{2\pi y(2y^2+1)}{3\epsilon(y^2-1)^{5/2}}\arcosh(y)\;, \\
%  c^{(2,2,2)}_{\text{1SF}} & = -\frac{\pi(2y^4-11y^2-10)}{3\epsilon(y^2-1)^3} + \frac{\pi y(4y^2-9)}{\epsilon(y^2-1)^{7/2}}\arcosh(y)\;, \nonumber \\
%  c^{(2,2,3)}_{\text{1SF}} & = -\frac{\pi(2y^4+3y^2-12)}{9\epsilon(y^2-1)^2} + \frac{\pi y(8y^2-11)}{3\epsilon(y^2-1)^{5/2}}\arcosh(y)\; \nonumber.
%
Note that the $1/\epsilon$ pole is combined with the factor $(-q^2)^{-2\epsilon}$ in \eqref{eq:QED_finite_sSquared}
to generate a Laurent expansion $1/\epsilon - 2 \log(-q^2)$, where the leading 
divergent term drops out of classical physics because it involves no non-analytic 
dependence on $q$.

%--------------------------------------------------------------------------------
\subsection{The Gravitational Case}
\label{sec:ScatteringAmplitudesGravity}
%-------------------------------------------------------------------------------

The finite remainder of the four-point tree-level, one-loop, and two-loop gravity amplitudes, 
written in a form for massive particles of any spin and truncated at the quadratic order in
spin, take the form
\begin{equation}
  \mathcal{M}_{\text{finite}}^{(L)} = \frac{G^{1+L}}{(-q^2)^{1-L/2+L\epsilon}} \sum_{n,i}^{} c^{(L,n,i)} \mathcal{J}^{(n,i)}\;.
 \label{eq:GR_finite_sSquared}
\end{equation}
To avoid cluttering the notation, here we use the same coefficient names
as used before for the electrodynamics results. The context should leave no confusion. 
The tree-level coefficients are
\begin{equation}
	\begin{split}
		& c^{(0,0,1)} = 16\pi(2y^2-1)m_1^2m_2^2\;, \qquad c^{(0,1,1)} = 32\pi y m_1^2m_2^2\;, \\
		& c^{(0,2,1)} = 8\pi(2y^2-1)m_1^2m_2^2\;, \qquad ~~ \hspace{-0.5mm}c^{(0,2,2)} = c^{(0,2,3)} = 0\;.
	\end{split}
\end{equation}
The one-loop coefficients read
\begin{equation}
	\begin{split}
		c^{(1,0,1)} & = 6\pi^2(5y^2-1)(m_1+m_2)m_1^2m_2^2\;, \\
		c^{(1,1,1)} & = 2\pi^2 y\frac{(5y^2-3)(4m_1 + 3m_2)m_1^2 m_2^2}{(y^2-1)}\;, \\
		c^{(1,2,1)} & = - c^{(1,2,3)} = \pi^2\frac{((95y^4 - 102y^2+15)m_1+4(15y^4 - 15y^2 + 2)m_2)m_1^2m_2^2}{4(y^2-1)}\;, \\
		c^{(1,2,2)} & = -\pi^2\frac{((65y^4 - 66y^2+9)m_1+2(15y^4 - 12y^2 + 1)m_2)m_1^2m_2^2}{2(y^2-1)^2}\;, 
	\end{split}
\end{equation}
Finally, the two-loop coefficients are organised again into probe structures, which 
dominate in either the $m_1 \gg m_2$ or $m_2 \gg m_1$ limits, 
as well as 1SF structures according to
\begin{equation}
c^{(2,n,i)} = c^{(2,n,i)}_{\text{probe}} + c^{(2,n,i)}_{\text{1SF,rational}} + c^{(2,n,i)}_{\text{1SF},\arcosh(y)}\;.
\end{equation}
Here we split rational and $\arcosh$ terms of the 1SF contributions. 
The individual terms are given by
\begin{itemize}
	\item Probe structures: 
		{\allowdisplaybreaks
		\begin{align}
			\hspace{-1.5cm} c^{(2,0,1)}_{\text{probe}} & = \frac{2\pi(64y^6 - 120y^4 + 60y^2 -5)(m_1^2+m_2^2)m_1^2m_2^2}{3  \epsilon(y^2-1)^2}\;, \nonumber \\
			\hspace{-1.5cm} c^{(2,1,1)}_{\text{probe}} & = 4\pi y\frac{(16y^4-20y^2+5)(3m_1^2+2m_2^2)m_1^2m_2^2}{3 \epsilon(y^2-1)^2}\;, \\
			\hspace{-1.5cm} c^{(2,2,1)}_{\text{probe}} & = - c^{(2,2,3)}_{\text{probe}} = \pi\frac{((128y^6-216y^4+96y^2-7)m_1^2+(64y^6-104y^4+44y^2-3)m_2^2)m_1^2m_2^2}{3\epsilon(y^2-1)^2}\;, \nonumber \\
			\hspace{-1.5cm} c^{(2,2,2)}_{\text{probe}} & = -2\pi\frac{((80y^6-132y^4+57y^2-4)m_1^2+(32y^6-48y^4+18y^2-1)m_2^2)m_1^2m_2^2}{3 \epsilon(y^2-1)^3}\;, \nonumber 
		\end{align}
		}
	\item 1SF rational terms:
	{\allowdisplaybreaks
	\begin{align}
		c^{(2,0,1)}_{\text{1SF,rational}} & =  4\pi y\frac{(36y^6-114y^4+132y^2-55)m_1^3 m_2^3}{3 \epsilon(y^2-1)^2}\;, \nonumber \\
		c^{(2,1,1)}_{\text{1SF,rational}} & = 4\pi \frac{(36y^6 - 156y^4 + 84y^2 +41)m_1^3 m_2^3}{3\epsilon(y^2-1)^2}\;, \nonumber \\
		c^{(2,2,1)}_{\text{1SF,rational}} & = 2\pi y\frac{(180y^6-1242y^4+104y^2+213)m_1^3 m_2^3}{15 \epsilon(y^2-1)^2}\;, \\
		c^{(2,2,2)}_{\text{1SF,rational}}  & = -2\pi y\frac{(138y^6-1769y^4+1394y^2+2122)m_1^3 m_2^3}{15 \epsilon(y^2-1)^3}\;,\nonumber  \\
		c^{(2,2,3)}_{\text{1SF,rational}}  & = -2\pi y\frac{(186y^6-1257y^4+446y^2+1005)m_1^3 m_2^3}{15 \epsilon(y^2-1)^2}\;. \nonumber
	\end{align}
	}
	\item 1SF $\arcosh(y)$ terms:
	{\allowdisplaybreaks
	\begin{align}
			c^{(2,0,1)}_{\text{1SF},\arcosh(y)} & = -8\pi \frac{(4y^4-12y^2-3)m_1^3 m_2^3}{ \epsilon\sqrt{y^2-1}}  \arcosh(y)\;, \nonumber\\
			c^{(2,1,1)}_{\text{1SF},\arcosh(y)} & = -16\pi y\frac{(2y^4-11y^2-6)m_1^3m_2^3}{ \epsilon(y^2-1)^{3/2}} \arcosh(y)\;, \nonumber \\
			c^{(2,2,1)}_{\text{1SF},\arcosh(y)} & = -4\pi y^2\frac{(4y^6 -36y^4+y^2+6)m_1^3m_2^3}{\epsilon(y^2-1)^{5/2}} \arcosh(y)\;, \\
			c^{(2,2,2)}_{\text{1SF},\arcosh(y)} & = 2\pi\frac{(8y^8 -104y^6+35y^4+162y^2+24)m_1^3m_2^3}{ \epsilon(y^2-1)^{7/2}} \arcosh(y)\;, \nonumber \\
			c^{(2,2,3)}_{\text{1SF},\arcosh(y)} & = 2\pi\frac{(8y^8 -72y^6+11y^4+66y^2+12)m_1^3m_2^3}{ \epsilon(y^2-1)^{5/2}} \arcosh(y)\;. \nonumber
	\end{align}
	}
\end{itemize}
As explained for the amplitudes in electrodynamics, the $1/\epsilon$ pole
is responsible for the finite $\log(-q^2)$ terms that
ultimately contribute to classical physics. 
%--------------------------------------------------------------------------------
\subsection{A Hidden Structure: Spin-Shift Symmetry}
\label{sec:SpinShiftSymmetry}
%-------------------------------------------------------------------------------

The spinning one-loop amplitude is well-known to exhibit a \textit{spin-shift symmetry} for 
gravity~\cite{Aoude:2022trd, Bern:2022kto} (as well as electromagnetism~\cite{Bern:2023ity}), in which 
$(q\cdot a)^2$ and $q^2 a^2$ always appear in the following combination 
\begin{equation}
  (q \cdot a)^2 - q^2 a^2\;,
\end{equation}
up to the quartic (quadratic) order in spin for gravity (electromagnetism). The immediate consequence is 
that the amplitude is invariant under a shift in the spin of the form
\begin{equation}
 a^\mu \rightarrow a^\mu +  \xi q^\mu\;,
\end{equation}
for arbitrary $\xi$ with a formal counting of $\mathcal{O}(1/|q|^2)$. In the case of $L=1$,
this property is explicitly manifest in our electrodynamics and gravity results as
\begin{equation}
  c^{(L,2,1)}_{\text{probe}} = -c^{(L,2,3)}_{\text{probe}}\;.
  \label{eq:coeffsHiddenRelation}
\end{equation}
which is expected. Furthermore, for the very first time, we observe
that the above relation also holds at two loops in either probe limit,
$m_1 \gg m_2$ or $m_2 \gg m_1$, up to quadratic order in spin. 
The spin-shift symmetry is in fact broken by the 1SF
corrections, which play a role at two loops and beyond. This is remarkable because 
it shows that this symmetry is not simply an accident, but rather there must be 
a deeper underlying cause closely tied to probe dynamics. We will revisit spin-shift symmetry in 
more detail in Chapter~\mysecref{sec:Chapter5}, although its exact origin is still an open question.

%--------------------------------------------------------------------------------
\section{Aligned-Spin Scattering Observables to $\mathcal{O}(G^3S^2)$}
\label{sec:AlignedSpinObservables}
%--------------------------------------------------------------------------------
We use the radial-action-like subtraction scheme to compute
aligned-spin observables for both electrodynamics and gravity, focusing on the
scattering angle. The aligned-spin limit fully captures the spin Casimir terms and
tests our method of spin interpolation for extracting such terms from
fixed-spin scattering amplitudes. The non-Casimir terms at $\mathcal{O}(G^3S^2)$ have
already been computed from scattering amplitudes in the earlier work of Ref.~\cite{FebresCordero:2022jts}.

The radial action is obtained from the finite remainder of the amplitude through the
following Fourier transformation\footnote{The delta functions and integration over two dimensions yield 
the factor of $4E|\boldsymbol{p}|$ appearing in \mysecref{sec:RadialAction}. See 
Appendix~\mysecref{sec:EvaluatingRankedIntegrals} to see this explicitly.}
\begin{equation}
 I_r^{(L)} = \int \hat{\mathrm{d}}^4 q\, \hat{\delta}(2 m_1 u_1 \cdot q)\hat{\delta}(2m_2 u_2 \cdot q) 
 e^{i b \cdot q} \mathcal{M}^{(L)}_{\text{finite}}\;,
 \label{eq:RadialActionFourierTransform}
\end{equation}
where $\mathcal{M}^{(L)}_{\text{finite}}$ is the finite remainder of 
the amplitude in terms of spin, which is given in \myeqref{eq:QED_finite_sSquared} for QED and in 
\myeqref{eq:GR_finite_sSquared} for gravity. As a consequence of the spin structures, we need to compute 
tensor integrals of the general form
\begin{equation}
    f^{\mu_1 \cdots \mu_n}_\beta = \int \hat{\delta}(2m_1 u_1 \cdot q)\hat{\delta}(2m_2 u_2 \cdot q)  
 	e^{i b \cdot q} q^{\mu_1}\ \cdots q^{\mu_n} (-q^2)^{-\beta}\;,
 	\label{eq:GeneralRankN_noI}
\end{equation} 
where the necessary integrals, alongside a discussion of how to evaluate them,
can be found in Appendix~\mysecref{sec:EvaluatingRankedIntegrals}. After the
relevant integrals are performed, we specialise to the aligned-spin case by
imposing
\begin{equation}
   a \cdot \bar{u}_i = a \cdot \hat{b} = 0\;, \quad 
   \epsilon^{\mu \nu \rho \sigma} u_{1\nu} u_{2\rho} a_\sigma =  |a| \sqrt{y^2-1} \ \hat{b}^\mu\;, \quad 
   a \cdot a = -|a|^2\;,
\end{equation}
where the $u_i$ span the equatorial plane (see \eqref{eq:velvectorsproperties}), we align the normalised 
spin vector, $a$, with the angular momentum, and we further define the normalised impact parameter 
$\hat{b}^\mu = b^\mu/|b|$. The aligned-spin scattering angle can then be computed
using \eqref{eq:angleradaction}, with $J = |\boldsymbol{p}| |\boldsymbol{b}|$ the angular momentum in the 
asymptotic past, together with
\begin{equation}
 |b|^2 = -b^2 = \boldsymbol{b}^2\;, \qquad 
 |\boldsymbol{p}| = \frac{m_1 m_2 \sqrt{y^2 - 1}}{\sqrt{m_1^2 + m_2^2 + 2m_1 m_2y}} 
 = \frac{m_1 m_2 \sqrt{y^2 - 1}}{\sqrt{s}}\;,
\end{equation} 
where $s = (p_1 + p_2)^2$ is the usual Mandelstam variable. The scattering angle in  
\eqref{eq:angleradaction} is expanded in the coupling constant
\begin{equation}
  \chi = g\,\chi^{(0)} + g^2\chi^{(1)} + g^3\chi^{(2)} + \mathcal{O}(g^4)\;, 
  \label{eq:ScatteringAngle_Expansion}
\end{equation}
with $g$ denoting the coupling constant, i.e.~$\alpha_{\text{eff}}$ for electrodynamics 
and $G$ for gravity.

In practice we need to perform derivatives of the radial action with 
respect to $J$ to acquire the scattering angle. To do this, we first realise that the action of the 
differentiation in \eqref{eq:angleradaction} can be recast as
\begin{equation}
	\frac{\partial}{\partial J} = \frac{1}{|\boldsymbol{p}|} \frac{\partial}{\partial |\boldsymbol{b}|} \iff \frac{1}{|\boldsymbol{p}|} \frac{b_\mu}{|\boldsymbol{b}|} \frac{\partial}{\partial b_\mu},
\end{equation}
such that we first differentiate \eqref{eq:RadialActionFourierTransform} and then evaluate 
the tensor integrals of the form \eqref{eq:GeneralRankN_noI}. To be more concrete, let us compute 
the electrodynamics scattering angle at the leading order. From the tree-level amplitude in scalar QED
\begin{align}
	\chi^{(0)}_{\text{SQED}} & = - \frac{\partial}{\partial J} I_{r,\text{SQED}}^{(0)} \nonumber \\
	& = \frac{16y\pi m_1 m_2}{|\boldsymbol{p}|} \frac{b_\mu}{|\boldsymbol{b}|} \frac{\partial}{\partial b_\mu}  \int \hat{\mathrm{d}}^4 q\, \hat{\delta}(2 m_1 u_1 \cdot q) \hat{\delta}(-2 m_2 u_2 \cdot q) 
 	e^{i b \cdot q} \frac{1}{(-q^2)} \nonumber \\
 	& = \frac{16y\pi m_1 m_2}{|\boldsymbol{p}|} \frac{b_\mu}{|\boldsymbol{b}|}\int \hat{\mathrm{d}}^4 q\, \hat{\delta}(2m_1 u_1 \cdot q) \hat{\delta}(-2m_2 u_2\cdot q) 
 	e^{i b \cdot q} \frac{i q^\mu}{(-q^2)} \\
 	& = -\frac{2y}{|\boldsymbol{p}||\boldsymbol{b}|\sqrt{y^2-1}} \nonumber\\
 	& = -\frac{2y}{J\sqrt{y^2-1}} \nonumber \;,
\end{align}
which is the well-known leading-order scattering angle for electrodynamics once multiplied by 
$\alpha_{\textrm{eff}}= q_\phi q_{\mbox{\tiny V}} \alpha$.
%-------------------------------------------------------------------------------
\subsection{Scattering Angle in Electrodynamics}
\label{sec:AlignedSpinScatteringAngleQED}
% -------------------------------------------------------------------------------
Here we give the scattering angle in electrodynamics up to two loops, 
i.e.~$\mathcal O(\alpha_{\text{eff}}^3)$, and quadratic order in spin. At leading order we have
\begin{equation}
 \chi^{(0)}_{\text{aligned-spin}}  = - \frac{2y}{J\sqrt{y^2-1}} + \frac{2 m_1m_2 \sqrt{y^2-1}}{J^2 \sqrt{s}}|a|   - \frac{2m_1^2 m_2^2 y \sqrt{y^2-1}}{J^3 s} |a|^2\;,
\end{equation}
at next-to-leading order we find
\begin{equation}
\begin{split}
    \chi^{(1)}_{\text{aligned-spin}}   = \frac{\pi (m_1 + m_2)}{2 J^2 \sqrt{s}} & - \frac{\pi y (2m_1+m_2)m_1m_2}{J^3 s} |a|  \\
    & + \frac{3\pi ((5y^2-3)m_1 + 2y^2 m_2)m_1^2m_2^2}{4 J^4 s^{3/2}} |a|^2\;,
\end{split}
\end{equation}
and at next-to-next-to-leading order we have 
\begin{equation}
\begin{split}
    \chi^{(2)}_{\text{aligned-spin}} = & -\frac{2y(2y^2-3)(m_1^2+m_2^2)}{3J^3s(y^2-1)^{3/2}} 
    + \frac{4(y^4-3y^2+3)m_1m_2}{3J^3  s (y^2-1)^{3/2}}\\
    & +\frac{2(2y^2-1)(3m_1^2 + m_2^2)m_1m_2}{J^4 s^{3/2} \sqrt{y^2-1}} |a|
    - \frac{4y(y^2-3)m_1^2m_2^2}{J^4 s^{3/2} \sqrt{y^2-1}}|a| \\
    & + \frac{12 m_1^2m_2^2}{J^4 s^{3/2}}  \arcosh(y) |a|
  	 -\frac{4y((10y^2-9)m_1^4 m_2^2 + (2y^2-1)m_1^2 m_2^4)}{J^5 s^2 \sqrt{y^2-1}}|a|^2 \\
  	& + \frac{8(y^4-y^2-1)m_1^3m_2^3}{J^5 s^2 \sqrt{y^2-1}}|a|^2 
      - \frac{48 m_1^3m_2^3 y }{J^5 s^2}\arcosh(y)|a|^2\;,
\end{split}
\end{equation}
%\begin{equation}
%\begin{split}
%    \chi^{(2)}_{\text{aligned-spin}} = & -\frac{2y(2y^2-3)(m_1^2+m_2^2)}{3J^3s(y^2-1)^{3/2}} 
%    + \frac{4(y^4-3y^2+3)m_1m_2}{3J^3  s (y^2-1)^{3/2}}\\
%    & +\frac{2(2y^2-1)(3m_1^2 + m_2^2)m_1m_2}{J^4 s^{3/2} \sqrt{y^2-1}} |a|
%    - \frac{4y(y^2-3)m_1^2m_2^2}{3J^4 s^{3/2} \sqrt{y^2-1}}|a| \\
%    & + \frac{4 m_1^2m_2^2}{J^4 s^{3/2}}  \arcosh(y) |a|
%  	 -\frac{4y((10y^2-9)m_1^4 m_2^2 + (2y^2-1)m_1^2 m_2^4)}{J^5 s^2 \sqrt{y^2-1}}|a|^2 \\
%  	& + \frac{8(y^4+15y^2-9)m_1^3m_2^3}{9J^5 s^2 \sqrt{y^2-1}}|a|^2 
%      - \frac{80 m_1^3m_2^3 y }{3J^5 s^2}\arcosh(y)|a|^2\;,
%\end{split}
%\end{equation}
%
where we identify the angular momentum at past infinity as 
\begin{equation}
 J = |\boldsymbol{b}\times\boldsymbol{p}| = |\boldsymbol{b}||\boldsymbol{p}|\;.
\end{equation}
To validate our results, we further specialise our analysis to the 
limiting case of a probe scalar ($m_1 \gg m_2$)
\begin{align}
    & \chi^{(0)}_{\text{aligned-spin,probe}}  = - \frac{2y}{J\sqrt{y^2-1}} + \frac{2m \sqrt{y^2-1}}{J^2} |a|   - \frac{2 m^2 y \sqrt{y^2-1}}{J^3}|a|^2 \;, \label{eq:QEDScatteringAmplitudeAngle} \nonumber \\ 
    & \chi^{(1)}_{\text{aligned-spin,probe}}   = \frac{\pi}{2 J^2} - \frac{2\pi y m }{J^3}|a|  + \frac{3 \pi (5y^2-3) m^2}{4 J^4}|a|^2 \;, \\ 
    & \chi^{(2)}_{\text{aligned-spin,probe}}   = -\frac{2y(2y^2-3)}{3J^3 (y^2-1)^{3/2}} +\frac{6 (2y^2-1)m}{J^4\sqrt{y^2-1}}|a|  -\frac{4 y (10y^2-9) m^2 }{J^5\sqrt{y^2-1}}|a|^2\;, \nonumber
\end{align}
where we relabel $m_2 = m$.

%--------------------------------------------------------------------------------
\subsubsection{Validation: Classical Aligned-Spin Probe Limit}
\label{sec:CrossCheckQED}
%--------------------------------------------------------------------------------
Observables for spinning two-body interactions in electrodynamics
have been computed using scattering amplitudes 
to $\mathcal{O}(\alpha S^2)$ in Ref.~\cite{Maybee:2019jus}, and to 
$\mathcal{O}(\alpha^2 S^2)$ in Ref.~\cite{Menezes:2022tcs} using the
massive spinor-helicity formalism. As there are no results in the literature for such
observables beyond one loop, the simplest cross-check we can make is to solve the classical
dynamics of a charged probe scalar mass in a root-Kerr 
background \cite{Monteiro:2014cda, Arkani-Hamed:2019ymq} which is the
electromagnetic field in the presence of a spinning disc of charge. The root-Kerr
four-vector potential solution relevant for the dynamics of a probe mass confined to the equatorial
plane, i.e., the $(x,y)$-plane, reads
\begin{equation}
    A_\mu \big \rvert_{z = 0} = \frac{\alpha}{e \tilde{r}} \left(1, \frac{x\tilde{r} + ay}{r^2}, \frac{y\tilde{r} - ax}{r^2}, 0\right)\;, 
    \label{eq: rootKerr}
\end{equation}
where $\tilde{r} = \sqrt{r^2 - a^2}$, $\alpha = e^2/4\pi$, and $a = |a|$ is the spin 
associated with the rotating ring of charge. The dynamics for the worldline of an 
electrically charged point particle of mass $m$ reads 
\begin{equation}
  \hspace{-0.5cm} S = \int d\tau \left(\frac{m}{2} \dot{X}_\mu \dot{X}^\mu + \sqrt{4\pi \alpha} A_\mu \dot{X}^\mu \right)\;, \quad 
 X^\mu = (t,x,y,z)\;, \quad 
 \dot{X}^\mu = \frac{dX^\mu}{d\tau}\;,
\end{equation}
which we find by gauge-fixing the usual polynomial point particle action. Then, one can determine
the Noether currents corresponding to time translation and rotation invariance,
which results in equations for energy and angular momentum conservation
\begin{equation}
  m \dot{X}^0 + eA^0 = \text{constant}\;, \qquad (m \dot{X}_i + e A_i) \Omega_{ij} X_j = \text{constant}\;,
\end{equation}
where
\begin{equation}
  \Omega_{ij} = \begin{pmatrix}
    0 & -1 \\
    1 & \phantom{-}0
  \end{pmatrix}\;,
\end{equation}
is the infinitesimal rotation matrix. Taking
the initial conditions to be those of the probe mass at past infinity with
velocity $v_{\infty}$, energy and angular momentum conservation allows us to
determine the total velocity, $v$, and angular velocity, $v_{\theta}$, as a
function of radial distance, respectively. More specifically, this is done by 
solving
\begin{equation}
  \sqrt{v^2 + 1} - \sqrt{v_{\infty}^2 + 1} + \frac{\alpha}{m \tilde{r}} = 0\;, \qquad 
  m (\boldsymbol{r} \times \boldsymbol{v})_z + e (\boldsymbol{r} \times \boldsymbol{A})_z = m b v_\infty\;,
\end{equation}
where $b = |\boldsymbol{b}|$ is related to the angular momentum at infinity 
via $J = m v_{\infty} b$. To this end, we find
\begin{align}
	v(r,a) & = v_{\infty} - \frac{(2r^2+a^2)\sqrt{1+v_{\infty}^2}}{2mr^3 v_{\infty}}\alpha 
			- \frac{(r^2+a^2)}{2m^2 r^4 v_{\infty}^3}\alpha^2 \nonumber \\
			& \hspace{4cm} - \frac{(2r^2+3a^2)\sqrt{1+v_{\infty}^2}}{4m^3 r^5 v_{\infty}^5} \alpha^3 + \mathcal{O}(a^3,\alpha^4)\;, \nonumber \\
		v_{\theta}(r,a) & = \frac{b v_\infty}{r} - \frac{a }{m r^2}\alpha + \mathcal{O}(a^3)\;,
\end{align}
where the radial velocity, $v_{r}$, is trivially fixed through 
$v^2 = v_{r}^2 + v_{\theta}^2$. Then, we can determine the scattering angle as
\begin{equation}
  \chi = 2\theta_{r_{\text{min}} \rightarrow \infty} - \pi\;,
\end{equation} 
together with
\begin{equation}
    \theta_{r_{\text{min}} \rightarrow \infty} = \int_{r_{\text{min}}}^{\infty} dr \frac{d\theta}{dr} = \int_{r_{\text{min}}}^{\infty} dr \frac{v_\theta}{r v_r}\;,
\end{equation}
where $r_{\text{min}}$ follows by solving $v_r = 0$. Finally, integrating and 
organising the scattering angle as in \eqref{eq:ScatteringAngle_Expansion}, we 
find
\begin{equation}
\begin{split}
    & \chi_{cl}^{(0)}  = - \frac{2y}{J\sqrt{y^2-1}} + \frac{2m \sqrt{y^2-1}}{J^2}a  - \frac{2 m^2 y \sqrt{y^2-1}}{J^3} a^2\;, \\ 
    & \chi_{cl}^{(1)}  = \frac{\pi}{2 J^2} - \frac{2\pi ym }{J^3}a + \frac{3\pi (5y^2-3) m^2 }{4 J^4}a^2\;, \\ 
    & \chi_{cl}^{(2)}  = -\frac{2y(2y^2-3)}{3J^3 (y^2-1)^{3/2}} + \frac{6 (2y^2-1)m}{J^4\sqrt{y^2-1}} a-\frac{4y (10y^2-9) m^2 }{J^5\sqrt{y^2-1}}a^2\;,
\end{split}
\end{equation}
where we identify
\begin{equation}
    v_\infty = \sqrt{y^2 - 1}\;, \qquad b = \frac{J}{m \sqrt{y^2-1}}\;.
\end{equation}
This is in exact agreement with \myeqref{eq:QEDScatteringAmplitudeAngle}, provided that we identify 
$\alpha$ with $\alpha_{\mathrm{eff}}$. Since the dynamics can be 
solved straightforwardly to higher orders in the coupling, we 
provide the scattering angle for a charged scalar probe through $\mathcal{O}(\alpha^9 S^2)$.
At each order in spin, the scattering angle reads
\begin{align}
	\hspace{-1.5cm} \chi_{\mathrm{cl}}^{a^0} & = - \frac{2y}{\sqrt{y^2-1}}\alpha_{\hspace{-0.05cm}J} + \frac{\pi}{2}\alpha_{\hspace{-0.05cm}J}^2 - \frac{2y(2y^2-3)}{3(y^2-1)^{3/2}}\alpha_{\hspace{-0.05cm}J}^3 + \frac{3 \pi }{8}\alpha_{\hspace{-0.05cm}J} ^4 -\frac{2  y \left(8 y^4-20 y^2+15\right)}{15 \left(y^2-1\right)^{5/2}}\alpha_{\hspace{-0.05cm}J} ^5 \nonumber \\
	& \hspace{0.5cm} + \frac{5 \pi  }{16}\alpha_{\hspace{-0.05cm}J} ^6 - \frac{2 y \left(16 y^6-56 y^4+70 y^2-35\right)}{35 \left(y^2-1\right)^{7/2}}\alpha_{\hspace{-0.05cm}J} ^7 +\frac{35 \pi  }{128}\alpha_{\hspace{-0.05cm}J} ^8\\
	& \hspace{0.5cm} -\frac{2  y \left(128 y^8-576 y^6+1008 y^4-840 y^2+315\right)}{315 \left(y^2-1\right)^{9/2}}\alpha_{\hspace{-0.05cm}J}^9 + \mathcal{O}(\alpha_{\hspace{-0.05cm}J}^{10})\;, \nonumber 
\end{align}
\begin{align}
	\hspace{-1.5cm} \chi_{\mathrm{cl}}^{a} & = \frac{2m \sqrt{y^2-1}}{J}\alpha_{\hspace{-0.05cm}J}- \frac{2 \pi m y}{J}\alpha_{\hspace{-0.05cm}J}^2 +\frac{6 m \left(2 y^2-1\right)}{J \sqrt{y^2-1}}\alpha_{\hspace{-0.05cm}J}^3-\frac{6 \pi m y}{J}\alpha_{\hspace{-0.05cm}J}^4 + \frac{10 m \left(8 y^4-12 y^2+3\right)}{3 J \left(y^2-1\right)^{3/2}}\alpha_{\hspace{-0.05cm}J}^5 \nonumber \\
	& \hspace{0.5cm} -\frac{45 \pi m y}{4 J}\alpha_{\hspace{-0.05cm}J}^6 +\frac{14 m \left(16 y^6-40 y^4+30 y^2-5\right)}{5 J \left(y^2-1\right)^{5/2}}\alpha_{\hspace{-0.05cm}J}^7 -\frac{35 \pi  m y}{2 J}\alpha_{\hspace{-0.05cm}J}^8 \\
	& \hspace{0.5cm} +\frac{18 m \left(128 y^8-448 y^6+560 y^4-280 y^2+35\right)}{35 J \left(y^2-1\right)^{7/2}}\alpha_{\hspace{-0.05cm}J}^9 + \mathcal{O}(\alpha_{\hspace{-0.05cm}J}^{10})\;, \nonumber
\end{align}
\begin{align}
	\hspace{-1.5cm} \chi_{\mathrm{cl}}^{a^2} & = -\frac{2  m^2 y \sqrt{y^2-1}}{J^2}\alpha_{\hspace{-0.05cm}J} +\frac{3 \pi m^2 \left(5 y^2-3\right)}{4 J^2}\alpha_{\hspace{-0.05cm}J}^2 - \frac{4  m^2 y \left(10 y^2-9\right)}{J^2 \sqrt{y^2-1}}\alpha_{\hspace{-0.05cm}J}^3 +\frac{15 \pi  m^2 \left(17 y^2-5\right)}{8 J^2} \alpha_{\hspace{-0.05cm}J}^4 \nonumber \\
	& \hspace{0.5cm} - \frac{2 m^2 y \left(104 y^4-180 y^2+75\right)}{J^2 \left(y^2-1\right)^{3/2}}\alpha_{\hspace{-0.05cm}J}^5 +\frac{105 \pi  m^2 \left(37 y^2-7\right)}{32 J^2}\alpha_{\hspace{-0.05cm}J}^6 \\
	& \hspace{0.5cm} -\frac{8 m^2 y \left(400 y^6-1064 y^4+910 y^2-245\right)}{5 J^2 \left(y^2-1\right)^{5/2}}\alpha_{\hspace{-0.05cm}J}^7 + \frac{315 \pi  m^2 \left(65 y^2-9\right)}{64 J^2}\alpha_{\hspace{-0.05cm}J}^8 \nonumber \\
	& \hspace{0.5cm} -\frac{2 m^2 y \left(5248 y^8-19008 y^6+25200 y^4-14280 y^2+2835\right)}{7 J^2 \left(y^2-1\right)^{7/2}}\alpha_{\hspace{-0.05cm}J}^9 + \mathcal{O}(\alpha_{\hspace{-0.05cm}J}^{10})\;, \nonumber
\end{align}
where we have defined $\alpha_{\hspace{-0.05cm}J} = \alpha/J$ to reduce clutter.
%-------------------------------------------------------------------------------
\subsection{Scattering Angle in Gravity}
\label{sec:AlignedSpinScatteringAngleGrav}
%-------------------------------------------------------------------------------

Here we provide the scattering angle in gravity up to two loops, i.e.~$\mathcal O(G^3)$, and to quadratic order 
in spin. At leading order we find
\begin{equation}
	\begin{split}
		\chi^{(0)}_{\text{aligned-spin}} = \frac{2(2y^2-1)m_1 m_2}{J\sqrt{y^2-1}} & - \frac{4y m_1^2 m_2^2 \sqrt{y^2-1}}{J^2 \sqrt{s}}|a| \\
		& + \frac{2m_1^3m_2^3 \sqrt{y^2-1}(2y^2-1)}{J^3 s} |a|^2\;,
	\end{split}
\end{equation}
at next-to-leading order we find
\begin{equation}
\begin{split}
    \chi^{(1)}_{\text{aligned-spin}} = & \ \frac{3\pi (5y^2-1)(m_1 + m_2)m_1^2 m_2^2}{4J^2 \sqrt{s}} - \frac{\pi y (5y^2-3)(4m_1+3m_2)m_1^3m_2^3}{2J^3 s} |a|  \\
    	& +\frac{3\pi ((95y^4-102y^2+15)m_1 + (60y^4-60y^2 + 8) m_2)m_1^4m_2^4}{16 J^4 s^{3/2}} |a|^2\;,
\end{split}
\end{equation}
in agreement with well-known results in the literature, e.g.~\cite{Kosmopoulos:2021zoq, Liu:2021zxr, 
Bern:2022kto, Aoude:2022trd, Aoude:2022thd}. Finally, at next-to-next-to-leading order we find
{\allowdisplaybreaks
\begin{align}
    & \chi^{(2)}_{\text{aligned-spin}} = \frac{2(64y^6-120y^4 + 60y^2 -5)(m_1^2+m_2^2)m_1^3 m_2^3}{3J^3 s (y^2-1)^{3/2}} \nonumber \\
    & + \frac{4y(36y^6-114y^4+132y^2-55)m_1^4m_2^4}{3 J^3 s(y^2-1)^{3/2}} - \frac{8(4y^4-12y^2-3)m_1^4m_2^4}{J^3 s}\arcosh(y) \nonumber \\
    & - \frac{4y(16y^4-20y^2+5)(3m_1^2 + 2 m_2^2)m_1^4 m_2^4}{J^4  s^{3/2}\sqrt{y^2-1}}|a| \nonumber \\
    & - \frac{4y(36y^6-156y^4+84y^2+41)m_1^5m_2^5}{J^4 s^{3/2}\sqrt{y^2-1}}|a| \\
    & + \frac{48y(2y^4 - 11y^2 - 6)m_1^5m_2^5}{J^4 s^{3/2}}\arcosh(y)|a| \nonumber \\
    & + \frac{4((128y^6 - 216y^4 + 96 y^2 - 7)m_1^2 + (64y^6 - 104y^4 + 44y^2 -3)m_2^2)m_1^5 m_2^5}{J^5 s^2 \sqrt{y^2-1}}|a|^2 \nonumber \\
    & + \frac{8 y(184 y^6 - 1252 y^4 + 332 y^2 + 741)m_1^6 m_2^6}{5 J^5 s^2 \sqrt{y^2-1}}|a|^2 \nonumber \\
    & - \frac{192(y^6 - 8 y^4 -7y^2 -1)m_1^6 m_2^6}{J^5 s^2}\arcosh(y)|a|^2\; \nonumber .
\end{align}
}

\noindent
This is in agreement with Ref.~\cite{Jakobsen:2022fcj} after converting their
result to the scattering angle in the centre-of-mass frame.

%--------------------------------------------------------------------------------
\chapter{Going Beyond the Quadratic-in-Spin Order}
\label{sec:Chapter4}
%--------------------------------------------------------------------------------

In the previous chapter we saw that modelling Kerr black holes as point particles in 
finite-spin representations inevitably introduces an ambiguity in spin Casimir terms. By 
developing the novel spin interpolation method we were able to resolve this ambiguity and 
therefore faithfully capture the scattering dynamics through $\mathcal{O}(S^2)$ in both 
electrodynamics and gravity. In this chapter, we wish to compute the dynamics at $\mathcal{O}(S^4)$ in 
Einstein gravity~\cite{Akpinar:2025bkt}, which is the current state-of-the-art for precision Kerr 
black hole scattering\footnote{Indeed, we do not consider electrodynamics as one would require 
coupling of a massive spin-$2$ field to the photon.}. In order to do this, we will make use of the spin 
interpolation method, albeit it no longer involves a simple subtraction between two amplitudes 
because there are five spin Casimir structures through the quartic in 
spin order. Indeed, to obtain results at the quartic order in spin we will require the scalar-scalar, 
scalar-vector, and scalar-tensor amplitudes. Given this fact, we will solve for coefficients in an 
ansatz using a linear system of equations which systematises this method for any spin order. 
%

%--------------------------------------------------------------------------------
\section{A Form Factor Decomposition for Spin-2}
\label{sec:Form_factor_decomp_tensor}
%--------------------------------------------------------------------------------

As for the case of scalar-vector scattering, we perform a form factor decomposition for the 
scalar-tensor scattering process
\begin{equation}
    H(-p_1,\varepsilon_1) + \phi(-p_2) \rightarrow \phi(p_3) + H(p_4,\varepsilon_4^*),
\end{equation}
which streamlines the computation. In this case, the decomposition is
\begin{equation}
    M_{\phi \tiny{\mbox{H}}} = \varepsilon^*_{4 \mu \nu} \mathcal{T}^{\mu \nu \alpha \beta}(\bar{p}_1,\bar{p}_2,q)\varepsilon_{1 \alpha \beta},
\end{equation}
where the general form of $\mathcal{T}^{\mu \nu \alpha \beta}$ reads
\begin{equation}
    \begin{split}
        \mathcal{T}^{\mu \nu \alpha \beta}(\bar{p}_1,\bar{p}_2,q) = \ & A^{(1)} g^{\mu \nu}g^{\alpha \beta}+A^{(2)} g^{\mu \alpha}g^{\nu \beta}+A^{(3)} g^{\mu \beta}g^{\alpha \nu} \\ 
        & \begin{split}
            + \sum_{i,j=1}^{3} \Big(& B^{(1)}_{ij} g^{\mu \nu} k^\alpha_i k^\beta_j +  B^{(2)}_{ij} g^{\mu \alpha} k^\nu_i k^\beta_j +  B^{(3)}_{ij} g^{\mu \beta} k^\nu_i k^\alpha_j \\
            & + B^{(4)}_{ij} g^{\alpha \beta} k^\mu_i k^\nu_j +  B^{(5)}_{ij} g^{\nu \alpha} k^\mu_i k^\beta_j +  B^{(6)}_{ij} g^{\nu \beta} k^\mu_i k^\alpha_j\Big)
        \end{split} \\
        & + \sum_{i,j,k,l=1}^{3} C_{ijkl}k^\mu_i k^\nu_j k^\alpha_k k^\beta_l,
    \end{split}
\end{equation}
with $k_i \in \{\bar{p}_1,\bar{p}_2,q\}$. Let us count to see how many terms there are: 
\begin{enumerate}
    \item 3 terms of the form $A^{(i)}$,
    \item 6 terms of the form $B^{(i)}$ with 9 terms each,
    \item 1 term of the form $C_{ijkl}$ which actually has 81 terms,
\end{enumerate}
with a total of $3 + 54 + 81 = 138$ terms. We may reduce the number of terms by 
imposing tracelessness and transversality
\begin{equation}
    \eta_{\mu \nu}\varepsilon_i^{\mu\nu} = 0, \ p_{i\mu}\varepsilon_i^{\mu\nu} =  p_{i\nu}\varepsilon_i^{\mu\nu} = 0,
\end{equation}
where the latter allows us to write 
\begin{equation}
    \begin{split}
        \bar{p}_{1\mu}\varepsilon_1^{\mu\nu} = \frac{1}{2}q_{\mu}\varepsilon_1^{\mu\nu}& , \ \bar{p}_{1\nu}\varepsilon_1^{\mu\nu} = \frac{1}{2}q_{\nu}\varepsilon_1^{\mu\nu}, \\
        \bar{p}_{1\mu}\varepsilon_4^{*\mu\nu} = -\frac{1}{2}q_{\mu}\varepsilon_4^{*\mu\nu}& , \ \bar{p}_{1\nu}\varepsilon_4^{*\mu\nu} = -\frac{1}{2}q_{\nu}\varepsilon_4^{*\mu\nu}.
    \end{split}
\end{equation}
As before, this reduces the total number of structures, which we find to be $14$. This is not the end of 
the story because we expect the amplitude to be invariant under crossing symmetry (exchange of 
$\varepsilon_1$ and $\varepsilon_4^*$ while flipping the sign of $q$), which further 
constrains the form of the tensor structures. Let us make crossing 
symmetry explicit and sort the tensors according to powers of $q$.

\begin{enumerate}
    \item There are 3 tensor structures at $\mathcal{O}(q^0)$:
    \begin{equation}
        \begin{split}
            & \mathcal{T}_1 = \text{Tr}\{\varepsilon_4^* \cdot \varepsilon_1 \}\;, \\
            & \mathcal{T}_2 = (\bar{p}_2 \cdot \varepsilon^*_4 \cdot \bar{p}_2)(\bar{p}_2 \cdot \varepsilon_1 \cdot \bar{p}_2)\;, \\
            & \mathcal{T}_3 = (\bar{p}_2 \cdot \varepsilon^*_4  \cdot \varepsilon_1 \cdot \bar{p}_2)\;.
        \end{split}
    \end{equation}
    \item There are 2 tensor structures at $\mathcal{O}(q^1)$:
    \begin{equation}
        \begin{split}
            & \mathcal{T}_4 = (\bar{p}_2 \cdot \varepsilon^*_4 \cdot \bar{p}_2)(\bar{p}_2\cdot \varepsilon_1 \cdot q)  - (q \cdot \varepsilon^*_4 \cdot \bar{p}_2)(\bar{p}_2 \cdot \varepsilon_1 \cdot \bar{p}_2) \;, \\
            & \mathcal{T}_5 = (\bar{p}_2 \cdot \varepsilon^*_4 \cdot \varepsilon_1 \cdot q  ) - (q \cdot \varepsilon^*_4 \cdot \varepsilon_1 \cdot \bar{p}_2)\;.
        \end{split}
    \end{equation}
    \item There are 2 tensor structures at $\mathcal{O}(q^2)$:
    \begin{equation}
        \begin{split}
            & \mathcal{T}_6 = (q \cdot \varepsilon_4^* \cdot \bar{p}_2) (q \cdot \varepsilon_1 \cdot \bar{p}_2) , \\
            & \mathcal{T}_7 = (q \cdot \varepsilon^*_4 \cdot \varepsilon_1 \cdot q), \\
        \end{split}
    \end{equation}
    \item There is 1 tensor structure at $\mathcal{O}(q^3)$:
    \begin{equation}
        \mathcal{T}_8 = (q \cdot \varepsilon^*_4 \cdot \bar{p}_2)  (q \cdot \varepsilon_1 \cdot q) -  (q \cdot \varepsilon^*_4 \cdot q) (\bar{p}_2 \cdot \varepsilon_1 \cdot q).
    \end{equation}
    \item There is 1 tensor structure at $\mathcal{O}(q^4)$:
    \begin{equation}
        \mathcal{T}_{9} =  (q \cdot \varepsilon^*_4 \cdot q) (q \cdot \varepsilon_1 \cdot q).
    \end{equation}
\end{enumerate}
In addition to these nine tensor structures there is one more crossing-symmetric structure appearing at 
$\mathcal{O}(q^2)$
\begin{equation}
	\mathcal{T}_{14} = (\bar{p}_2 \cdot \varepsilon^*_4 \cdot \bar{p}_2) (q \cdot \varepsilon_1 \cdot q) +  (q \cdot \varepsilon^*_4 \cdot q)(\bar{p}_2 \cdot \varepsilon_1 \cdot \bar{p}_2)\;,	
\end{equation}
which is evanescent, i.e., it is linearly dependent on the other crossing 
symmetric structures in four spacetime dimensions. This means that the projection matrix one 
would obtain is singular in four spacetime dimensions as there is a redundancy in the system. To 
circumvent this, we simply discard this evanescent structure and keep the other nine. There are four additional structures which are not crossing symmetric, and so should have 
vanishing coefficients (after IBP reduction at loop level)
\begin{equation}
    \begin{split}
        & \mathcal{T}_{10} =  (\bar{p}_2 \cdot \varepsilon^*_4 \cdot \bar{p}_2)  ( q \cdot \varepsilon_1  \cdot  \bar{p}_2 ) + (q \cdot \varepsilon^*_4 \cdot \bar{p}_2)(\bar{p}_2 \cdot \varepsilon_1 \cdot \bar{p}_2) , \\ 
        & \mathcal{T}_{11} = (\bar{p}_2 \cdot \varepsilon^*_4 \cdot \varepsilon_1  \cdot q) + (q \cdot \varepsilon^*_4 \cdot \varepsilon_1 \cdot \bar{p}_2), \\
        & \mathcal{T}_{12} = (\bar{p}_2 \cdot \varepsilon^*_4 \cdot \bar{p}_2)(q \cdot \varepsilon_1 \cdot q)  - (q \cdot \varepsilon^*_4 \cdot q) (\bar{p}_2 \cdot \varepsilon_1 \cdot \bar{p}_2) , \\
        & \mathcal{T}_{13} = (q \cdot \varepsilon^*_4 \cdot \bar{p}_2) (q \cdot \varepsilon_1 \cdot q)  + (q \cdot \varepsilon^*_4 \cdot q) (q \cdot \varepsilon_1 \cdot \bar{p}_2) .
    \end{split}
\end{equation}
Although the coefficients of these structures are expected to vanish, we do not assume this 
a priori. As a stringent consistency check, we include these non-crossing-symmetric 
structures explicitly in our computations and find that their coefficients indeed vanish after integration. The 
amplitude therefore decomposes as 
\begin{equation}
    M_{\phi \tiny{\mbox{H}}} = \sum_{n = 1}^{13} \mathbb{M}_n \mathcal{T}_n,
    \label{eq:AmplitudeDecomp}
\end{equation}
where $\mathcal{T}_n = \varepsilon^*_{4 \mu \nu} \mathcal{T}_n^{\mu \nu \alpha \beta
}\varepsilon_{1 \alpha \beta}$ together with
\allowdisplaybreaks
\begin{equation}
\begin{split}
&\mathcal{T}_1^{\mu \nu \alpha \beta} = \eta^{\mu \alpha}\eta^{\nu\beta}\;, \\
&\mathcal{T}_2^{\mu \nu \alpha \beta} = \bar{p}_2^\mu\bar{p}_2^\nu\bar{p}_2^\alpha\bar{p}_2^\beta\;, \\
&\mathcal{T}_3^{\mu \nu \alpha \beta} = \bar{p}_2^\mu\eta^{\nu\alpha}\bar{p}_2^\beta\;, \\
&\mathcal{T}_4^{\mu \nu \alpha \beta} = 2\bar{p}_2^\nu\bar{p}_2^\alpha \bar{p}_2^{[\mu} q_{\phantom{2}}^{\beta]}\;, \\
&\mathcal{T}_5^{\mu \nu \alpha \beta} = 2\eta^{\nu \alpha} \bar{p}_2^{[\mu} q_{\phantom{2}}^{\beta]}\;, \\
&\mathcal{T}_6^{\mu \nu \alpha \beta} = q^\mu \bar{p}_2^\nu\bar{p}_2^\alpha q^\beta\;, \\
&\mathcal{T}_7^{\mu \nu \alpha \beta} = q^\mu \eta^{\nu \alpha} q^\beta\;, \\
\end{split}
\hspace{0.2cm}
\begin{split}
&\mathcal{T}_8^{\mu \nu \alpha \beta} = 2q^\mu \bar{p}_2^{[\nu} q_{\phantom{2}}^{\alpha]}q^\beta\;, \\
&\mathcal{T}_9^{\mu \nu \alpha \beta} = q^\mu q^\nu q^\alpha q^\beta\;, \\ 
&\mathcal{T}_{10}^{\mu \nu \alpha \beta} = 2\bar{p}_2^\nu \bar{p}_2^\beta \bar{p}_2^{(\mu} q_{\phantom{2}}^{\alpha)} \;, \\
&\mathcal{T}_{11}^{\mu \nu \alpha \beta} = 2 \eta^{\nu \alpha} \bar{p}_2^{(\mu} q_{\phantom{2}}^{\beta)} \;, \\
&\mathcal{T}_{12}^{\mu \nu \alpha \beta} = \bar{p}_2^\mu \bar{p}_2^\nu q^\alpha q^\beta - q^\mu q^\nu \bar{p}_2^\alpha \bar{p}_2^\beta \;, \\
&\mathcal{T}_{13}^{\mu \nu \alpha \beta} = 2q^\mu q^\alpha  \bar{p}_2^{(\nu} q_{\phantom{2}}^{\beta)}\;.
\end{split}
\label{eq:TensorStructures}
\end{equation}
Again, the projection matrix is constructed by following the same steps as in 
\mysecref{sec:Form_factor_decomp}. More explicitly, we write (again suppressing the subscript 
$\phi \mbox{H}$)
\begin{equation}
    \sum_{\text{spin}} M \mathcal{T}^\dagger_m = \sum_{n = 1}^{13} \mathbb{M}_n \sum_{\text{spin}}\mathcal{T}_n \mathcal{T}^\dagger_m \equiv \sum_{n = 1}^{13} \mathbb{M}_n \mathcal{P}_{nm},
\end{equation}
and define our projector as
\begin{equation}
    \begin{split}
        \mathcal{P}_{nm} = \mathcal{T}_n \mathcal{T}^\dagger_m = P_{4 \mu \nu \rho \sigma} \mathcal{T}_n^{\mu \nu \alpha \beta} P_{1 \lambda \gamma \alpha \beta} \mathcal{T}_m^{\dagger \rho \sigma \lambda \gamma} = \operatorname{Tr}[P_4 \mathcal{T}_n P_1 \mathcal{T}^\dagger_m]\;,
    \end{split}
    \label{eq:ProjectionMatrix}
\end{equation}
where $P_{i}^{\mu \nu \rho \sigma}$ is the polarisation sum 
\myeqref{eq:spin2-transverse-proj} for the tensor particle legs $p_1$ and $p_4$ both 
with mass $m_1$. Then, we project the amplitude into a vector of coefficients and 
extract the form factors in exactly the same way as \mysecref{sec:Form_factor_decomp}.

%--------------------------------------------------------------------------------
\section{Resolving Spin Structures}
\label{sec:ResolvingSpinStructures}
%--------------------------------------------------------------------------------

Having factorised the spin-dependent information into tensor structures, we can now think about translating 
these polarisation structures into explicit spin vector structures. The first thing to do is to investigate
the expectation value of the Pauli-Lubanski vector in the spin-$2$ representation, and derive the spin-$2$ analogue 
of \myeqref{eq:Spin1Decomposition}. Although we will not use this formula explicitly, it is nice to 
derive it for the sake of completeness. Then, we will determine a boost matrix that maps tensor structures 
written in terms of polarisations to tensor structures written in terms of boosted 
polarisations. Indeed, we will do this for both the spin-$1$ and spin-$2$ representations which will be crucial in 
performing the spin interpolation. To finalise, we will outline a systematised strategy for the spin interpolation 
method in detail.

%--------------------------------------------------------------------------------
\subsection{Spin Decomposition}
\label{sec:Spin2Decomposition}
%--------------------------------------------------------------------------------

Here we will follow a slightly different approach to determine the spin-$2$ analogue of 
\eqref{eq:Spin1Decomposition}. Let us start by writing an ansatz for the spin decomposition of 
tensor polarisations
\begin{align}
        \bar \varepsilon^{\star \mu_1 \mu_2}_{i} \bar \varepsilon^{\nu_1 \nu_2}_{j} & = P^{\mu_1 \mu_2 \nu_1 \nu_2}_{(2)} \delta_{ij} + \frac{\alpha_1}{4} \Big[P^{\mu_1 \nu_1}_{(1)} \epsilon_{\phantom{(1)}}^{\mu_2 \nu_2 \kappa \rho} + P^{\mu_2 \nu_1}_{(1)}  \epsilon_{\phantom{(1)}}^{\mu_1 \nu_2 \kappa \rho} + P^{\mu_1 \nu_2}_{(1)} \epsilon_{\phantom{(1)}}^{\mu_2 \nu_1 \kappa \rho} \nonumber \\[0.25cm]
        & + P^{\mu_2 \nu_2}_{(1)} \epsilon_{\phantom{(1)}}^{\mu_1 \nu_1 \kappa \rho}\Big] \bar p_{1\kappa}^{\prime} \mathds{S}_{(2)\rho ij} + \alpha_2 P^{\mu_1 \mu_2}_{(1)}\big(\mathds{S}^{(\nu_1}_{(2)}\mathds{S}^{\nu_2)}_{(2)}\big)_{ij} + \alpha_3 P^{\nu_1 \nu_2}_{(1)}\big(\mathds{S}^{(\mu_1}_{(2)}\mathds{S}^{\mu_2)}_{(2)}\big)_{ij} \nonumber  \\[0.25cm]
        & + \frac{\alpha_4}{4} \Big[P^{\mu_1 \nu_1}_{(1)}\big(\mathds{S}^{(\mu_2}_{(2)}\mathds{S}^{\nu_2)}_{(2)}\big)_{ij} + P^{\mu_2 \nu_1}_{(1)}\big(\mathds{S}^{(\mu_1}_{(2)}  \mathds{S}^{\nu_2)}_{(2)}\big)_{ij} + P^{\mu_1 \nu_2}_{(1)}\big(\mathds{S}^{(\mu_2}_{(2)}\mathds{S}^{\nu_1)}_{(2)}\big)_{ij}  \nonumber  \\[0.25cm]
        & + P^{\mu_2 \nu_2}_{(1)}\big(\mathds{S}^{(\mu_1}_{(2)}\mathds{S}^{\nu_1)}_{(2)}\big)_{ij}\Big] + \frac{\alpha_5}{4}\Big[\epsilon^{\mu_1 \nu_1 \kappa}_{~~~~~~\rho} \bar p_{1\kappa}^{\prime} \big(\mathds{S}^{(\mu_2}_{(2)}\mathds{S}^{\nu_2}_{(2)}\mathds{S}^{\rho)}_{(2)}\big)_{ij}  \\[0.25cm]
        & + \epsilon^{\mu_2 \nu_1 \kappa}_{~~~~~~\rho} \bar p_{1\kappa}^{\prime} \big(\mathds{S}^{(\mu_1}_{(2)}\mathds{S}^{\nu_2}_{(2)}\mathds{S}^{\rho)}_{(2)}\big)_{ij} + \epsilon^{\mu_1 \nu_2 \kappa}_{~~~~~~\rho} \bar p_{1\kappa}^{\prime} \big(\mathds{S}^{(\mu_2}_{(2)}\mathds{S}^{\nu_1}_{(2)}\mathds{S}^{\rho)}_{(2)}\big)_{ij} \nonumber \\[0.25cm]
        & + \epsilon^{\mu_2 \nu_2 \kappa}_{~~~~~~\rho} \bar p_{1\kappa}^{\prime} \big(\mathds{S}^{(\mu_1}_{(2)}\mathds{S}^{\nu_1}_{(2)}\mathds{S}^{\rho)}_{(2)}\big)_{ij}\Big] + \alpha_6 \big(\mathds{S}^{(\mu}_{(2)}\mathds{S}^\nu_{(2)}\mathds{S}^{\rho}_{(2)}\mathds{S}^{\sigma)}_{(2)}\big)_{ij}\;,\nonumber 
    \label{eq:SpinDecomp_Spin2Ansatz_FreeCoeff}
\end{align}
where we have symmetrised our ansatz in $(\mu_1, \mu_2)$ and $(\nu_1,\nu_2)$. The right-hand side is simply all possible 
spin vector structures that can be constructed through quartic order in spin, where the $\alpha_i$ are arbitrary coefficients that 
must be fixed.

In order to fix the free coefficients in our ansatz, we will need to make use of the tensor basis 
into which we project our amplitudes. Note that the form factor basis that we constructed involves the unboosted 
polarisation tensors, whereas we require the basis written in terms of boosted polarisations. This is 
determined by simply contracting \myeqref{eq:TensorStructures} with the boosted polarisation tensors. At this point 
of our discussion, how the bases are related by Lorentz transformations is irrelevant. Then, the strategy we take 
is to contract both sides with external momenta and/or the metric, substitute the explicit spin-$2$ representation 
spin operator in order to translate everything into polarisation structures, and then project onto the 
tensor basis. The spin-$2$ representation of the spin operator reads 
\begin{align}
    \mathds{S}^\mu_{(2)ij} & = \frac{1}{2m}\epsilon^{\mu \lambda \alpha \beta} p_\lambda \varepsilon^\star_{i \gamma \kappa}(p)\big(\Sigma_{(2)\alpha \beta}\big)^{\gamma \kappa}_{~~\rho \sigma}\varepsilon^{\rho \sigma}_j(p)\;, \\
    & = \frac{2i}{m_1} \epsilon^{\mu \nu \alpha \beta} \bar{p}^\prime_{1\nu} \bar \varepsilon^\star_{i \alpha \rho}(\bar{p}^\prime_{1}) \bar \varepsilon^{\rho}_{j\beta}(\bar{p}^\prime_{1})\;, \nonumber
\end{align}
where we have substituted the explicit spin-$2$ representation of the infinitesimal Lorentz transformation to 
acquire the second equality; see \mysecref{sec:SpinVector_Spin2} and \mysecref{sec:LorentzGeneratorSpin2} for more 
details. Indeed, requiring that the equality holds will impose constraints that determine the explicit form of the 
coefficients.

It turns out that, in practice, we actually need only the first four tensor structures. The remaining structures impose 
no additional constraints. In light of this, the first tensor structure leads to the constraint
\begin{equation}
    0 = 4 + 6(\alpha_2 + \alpha_3 + 2\alpha_4) + 34 \alpha_6 \label{eq:Constraint1}\;.
\end{equation}
The second structure gives the constraints
\begin{align}
    & 0 = 1 + 6(\alpha_2 + \alpha_3 + \alpha_4) + 24 \alpha_6 \label{eq:Constraint2}\;,\\
    & 0 = 1 - 6\alpha_6 \label{eq:Constraint3}\;, \\
    & 0 = \alpha_2 + \alpha_3 + \alpha_4 + 5 \alpha_6 \label{eq:Constraint4}\;.
\end{align}
Immediately we see that \eqref{eq:Constraint3} leads to $\alpha_6 = 1/6$. Isolating $\alpha_2$ in 
\myeqref{eq:Constraint4} gives $\alpha_2 = -5/6 - \alpha_3 - \alpha_4$, which, when substituted into 
\myeqref{eq:Constraint1}, leads to $\alpha_4 = -7/9$. Putting this information together, we find that 
\myeqref{eq:Constraint2} gives $\alpha_2 + \alpha_3 = -1/18$.
The third tensor structure provides no additional constraints, as the equations are trivially satisfied. So we move 
onto the fourth tensor structure, which yields
\begin{align}
    & 0 = \alpha_2 - \alpha_3  \label{eq:Constraint5}\;, \\
    & 0 = 1 - 6i m_1 \alpha_5\label{eq:Constraint6}\;,\\
    & 0 = 6i m_1 \alpha_1 + 9 \alpha_2 - 9\alpha_3 + 14i m_1 \alpha_5 \label{eq:Constraint7}\;.
\end{align}
In this way, \eqref{eq:Constraint5} gives $\alpha_2 = \alpha_3 = -1/36$, \eqref{eq:Constraint6} leads to 
$\alpha_5 = -i/6m_1$, and, finally, using all fixed coefficients, \eqref{eq:Constraint7} yields $\alpha_1 = 7i/18m_1$. 
Notice that we obtain factors of $m_1$ because we are working with 
$\bar p_1^{\,\prime \mu} = \bar p_1^{\,\mu}(m_1/\bar m_1)$. 

Thus, the spin decomposition of boosted polarisation tensors for spin-$2$ reads
\begin{align}
        \bar \varepsilon^{\star \mu_1 \mu_2}_{i} \bar \varepsilon^{\nu_1 \nu_2}_{j} & = P^{\mu_1 \mu_2 \nu_1 \nu_2}_{(2)} \delta_{ij} + \frac{7i}{72 m_1} \Big[P^{\mu_1 \nu_1}_{(1)} \epsilon_{\phantom{(1)}}^{\mu_2 \nu_2 \kappa \rho} + P^{\mu_2 \nu_1}_{(1)}  \epsilon_{\phantom{(1)}}^{\mu_1 \nu_2 \kappa \rho} + P^{\mu_1 \nu_2}_{(1)} \epsilon_{\phantom{(1)}}^{\mu_2 \nu_1 \kappa \rho} \nonumber \\[0.25cm]
        & + P^{\mu_2 \nu_2}_{(1)} \epsilon_{\phantom{(1)}}^{\mu_1 \nu_1 \kappa \rho}\Big] \bar p_{1\kappa}^{\prime} \mathds{S}_{(2)\rho ij} - \frac{1}{36} P^{\mu_1 \mu_2}_{(1)}\big(\mathds{S}^{(\nu_1}_{(2)}\mathds{S}^{\nu_2)}_{(2)}\big)_{ij} - \frac{1}{36} P^{\nu_1 \nu_2}_{(1)}\big(\mathds{S}^{(\mu_1}_{(2)}\mathds{S}^{\mu_2)}_{(2)}\big)_{ij} \nonumber \\[0.25cm]
        & - \frac{7}{36} \Big[P^{\mu_1 \nu_1}_{(1)}\big(\mathds{S}^{(\mu_2}_{(2)}\mathds{S}^{\nu_2)}_{(2)}\big)_{ij} + P^{\mu_2 \nu_1}_{(1)}\big(\mathds{S}^{(\mu_1}_{(2)}  \mathds{S}^{\nu_2)}_{(2)}\big)_{ij} + P^{\mu_1 \nu_2}_{(1)}\big(\mathds{S}^{(\mu_2}_{(2)}\mathds{S}^{\nu_1)}_{(2)}\big)_{ij} \nonumber  \\[0.25cm]
        & + P^{\mu_2 \nu_2}_{(1)}\big(\mathds{S}^{(\mu_1}_{(2)}\mathds{S}^{\nu_1)}_{(2)}\big)_{ij}\Big] - \frac{i}{24 m_1}\Big[\epsilon^{\mu_1 \nu_1 \kappa}_{~~~~~~\rho} \bar p_{1\kappa}^{\prime} \big(\mathds{S}^{(\mu_2}_{(2)}\mathds{S}^{\nu_2}_{(2)}\mathds{S}^{\rho)}_{(2)}\big)_{ij}  \\[0.25cm]
        & + \epsilon^{\mu_2 \nu_1 \kappa}_{~~~~~~\rho} \bar p_{1\kappa}^{\prime} \big(\mathds{S}^{(\mu_1}_{(2)}\mathds{S}^{\nu_2}_{(2)}\mathds{S}^{\rho)}_{(2)}\big)_{ij} + \epsilon^{\mu_1 \nu_2 \kappa}_{~~~~~~\rho} \bar p_{1\kappa}^{\prime} \big(\mathds{S}^{(\mu_2}_{(2)}\mathds{S}^{\nu_1}_{(2)}\mathds{S}^{\rho)}_{(2)}\big)_{ij}\nonumber  \\[0.25cm]
        & + \epsilon^{\mu_2 \nu_2 \kappa}_{~~~~~~\rho} \bar p_{1\kappa}^{\prime} \big(\mathds{S}^{(\mu_1}_{(2)}\mathds{S}^{\nu_1}_{(2)}\mathds{S}^{\rho)}_{(2)}\big)_{ij}\Big] + \frac{1}{6} \big(\mathds{S}^{(\mu}_{(2)}\mathds{S}^\nu_{(2)}\mathds{S}^{\rho}_{(2)}\mathds{S}^{\sigma)}_{(2)}\big)_{ij}\;,\nonumber 
    \label{eq:SpinDecomp_Spin2Operator}
\end{align}
which is in exact agreement with Refs.~\cite{Cangemi:2022abk,Gatica:2025uhx}\footnote{As a test, we have 
also re-derived the spin-$1$ spin decomposition using this approach.}. The final step is to perform the weighted sum 
over polarisation tensors of different spin states
\begin{equation}
    \bar{\varepsilon}^{\star \mu \nu} = \sum_j \omega_j \bar{\varepsilon}^{\star \mu \nu}_j\;, \quad \sum_i \omega_i^\star \omega_i = 1\;,
\end{equation}
such that the expectation value of the product of spin vector operators becomes products of classical spin vectors. More 
explicitly, the final formula is 
\begin{align}
        \bar \varepsilon^{\star \mu_1 \mu_2} \bar \varepsilon^{\nu_1 \nu_2} & = P^{\mu_1 \mu_2 \nu_1 \nu_2}_{(2)}  + \frac{7i}{72 m_1} \Big[P^{\mu_1 \nu_1}_{(1)} \epsilon_{\phantom{(1)}}^{\mu_2 \nu_2 \kappa \rho} + P^{\mu_2 \nu_1}_{(1)}  \epsilon_{\phantom{(1)}}^{\mu_1 \nu_2 \kappa \rho} + P^{\mu_1 \nu_2}_{(1)} \epsilon_{\phantom{(1)}}^{\mu_2 \nu_1 \kappa \rho} \nonumber \\[0.25cm]
        & + P^{\mu_2 \nu_2}_{(1)} \epsilon_{\phantom{(1)}}^{\mu_1 \nu_1 \kappa \rho}\Big] \bar p_{1\kappa}^{\prime} S_{\rho} - \frac{1}{36} P^{\mu_1 \mu_2}_{(1)}S^{\nu_1}S^{\nu_2} - \frac{1}{36} P^{\nu_1 \nu_2}_{(1)}S^{\mu_1}S^{\mu_2} \nonumber \\[0.25cm]
        & - \frac{7}{36} \Big[P^{\mu_1 \nu_1}_{(1)}S^{\mu_2}S^{\nu_2} + P^{\mu_2 \nu_1}_{(1)}S^{\mu_1}  S^{\nu_2} + P^{\mu_1 \nu_2}_{(1)} S^{\mu_2}S^{\nu_1} + P^{\mu_2 \nu_2}_{(1)}S^{\mu_1}S^{\nu_1}\Big] \nonumber \\[0.25cm]
        &  - \frac{i}{24 m_1}\Big[\epsilon^{\mu_1 \nu_1 \kappa}_{~~~~~~\rho} \bar p_{1\kappa}^{\prime} S^{\mu_2}S^{\nu_2}S^{\rho} + \epsilon^{\mu_2 \nu_1 \kappa}_{~~~~~~\rho} \bar p_{1\kappa}^{\prime}S^{\mu_1}S^{\nu_2}S^{\rho} + \epsilon^{\mu_1 \nu_2 \kappa}_{~~~~~~\rho} \bar p_{1\kappa}^{\prime}S^{\mu_2}S^{\nu_1}S^{\rho} \nonumber \\[0.25cm]
        & + \epsilon^{\mu_2 \nu_2 \kappa}_{~~~~~~\rho} \bar p_{1\kappa}^{\prime} S^{\mu_1}S^{\nu_1}S^{\rho}\Big] + \frac{1}{6} S^{\mu}S^{\nu} S^{\rho}S^{\sigma}\;,
    \label{eq:SpinDecomp_Spin2Classical}
\end{align}
where products of classical spin vectors are mapped to symmetrised products of spin operators in 
the spin-$2$ representation
\begin{equation}
	\label{eq:prodOfSpin}
	S^\mu S^\rho \ldots ~S^\sigma S^\nu   = \sum_{i,k,...}\omega_i^\star \mathds{S}^{(\mu}_{(2)ik}\mathds{S}^{\rho}_{(2)k\ell} \ldots \mathds{S}^{\sigma}_{(2)rm} \mathds{S}^{\nu)}_{(2)mj}\omega_j\;.
\end{equation}

%--------------------------------------------------------------------------------
\subsection{Boosting Polarisations}
\label{sec:BoostingPolarisations}
%--------------------------------------------------------------------------------

As before, the polarisations that appear in our scattering amplitudes are related to the polarisations that 
play a role in the definition of the spin vector via Lorentz transformations. For the case of polarisation vectors, 
we have already discussed the explicit transformations in \mysecref{sec:ExtractingSpin}. Specifically, the explicit 
boosts of the tensor structures required for the computation through the quadratic order in spin were explicitly given in 
\myeqref{eq:Spin1TensorBoostToSpin}, albeit we simultaneously traded polarisation structures for spin structures. 
For computations through the quartic order in spin, we must perform boosts of these tensor structures to higher orders in 
the $|q|$ expansion, namely the $|q|^4$ order. Note that we will not be explicitly translating to polarisation 
structures. Of course, the same will be done for the tensor structures describing scalar-tensor scattering. 

\textbf{Spin-1 Structures.}~Performing the Lorentz transformation given by \myeqref{eq:polarisationBoost} for the crossing 
symmetric tensor structures in \myeqref{eq:Spin1TensorStructures} and expanding consistently to $|q|^4$ yields 
\begin{equation}
	\bar{T}_i = B_{ij} T_j\;,
\end{equation}
where $\bar{T}_i$ are the boosted tensor structures written in terms of $\bar{\varepsilon}$ and 
$\bar{\varepsilon}^\star$, while $B_{ij}$ is the boost matrix truncated to the $\mathcal{O}(|q|^4)$: 
\begin{equation}
  B_{ij} = \begin{pmatrix}
    1 & -\frac{1}{2\bar{m}_1^2} + \frac{q^2}{8\bar{m}_1^4}-\frac{q^4}{32 \bar{m}_1^6} & 0 & 0 \\[0.2cm]
    0 & 1 - \frac{q^2}{4\bar{m}_1^2} + \frac{q^4}{16 \bar{m}_1^4} & 0 & 0 \\[0.2cm]
    0 & - \frac{y^2 \bar{m}_2^2  q^2}{4 \bar{m}_1^2} + \frac{y^2 \bar{m}_2^2 q^4}{16 \bar{m}_1^4} & 1 & - \frac{y \bar{m}_2  q^2}{2 \bar{m}_1} + \frac{y \bar{m}_2 q^4}{16 \bar{m}_1^3}\\[0.2cm]
    0 & \frac{y \bar{m}_2}{\bar{m}_1} - \frac{y \bar{m}_2 q^2}{4\bar{m}_1^3} +\frac{y \bar{m}_2 q^4}{16\bar{m}_1^5} & 0 & 1 - \frac{q^2}{8 \bar{m}_1^2} + \frac{3 q^4}{128 \bar{m}_1^4} \\[0.2cm]
  \end{pmatrix}\;.
\end{equation}
It is important to note that some tensor structures contain explicit powers of $q$, which we must take into 
account when consistently expanding to the $|q|^4$ order. 

\textbf{Spin-2 Structures.} To perform exactly the same thing for spin tensors we begin with the 
Lorentz transformations derived in Appendix~\mysecref{sec:CovariantBoostedpolarisations}
\begin{equation}
    \begin{split}
        & \Lambda(\psi)^{\mu}_{~\nu} = \delta^{\mu}_{~\nu} + \Lambda_1(\bar{p}_1^\mu q_\nu - q^\mu \bar{p}_{1\nu}) + \frac{\Lambda_2}{\bar{m}_1^2}(q^2 \bar{p}_1^\mu\bar{p}_{1 \nu} +\bar{m}_1^2 q^\mu q_\nu)\;, \\
        & \Lambda(-\psi)^{\mu}_{~\nu} = \delta^{\mu}_{~\nu} - \Lambda_1(\bar{p}_1^\mu q_\nu - q^\mu \bar{p}_{1\nu}) + \frac{\Lambda_2}{\bar{m}_1^2}(q^2 \bar{p}_1^\mu\bar{p}_{1 \nu} +\bar{m}_1^2 q^\mu q_\nu)\;,
    \end{split}
\end{equation}
where $\Lambda_1, \Lambda_2$ are defined in \myeqref{eq:BoostCoeffDefns} and $\psi$ is defined 
in \myeqref{eq:BoostAngleDefns}. Explicitly, performing the boosts yields
\begin{equation}
    \begin{split}
        \varepsilon_1^{\mu \nu} = \Lambda(\psi)^{\mu}_{~\alpha} \Lambda(\psi)^{\nu}_{~\beta} \bar{\varepsilon}^{\alpha \beta} = \bar{\varepsilon}^{\mu \nu} + 2 \Lambda_1 q_\alpha \bar{\varepsilon}_{\phantom{1}}^{\alpha (\mu} \bar{p}_1^{\nu)} & + 2 \Lambda_2 q_\alpha \bar{\varepsilon}^{\alpha (\mu} q^{\nu)} + 2 \Lambda_1 \Lambda_2 q_\alpha \bar{\varepsilon}^{\alpha \beta} q_\beta \bar{p}_1^{(\mu} q^{\nu)}_{\phantom{1}} \\
        & \hspace{-1cm}+ q_\alpha \bar{\varepsilon}^{\alpha \beta} q_\beta ( \Lambda_1^2 \bar{p}_1^{\mu} \bar{p}_1^{\nu} + \Lambda_2^2 q^\mu q^\nu)\;,
    \end{split}
\end{equation}
\begin{equation}
    \begin{split}
        \varepsilon_4^{\star \mu \nu} = \Lambda(-\psi)^{\mu}_{~\alpha} \Lambda(-\psi)^{\nu}_{~\beta} & \bar{\varepsilon}^{\star \alpha \beta} = \bar{\varepsilon}^{\star \mu \nu}_{\phantom{1}} - 2 \Lambda_1 q_\alpha \bar{\varepsilon}^{\star \alpha (\mu} \bar{p}_1^{\nu)} + 2 \Lambda_2 q_\alpha \bar{\varepsilon}^{\star \alpha (\mu} q^{\nu)} \\
        & \hspace{2cm} - 2 \Lambda_1 \Lambda_2 q_\alpha \bar{\varepsilon}^{\star \alpha \beta} q_\beta \bar{p}_1^{(\mu} q^{\nu)}_{\phantom{1}} \\
        & \hspace{2cm} + q_\alpha \bar{\varepsilon}^{\star \alpha \beta} q_\beta ( \Lambda_1^2 \bar{p}_1^{\mu} \bar{p}_1^{\nu} + \Lambda_2^2 q^\mu q^\nu)\;,
    \end{split}
\end{equation}
which preserves the symmetric tracelessness of the polarisation tensors. Expanding this consistently to $|q|^4$ and 
projecting onto the tensor basis yields
\begin{equation}
	\bar{\mathcal{T}}_i = \mathcal{B}_{ij} \mathcal{T}_j\;.
\end{equation}
Here $\bar{\mathcal{T}}_i$ are the boosted tensor structures and, for visual purposes, we split the boost matrix as
\begin{equation}
	\mathcal{B}_{ij} = \mathcal{B}_{ij}^{(0)} + q^2 \mathcal{B}_{ij}^{(1)} + q^4 \mathcal{B}_{ij}^{(2)} + \ldots
\end{equation}
where $\ldots$ denote higher corrections that we 
neglect. The matrices appearing here take the form
\vspace{1cm}
\begin{equation}
\mathcal{B}_{ij}^{(0)} = \begin{pmatrix}
    1 & 0 & 0 & 0 & 0 & -\frac{1}{\bar{m}_1^2} & 0 & 0 & \frac{1}{4\bar{m}_1^4}\\[0.2cm]
    0 & 1 & 0  & \frac{y\bar{m}_2}{\bar{m}_1} & 0 & -\frac{y^2 \bar{m}_2^2}{2\bar{m}_1^2} & \frac{y^2(y^2-1) \bar{m}_2^4}{2\bar{m}_1^2} & -\frac{y^3\bar{m}_2^3}{4\bar{m}_1^3} & \frac{y^4\bar{m}_2^4}{16\bar{m}_1^4}\\[0.2cm]
    0 & 0 & 1 & 0 & \frac{y\bar{m}_2}{2\bar{m}_1}  & -\frac{1}{2 \bar{m}_1^2} & -\frac{y^2\bar{m}_2^2}{4 \bar{m}_1^2} & -\frac{y\bar{m}_2}{4 \bar{m}_1^3} & \frac{y^2\bar{m}_2^2}{8 \bar{m}_1^4}\\[0.2cm]
    0 & 0 & 0 & 1 & 0 & -\frac{y \bar{m}_2}{\bar{m}_1}  & \frac{y(y^2-1)\bar{m}_2^3}{\bar{m}_1} & -\frac{3 y^2 \bar{m}_2^2}{4 \bar{m}_1^2} & \frac{y^3 \bar{m}_2^3}{4 \bar{m}_1^3} \\[0.2cm]
    0 & 0 & 0 & 0 & 1  & 0 & -\frac{y \bar{m}_2}{\bar{m}_1} & -\frac{1}{2\bar{m}_1^2}  & \frac{y \bar{m}_2}{2\bar{m}_1^3} \\[0.2cm]
    0 & 0 & 0 & 0 & 0 & 1 & 0 & \frac{y \bar{m}_2}{2\bar{m}_1} & -\frac{y^2 \bar{m}_2^2}{4\bar{m}_1^2}\\[0.2cm]
    0 & 0 & 0 & 0 & 0 & 0 & 1 & 0 & -\frac{1}{2\bar{m}_1^2}\\[0.2cm]
    0 & 0 & 0 & 0 & 0 & 0 & 0 & 1 & -\frac{y \bar{m}_2}{\bar{m}_1}\\[0.2cm]
    0 & 0 & 0 & 0 & 0 & 0 & 0 & 0 & 1\\[0.2cm]
  \end{pmatrix},
\end{equation}
\vspace{2cm}
\begin{equation}
\hspace{-1.5cm}
\mathcal{B}_{ij}^{(1)} = \begin{pmatrix}
    0 & 0 & 0 & 0 & 0 & 0 & \frac{1}{4\bar{m}_1^4} & 0 & -\frac{1}{8\bar{m}_1^6}\\[0.2cm]
    -\frac{y^2(y^2-1)\bar{m}_2^4}{4 \bar{m}_1^2} & 0 & -\frac{y^2\bar{m}_2^2}{2\bar{m}_1^2} & -\frac{y \bar{m}_2}{8\bar{m}_1^3} & 0 & \frac{y^2\bar{m}_2^2}{8\bar{m}_1^4} &  -\frac{y^2(y^2-1)\bar{m}_2^4}{8 \bar{m}_1^4} & \frac{3y^3\bar{m}_2^3}{32\bar{m}_1^5} & -\frac{y^4\bar{m}_2^4}{32\bar{m}_1^6} \\[0.2cm]
    0 & 0 & 0 & 0 & -\frac{y \bar{m}_2}{16 \bar{m}_1^3} & \frac{1}{8 \bar{m}_1^4} & \frac{y^2 \bar{m}_2^2}{16 \bar{m}_1^4} & \frac{3 y \bar{m}_2}{32 \bar{m}_1^5} & -\frac{y^2 \bar{m}_2^2}{16 \bar{m}_1^6}\\[0.2cm]
    -\frac{y(y^2-1)\bar{m}_2^3}{2 \bar{m}_1} & 0 & -\frac{y \bar{m}_2}{\bar{m}_1} & -\frac{1}{8 \bar{m}_1^2} & 0 & \frac{y \bar{m}_2}{4 \bar{m}_1^3} & -\frac{y(y^2-1)\bar{m}_2^3}{4 \bar{m}_1^3} & \frac{9y^2 \bar{m}_2^2}{32\bar{m}_1^4} & -\frac{y^3 \bar{m}_2^3}{8\bar{m}_1^5}\\[0.2cm]
    0 & 0 & 0 & 0 & -\frac{1}{8 \bar{m}_1^2} & 0 & \frac{y \bar{m}_2}{4\bar{m}_1^3} & \frac{3}{16 \bar{m}_1^4} & -\frac{y \bar{m}_2}{4\bar{m}_1^5}\\[0.2cm]
    0 & 0 & 0 & 0 & 0 & -\frac{1}{4\bar{m}_1^2} & 0 & -\frac{3 y \bar{m}_2}{16\bar{m}_1^3} & \frac{y^2 \bar{m}_2^2}{8\bar{m}_1^4}\\[0.2cm]
    0 & 0 & 0 & 0 & 0 & 0 & -\frac{1}{4\bar{m}_1^2} & 0 & \frac{1}{4\bar{m}_1^4}\\[0.2cm]
    0 & 0 & 0 & 0 & 0 & 0 & 0 & -\frac{3}{8\bar{m}_1^2} & \frac{y \bar{m}_2}{2\bar{m}_1^3}\\[0.2cm]
    0 & 0 & 0 & 0 & 0 & 0 & 0 & 0 & -\frac{1}{2\bar{m}_1^2}\\[0.2cm]  
    \end{pmatrix},
\end{equation}
\begin{equation}
\hspace{-1.5cm}
\mathcal{B}_{ij}^{(2)} = \begin{pmatrix}
    0 & 0 & 0 & 0 & 0 & 0 & -\frac{1}{16 \bar{m}_1^6} & 0 & \frac{3}{64 \bar{m}_1^8}\\[0.2cm]
    \frac{y^2(y^2-1)\bar{m}_2^4}{16 \bar{m}_1^4} & 0 &  \frac{y^2 \bar{m}_2^2}{8 \bar{m}_1^4} & \frac{3 y \bar{m}_2}{128 \bar{m}_1^5} & 0 & -\frac{y^2 \bar{m}_2^2}{32 \bar{m}_1^6} & \frac{y^2(y^2-1)\bar{m}_2^4}{32 \bar{m}_1^6} & -\frac{15 y^3 \bar{m}_2^3}{512 \bar{m}_1^7} & \frac{3 y^4 \bar{m}_2^4}{256 \bar{m}_1^8}\\[0.2cm]
    0 & 0 & 0 & 0 & \frac{3 y \bar{m}_2}{256\bar{m}_1^5} & -\frac{1}{32\bar{m}_1^6} & -\frac{y^2 \bar{m}_2^2}{64 \bar{m}_1^6} & -\frac{15 y \bar{m}_2}{512 \bar{m}_1^7} & \frac{3 y^2 \bar{m}_2^2}{128 \bar{m}_1^8}\\[0.2cm]
     \frac{y(y^2-1)\bar{m}_2^3}{8 \bar{m}_1^3} & 0 & \frac{y \bar{m}_2}{4 \bar{m}_1^3} & \frac{3}{128 \bar{m}_1^4} & 0 & -\frac{y \bar{m}_2}{16\bar{m}_1^5} & \frac{y(y^2-1)\bar{m}_2^3}{16\bar{m}_1^5} & -\frac{45 y^2 \bar{m}_2^2}{512\bar{m}_1^6} & \frac{3 y^3 \bar{m}_2^3}{64\bar{m}_1^7}\\[0.2cm]
    0 & 0 & 0 & 0 & \frac{3}{128 \bar{m}_1^4} & 0 & -\frac{y \bar{m}_2}{16\bar{m}_1^5} & -\frac{15}{256 \bar{m}_1^6} & \frac{3 y \bar{m}_2}{32\bar{m}_1^7} \\[0.2cm]
    0 & 0 & 0 & 0 & 0 & \frac{1}{16 \bar{m}_1^4} & 0 & \frac{15 y \bar{m}_2}{256 \bar{m}_1^5} & - \frac{3 y^2 \bar{m}_2^2}{64\bar{m}_1^6}\\[0.2cm]
    0 & 0 & 0 & 0 & 0 & 0 & \frac{1}{16 \bar{m}_1^4} & 0 & -\frac{3}{32 \bar{m}_1^6}\\[0.2cm]
    0 & 0 & 0 & 0 & 0 & 0 & 0 & \frac{15}{128 \bar{m}_1^4} & -\frac{3 y \bar{m}_2}{16\bar{m}_1^5}\\[0.2cm]
    0 & 0 & 0 & 0 & 0 & 0 & 0 & 0 & \frac{3}{16 \bar{m}_1^4}\\[0.2cm]  \end{pmatrix}\;.
\end{equation}

%--------------------------------------------------------------------------------
\subsection{Spin Interpolation: A Systematised Strategy}
\label{sec:SpinInterpolation_spin2}
%--------------------------------------------------------------------------------

All of our discussions thus far have led us to this moment: we now have all of the necessary 
knowledge to perform the spin interpolation through the quartic order in spin. More specifically,
we have the scattering amplitudes through two loops for the three types of processes that are required: 
scalar-scalar, scalar-vector, and scalar-tensor. For processes that involve 
external massive spinning legs we have further performed a form factor decomposition to 
essentially factorise the spin information from the theory-dependent information. Then, in order 
to make a connection with polarisations that appear in the definition of the spin vector, in 
\mysecref{sec:BoostingPolarisations} we have determined the necessary boost matrices that relate 
unboosted and boosted tensor structures. On the side of constructing scattering amplitudes, this 
is everything that we need.

The spin interpolation is performed as follows. The very first step is to construct an ansatz for 
the finite part of the amplitude through the quartic order in spin:
\begin{equation}
	\mathcal{M}_{\text{ansatz}}^{(L)} = \frac{\kappa^{2+2L}}{(-q^2)^{1-L/2+L\epsilon}}\sum_{n = 0}^{4}\sum_{m } \alpha^{(L, n,m)} \mathcal{J}^{(n,m)}\;,
	\label{eq:spinningAnsatz}
\end{equation}
where the spin structures $\mathcal{J}^{(n,m)}$ are summarised in Table~\ref{tab:spinStructures}. The spin structures 
are written in a convenient basis in terms of 
\begin{equation}
 	\Omega_\pm = (q\cdot S)^2 \pm q^2 S^2\;,
 	\label{eq:omega}
\end{equation}
which is useful for investigating spin-shift symmetry; in fact, $\Omega_-$ preserves the spin-shift 
symmetry, whereas $\Omega_+$ breaks this symmetry. In this ansatz, the spin coefficients $\alpha^{(L, n, m)}$ are 
functions of kinematic variables and admit an expansion in $|q|$
\begin{equation}
	\alpha^{(L, n, m)}(q) \equiv \alpha^{(L, n, m)} + q^2 \alpha^{(L, n, m)}_{q^2} + q^4 \alpha^{(L, n, m)}_{q^4} + \ldots\;,
\end{equation}
where the leading contribution is understood to be the classical coefficients while all subleading coefficients 
are quantum. It is important to note that we will have to consistently expand to $\mathcal{O}(|q|^4)$ such 
that $\ldots$ denotes higher-order terms that are irrelevant for our discussion. The crucial assumption in 
the construction of this ansatz is that it is said to be valid for \textit{any} spin representation. This 
means that the spin coefficients are the same no matter what spin representation we evaluate this ansatz in. 
Indeed this is an explicit manifestation of spin universality which we understand to hold also at the quantum 
orders. It is worth stressing that, although we assume that spin universality holds at the quantum orders, if it 
were not true then we would never be able to faithfully extract all spin Casimir terms. 

The next step is to fix all of the coefficients appearing in \myeqref{eq:spinningAnsatz}. To this end, we 
evaluate the ansatz in the $s=0, 1$, and $2$ representations and expand consistently to 
$\mathcal{O}(|q|^4)$. In doing so, we trade all spin structures for symmetrised products of spin operators as 
in \myeqref{eq:prodOfSpin}, where the $\mathds{S}_{ij}$ are evaluated in the respective representation. The outcome is that all spin 
structures are traded in favour of polarisation structures. Indeed for $s=1$ and $s=2$ we perform a form factor 
decomposition onto the appropriate basis of tensor structures written in terms of boosted polarisations 
$\bar{\varepsilon}$ and $\bar{\varepsilon}^\star$. As the steps that follow are independent of the loop order, we will 
strip the loop-dependent prefactor in \myeqref{eq:spinningAnsatz} by writing 
\begin{equation}
	\mathcal{M}_{\text{ansatz}}^{(L)} = \frac{\kappa^{2+2L}}{(-q^2)^{1-L/2+L\epsilon}}\bar{\mathcal{M}}_{\text{ansatz}}^{(L)}\;.
\end{equation}
Explicitly, for $s = 0$ we have
\begin{equation}
	\bar{\mathcal{M}}_{\text{ansatz}}^{(L)} \Big\rvert_{s = 0} = \alpha^{(L, 0, 0)} + \alpha^{(L, 0, 0)}_{q^2} q^2 + \alpha^{(L, 0, 0)}_{q^4} q^4 + \ldots\;,
\end{equation}
where $\ldots$ denote higher-order corrections in $|q|$ that we ignore. For $s=1$ we find
\begin{align}
	\bar{\mathcal{M}}_{\text{ansatz}}^{(L)} \Big\rvert_{s = 1} = \sum_{i = 1}^{4} \beta_i \bar{T}_i + \ldots\;,
\end{align}
\begin{table}[t]
\hspace*{-2.8cm}
\centering
\renewcommand{\arraystretch}{1.2}
\setlength{\tabcolsep}{8pt}
\begin{tabular}{c|cccccc}
\toprule
$\mathcal{J}_{(n,m)}$ 
& $m=0$ 
& $m=1$ 
& $m=2$ 
& $m=3$ 
& $m=4$ 
& $m=5$ \\
\midrule
$n=0$ 
& $\mathds{1}$ 
& -- 
& -- 
& -- 
& -- 
& -- \\

$n=1$ 
& $i(n \cdot S)$ 
& -- 
& -- 
& -- 
& -- 
& -- \\

$n=2$ 
& $\Omega_-$ 
& $\Omega_+$ 
& $q^2 (\bar{p}_2 \cdot S)^2$ 
& -- 
& -- 
& -- \\

$n=3$ 
& $i(n \cdot S)\Omega_-$ 
& $i(n \cdot S)\Omega_+$ 
& $i q^2(n \cdot S)(\bar{p}_2 \cdot S)^2$ 
& -- 
& -- 
& -- \\

$n=4$ 
& $\Omega_-^2$ 
& $\Omega_+^2$ 
& $\Omega_+\Omega_-$ 
& $q^2 (\bar{p}_2 \cdot S)^2\Omega_-$ 
& $q^2 (\bar{p}_2 \cdot S)^2\Omega_+$ 
& $q^4 (\bar{p}_2 \cdot S)^4$ \\
\bottomrule
\end{tabular}

\caption{Spin structures for the ansatz~\eqref{eq:spinningAnsatz}, with 
$n^\mu \equiv -\epsilon^{\mu\nu\alpha\beta}q_\nu \bar{p}_{1\alpha}\bar{p}_{2\beta}$ 
and the $\Omega_\pm$ structures defined in \eqref{eq:omega}.}
\label{tab:spinStructures}
\end{table}
with 
\begin{align}
	\beta_1 & = - \alpha^{(L, 0, 0)} + q^2\Big[-\alpha^{(L, 2, 0)} + 3 \alpha^{(L, 2, 1)} - (y^2-1)\bar{m}_2^2 \alpha^{(L, 2, 2)} - \alpha^{(L, 0, 0)}_{q^2}\Big] \\
	& + \frac{q^4}{6}\Big[-8 \alpha^{(L, 4, 0)} - 44 \alpha^{(L, 4, 1)} + 14 \alpha^{(L, 4, 2)} - 6 \Big(\alpha^{(L, 0, 0)}_{q^4} + \alpha^{(L, 2, 0)}_{q^2} - 3\alpha^{(L, 2, 1)}_{q^2} \Big) \nonumber \\
	& + (y^2-1)\bar{m}_2^2 \Big( -7 \alpha^{(L, 4, 3)} + 11 \alpha^{(L, 4, 4)} - 6 (y^2-1)\bar{m}_2^2 \alpha^{(L, 4, 5)} -6 \alpha^{(L, 2, 2)}_{q^2}\Big) \Big] \nonumber \;,
\end{align}
\begin{align}
	\beta_2 & =  -\alpha^{(L, 2, 0)} -\alpha^{(L, 2, 1)} + \frac{q^2}{6} \Big[-8 \alpha^{(L, 4, 0)} + 20 \alpha^{(L, 4, 1)} + 6 \alpha^{(L, 4, 2)} \\
	& - (y^2-1)\bar{m}_2^2 \Big( \alpha^{(L, 4, 3)} + \alpha^{(L, 4, 4)}\Big) - 6\Big( \alpha^{(L, 2, 0)}_{q^2} + \alpha^{(L, 2, 1)}_{q^2}\Big)\Big] \nonumber \;,
\end{align}
\begin{align}
	\hspace{-0.25cm}
	\beta_3 & =  -\alpha^{(L, 2, 3)} + q^2\Big[ - \alpha^{(L, 4, 3)} + \frac{4}{3} \alpha^{(L, 4, 4)} - (y^2-1)\bar{m}_2^2 \alpha^{(L, 4, 5)} - \alpha^{(L, 2, 2)}_{q^2}\Big]\;,
\end{align}
\begin{align}
	\beta_4 & = -\bar{m}_1 \alpha^{(L, 1, 0)} + \frac{\bar{m}_1 q^2}{3}\Big[- 4 \alpha^{(L, 3, 0)} + 6 \alpha^{(L, 3, 1)} \\
	& \hspace{3cm} - (y^2-1)\bar{m}_2^2 \alpha^{(L, 3, 2)} - 3\alpha^{(L, 1, 0)}_{q^2}\Big]\;. \nonumber
\end{align}
Similarly, for the $s = 2$ case we find
\begin{align}
	\bar{\mathcal{M}}_{\text{ansatz}}^{(L)} \Big\rvert_{s = 2} = \sum_{i = 1}^{9} \gamma_i \bar{\mathcal{T}}_i + \ldots\;,
\end{align}
with 
\begin{align}
	\hspace{-1cm}
	\gamma_1 & = \alpha^{(L, 0, 0)} + q^2 \Big[2 \alpha^{(L, 2, 0)} - 10 \alpha^{(L, 2, 1)} + 4(y^2-1)\bar{m}_2^2 \alpha^{(L, 2, 2)} + \alpha^{(L, 0, 0)}_{q^2}\Big] \\
	& + \frac{q^4}{3} \Big[ 20\alpha^{(L, 4, 0)} + 3 \alpha^{(L, 4, 1)} - 108 \alpha^{(L, 4, 2)} + 3\alpha^{(L, 0, 0)}_{q^4} + 6\alpha^{(L, 2, 0)}_{q^2} - 30 \alpha^{(L, 2, 1)}_{q^2} \nonumber \\
	&  + 2 (y^2-1) \bar{m}_2^2 \Big( 23  \alpha^{(L, 4, 3)} + 6 \Big(-7 \alpha^{(L, 4, 4)} + 4(y^2-1)\bar{m}_2^2  \alpha^{(L, 4, 5)} +  \alpha^{(L, 2, 2)}_{q^2}\Big)\Big)\Big]\;, \nonumber
\end{align}
\begin{align}
	\gamma_2 = 6 q^4 \alpha^{(L, 4, 5)}\;,
\end{align}
\begin{align}
	\hspace{-1cm}
	\gamma_3 = 6 q^2 \alpha^{(L, 2, 2)} + q^4 \Big[24 \alpha^{(L, 4, 3)} - 38 \alpha^{(L, 4, 4)} + 30 (y^2-1) \bar{m}_2^2 \alpha^{(L, 4, 5)} + 5 \alpha^{(L, 2, 2)}_{q^2}\Big]\;,
\end{align}
\begin{align}
	\gamma_4 = - 6 \bar{m}_1 q^2 \alpha^{(L, 3, 2)}\;,
\end{align}
\begin{align}
	\hspace{-1cm}
	\gamma_5 = - 2 \bar{m}_1 \alpha^{(L, 1, 0)} - \frac{2}{3}\bar{m}_1 q^2 \Big[10 \alpha^{(L, 3, 0)} -24 \alpha^{(L, 3, 1)} + 7 (y^2-1) \bar{m}_2^2 \alpha^{(L, 3, 2)} + \alpha^{(L, 1, 0)}_{q^2} \Big]\;,
\end{align}
\begin{align}
	\gamma_6 = 6 q^2 \Big[ \alpha^{(L, 3, 0)} +  \alpha^{(L, 3, 1)}\Big]\;,
\end{align}
\begin{align}
	\gamma_7 & = 6 \Big[\alpha^{(L, 2, 0)} +  \alpha^{(L, 2, 1)} \Big] + q^2 \Big[32  \alpha^{(L, 4, 0)} - 92  \alpha^{(L, 4, 1)} - 30  \alpha^{(L, 4, 2)} \\ 
	& + 7(y^2-1) \bar{m}_2^2 \Big( \alpha^{(L, 4, 3)} + \alpha^{(L, 4, 4)}\Big) + 6 \Big(\alpha^{(L, 2, 0)}_{q^2} + \alpha^{(L, 2, 1)}_{q^2}\Big)\Big] \nonumber \;,
\end{align}
\begin{align}
	\gamma_8 = - 6\bar{m}_1 \Big[\alpha^{(L, 3, 0)} + \alpha^{(L, 3, 1)}\Big]\;,
\end{align}
\begin{align}
	\gamma_9 = 6 \Big[\alpha^{(L, 4, 0)} + \alpha^{(L, 4, 1)} + \alpha^{(L, 4, 2)}\Big]\;.
\end{align}
Notice that there are $20$ spin coefficients in total, of which $9$ are classical while the 
remaining $6$ are quantum suppressed. To make a connection with our scattering amplitudes, we 
subsequently perform the inverse boost of the tensor structures
\begin{equation}
	\bar{T}_{i} = (B_{ij})^{-1} T_j\;, \quad \bar{\mathcal{T}}_{i} = (\mathcal{B}_{ij})^{-1} \mathcal{T}_j\;,
\end{equation}
while expanding consistently to $\mathcal{O}(|q|^4)$. These $20$ spin coefficients are then fixed by 
demanding equality with the corresponding spinning amplitude order by order in $|q|$. Crucially, we 
may demand equality with scattering amplitudes written in terms of general functions of kinematic 
information. The advantage is that the spin coefficients are then fixed in terms of general functions that 
are independent of the loop order. We write finite 
the part of the classical amplitude as
\begin{equation}
	\bar{\mathcal{M}}^{(L)}_{\phi \phi} = \mathrm{a}_1 + \mathrm{a}_{2} q^2 + \mathrm{a}_{3} q^4 + \ldots \;,
\end{equation}
\begin{align}
	\bar{\mathcal{M}}^{(L)}_{\phi \tiny{\mbox{V}}} = & \ (\mathrm{b}_1 + \mathrm{b}_{2} q^2 + \mathrm{b}_{3} q^4) T_1  + (\mathrm{b}_4 + \mathrm{b}_5 q^2) T_2 \\
	& \hspace{2cm} + (\mathrm{b}_6 + \mathrm{b}_7 q^2) T_3 + (\mathrm{b}_8 + \mathrm{b}_9 q^2) T_4 + \ldots\;, \nonumber
\end{align}
\begin{align}
	\bar{\mathcal{M}}^{(L)}_{\phi \tiny{\mbox{H}}} & = (\mathrm{c}_1 + \mathrm{c}_{2} q^2 + \mathrm{c}_{3} q^4) \mathcal{T}_1  + (\mathrm{c}_4 + \mathrm{c}_5 q^2 + \mathrm{c}_6 q^4) \mathcal{T}_2 \\
	& + (\mathrm{c}_7 + \mathrm{c}_8 q^2 + \mathrm{c}_9 q^4) \mathcal{T}_3 + (\mathrm{c}_{10} + \mathrm{c}_{11} q^2) \mathcal{T}_4 + (\mathrm{c}_{12} + \mathrm{c}_{13} q^2) \mathcal{T}_5\nonumber \\
	&  + (\mathrm{c}_{14} + \mathrm{c}_{15} q^2) \mathcal{T}_6 + (\mathrm{c}_{16} + \mathrm{c}_{17} q^2) \mathcal{T}_7 + \mathrm{c}_{18} \mathcal{T}_8 + \mathrm{c}_{19} \mathcal{T}_9 + \ldots \;, \nonumber 
\end{align}
where the coefficients $(\mathrm{a}_i, \mathrm{b}_i, \mathrm{c}_i)$ are generic functions of 
kinematic variables and we have stripped off the loop-order-dependent prefactor. 

Then, by demanding equality between our ansätze and amplitudes, we proceed with solving a 
system of linear equations
\begin{equation}
	\begin{pmatrix}
		\bar{\mathcal{M}}^{(L)}_{\phi \phi} - \bar{\mathcal{M}}_{\text{ansatz}}^{(L)} \Big\rvert_{s = 0}\\[0.3cm]
		\bar{\mathcal{M}}^{(L)}_{\phi \tiny{\mbox{V}}} - \bar{\mathcal{M}}_{\text{ansatz}}^{(L)} \Big\rvert_{s = 1}\\[0.3cm]
		\bar{\mathcal{M}}^{(L)}_{\phi \tiny{\mbox{H}}} - \bar{\mathcal{M}}_{\text{ansatz}}^{(L)} \Big\rvert_{s = 2}\\[0.3cm]
	\end{pmatrix} = \boldsymbol{0}\;,
\end{equation}
which yields $31$ linear relations. As our goal is to fix the spin coefficients in terms of the 
amplitude coefficients, we will simply use linear algebra. We start with
\begin{equation}
	\mathrm{u}_i = \mathrm{A}_{ij} \mathrm{v}_j\;,
\end{equation}
where $\mathrm{u}_i$ and $\mathrm{v}_i$ are vectors constructed from the amplitude 
coefficients and spin coefficients, respectively, and $\mathrm{A}_{ij}$ is a $31 \times 20$ 
matrix which we want to invert. To be as explicit as possible, the vectors are 
\begin{equation}
	\mathrm{u}_i = 
	\begin{pmatrix}
		\mathrm{a}_1\\
		\mathrm{a}_2\\
		\mathrm{a}_3\\
		\mathrm{b}_1\\
		\mathrm{b}_2\\
		\mathrm{b}_3\\
		\mathrm{b}_4\\
		\mathrm{b}_5\\
		\mathrm{b}_6\\
		\mathrm{b}_7\\
		\mathrm{b}_8\\
		\mathrm{b}_9\\
		\mathrm{c}_1\\
		\mathrm{c}_2\\
		\mathrm{c}_3\\
		\mathrm{c}_4\\
		\mathrm{c}_5\\
		\mathrm{c}_6\\
		\mathrm{c}_7\\
		\mathrm{c}_8\\
		\mathrm{c}_9\\
		\mathrm{c}_{10}\\
		\mathrm{c}_{11}\\
		\mathrm{c}_{12}\\
		\mathrm{c}_{13}\\
		\mathrm{c}_{14}\\
		\mathrm{c}_{15}\\
		\mathrm{c}_{16}\\
		\mathrm{c}_{17}\\
		\mathrm{c}_{18}\\
		\mathrm{c}_{19}
	\end{pmatrix}\;,
	\quad
	\mathrm{v}_i = 
	\begin{pmatrix}
		\alpha^{(L, 0, 0)}\\[0.15cm]
		\alpha^{(L, 1, 0)}\\[0.15cm]
		\alpha^{(L, 2, 0)}\\[0.15cm]
		\alpha^{(L, 2, 1)}\\[0.15cm]
		\alpha^{(L, 2, 2)}\\[0.15cm]
		\alpha^{(L, 3, 0)}\\[0.15cm]
		\alpha^{(L, 3, 1)}\\[0.15cm]
		\alpha^{(L, 3, 2)}\\[0.15cm]
		\alpha^{(L, 4, 0)}\\[0.15cm]
		\alpha^{(L, 4, 1)}\\[0.15cm]
		\alpha^{(L, 4, 2)}\\[0.15cm]
		\alpha^{(L, 4, 3)}\\[0.15cm]
		\alpha^{(L, 4, 4)}\\[0.15cm]
		\alpha^{(L, 4, 5)}\\[0.15cm]
		\alpha^{(L, 0, 0)}_{q^2}\\[0.15cm]
		\alpha^{(L, 0, 0)}_{q^4}\\[0.15cm]
		\alpha^{(L, 1, 0)}_{q^2}\\[0.15cm]
		\alpha^{(L, 2, 0)}_{q^2}\\[0.15cm]
		\alpha^{(L, 2, 1)}_{q^2}\\[0.15cm]
		\alpha^{(L, 2, 2)}_{q^2}\\[0.15cm]
	\end{pmatrix}\;,
\end{equation}
while the matrix $\mathrm{A}_{ij}$ is simply a function of $\bar{m}_i$ and $y$. By virtue of the 
matrix being non-square, we cannot in general invert it to solve for $\mathrm{v}_i$ in 
terms of $\mathrm{u}_i$. Towards this goal, we find that the transpose of this matrix has a 
nullspace of dimension $11$, which is a subspace of all length-$31$ column vectors annihilated by the 
matrix. The nullspace is represented by a $11 \times 31$ matrix and constitutes 
the linear dependencies of the system. Then, acting this matrix on the vector of amplitude 
coefficients yields the following constraints
\begin{align}
	6 \mathrm{b}_4 + \mathrm{c}_{16} + \frac{2}{\bar{m}_1^2}\mathrm{a}_1 + 
	\frac{4y \bar{m}_2}{\bar{m}_1}\mathrm{b}_8 = 0 \;,& \\[0.29cm]
	\mathrm{c}_{14} = 0\;, & \\[0.29cm]
	2 \mathrm{b}_8 - \mathrm{c}_{12} = 0\;, & \\[0.29cm]
	\mathrm{c}_{10} = 0\;, & \\[0.29cm]
	6 \mathrm{b}_8 - \mathrm{c}_8 = 0\;, & \\[0.29cm]
	\mathrm{c}_{7} = 0\;, & \\[0.29cm]
	\mathrm{c}_{5} = 0\;, & \\[0.29cm]
	\mathrm{c}_{4} = 0\;, & \\[0.29cm]
	2 \mathrm{a}_{2} + 3 \mathrm{b}_{2} + \mathrm{b}_{4} + \mathrm{c}_{2} + \frac{\mathrm{a}_{1}}{2\bar{m}_1^2} + \frac{y \bar{m}_2}{\bar{m}_1} \mathrm{b}_8 - (y^2-1) \bar{m}_2^2 \mathrm{b}_{6} =0\;, & \\[0.29cm]
	\mathrm{a}_1 - \mathrm{c}_1 = 0\;, & \\[0.29cm]
	\mathrm{a}_1 + \mathrm{b}_1 = 0\;. & 
\end{align}
Crucially, this linear dependence is an explicit manifestation of spin universality, since terms 
fixed by lower-spin amplitudes, such as spinless terms and spin-orbit interaction terms, reappear
in higher-spin amplitudes. Following this, we impose the constraints at the level of our 
amplitudes, reconstruct the linear system, and subsequently find all of the linearly 
dependent relations by deleting rows while ensuring that the rank of the (rank $20$) matrix 
remains unchanged. Keep in mind that imposing the constraints will alter the form 
of $\mathrm{u}_{i}$, which we will now denote as $\mathrm{w}_{i}$. In fact, keeping track of which 
rows we have deleted allows us to construct a $20 \times 31$ projection matrix $\mathrm{R}_{ij}$ by 
deleting the corresponding rows of the $31 \times 31$ identity matrix. Then, the task is to solve the 
system 
\begin{equation}
	\mathrm{w}'_i \equiv \mathrm{R}_{ij} \mathrm{w}_j =  \mathrm{R}_{ik}\mathrm{A}_{kj}\mathrm{v}_j\;,
\end{equation}
where $\mathrm{w}'_i$ is now a $20$-dimensional vector and $\mathrm{R}_{ik}\mathrm{A}_{kj}$ is a 
$20 \times 20$ matrix. Inverting the system gives
\begin{equation}
	\mathrm{v}_i = (\mathrm{R}_{ik}\mathrm{A}_{kj})^{-1}\mathrm{w}'_j\;,
\end{equation}
which, finally, yields the spin coefficients in terms of the amplitude coefficients\footnote{For 
the sake of sanity we will omit the explicit form of $\mathrm{v}_i$.}. Now, to obtain the final 
answer one simply needs to extract the explicit coefficients from the amplitudes (at any loop 
order) and substitute them into $\mathrm{v}_i$. 

An important point is that this method allows us to directly solve for both the classical and quantum 
coefficients, however we will only present the classical coefficients as they are relevant for
classical black hole scattering. 

%--------------------------------------------------------------------------------
\section{Classical Scattering Dynamics to $\mathcal{O}(G^3S^4)$}
\label{sec:DynamicsThroughS4}
%--------------------------------------------------------------------------------

Now that we have completely fixed the spin coefficients in terms of amplitude coefficients using 
the spin interpolation method, we are ready to present entirely novel results for the scattering of a spinless and 
spinning black hole to $\mathcal{O}(G^3 S^4)$. The first thing to do is to specify the spin coefficients through the 
two-loop order, which remarkably exhibits spin-shift symmetry for probe dynamics through the quartic order in spin. 
Using these new results, we extract the radial action and compute observables through $\mathcal{O}(G^3 S^4)$ using the 
Dirac bracket formalism. Since the non-aligned spin observables, such as the impulse and spin-kick, are very long, we 
will omit their explicit forms (as well as the form of the radial action) and instead present the scattering angle for 
aligned-spin configurations. It is very well known that the 
conservative two-loop dynamics exhibit a logarithmic divergence in the ultra-relativistic (high-energy) 
limit~\cite{Bern:2019crd, Bern:2019nnu}, which we point out through the quartic-in-spin order. In fact, by 
incorporating the radiation-reaction results of Ref.~\cite{Alessio:2022kwv} we will show that these divergences 
cancel, as expected at this loop order~\cite{DiVecchia:2020ymx,Rothstein:2024nlq,Damour:2020tta}. Importantly, this 
cancellation had already been established to the quadratic order in spin~\cite{Jakobsen:2022zsx}. Using this result, 
we compute the non-aligned spin radiation-reaction corrections to the aforementioned observables to all orders in 
spin, but, for the same reason as before, we will only present the aligned-spin scattering angle through the 
quartic order in spin. Together, this constitutes the full non-dissipative dynamics for the scattering of a spinless 
and spinning black hole through the quartic-in-spin order and two loops. 

As in \mysecref{sec:ScatteringAmplitudes} we will homogenise the mass dependence in all spin structures by using 
$u_i = \bar{p}_i/\bar{m}_i$ and $a = S/\bar{m}_1$, and we drop the barred notation since $\bar{m}_i = m_i +
\mathcal{O}(q)$.

%--------------------------------------------------------------------------------
\subsection{Scattering Amplitudes and Observables}
\label{sec:ScatteringAmplitudesAndObservables}
%--------------------------------------------------------------------------------

We begin by presenting the final classical amplitude through the 
two-loop order. The finite part of the classical amplitudes takes the form 
of \eqref{eq:spinningAnsatz} where the spin coefficients are truncated to the classical order:
\begin{equation}
	\mathcal{M}_{\text{finite}}^{(L)} = \frac{\kappa^{2+2L}}{(-q^2)^{1-L/2+L\epsilon}}\sum_{n = 0}^{4}\sum_{m } \alpha^{(L, n,m)} \mathcal{J}^{(n,m)}\;,
\end{equation}
with the spin structures now written in terms of $u_i$ and $a$.

\textbf{Tree-level.} The tree-level amplitude is well known and takes a very simple form
\begin{align}
	\mathcal{M}_{\text{finite}}^{(0)} = \frac{\kappa^2}{(-q^2)}\Big\{& \frac{(2y^2-1)}{2} m_1^2 m_2^2 + i y m^2_1 m^2_2 (n \cdot a) + \frac{(2y^2-1) m^2_1 m^2_2}{4}(q \cdot a)^2 \nonumber \\
	& + \frac{i y m^2_1 m^2_2}{6}(n \cdot a)(q \cdot a)^2 + \frac{(2y^2-1) m^2_1 m^2_2}{48}(q \cdot a)^4\Big\},
\end{align}
where we drop terms that are proportional to $q^2$ as they cancel the $t$-channel pole.

\textbf{One-loop.} The coefficients for the one-loop amplitude are
\allowdisplaybreaks{
\begin{align}
	\alpha^{(1,0,0)} & = \frac{3m_1^2 m_2^2(m_1 + m_2)}{512}(5y^2-1)\;, \nonumber \\[0.2cm]
	\alpha^{(1,1,0)} &= \frac{m_1^2 m_2^2(4 m_1 + 3 m_2)}{512(y^2-1)}y(5y^2-3)\;, \nonumber\\[0.2cm]
	\alpha^{(1,2,0)} & = \frac{m_1^2 m_2^2}{4096 (y^2-1)}((95 y^4 - 102 y^2 + 15) m_1 +  4 (15 y^4 - 15 y^2 + 2) m_2)\;, \nonumber \\[0.2cm]
	\alpha^{(1,2,1)} & = 0\;, \nonumber\\[0.2cm] 
	\alpha^{(1,2,2)} & = -\frac{m_1^2 m_2^2}{2048 (y^2-1)^2}((65 y^4 - 66 y^2 + 9) m_1 +  2 (15 y^4 - 12 y^2 + 1) m_2)\;, \nonumber\\[0.2cm]
	\alpha^{(1,3,0)} & = \frac{m_1^2 m_2^2}{1024 (y^2-1)}y((9 y^2 - 5) m_1 + (5 y^2 -2) m_2)\;, \nonumber\\[0.2cm]  
	\alpha^{(1,3,1)} & = 0\;, \\[0.2cm]
	\alpha^{(1,3,2)} & = -\frac{m_1^2 m_2^2}{1024 (y^2-1)^2}y(8(2y^2 - 1) m_1 + (5 y^2 + 1) m_2)\;, \nonumber\\[0.2cm]
	\alpha^{(1,4,0)} & = \frac{m_1^2 m_2^2}{98304 (y^2-1)}((239 y^4- 250 y^2+35 ) m_1 + 24 y^2 (5 y^2 - 4) m_2)\;, \nonumber\\[0.2cm]
	\alpha^{(1,4,1)} & = 0\;, \nonumber \\[0.2cm]
	\alpha^{(1,4,2)} & = 0\;, \nonumber\\[0.2cm]
	\alpha^{(1,4,3)} & = -\frac{m_1^2 m_2^2}{24576 (y^2-1)^2}((193 y^4 - 194 y^2 + 25) m_1 + 12 (5 y^4 - y^2 - 2)m_2)\;, \nonumber\\[0.2cm] 
	\alpha^{(1,4,4)} & = 0\;, \nonumber\\[0.2cm]
	\alpha^{(1,4,5)} & =  \frac{m_1^2 m_2^2}{12288 (y^2-1)^3}(8 (8 y^4 - 8 y^2 + 1) m_1 + (15 y^4 + 6 y^2 -13) m_2)\nonumber\;,
\end{align}
}%
in perfect agreement with known results, for example 
Refs.~\cite{Aoude:2022trd,Chen:2021qkk,Bern:2022kto}. 
Now the convenience of our basis becomes apparent: vanishing coefficients 
correspond to spin-shift symmetry violating spin structures. Thus, we immediately see that the 
one-loop amplitude is spin-shift symmetric, as expected.

\textbf{Two-loops.} Finally, at two loops the coefficients are organised as
\begin{equation}
	\alpha^{(2,n,m)} = \alpha^{(2,n,m)}_{\text{probe}} + m_1^3 m_2^3 \, \alpha^{(2,n,m)}_{\text{1SF}}\;.
\end{equation}
The probe contributions read
\allowdisplaybreaks{
\begin{align}
	& \alpha_{\text{probe}}^{(2,0,0)} = \frac{
	m_1^2 m_2^2 \left(m_1^2+m_2^2\right) \left(64 y^6-120 y^4+60 y^2-5\right)}{49152 \pi ^2 \epsilon \left(y^2-1\right)^2} \;,\nonumber \\[0.2cm]
	& \alpha_{\text{probe}}^{(2,1,0)} = \frac{m_1^2m_2^2(3m_1^2+2m_2^2)y \left(16 y^4-20 y^2+5\right)}{24576 \pi ^2 \epsilon \left(y^2-1\right)^2 }\;, \nonumber\\[0.2cm]
	& \alpha_{\text{probe}}^{(2,2,0)} = \frac{m_1^4 m_2^2 \left(128 y^6-216 y^4+96 y^2-7\right)+ m_1^2 m_2^4 \left(64 y^6-104 y^4+44 y^2-3\right)}{98304 \pi ^2\epsilon  \left(y^2-1\right)^2 }\;, \nonumber\\[0.2cm]
	& \alpha_{\text{probe}}^{(2,2,1)} = 0 \;,\nonumber \\[0.2cm]
	& \alpha_{\text{probe}}^{(2,2,2)} = -\frac{m_1^4 m_2^2 \left(80 y^6-132 y^4+57 y^2-4\right)+ m_1^2 m_2^4 \left(32 y^6-48 y^4+18 y^2-1\right)}{49152 \pi ^2\epsilon \left(y^2-1\right)^3 }\;, \nonumber\\[0.2cm]
	& \alpha_{\text{probe}}^{(2,3,0)} = y \frac{ m_1^4 m_2^2\left(80 y^4-92 y^2+21\right)+2  m_1^2 m_2^4\left(16 y^4-16 y^2+3\right)}{147456 \pi ^2 \epsilon \left(y^2-1\right)^2 }\;, \nonumber\\[0.2cm]
	& \alpha_{\text{probe}}^{(2,3,1)} = 0\;, \\[0.2cm]
	& \alpha_{\text{probe}}^{(2,3,2)} = -\frac{ m_1^4 m_2^2y \left(64 y^4-70 y^2+15\right)+2 m_1^2 m_2^4 y^3 \left(8 y^2-5\right)}{73728 \pi ^2 \epsilon \left(y^2-1\right)^3 }\;, \nonumber\\[0.2cm]
	& \alpha_{\text{probe}}^{(2,4,0)} = \frac{3  m_1^4 m_2^2\left(64 y^6-104 y^4+44 y^2-3\right)+m_1^2 m_2^4 \left(64 y^6-88 y^4+28 y^2-1\right)}{1179648 \pi ^2 \epsilon \left(y^2-1\right)^2 }\;, \nonumber\\[0.2cm]
	& \alpha_{\text{probe}}^{(2,4,1)} = 0\;, \nonumber\\[0.2cm]
	& \alpha_{\text{probe}}^{(2,4,2)} = 0\;, \nonumber\\[0.2cm]
	& \alpha_{\text{probe}}^{(2,4,3)} = -\frac{3 m_1^4 m_2^2\left(48 y^6-76 y^4+31 y^2-2\right)+ m_1^2m_2^4 \left(32 y^6-32 y^4+2 y^2+1\right)}{294912 \pi ^2 \epsilon \left(y^2-1\right)^3 }\;, \nonumber\\[0.2cm]
	& \alpha_{\text{probe}}^{(2,4,4)} = 0\;, \nonumber\\[0.2cm]
	& \alpha_{\text{probe}}^{(2,4,5)} = \frac{3m_1^4  m_2^2 \left(128 y^6-200 y^4+80 y^2-5\right)+m_1^2 m_2^4 \left(64 y^6-40 y^4-20 y^2+5\right)}{1179648 \pi ^2 \epsilon \left(y^2-1\right)^4 }\;,\nonumber
\end{align}
}%
where the vanishing coefficients directly imply that the amplitude in both probe
limits satisfies spin-shift symmetry. The non-probe, 1SF
contributions read
\allowdisplaybreaks{
\begin{align}
	\hspace{-5cm}
	& \alpha_{\text{1SF}}^{(2,0,0)} = \frac{y \left(36 y^6-114 y^4+132 y^2-55\right)}{24576 \pi ^2  \epsilon \left(y^2-1\right)^2} - \frac{\left(4 y^4-12 y^2-3\right) \operatorname{arcosh}(y)}{4096 \pi ^2  \epsilon \sqrt{y^2-1}}\;, \nonumber \\[0.2cm]
	& \alpha_{\text{1SF}}^{(2,1,0)} = \frac{\left(36 y^6-156 y^4+84 y^2+41\right)}{24576 \pi ^2 \epsilon \left(y^2-1\right)^2 } - \frac{y \left(2 y^4-11 y^2-6\right) \operatorname{arcosh}(y)}{2048 \pi ^2  \epsilon \left(y^2-1\right)^{3/2}}\;, \nonumber\\[0.2cm]
	& \alpha_{\text{1SF}}^{(2,2,0)} = \frac{y \left(366 y^6-2499 y^4+550 y^2+1218\right)}{491520 \pi ^2 \epsilon \left(y^2-1\right)^2 } \nonumber \\
	& \hspace{4cm} - \frac{\left(16 y^8-144 y^6+13 y^4+78 y^2+12\right) \operatorname{arcosh}(y)}{32768 \pi ^2 \epsilon \left(y^2-1\right)^{5/2}}\;, \nonumber\\[0.2cm] 
	& \alpha_{\text{1SF}}^{(2,2,1)} = -\frac{y \left(2 y^6-5 y^4+114 y^2+264\right)}{163840 \pi ^2 \epsilon  \left(y^2-1\right)^2} +\frac{3\left(3 y^4+18 y^2+4\right) \operatorname{arcosh}(y)}{32768 \pi ^2  \epsilon \left(y^2-1\right)^{5/2}}\;, \\[0.2cm]
	& \alpha_{\text{1SF}}^{(2,2,2)} = -\frac{y \left(138 y^6-1769 y^4+1394 y^2+2122\right)}{245760 \pi ^2 \epsilon \left(y^2-1\right)^3 } \nonumber \\
	& \hspace{4cm} +\frac{\left(8 y^8-104 y^6+35 y^4+162 y^2+24\right) \operatorname{arcosh}(y)}{16384 \pi ^2  \epsilon \left(y^2-1\right)^{7/2}}\;, \nonumber\\[0.2cm]
	& \alpha_{\text{1SF}}^{(2,3,0)} = \frac{\left(372 y^8-3524 y^6-653 y^4+1945 y^2-30\right)}{1474560 \pi ^2 \epsilon \left(y^2-1\right)^3}\nonumber \\
	& \hspace{4cm} -\frac{y \left(16 y^8-184 y^6-50 y^4+57 y^2+35\right) \operatorname{arcosh}(y)}{98304 \pi ^2 \epsilon \left(y^2-1\right)^{7/2}}\;, \nonumber\\[0.2cm]
	& \alpha_{\text{1SF}}^{(2,3,1)} = -\frac{\left(4 y^8-12 y^6+385 y^4+1305 y^2+208\right)}{491520 \pi ^2 \epsilon \left(y^2-1\right)^3}+\frac{y \left(10 y^4+75 y^2+41\right)\operatorname{arcosh}(y)}{32768 \pi ^2 \epsilon \left(y^2-1\right)^{7/2}}\;, \nonumber\\[0.2cm]
	& \alpha_{\text{1SF}}^{(2,3,2)} = -\frac{\left(96 y^8-3724 y^6+2687 y^4+12809 y^2+1362\right)}{737280 \pi ^2 \epsilon\left(y^2-1\right)^4}\nonumber \\
	& \hspace{4cm} +\frac{y \left(8 y^8-188 y^6-10 y^4+741 y^2+331\right) \operatorname{arcosh}(y)}{49152 \pi ^2 \epsilon \left(y^2-1\right)^{9/2}}\;, \nonumber\\[0.2cm]
	& \alpha_{\text{1SF}}^{(2,4,0)} = \frac{y \left(10596 y^{10}-139906 y^8+50884 y^6+120462 y^4-38972 y^2+25601\right)}{165150720 \pi ^2 \epsilon \left(y^2-1\right)^4} \nonumber\\[0.2cm] 
	& \hspace{2cm} - \frac{\left(64 y^{12}-960 y^{10}+314 y^8+280 y^6+534 y^4-14 y^2+55\right) \operatorname{arcosh}(y)}{1572864 \pi ^2 \epsilon \left(y^2-1\right)^{9/2}}\;, \nonumber\\[0.2cm]
	& \alpha_{\text{1SF}}^{(2,4,1)} = -\frac{y \left(4 y^{10}-18 y^8-108 y^6+1022 y^4+13732 y^2+11093\right)}{55050240 \pi ^2 \epsilon \left(y^2-1\right)^4} \nonumber \\
	& \hspace{4cm} - \frac{\left(2 y^8-8 y^6-234 y^4-446 y^2-49\right) \operatorname{arcosh}(y)}{1572864 \pi ^2 \epsilon \left(y^2-1\right)^{9/2}}\;, \nonumber\\[0.2cm]
	& \alpha_{\text{1SF}}^{(2,4,2)} = -\frac{y \left(12 y^{10}-54 y^8+1636 y^6+6126 y^4-4604 y^2-2591\right)}{3932160 \pi ^2 \epsilon \left(y^2-1\right)^4} \nonumber \\
	& \hspace{4cm} +\frac{\left(118 y^8+1064 y^6-90 y^4-934 y^2-53\right) \operatorname{arcosh}(y)}{786432 \pi ^2 \epsilon \left(y^2-1\right)^{9/2}}\;, \nonumber\\[0.2cm]
	& \alpha_{\text{1SF}}^{(2,4,3)} = -\frac{y \left(3312 y^{10}-129464 y^8+146636 y^6+657312 y^4-341941 y^2-157250\right)}{41287680 \pi ^2 \epsilon \left(y^2-1\right)^5} \nonumber\\[0.2cm] 
	& \hspace{2cm} +\frac{\left(32 y^{12}-864 y^{10}+312 y^8+4944 y^6+450 y^4-3039 y^2-134\right) \operatorname{arcosh}(y)}{393216 \pi ^2 \epsilon \left(y^2-1\right)^{11/2}}\;, \nonumber\\[0.2cm]
	& \alpha_{\text{1SF}}^{(2,4,4)} = \frac{y \left(8 y^{10}-36 y^8+1156 y^6+1624 y^4-15747 y^2-6850\right)}{2752512 \pi ^2 \epsilon \left(y^2-1\right)^5} \nonumber \\
	& \hspace{4cm} - \frac{\left(20 y^8+128 y^6-474 y^4-581 y^2-38\right) \operatorname{arcosh}(y)}{131072 \pi ^2 \epsilon \left(y^2-1\right)^{11/2}} \;, \nonumber\\[0.2cm]
	& \alpha_{\text{1SF}}^{(2,4,5)} = \frac{y \left(980 y^{10}-98474 y^8+122716 y^6+626512 y^4-602806 y^2-267223\right)}{41287680 \pi ^2 \epsilon \left(y^2-1\right)^6} \nonumber \\[0.2cm]
	& \hspace{2cm} -\frac{\left(16 y^{12}-624 y^{10}+74 y^8+5048 y^6-1474 y^4-4860 y^2-259\right) \operatorname{arcosh}(y)}{393216 \pi ^2 \epsilon \left(y^2-1\right)^{13/2}} \;. \nonumber
\end{align}
}
These results are completely new and advance the state-of-the-art for precision Kerr black hole scattering.
%--------------------------------------------------------------------------------
\subsection*{Observables}
\label{sec:observablesS4}
%--------------------------------------------------------------------------------
Given the finite part of the classical amplitude through $\mathcal{O}(G^3 S^4 )$ in hand, we may go 
ahead and obtain the radial action using \myeqref{eq:RadialActionFourierTransform}. As the 
expressions are very long we choose to omit their explicit forms. But a crucial point is 
that we find exact agreement with the results of Ref.~\cite{Gonzo:2024zxo} through the 3PM 
order, in the limit of a spinless probe moving in a Kerr background. This provides strong 
evidence that the amplitude-action formalism \myeqref{eq:aarelation} truly does extract the 
radial action for the scattering process in the presence of spin. The radial action can be found 
in the ancillary file of Ref.~\cite{Akpinar:2025tct}.

Together with the Dirac bracket formalism described in \mysecref{sec:CovariantDiracBrackets}, 
the radial action provides a powerful approach that allows for the computation of the 
conservative linear and angular impulses. These observables are 
written in the basis \myeqref{eq:ObservableBasis} and are denoted by $\Delta u_1^\mu$ and 
$\Delta a^\mu$, respectively. Indeed, they satisfy \myeqref{eq:consv_mon_constraint} and
\myeqref{eq:u_dot_a_constraint}, and are in agreement with existing results through the quadratic order in 
spin~\cite{Akpinar:2024meg, FebresCordero:2022jts, Jakobsen:2022fcj}. 
As before, we omit the explicit expressions here, but the results 
through $\mathcal{O}(G^3 S^4)$ may be found in the ancillary file of Ref.~\cite{Akpinar:2025bkt}.

To compute the scattering angle we take the aligned-spin configuration described in 
\mysecref{sec:AlignedSpinObservables} and follow Appendix~\mysecref{sec:ScatteringAngleAndImpulse}. Although these 
results overlap with \mysecref{sec:AlignedSpinScatteringAngleGrav} through the quadratic order in spin, we 
nonetheless present the full result for completeness. At leading order we have
\begin{align}
	\chi^{(0)}_{\text{aligned-spin}} = & \ \frac{2(2y^2-1)m_1 m_2}{J\sqrt{y^2-1}} - \frac{4y m_1^2 m_2^2 \sqrt{y^2-1}}{J^2 \sqrt{s}}|a| \nonumber \\
	& + \frac{2m_1^3m_2^3 \sqrt{y^2-1}(2y^2-1)}{J^3 s} |a|^2 - \frac{4 y m_1^4 m_2^4 (y^2-1)^{3/2}}{J^4 s^{3/2}}|a|^3 \\
	& + \frac{2m_1^5 m_2^5 (y^2-1)^{3/2}(2y^2-1)}{J^5 s^2}|a|^4\;. \nonumber
\end{align}
At next-to-leading order we find
\begin{align}
    \chi^{(1)}_{\text{aligned-spin}} = & \ \frac{3\pi (5y^2-1)(m_1 + m_2)m_1^2 m_2^2}{4J^2 \sqrt{s}} - \frac{\pi y (5y^2-3)(4m_1+3m_2)m_1^3m_2^3}{2J^3 s} |a| \nonumber \\
    	& +\frac{3\pi ((95y^4-102y^2+15)m_1 + (60y^4-60y^2 + 8) m_2)m_1^4m_2^4}{16 J^4 s^{3/2}} |a|^2 \nonumber \\
    	& - \frac{3\pi y (y^2-1)((9y^2 - 5)m_1 + (5y^2-2)m_2)m_1^5 m_2^5}{J^5 s^2}|a|^3 \\
    	& + \frac{5 \pi (y^2-1)((239 y^4 - 250 y^2 + 35)m_1 + 24y^2(5y^2 - 4)m_2)m_1^6 m_2^6}{32 J^6 s^{5/2}}|a|^4\;. \nonumber
\end{align}
Finally, at next-to-next-to-leading order we find
{\allowdisplaybreaks
\begin{align}
    & \chi^{(2)}_{\text{aligned-spin}} = \frac{2(64y^6-120y^4 + 60y^2 -5)(m_1^2+m_2^2)m_1^3 m_2^3}{3J^3 s (y^2-1)^{3/2}} \nonumber \\
    & + \frac{4y(36y^6-114y^4+132y^2-55)m_1^4m_2^4}{3 J^3 s (y^2-1)^{3/2}} - \frac{8(4y^4-12y^2-3)m_1^4m_2^4}{J^3 s}\arcosh(y) \nonumber \\
    & - \frac{4y(16y^4-20y^2+5)(3m_1^2 + 2 m_2^2)m_1^4 m_2^4}{J^4  s^{3/2}\sqrt{y^2-1}}|a| - \frac{4y(36y^6-156y^4+84y^2+41)m_1^5m_2^5}{J^4 s^{3/2}\sqrt{y^2-1}}|a| \nonumber \\
    & + \frac{48y(2y^4 - 11y^2 - 6)m_1^5m_2^5}{J^4 s^{3/2}}\arcosh(y)|a|  \\
    & + \frac{4((128y^6 - 216y^4 + 96 y^2 - 7)m_1^2 + (64y^6 - 104y^4 + 44y^2 -3)m_2^2)m_1^5 m_2^5}{J^5 s^2 \sqrt{y^2-1}}|a|^2 \nonumber \\
    & + \frac{8 y(184 y^6 - 1252 y^4 + 332 y^2 + 741)m_1^6 m_2^6}{5 J^5 s^2 \sqrt{y^2-1}}|a|^2 \nonumber \\
    & - \frac{192(y^6 - 8 y^4 -7y^2 -1)m_1^6 m_2^6}{J^5 s^2}\arcosh(y)|a|^2 \nonumber \\
    & - \frac{40y \sqrt{y^2-1}((80 y^4 - 92 y^2 + 21)m_1^2 + 2(16 y^4 - 16y^2 + 3)m_2^2)m_1^6 m_2^6}{3 J^6 s^{5/2}}|a|^3 \nonumber \\
    & - \frac{8(188 y^8-1768 y^6-134 y^4+1625 y^2+89)m_1^7 m_2^7}{3 J^6 s^{5/2} \sqrt{y^2-1}}|a|^3 \nonumber \\
    & + \frac{80 y(4 y^6-42 y^4-52 y^2-19)m_1^7 m_2^7}{J^6 s^{5/2}}\arcosh(y)|a|^3 \nonumber \\
    & + \frac{10 \sqrt{y^2-1}(2(64 y^6-104 y^4+44 y^2-3)m_1^2 + (64 y^6-88 y^4+28 y^2-1)m_2^2)m_1^6 m_2^6}{ J^7 s^3}|a|^4 \nonumber \\
    & + \frac{4 y (6724 y^8-81174 y^6-34962 y^4+93035 y^2+16377)m_1^8 m_2^8}{35 J^7 s^3 \sqrt{y^2-1}}|a|^4 \nonumber \\
    & - \frac{24(20 y^8-260 y^6-417 y^4-266 y^2-1)m_1^8 m_2^8}{J^7 s^3}\arcosh(y)|a|^4\;. \nonumber 
\end{align}
}
This is in agreement with post-Newtonian results where there is overlap~\cite{Bautista:2024agp}. 
%--------------------------------------------------------------------------------
\subsection{The Ultra-Relativistic Limit and Radiation-Reaction}
\label{sec:IncludingRR}
%--------------------------------------------------------------------------------

The ultra-relativistic limit of the conservative two-loop scattering dynamics of two massive objects exhibits 
a logarithmic divergence. It is well-established that such a feature must cancel at this order 
for physical observables---like the  scattering angle---once the effects of radiation-reaction are accounted 
for~\cite{DiVecchia:2020ymx,Rothstein:2024nlq,Damour:2020tta}. 

Here we will show that this logarithmic divergence persists through quartic order in spin, and that it cancels 
upon inclusion of radiation-reaction effects. As the radial action arises directly from the finite part of the 
scattering amplitude and is closely connected to physical observables, we will choose to analyse its structure in this 
limit. While the ultra-relativistic limit involves sending $y \rightarrow \infty$, to clean up the expression we 
will instead work with $y = (1+x^2)/2x$ and take the limit $x \rightarrow 0$. One should also note that 
$\arcosh(y) = -\operatorname{log}(x)$ together with $y \geq 1$ and $0 < x <1$. Then, in the ultra-relativistic limit 
the 3PM radial action for generic spin configurations reads
\begin{align}
	& \lim_{x \to 0} I_r^{(2)} \Big\rvert_{S^0} =  - \frac{2G^3 m_1^2 m_2^2}{|b|^2 x^2} \Big(2\operatorname{log}(x) + 3\Big) + \mathcal{O}(x^{-1})\;, \\
	& \lim_{x \to 0} I_r^{(2)} \Big\rvert_{S^1} = - \frac{8 G^3 m_1^2 m_2^2}{|b|^4 x} \Big(2\operatorname{log}(x) + 3\Big)(l \cdot a) + \mathcal{O}(x^{0})\;, \\
	& \lim_{x \to 0} I_r^{(2)} \Big\rvert_{S^2} = \frac{8 G^3 m_1^2 m_2^2}{|b|^6 x^2} \Big(2\operatorname{log}(x) + 3\Big)(b \cdot a)^2 \\
	& \hspace{3cm} - \frac{4 G^3 m_1^2 m_2^2}{5|b|^4 x^2} \Big(15\operatorname{log}(x) + 23\Big)a^2 + \mathcal{O}(x^{-1})\;, \nonumber \\
	& \lim_{x \to 0} I_r^{(2)} \Big\rvert_{S^3} = \frac{32 G^3 m_1^2 m_2^2}{|b|^8 x}\Big(2\operatorname{log}(x) + 3\Big)(l \cdot a)(b \cdot a)^2\;, \\
	& \hspace{3cm} - \frac{16 G^3 m_1^2 m_2^2}{15|b|^6 x} \Big(30\operatorname{log}(x) + 47\Big)(l \cdot a)a^2 + \mathcal{O}(x^{0})\;, \nonumber \\
	& \lim_{x \to 0} I_r^{(2)} \Big\rvert_{S^4} = - \frac{32 G^3 m_1^2 m_2^2}{|b|^{10} x^2}\Big(2\operatorname{log}(x) + 3\Big)(b \cdot a)^4\;, \\
	& \hspace{3cm} + \frac{16 G^3 m_1^2 m_2^2}{35|b|^8 x^2} \Big(175\operatorname{log}(x) + 268\Big)(b \cdot a)^2 a^2 \nonumber  \\
	& \hspace{3cm} - \frac{2G^3 m_1^2 m_2^2}{105|b|^6 x^2}\Big(1050\operatorname{log}(x) + 1681\Big)a^4 + \mathcal{O}(x^{-1})\;, \nonumber 
\end{align}
where $l^\mu = \epsilon^{\mu \nu \rho \sigma} b_\nu u_{1\rho} u_{2\sigma}$. Notice that these expressions are clearly 
logarithmically divergent as $x \rightarrow 0$. 

The radiation-reaction results of Ref.~\cite{Alessio:2022kwv}, following from the real 
part of the Eikonal phase, are valid to all orders in spin. Since this is a unique and finite contribution at 
$\mathcal{O}(G^3)$---meaning that there is no freedom associated to subtracting off iterations of lower-order 
contributions---we naively treat their results as a radiation-reaction contribution to the radial action. This 
contribution to the radial action is denoted as $I_{r,\mathrm{RR}}$. Therefore, we are able to obtain the 
radiation-reaction correction to the impulse and spin-kick to all orders in spin and for generic spin orientations, using 
the covariant Dirac bracket formalism\footnote{As before, these results can be found in the ancillary file of 
Ref.~\cite{Akpinar:2025bkt}.}. Taking the sum of the conservative and radiation-reaction contributions we find 
an exact cancellation of the logarithmic divergence in the ultra-relativistic limit through the quartic order in spin:
\begin{align}
	& \lim_{x \to 0} \Big(I_r^{(2)} + I_{r,\mathrm{RR}}^{(2)}\Big) \Big\rvert_{S^0} =  - \frac{8G^3 m_1^2 m_2^2}{|b|^2 x^2} + \mathcal{O}(x^{-1})\;, \\
	& \lim_{x \to 0} \Big(I_r^{(2)} + I_{r,\mathrm{RR}}^{(2)}\Big)  \Big\rvert_{S^1} = - \frac{32 G^3 m_1^2 m_2^2}{3|b|^4 x} + \mathcal{O}(x^{0})\;, \\
	& \lim_{x \to 0} \Big(I_r^{(2)} + I_{r,\mathrm{RR}}^{(2)}\Big)  \Big\rvert_{S^2} = \frac{32 G^3 m_1^2 m_2^2}{3|b|^6 x^2} (b \cdot a)^2 - \frac{42 G^3 m_1^2 m_2^2}{5|b|^4 x^2}a^2 + \mathcal{O}(x^{-1})\;, \nonumber \\
	& \lim_{x \to 0} \Big(I_r^{(2)} + I_{r,\mathrm{RR}}^{(2)}\Big)  \Big\rvert_{S^3} = \frac{128 G^3 m_1^2 m_2^2}{3|b|^8 x}(l \cdot a)(b \cdot a)^2 \\ 
	& \hspace{4cm} - \frac{352 G^3 m_1^2 m_2^2}{15|b|^6 x} (l \cdot a)^2 a^2 + \mathcal{O}(x^{0})\;, \nonumber \\
	& \lim_{x \to 0} \Big(I_r^{(2)} + I_{r,\mathrm{RR}}^{(2)}\Big)  \Big\rvert_{S^4} = - \frac{128 G^3 m_1^2 m_2^2}{3|b|^{10} x^2}(b \cdot a)^4 \\ 
	& \hspace{4cm} + \frac{5864 G^3 m_1^2 m_2^2}{105|b|^8 x^2} (b \cdot a)^2 a^2  - \frac{1612G^3 m_1^2 m_2^2}{105|b|^6 x^2}a^4 + \mathcal{O}(x^{-1})\;. \nonumber 
\end{align}
To conclude our discussion regarding precision computations and the incorporation of radiation-reaction effects, 
we now present the radiation-reaction correction to the aligned-spin scattering angle (for which we revert back to 
using the variable $y$):
\begin{align}
    \chi^{(2)}_{\text{aligned-spin}, \mathrm{RR}} = & - \frac{4(2y^2-1)^2(5y^2-8) m_1^4 m_2^4}{3J^3 s (y^2-1)} + \frac{4 y (2y^2-1)(2y^2-3)m_1^4 m_2^4}{J^3 s(y^2-1)^{3/2}}\arcosh(y) \nonumber \\
    & + \frac{8y(10 y^4-21 y^2+8)m_1^5 m_2^5}{J^4 s^{3/2}}|a| \nonumber \\ 
    & - \frac{24 y^2(4 y^4-8 y^2+3)m_1^5 m_2^5}{J^4s^{3/2}\sqrt{y^2-1}}\arcosh(y)|a| \\
    & - \frac{16(30 y^6-78 y^4+53 y^2-8)m_1^6 m_2^6}{3 J^5 s^2}|a|^2 \nonumber \\ 
    & + \frac{16y(12 y^6-30 y^4+20 y^2-3)m_1^6 m_2^6}{J^5 s^2 \sqrt{y^2-1}}\arcosh(y)|a|^2 \nonumber \\
    & + \frac{80y (y^2-1)(10 y^4-21 y^2+8)m_1^7 m_2^7}{3 J^6 s^{5/2}}|a|^3 \nonumber \\ 
    & - \frac{80y^2\sqrt{y^2-1}(4 y^4-8 y^2+3)m_1^7 m_2^7}{J^6 s^{5/2}}\arcosh(y)|a|^3 \nonumber \\
    & - \frac{4(y^2-1)(100 y^6-260 y^4+175 y^2-24)m_1^8 m_2^8}{J^7 s^{3}}|a|^4 \nonumber \\ 
    & + \frac{12y\sqrt{y^2-1}(40 y^6-100 y^4+66 y^2-9)m_1^8 m_2^8}{J^7 s^{3}}\arcosh(y)|a|^4\;. \nonumber 
\end{align}
%

%--------------------------------------------------------------------------------
\chapter{Symmetries, Conserved Quantities, and Asymptotic Integrability}
\label{sec:Chapter5}
%--------------------------------------------------------------------------------

In this chapter, we will shift our attention to the study of symmetries, conserved 
quantities, and integrability for the motion of a spinning probe particle in a Kerr 
background, using on-shell techniques. Our goal is to leverage two 
complementary advances---existing PM data for the scattering dynamics of spinning 
black holes and the Dirac bracket formalism---for exactly this purpose. In fact, this 
approach provides a natural setting to explore questions, such as:
\begin{itemize}
\item Can conserved quantities be studied in an on-shell, gauge-invariant manner?
\item What role does spin-shift symmetry play in Kerr black hole scattering?
\item Does there exist a conserved quantity that generates spin-shift symmetry?
\end{itemize}
The primary system of interest in this chapter is a spinning probe particle moving in 
a Kerr background. Of particular interest are the symmetries and integrability 
properties of the spinning probe particle’s motion in the Kerr 
spacetime~\cite{Carter:1968ks,Carter:1968rr}. In the absence of spin, the conservation 
of energy, azimuthal angular momentum, and the Carter constant are usually introduced 
as a consequence of the presence of Killing vector~\cite{Dixon:1970zza} and 
tensor~\cite{Walker:1970un} symmetries of the background spacetime. Interestingly, 
such conservation laws have been extended to the case of a spinning probe in Kerr, 
through the introduction of the R\"udiger constant at linear-in-spin 
order~\cite{rudiger1981conserved,rudiger1983conserved} and a suitable extension of the 
Carter constant at the quadrupolar order~\cite{Compere:2023alp}; see 
Ref.~\cite{Compere:2021kjz} for a nice review. It is important to stress that these 
conservation laws hold only for probes whose spin-induced multipole moments match 
those of a Kerr black hole, while for generic compact objects with generic Wilson 
coefficients---most notably neutron stars---integrability is expected to be broken 
already at quadratic order in spin.

In light of this, the overarching goal of this chapter is to combine existing PM 
results for the radial action, together with the Dirac bracket formalism, to shed 
light on the deeper structure of these symmetries, at least at asymptotic infinity. 
Towards this direction, we will begin with a brief review of the conservation laws 
defined in the bulk, from which we obtain conservation laws from the asymptotic 
scattering perspective by taking $r \rightarrow \infty$. After extracting from the 
literature the relevant spinning radial actions in the probe limit and beyond---up to 
3PM at the quartic order in spin---we will then establish the validity of the 
aforementioned symmetries, further comparing them with a new position-space 
reformulation of the spin-shift symmetry in our phase space. Equipped with this 
knowledge, we define a novel notion of asymptotic integrability, which we find to be 
satisfied by a spinning probe in Kerr to all PM orders and for generic spinning 
binaries at low PM orders. Remarkably, the conservation of these symmetries, 
alongside spin-shift symmetry, enforces powerful constraints on the radial action, 
allowing us to bootstrap the radial action for a spinning probe particle in a Kerr 
background at any PM order and through the quartic order in spin. 

%--------------------------------------------------------------------------------
\section{Conserved Quantities for a Quadrupolar Probe in Kerr Spacetime}
\label{sec:BulkDynamicsQuadrupoleProbeInKerr}
%--------------------------------------------------------------------------------

The first things to discuss are the conserved quantities associated to a quadrupolar 
test body moving in Kerr spacetime. To illustrate the main ideas we will consider a 
simplified system involving a spinless test body in Kerr spacetime, which 
follows geodesic motion. This involves investigating the conserved quantities 
associated with the Kerr metric itself. Once we have understood this, 
we will endow the test body with spin up to the quadrupolar order in the spin 
multipole expansion. It is important to note that such conserved quantities are 
defined in the bulk. These quantities are then evaluated at asymptotic infinity to 
acquire conservation laws for scattering dynamics.

%--------------------------------------------------------------------------------
\subsection{Symmetries of the Kerr Metric}
\label{sec:SymmetriesOfTheKerrMetric}
%--------------------------------------------------------------------------------

Our starting point is the Kerr metric, describing astrophysical rotating black holes 
of mass $M$ and angular momentum $J = aM$. Here $a$ is referred to as the ring 
radius and it constitutes the normalised angular momentum. In Boyer-Lindquist 
coordinates $(t, r, \theta, \phi)$ and natural units $G = c = 1$, the Kerr line 
element is
\begin{equation}
    \begin{split}
    \mathrm{d}s^2 = \left(1-\frac{2Mr}{\Sigma}\right)\mathrm{d}t^2 & + \frac{4Mar \operatorname{sin}^2(\theta)}{\Sigma}\mathrm{d}t \mathrm{d}\phi - \frac{\Sigma}{\Delta}\mathrm{d}r^2 - \Sigma \mathrm{d}\theta^2 \\
    & - \left(r^2 + a^2 + \frac{2Mr}{\Sigma}a^2\operatorname{sin}^2(\theta)\right)\operatorname{sin}^2(\theta)\mathrm{d}\phi^2\;,
    \end{split}
\end{equation}
together with
\begin{equation}
    \Sigma \equiv r^2 + a^2 \operatorname{cos}^2(\theta)\;, \quad \Delta \equiv r^2 + a^2 - 2Mr.
\end{equation}
From this we determine the metric as $g_{\mu \nu} dx^\mu dx^\nu$, 
where $x^\mu$ is the coordinate of the spacetime, i.e.~$x^\mu = (t, r, \theta, \phi)
$. The explicit metric components are 
\begin{equation}
    \begin{split}
        & g_{tt} = \left(1-\frac{2Mr}{\Sigma}\right)\;, \quad g_{t\phi} = g_{\phi t} =  \frac{2Mar \operatorname{sin}^2(\theta)}{\Sigma}\;, \\
        & g_{rr} = -\frac{\Sigma}{\Delta}\;, \quad g_{\theta \theta} = -\Sigma\;, \quad g_{\phi \phi} = -\left(r^2 + a^2 + \frac{2Mr}{\Sigma}a^2\operatorname{sin}^2(\theta)\right)\operatorname{sin}^2(\theta).
    \end{split}
\end{equation}
The inverse metric is defined as $g^{\mu \nu}$ such that 
$g^{\mu \alpha} g_{\alpha \nu} = \delta^{\mu}_{~\nu}$, with components
\begin{equation}
    \begin{split}
        & g^{tt} = \frac{1}{\Sigma \Delta }\left(r^2 + a^2 - \Delta a^2\operatorname{sin}^2(\theta)\right)\;, \quad g^{t\phi} = g^{\phi t} =  \frac{2Mar}{\Sigma \Delta}\;, \\
        & g^{rr} = -\frac{\Delta}{\Sigma}\;, \quad g^{\theta \theta} = -\frac{1}{\Sigma}\;, \quad g^{\phi \phi} = -\frac{\Delta - a^2 \operatorname{sin}^2(\theta)}{\Sigma \Delta \operatorname{sin}^2(\theta)}.
    \end{split}
    \label{eq:InvKerrMetric}
\end{equation}
Towards the direction of conserved quantities, the Kerr metric admits 
two Killing vectors 
\begin{equation}
    \xi^\mu_{(t)} := (\partial_t)^\mu = (1,0,0,0)\;, \quad \xi^\mu_{(\phi)} : = (\partial_\phi)^\mu = (0,0,0,1)\;,
\end{equation}
associated with stationarity and axisymmetry. This follows from the metric having no 
explicit dependence on $(t,\phi)$. Since these vectors generate an isometry of the 
spacetime, they satisfy the Killing equation
\begin{equation}
	\nabla_{(\mu} \xi_{\nu)} = 0\;,
	\label{eq:KillingEquation}
\end{equation}
following from the vanishing of the Lie derivative of the metric under the flow 
of $\xi$. Here $\nabla_\mu$ is a covariant derivative on the Kerr spacetime. Alongside 
this, there exists a Killing tensor 
$K_{\mu \nu} = Y_{\mu \rho}Y_{\nu}^{~\rho}$ which is conveniently constructed from 
an antisymmetric Killing-Yano tensor $Y_{\mu \nu} = - Y_{\nu \mu}$ that encodes a 
hidden symmetry of the background. The Killing-Yano tensor follows by solving 
$\nabla_{(\rho} Y_{\mu) \nu} = 0$. As one might expect, the Killing tensor satisfies a 
rank-$2$ version of \myeqref{eq:KillingEquation}
\begin{equation}
    \nabla_{(\rho} K_{\mu \nu)} = 0.
\end{equation}
More explicitly, the Killing-Yano tensor takes the form~\cite{Gonzo:2024zxo}
\begin{equation}
	Y_{\mu \nu} = a \operatorname{cos}(\theta)( \ell_\mu n_\nu - \ell_\nu n_\mu) + r \epsilon_{\mu \nu \rho \sigma}\ell^\rho n^\sigma \sqrt{- \operatorname{det}g_{\mu \nu}}\;,
\end{equation}
expressed in terms of principal null directions
\begin{align}
	\ell^\mu = \frac{1}{\Delta}(r^2 + a^2, \Delta, 0, a)\;, \quad n^\mu = \frac{1}{2\Sigma}(r^2 + a^2, - \Delta, 0, a)\;,
\end{align}
together with the flat-space Levi-Civita tensor $\epsilon_{\mu \nu \rho \sigma}$ and 
$\operatorname{det}g_{\mu \nu} = -\Sigma^2 \operatorname{sin}^2(\theta)$. 

Now consider a spinless probe moving in a Kerr spacetime. The dynamics of the probe 
with four-momentum $p^\mu$ is given by the usual geodesic motion
\begin{equation}
	\frac{\mathrm{D} p^\mu}{\mathrm{d} \tau} \equiv p^\nu \nabla_\nu p^\mu = 0 \;.
\end{equation}
A particularly useful property of the Killing-Yano tensor is that it defines a vector 
$\Upsilon^\mu = Y^{\mu \nu}p_\nu$ that is parallel transported along the 
trajectory, i.e.~$p^\nu \nabla_\nu \Upsilon^\mu = 0$. Thus, for such a system the 
following quantities are conserved~\cite{Carter:1968ks}
\begin{align}
	& \text{Energy}: E = - p_\mu \xi^{\mu}_{(t)}\;, \\
	& \text{Azimuthal angular momentum}: L = p_\mu \xi^{\mu}_{(\phi)}\;, \\ 
	& \text{The Carter constant}: \mathcal{K} = K^{\mu \nu}p_\mu p_\nu\;, \\
	& \text{Mass invariant}: m^2 = g^{\mu \nu}p_\mu p_\nu\;, 
\end{align}
where $m$ is the mass of the probe. Importantly, conservation 
means that along the trajectory we have
\begin{equation}
	\frac{\mathrm{D}}{\mathrm{d} \tau}\mathcal{Q} = 0\;, \quad \mathcal{Q} \in \{E, L, \mathcal{K}, m^2\}\;.
\end{equation}
This means there are four Poisson-commuting constants of motion, ensuring that 
the geodesic motion in the Kerr spacetime is a fully integrable dynamical system.
Note that one may also derive the Carter constant by assuming separable solutions 
of the Hamilton-Jacobi equation; see Appendix~\mysecref{sec:CarterConstFromHJEquations} for details.
%--------------------------------------------------------------------------------
\subsection{Symmetries for a Quadrupolar Probe in Kerr Spacetime}
\label{sec:SymmetriesQuadrupoleProbeInKerr}
%--------------------------------------------------------------------------------

Now we endow the probe with spin. In this case, the motion of an extended test body 
over a curved background is described by the Mathisson-Papapetrou-Dixon (MPD) 
equations~\cite{Mathisson:1937zz, Papapetrou:1951pa, Dixon:1970zza,Compere:2023alp}
\begin{align}
	\frac{\mathrm{D} p^\mu}{\mathrm{d} \tau} & = - \frac{1}{2} R^{\mu}_{~\nu \alpha \beta}u^\nu \mathcal{S}^{\alpha \beta} - \underbrace{\frac{1}{6}J^{\alpha \beta \gamma \delta} \nabla^{\mu} R_{\alpha \beta \gamma \delta}}_{-\mathcal{F}^\mu} + \ldots\;, \\
	\frac{\mathrm{D} \mathcal{S}^{\mu \nu}}{\mathrm{d} \tau} & = 2 p^{[\mu} u^{\nu]} + \underbrace{\frac{4}{3} R^{[\mu}_{~~\alpha \beta \gamma} J^{\nu]\alpha \beta \gamma}}_{\mathcal{L}^{\mu \nu}} + \ldots\;,
\end{align}
written up to the quadrupolar order in the test body's multipole expansion. Here 
$u^\mu$ defines the four-velocity of the test body, with four-momentum $p^\mu$, and 
spin dipole tensor $\mathcal{S}^{\mu \nu}$, while the $\ldots$ represent octupolar 
and higher corrections. The quadrupolar dynamics is encoded via 
the quadrupole tensor~\cite{Compere:2023alp}
\begin{equation}
	J^{\mu \nu \rho \sigma} = k \frac{3 (p \cdot u)}{(p^2)^2} p^{[\mu}\mathcal{S}^{\nu]\lambda} \mathcal{S}^{[\rho}_{~~\lambda} p^{\sigma]}\;,
\end{equation}
where $k$ is a response coefficient that controls the magnitude of the quadrupolar 
deformation. Typical values for a neutron star range from $k = 4$ to 
$k = 8$~\cite{Compere:2023alp}, while 
for Kerr black holes we have $k_{\mathrm{BH}} = 1$. In addition to the MPD equations, 
the motion of the test body is further supplemented with the SSC 
$p_\mu \mathcal{S}^{\mu \nu} = 0$, which is often referred to as the 
\textit{Tulczyjew-Dixon} condition~\cite{Tulczyjew1957, Tulczyjew1959}. Not only does 
the SSC condition allow us to fix the body's centre-of-mass~\cite{Compere:2021kjz}, it 
conveniently allows us to express the dipole spin tensor in terms of a spin vector 
$a^\mu$, which is orthogonal to the four-momentum, i.e., $p_\mu a^\mu = 0$. This is 
exactly the same as in our amplitude discussions. 

Before explicitly presenting the conserved quantities for the motion of a quadrupolar 
test body in a Kerr background, there are two important points to touch upon. 
The first is that, in the case of a test body, $p^\mu$ is no longer aligned 
with $u^\mu$, i.e. $p^\mu \neq m u^\mu$ as in the spinless case. In fact, they begin 
to differ at the quadrupolar order. To see this explicitly, consider the evolution of 
the SSC 
\begin{align}
	\frac{\mathrm{D}}{\mathrm{d} \tau} (\mathcal{S}^{\mu \nu} p_\nu) & = \frac{\mathrm{D} \mathcal{S}^{\mu \nu}}{\mathrm{d} \tau} p_\nu + \mathcal{S}^{\mu \nu} \frac{\mathrm{D} p_\nu}{\mathrm{d} \tau} \\
	& = (p \cdot u) p^\mu - m^2 u^\mu - \frac{1}{2}\mathcal{S}^{\mu \nu} R_{\nu \alpha \beta \gamma} u^\alpha \mathcal{S}^{\beta \gamma}  + \mathcal{L}^{\mu \nu} p_\nu + \mathcal{O}(\mathcal{S}^3) \nonumber \\
	& = 0 + \mathcal{O}(\mathcal{S}^3)\;, \nonumber
\end{align}
such that 
\begin{align}
	m^2 u^\mu - (p \cdot u) p^\mu =  -\frac{1}{2}\mathcal{S}^{\mu \nu} R_{\nu \alpha \beta \gamma} u^\alpha \mathcal{S}^{\beta \gamma}  + \mathcal{L}^{\mu \nu} p_\nu + \mathcal{O}(\mathcal{S}^3)\;,
	\label{eq:MassSquaredQuadrupole}
\end{align}
where we identify $\mathcal{S}^{\mu \nu} \mathcal{F}_{\nu} \sim \mathcal{O}(\mathcal{S}^3)$. Contracting with $u_\mu$ gives
\begin{equation}
	m^2 = (p \cdot u)^2 -\frac{1}{2}u_\mu \mathcal{S}^{\mu \nu} R_{\nu \alpha \beta \gamma} u^\alpha \mathcal{S}^{\beta \gamma}  + u_\mu\mathcal{L}^{\mu \nu} p_\nu + \mathcal{O}(\mathcal{S}^3)\;.
\end{equation}
Then, if we solve \myeqref{eq:MassSquaredQuadrupole} for $u_\mu$ and plug it into 
the above equation, we arrive at 
\begin{align}
	m^2 & = (p \cdot u)^2 -\frac{1}{2}\frac{(p \cdot u)}{m^2} p_\mu\left[ \mathcal{S}^{\mu \nu} R_{\nu \alpha \beta \gamma} u^\alpha \mathcal{S}^{\beta \gamma} - 2\mathcal{L}^{\mu \nu} p_\nu \right] + \mathcal{O}(\mathcal{S}^3) \nonumber  \\
	& = (p \cdot u)^2 + \mathcal{O}(\mathcal{S}^3)\;, 
\end{align}
by virtue of the SSC and antisymmetry of $\mathcal{L}^{\mu \nu}$. Note that we denote
the higher-spin corrections as $\mathcal{O}(\mathcal{S}^3)$ because we have not 
included octupolar corrections in the MPD equations. Thus, the analysis yields
\begin{equation}
	p^\mu = m u^\mu + \frac{1}{2m} \mathcal{S}^{\mu \nu} R_{\nu\alpha\beta\gamma}u^\alpha\mathcal{S}^{\beta\gamma} - \mathcal{L}^{\mu \nu} u_\nu + \mathcal{O}(\mathcal{S}^3)\;,
\end{equation}
showing that, indeed, $p_\mu$ and $u_\mu$ differ starting from the quadrupolar order. 

The second point to mention is that, as we are interested in studying 
conserved quantities, the Hamiltonian formulation is particularly powerful. In
this context, one usually works with the canonical four-momentum for the probe, defined as\footnote{Global spacetime indices are denoted with Greek letters, 
whereas the local spacetime indices are denoted with Latin letters.}
\begin{equation}
	\pi_\mu = p_\mu - \frac{1}{2}\omega_{\mu ab} \mathcal{S}^{ab}\;,
\end{equation}
where $\mathcal{S}^{ab}$ is the dipole spin tensor in the local Lorentz frame and 
$\omega_{\mu}^{~ab}$ is the spin connection defined in terms of the vierbein:
\begin{equation}
	\omega_{\mu}^{~ab} = e_{\nu}^{~a} \nabla_{\mu} e^{b \nu} = e_{\nu}^{~a}(\partial_\mu e^{b \nu} + \Gamma^{\nu}_{~\mu \lambda}e^{b \lambda})\;.
\end{equation}
The vierbein allows us to embed objects defined in the local spacetime into the global 
spacetime. In light of this, we will choose to write our conserved quantities in terms 
of $\pi_{\mu}$ where possible.

Given our discussion, the conserved quantities for a quadrupolar probe in a Kerr background are as follows. 
First, we have the spin-corrected energy and azimuthal angular momentum:
\begin{align}
E = -\pi_\mu \xi_{(t)}^\mu + \frac{1}{2} \mathcal{S}^{\alpha \beta} \nabla_\beta \xi_{(t)\alpha}\;, \quad L = \pi_\mu \xi_{(\phi)}^\mu - \frac{1}{2} \mathcal{S}^{\alpha \beta} \nabla_\beta \xi_{(\phi)\alpha}\;.
\end{align}
Second, the Killing tensor gives rise to the spin-corrected Carter 
constant~\cite{Carter:1968ks,Compere:2021kjz,Compere:2023alp}: 
\begin{align}
\label{eq:carterQ}
\mathcal{K} &= Y_{\mu \rho} Y^\rho{}_\nu p^\mu p^\nu + 4 \xi_{(t)}^\lambda \epsilon_{\lambda \mu \sigma[\rho} Y_{\nu]}{}^\sigma \mathcal{S}^{\mu \nu} p^\rho - \Big[ g_{\mu \rho} \left( (\xi_{(t)})_\nu (\xi_{(t)})_\sigma - \frac{1}{2} g_{\nu \sigma} \xi_{(t)}^2 \right) \nonumber \\
& \qquad - \frac{1}{2} Y_\mu{}^\lambda \left( Y_\rho{}^\kappa R_{\lambda \nu \kappa \sigma} + \frac{1}{2} Y_\lambda{}^\kappa R_{\kappa \nu \rho \sigma} \right) \Big] \mathcal{S}^{\mu \nu} \mathcal{S}^{\rho \sigma}\;. \nonumber
\end{align}
Finally, there is another hidden symmetry that yields a conserved projection of the 
spin along the principal plane of the Kerr geometry, namely the R\"udiger linear 
invariant:
\begin{align}
Q_Y =\frac{1}{2} \epsilon^{\mu \nu \alpha \beta} \mathcal{S}_{\mu \nu} Y_{\alpha \beta} \;.
\end{align}
These four quantities are understood to be conserved up to octupolar corrections (with 
$k = k_{\mathrm{BH}}$)
\begin{equation}
	\frac{\mathrm{D}}{\mathrm{d} \tau}\mathcal{Q} = \mathcal{O}(\mathcal{S}^3)\;, \quad \mathcal{Q} \in \{E, L, \mathcal{K}, Q_Y\}\;,
\end{equation}
and they fully characterise the integrable structure of spinning particle dynamics in 
Kerr at the quadrupolar order in probe spin~\cite{Compere:2023alp}. It is important to 
emphasise that this conservation is non-perturbative in Newton's constant and holds to 
all orders in the background spin. Another observation to 
note is that $\{E, L, Q_Y\}$ are conserved for a quadrupolar probe 
\textit{without} adding quadrupolar corrections to the quantities themselves. In other 
words, the dipolar correction is sufficient for conservation of $E$, $L$, and $Q_Y$\footnote{See 
Ref.~\cite{deFirmian:2026mln} for a discussion on conserved quantities for a quadrupolar probe in root-Kerr.}.

%--------------------------------------------------------------------------------
\subsection{Symmetries for Scattering Dynamics}
\label{sec:SymmetriesforScattering}
%--------------------------------------------------------------------------------

To make a connection with the asymptotic radial action, we evaluate these four 
conserved quantities in the asymptotic limit, i.e.~as the separation goes to 
infinity $r \to +\infty$. In this limit, the conserved quantities reduce to 
very clean expressions that depend only on the two-particle 
kinematics~\cite{Gonzo:2024zxo,Vinesslides}:
\begin{align}
	\label{eq:conserved-quantities-asymptotic}
	y & = \frac{E}{m_1}\;, \\
	\tilde{L} & \equiv \frac{|a_2|}{m_1}L =l\cdot a_2 -y (a_1\cdot a_2)+(u_2 \cdot a_1)(u_1 \cdot a_2)\;, \\
    Q_Y&= l\cdot a_1 + y (a_1\cdot a_2) - (u_2 \cdot a_1)(u_1 \cdot a_2)\;, \\[0.1cm]
    Q &\equiv \frac{\mathcal{K}}{m_1^2} - a_1^2 - a_2^2 \\
    & =  -l^2-2 y (l\cdot a_2) - (y^2-1)a_2^2 - (u_1\cdot a_2)^2 +2y (l\cdot a_1) \nonumber \\
    & +2 (y^2+1)(a_1\cdot a_2) -2y (u_1\cdot a_2)(u_2 \cdot a_1) - (y^2-1)a_1^2-(u_2\cdot a_1)^2\;, \nonumber
\end{align}
with 
\begin{equation}
	l^\mu = \epsilon^{\mu \nu \rho \sigma} b_{\nu}u_{1\rho}u_{2\sigma}\;, \quad 
	\mathcal{S}_i^{\mu \nu}= m_i \epsilon^{\mu \nu}{ }_{\rho \sigma} u_i^\rho a_i^\sigma\;,
\end{equation}
and we choose particle 1 as the probe while treating particle 2 as the 
background. The conserved quantities here are understood to simply be the asymptotic 
limits of the bulk conserved quantities. More specifically, $E$ and $L$ are the energy 
and axial angular momentum of particle $1$ in the rest frame of $2$, respectively, 
while $Q_Y$ is the R\"udiger linear invariant and $Q$ is the quadrupolar extension of 
the Carter constant. 

Now we may proceed with defining exactly what we mean by conserved quantities for our 
scattering dynamics. To begin, let us specify the relevant kinematic phase space for 
the classical gravitational scattering of two spinning bodies, given by $\mathcal{H}_{\rm 
2B} = \{u_1^{\mu},u_2^{\mu}, a_1^{\mu}, a_2^{\mu}, b^{\mu}\}$. Here, $2\mathrm{B}$ stands 
for two-body, $u_i^\mu$ denotes the particle velocities, $a_i^\mu$ their normalised spin 
vectors, and $b^\mu = b_2^\mu - b_1^\mu$ is the initial impact parameter of the system. 
Then, following the discussion of \mysecref{sec:CommutatorsToCovariantDiracBrackets}, given 
a quantity $\mathcal{Q}(\{u_i, a_i, b\})$ in our phase space $\mathcal{H}_{\rm 2B}$, we say 
that such a quantity is conserved if 
\begin{equation}
	 \{\mathcal{Q}, I_r\}_{\mathrm{D.B}} = 0 \implies  \langle \Delta \mathcal{Q} \rangle = 0\;,
	 \label{eq:vanishing_bracket}
\end{equation}
just as we would do in a Hamiltonian formulation\footnote{Unlike the usual Hamiltonian flow, 
the Dirac brackets used here are gauge-invariant. It should be possible to extend this 
definition of conserved quantity to the generic dissipative case upon inclusion of the 
radiative kernels, as suggested in Ref.~\cite{Alessio:2025flu}.}. 

A brief word of caution is necessary here regarding the distinction between local and 
asymptotic conservation laws in the sense of \myeqref{eq:vanishing_bracket}: while the 
former implies the latter, the converse need not hold true. In particular, starting at the 
quadratic order in spin, curvature-dependent contributions generate subleading 
corrections in the $1/r$ expansion to the asymptotic quantities in 
\myeqref{eq:conserved-quantities-asymptotic}. For example, the 
kinematic mass $m_1^2 = \eta_{\mu \nu}p_1^\mu p_1^\nu$ is not conserved off-shell at $
\mathcal{O}(\mathcal{S}_1^2)$ in the worldline setup. The conserved quantity at this 
order is instead the dynamical mass~\cite{Jakobsen:2021zvh}
\begin{equation}
    p_\mu p^\mu \equiv \mu_1^2 = m_1^2 - \frac{1}{4} R_{\mu\nu\rho\sigma}\; \mathcal{S}_1^{\mu\nu} \mathcal{S}_1^{\rho\sigma}
    + \mathcal{O}(\mathcal{S}_1^3),
\end{equation}
whose curvature-dependent correction vanishes in the $r\to\infty$ limit, effectively 
leading to $\mu_1 \to m_1$ in our on-shell approach. Similar curvature terms appear on 
the last line of the quadratic-in-spin correction to the Carter constant, 
i.e.,~\myeqref{eq:carterQ}, but likewise drop out asymptotically.\footnote{Indeed these 
terms become important when extending the conservation laws into the bulk or 
describing the local off-shell dynamics in the strong field regime, where curvature 
effects cannot be discarded.} Nevertheless, since our primary goal is to analyse the 
simplicity of the radial action and other on-shell quantities relevant for Kerr black 
hole scattering, the asymptotic formulation adopted here provides precisely the 
gauge-invariant notion of conservation laws that are appropriate for our purposes.

%--------------------------------------------------------------------------------
\section{Unexpected Symmetries of Kerr Black Hole Scattering}
\label{sec:UnexpectedSymmsOfBHScattering}
%--------------------------------------------------------------------------------

In this section, our goal is to test the symmetries of the radial action using data available in 
the literature:
\begin{enumerate}
	\item 1PM and all orders in spin~\cite{Guevara:2019fsj}, 
    \item 2PM through $\mathcal{O}(a^{11})$~\cite{Bohnenblust:2024hkw}\footnote{It is still unknown if the dynamics at $\mathcal{O}(a^{5})$ and beyond reproduces the dynamics of a Kerr black hole.}, 
    \item 2PM and 3PM through $\mathcal{O}(a^2)$~\cite{Jakobsen:2022fcj}, 
    \item 2PM and 3PM through $\mathcal{O}(a_1^4 a_2^0)$~\cite{Akpinar:2025bkt}, 
    \item $\infty$PM through $\mathcal{O}(a_1^1 a_2^\infty)$ in the probe limit~\cite{Gonzo:2024zxo}\footnote{Note that the results of Ref.~\cite{Gonzo:2024zxo} are already written in terms of conserved quantities and so automatically obey \eqref{eq:vanishing_bracket}.},
    \item 3PM through $\mathcal{O}(a^4)$, 4PM through $\mathcal{O}(a^3)$, and 5PM through $\mathcal{O}(a^2)$ in the probe limit~\cite{Hoogeveen:2025tew}, 
\end{enumerate}
where the total power in spin is denoted by $\mathcal{O}(a^n)$ (for example, 
$\mathcal{O}(a^2)=\mathcal{O}(a_1^2)+\mathcal{O}(a_1 a_2)+\mathcal{O}(a_2^2)$). For 
Refs.~\cite{Jakobsen:2022fcj,Hoogeveen:2025tew}, we extracted the radial action 
uniquely via an ansatz matched to their impulse. The generic expression for the spinning radial action 
at order $G^n$, including terms up to $a_1 + a_2 \leq 4$, reads
\begin{align}
    \mathcal{I}^{(n)}_{r}(b) = \frac{1}{|b|^{(n-1)(1-2\epsilon)}}\sum_{a_1,a_2}\sum_{i}\alpha^{(a_1,a_2,i)}\mathcal{J}^{(a_1,a_2,i)}\;,
    \label{eq:IrGeneralAnsatz}
\end{align}
with $|b|^2 = -b^2$ and $\epsilon = (4 - D)/2$. The integers $a_1$ and $a_2$ label the spin orders of the two 
bodies, while $i$ indexes independent structures at each spin 
level. The coefficients $\alpha$ are general functions of the kinematic variables, and $\mathcal{J}$ represent the 
corresponding independent spin structures. Now we list all of the spin structures with overall factors of $|b|$ 
stripped off. In other words, every factor 
of $b^\mu$ or $a_i^\mu$ that appears in our list must be compensated by a factor of $|b|$, 
for example, $(l \cdot a_1)$ actually appears as $(l \cdot a_1)/|b|^2$. Moreover, the spin-independent contribution 
for $n = 1$ should have an overall factor of $\operatorname{log}(|b|^2)$. At $\mathcal{O}(a_1^0 a_2^0)$ we have 
\begin{equation}
     \mathcal{J}^{(0,0,0)} = \mathds{1}_{a_1} \mathds{1}_{a_2}\;.
\end{equation}
For $\mathcal{O}(a_1^1 a_2^0)$ and $\mathcal{O}(a_1^0 a_2^1)$ we have 
\begin{align}
    \mathcal{J}^{(1,0,0)} = (l \cdot a_1) \mathds{1}_{a_2}\;, \qquad \mathcal{J}^{(0,1,0)} = (l \cdot a_2) \mathds{1}_{a_1}\;.
\end{align}
For the quadratic order in spin we have terms at $\mathcal{O}(a_1^2 a_2^0)$ and $\mathcal{O}(a_1^0 a_2^2)$
\begin{align}
    & \mathcal{J}^{(2,0,0)} = (b \cdot a_1)^2 \mathds{1}_{a_2}\;, \ \mathcal{J}^{(2,0,1)} = a_1^2 \mathds{1}_{a_2}\;, \ \mathcal{J}^{(2,0,2)} = (u_2 \cdot a_1)^2 \mathds{1}_{a_2}\;, \\
    & \mathcal{J}^{(0,2,0)} = (b \cdot a_2)^2 \mathds{1}_{a_1}\;, \ \mathcal{J}^{(0,2,1)} = a_2^2 \mathds{1}_{a_1}\;, \ \mathcal{J}^{(0,2,2)} =  (u_1 \cdot a_2)^2 \mathds{1}_{a_1}\;, 
\end{align}
as well as $\mathcal{O}(a_1^1 a_2^1)$ 
\begin{equation}
	 \mathcal{J}^{(1,1,0)} = (b \cdot a_1)(b \cdot a_2)\;, \  \mathcal{J}^{(1,1,1)} = (a_1 \cdot a_2) \;, \  \mathcal{J}^{(1,1,2)} = (u_2 \cdot a_1)(u_1 \cdot a_2)\;.
\end{equation}
At cubic order in spin we have terms at $\mathcal{O}(a_1^3 a_2^0)$ and $\mathcal{O}(a_1^0 a_2^3)$ 
\begin{align}
	 \mathcal{J}^{(3,0,0)} = (l \cdot a_1)(b \cdot a_1)^2 \mathds{1}_{a_2}\;, \ & \mathcal{J}^{(3,0,1)} = (l \cdot a_1) a_1^2 \mathds{1}_{a_2}\;, \\
	 & \hspace{-3.5cm} \mathcal{J}^{(3,0,2)} = (l \cdot a_1)(u_2 \cdot a_1)^2 \mathds{1}_{a_2}\;, \nonumber
\end{align}
\begin{align}
	\mathcal{J}^{(0,3,0)} = (l \cdot a_2)(b \cdot a_2)^2 \mathds{1}_{a_1}\;, \  & \mathcal{J}^{(0,3,1)} = (l \cdot a_2)a_2^2 \mathds{1}_{a_1}\;, \\
	& \hspace{-3.5cm} \mathcal{J}^{(0,3,2)} = (l \cdot a_2)(u_1 \cdot a_2)^2 \mathds{1}_{a_1}\;, \nonumber
\end{align}
as well as $\mathcal{O}(a_1^2 a_2^1)$ and $\mathcal{O}(a_1^1 a_2^2)$ 
\begin{align}
   & \hspace{-0.25cm} \mathcal{J}^{(2,1,0)} = (l \cdot a_2) (b \cdot a_1)^2\;, \ \mathcal{J}^{(2,1,1)} = (l \cdot a_2) a_1^2\;, \ \mathcal{J}^{(2,1,2)} = (l \cdot a_2) (u_1 \cdot a_2)^2\;, \nonumber \\
   & \hspace{0.5cm} \mathcal{J}^{(2,1,3)} = (l \cdot a_1) (b \cdot a_1) (b \cdot a_2)\;, \ \mathcal{J}^{(2,1,4)} = (l \cdot a_1) (u_2 \cdot a_1)(u_1 \cdot a_2)\;, \\
   & \hspace{-0.25cm} \mathcal{J}^{(1,2,0)} = (l \cdot a_1) (b \cdot a_2)^2\;, \ \mathcal{J}^{(1,2,1)} = (l \cdot a_1) a_2^2\;, \ \mathcal{J}^{(1,2,2)} = (l \cdot a_1) (u_2 \cdot a_1)^2\;, \nonumber \\
   & \hspace{0.5cm} \mathcal{J}^{(1,2,3)} = (l \cdot a_2) (b \cdot a_1) (b \cdot a_2)\;, \ \mathcal{J}^{(1,2,4)} = (l \cdot a_2) (u_2 \cdot a_1)(u_1 \cdot a_2)\;.
\end{align}
Finally, at quartic order in spin, we have terms at $\mathcal{O}(a_1^4 a_2^0)$ and $\mathcal{O}(a_1^0 a_2^4)$
\begin{align}
     & \hspace{1cm} \mathcal{J}^{(4,0,0)} = (b \cdot a_1)^4 \mathds{1}_{a_2}\;, \ \mathcal{J}^{(4,0,1)} = (b \cdot a_1)^2 a_1^2 \mathds{1}_{a_2}\;, \ \mathcal{J}^{(4,0,2)} = a_1^4 \mathds{1}_{a_2}\;,  \\
     & \mathcal{J}^{(4,0,3)} = (b \cdot a_1)^2 (u_2 \cdot a_1)^2 \mathds{1}_{a_2}\;, \ \mathcal{J}^{(4,0,4)} = a_1^2 (u_2 \cdot a_1)^2 \mathds{1}_{a_2}\;, \ \mathcal{J}^{(4,0,5)} = (u_2 \cdot a_1)^4 \mathds{1}_{a_2}\;, \nonumber \\
     & \hspace{1cm} \mathcal{J}^{(0,4,0)} = (b \cdot a_2)^4\mathds{1}_{a_1}\;, \ \mathcal{J}^{(0,4,1)} = (b \cdot a_2)^2 a_2^2 \mathds{1}_{a_1}\;, \ \mathcal{J}^{(0,4,2)} = a_2^4\mathds{1}_{a_1}\;,  \\
     & \mathcal{J}^{(0,4,3)} = (b \cdot a_2)^2 (u_1 \cdot a_2)^2\mathds{1}_{a_1}\;, \ \mathcal{J}^{(0,4,4)} = a_2^2 (u_1 \cdot a_2)^2\mathds{1}_{a_1}\;, \ \mathcal{J}^{(0,4,5)} = (u_1 \cdot a_2)^4\mathds{1}_{a_1}\;, \nonumber
\end{align}
at $\mathcal{O}(a_1^3 a_2^1)$ and $\mathcal{O}(a_1^1 a_2^3)$ 
\begin{align}
     & \mathcal{J}^{(3,1,0)} = (b \cdot a_1)^3(b \cdot a_2)\;, \ \mathcal{J}^{(3,1,1)} = (b \cdot a_1)(b \cdot a_2) a_1^2\;, \ \mathcal{J}^{(3,1,2)} = (b \cdot a_1)^2 (a_1 \cdot a_2)\;,\nonumber \\
     & \hspace{2cm} \mathcal{J}^{(3,1,3)} = a_1^2 (a_1 \cdot a_2)\;, \ \mathcal{J}^{(3,1,4)} = (b \cdot a_1)(b \cdot a_2)(u_2 \cdot a_1)^2\;, \nonumber \\
     & \hspace{1.25cm} \mathcal{J}^{(3,1,5)} = (a_1 \cdot a_2)(u_2 \cdot a_1)^2\;, \ \mathcal{J}^{(3,1,6)} = (b \cdot a_1)^2(u_2 \cdot a_1)(u_1 \cdot a_2)\;,\\
     & \hspace{2cm} \mathcal{J}^{(3,1,7)} = a_1^2 (u_2 \cdot a_1)(u_1 \cdot a_2)\;, \ \mathcal{J}^{(3,1,8)} = (u_2 \cdot a_1)^3(u_1 \cdot a_2)\;, \nonumber 
\end{align}
\begin{align}
     & \mathcal{J}^{(1,3,0)} = (b \cdot a_2)^3(b \cdot a_1)\;, \ \mathcal{J}^{(1,3,1)} = (b \cdot a_2)(b \cdot a_1) a_2^2\;, \ \mathcal{J}^{(1,3,2)} = (b \cdot a_2)^2 (a_1 \cdot a_2)\;,\nonumber \\
     & \hspace{2cm} \mathcal{J}^{(1,3,3)} = a_2^2 (a_1 \cdot a_2)\;, \ \mathcal{J}^{(1,3,4)} = (b \cdot a_1)(b \cdot a_2)(u_1 \cdot a_2)^2\;, \nonumber \\
     & \hspace{1.25cm} \mathcal{J}^{(1,3,5)} = (a_1 \cdot a_2)(u_1 \cdot a_2)^2\;, \ \mathcal{J}^{(1,3,6)} = (b \cdot a_2)^2(u_2 \cdot a_1)(u_1 \cdot a_2)\;,\\
     & \hspace{2cm} \mathcal{J}^{(1,3,7)} = a_2^2 (u_2 \cdot a_1)(u_1 \cdot a_2)\;, \ \mathcal{J}^{(1,3,8)} = (u_1 \cdot a_2)^3(u_2 \cdot a_1)\;, \nonumber 
\end{align}
and also at $\mathcal{O}(a_1^2 a_2^2)$
\begin{align}
    & \mathcal{J}^{(2,2,0)} = (b \cdot a_1)^2(b \cdot a_2)^2\;, \ \mathcal{J}^{(2,2,1)} = (b \cdot a_1)^2(u_1 \cdot a_2)^2\;, \ \mathcal{J}^{(2,2,2)} = (b \cdot a_1)^2 a_2^2\;, \nonumber \\ 
    & \mathcal{J}^{(2,2,3)} = (u_2 \cdot a_1)^2(b \cdot a_2)^2\;, \ \mathcal{J}^{(2,2,4)} = (u_2 \cdot a_1)^2(u_1 \cdot a_2)^2\;, \ \mathcal{J}^{(2,2,5)} = (u_2 \cdot a_1)^2 a_2^2\;, \nonumber \\
    & \mathcal{J}^{(2,2,6)} = a_1^2 (b \cdot a_2)^2\;, \ \mathcal{J}^{(2,2,7)} = a_1^2 (u_1 \cdot a_2)^2\;, \ \mathcal{J}^{(2,2,8)} = a_1^2 a_2^2\;,  \\ & \mathcal{J}^{(2,2,9)} = (b \cdot a_1) (b \cdot a_2) (a_1 \cdot a_2)\;, \ \mathcal{J}^{(2,2,10)} = (u_2 \cdot a_1)(u_1 \cdot a_2) (a_1 \cdot a_2)\;, \nonumber\\
    & \mathcal{J}^{(2,2,11)} = (b \cdot a_1) (b \cdot a_2) (u_2 \cdot a_1)(u_1 \cdot a_2)\;. \nonumber
\end{align}
This yields a total of 70 spin structures~\cite{Bern:2022kto}. Then we substitute this ansatz of the radial action 
into \myeqref{eq:obsFormula}, compute all of the brackets, and match to the 
impulse. This allows us to uniquely fix the coefficients in \myeqref{eq:IrGeneralAnsatz}. 

We have checked that the radial actions considered in this analysis are in agreement where they 
overlap (up to metric signature and Levi-Civita conventions), and we have included 
their expressions in the ancillary file of Ref.~\cite{Akpinar:2025tct}. Note that, because of vanishing Gram 
determinants, the basis of spin vectors is not unique in $D = 4$ dimensions. In fact, this causes particular 
annoyance at the cubic-in-spin order and beyond, since comparing different results requires expressing everything 
in the same basis. The comparison can be streamlined by simply choosing a specific four-dimensional frame. A 
convenient choice is 
\begin{align}
	& u_1^\mu = (1,0,0,0)\;, \quad u_2^\mu = (y, 0, 0, \sqrt{y^2 - 1})\;, \quad b^\mu = (0, 0, |b|, 0)\;, \\
	& \hspace{0.25cm} a_1^\mu = (0, a_{1}^{(x)}, a_{1}^{(y)}, a_{1}^{(z)})\;, \quad a_2^\mu = (\frac{\sqrt{y^2-1}}{y} a_{2}^{(z)}, a_{2}^{(x)}, a_{2}^{(y)}, a_{2}^{(z)})\;, \nonumber
\end{align}
which satisfies all of the constraints of the dynamics.

Recall from \myeqref{eq:vanishing_bracket} that the radial action exhibits an (asymptotic) symmetry if its Dirac 
bracket with the corresponding conserved quantity vanishes. Since available results hold only up to a fixed order 
in the spins, conservation is likewise expected only at that order. More specifically, for a conserved quantity 
$\mathcal{Q}$ and a radial action $I_r^{n,m}$ including terms up to $\mathcal{O}(a_1^n a_2^m)$, conservation 
requires
\begin{equation}\label{eq:vanishing_bracket_spin}
    \{\mathcal{Q}, I_r^{n,m}\}_{\mathrm{D.B}} = 0 +\mathcal{O}(a_1^{n+1}a_2^{m})+\mathcal{O}(a_1^{n}a_2^{m+1})\;.
\end{equation}
The Dirac brackets do not mix loop orders but do mix spin orders, so all spin terms up to the desired order must 
be included in $I_r^{n,m}$. Moreover, because the brackets involve explicit factors of $1/m_i$, they also mix 
powers in the self-force expansion. In fact, in the probe limit $m_2\rightarrow \infty$, one must therefore use 
the probe limit of the radial action and discard any terms in \myeqref{eq:FundamentalDBs} proportional to 
$1/m_2$.

Our main results are summarised in Table~\ref{tab:core_conservation_results}, but there are some general features 
that we wish to mention. First, in the probe limit $m_2\rightarrow\infty$: 
\begin{itemize}
    \item The $\mathcal{O}(G^{\infty} a_1^1  a_2^\infty)$ radial action commutes with all conserved quantities.
    \item The 1PM radial action commutes with all of the conserved quantities to all orders in spin.
    \item The 2PM and 3PM radial action commute with all conserved quantities through $\mathcal{O}(a^4)$; likewise for the 4PM radial action through $\mathcal{O}(a^3)$ and the 5PM radial action through $\mathcal{O}(a^2)$.
\end{itemize}
Beyond the probe limit, conservation continues to hold for the spin-orbit sector (that is, for terms of 
order $\mathcal{O}(a_{1}^{0}a_{2}^{1})$ or $\mathcal{O}(a_{1}^{1}a_{2}^{0})$) since these couplings are 
universal. However, once we go beyond the spin-orbit level and include, for example, 
$\mathcal{O}(a_1^1 a_2^1)$ contributions, the conservation laws are generically broken. Furthermore, we also 
discuss checks of conservation laws for terms beyond $\mathcal{O}(a_1^n a_2^m)$ for $n+m \leq 4$ including 
non-minimal couplings in \mysecref{sec:BeyondProbe&Kerr}. 

\begin{table}[htbp]
\centering

% -------------------- First table --------------------
\begin{tabular}{|c|c|c|c|c|c|}
\hline
\multicolumn{6}{|c|}{\textbf{Probe limit}} \\ \hline
$\tilde{L}, Q_Y, Q$ 
& $\mathcal{O}(a^2)$ 
& $\mathcal{O}(a^3)$ 
& $\mathcal{O}(a^4)$ 
& $\mathcal{O}(a^5)$ 
& $\mathcal{O}(a^\infty)$ \\ \hline

$G^1$ & \tickg & \tickg & \tickg & \tickg & \tickg \\ \hline
$G^2$ & \tickg & \tickg & \tickg & \crossr & ? \\ \hline
$G^3$ & \tickg & \tickg & \tickg & ? & ? \\ \hline
$G^4$ & \tickg & \tickg & ? & ? & ? \\ \hline
$G^5$ & \tickg & ? & ? & ? & ? \\ \hline
\end{tabular}

\vspace{0.5cm}

% -------------------- Second table --------------------
\begin{tabular}{|c|c|c|c|c|c|c|c|}
\hline
\multicolumn{8}{|c|}{\textbf{Beyond probe limit (conservative)}} \\ \hline

$\tilde{L}, Q_Y, Q$ 
& $\mathcal{O}(a^1)$ 
& $\mathcal{O}(a_1^2 a_2^0)$
& $\mathcal{O}(a_1^3 a_2^0)$
& $\mathcal{O}(a_1^4 a_2^0)$
& $\mathcal{O}(a_1^5 a_2^0)$
& $\mathcal{O}(a_1^\infty a_2^0)$
& $\mathcal{O}(a_1^1 a_2^1)$ \\ \hline

$G^1$ & \tickg & \tickg & \tickg & \tickg & \tickg & \tickg & \crossr \\ \hline
$G^2$ & \tickg & \tickg & \tickg & \tickg & \crossr & ? & \crossr \\ \hline
$G^3$ & \tickg & \crossr & \crossr & \crossr & ? & ? & \crossr \\ \hline

\end{tabular}

\caption{Conservation results in the probe limit (top) and beyond the probe limit (bottom), restricted to the conservative sector. A `?' indicates orders in the radial action that are currently unknown. In the probe limit, these conservation laws enhance to asymptotic integrability. The pattern shown applies to all quantities $\tilde{L}, Q, Q_Y$ (and similarly to the $a_1 \leftrightarrow a_2$ contributions beyond the probe limit), with the sole exception of the $\mathcal{O}(G^2 a^5)$ term in the probe limit which is discussed further in \mysecref{sec:BeyondProbe&Kerr}. We find that spin-shift symmetry holds precisely in the same regimes in which integrability is realised, both in the probe limit and beyond.}

\label{tab:core_conservation_results}

\end{table}

%--------------------------------------------------------------------------------
\section{Reinterpreting Spin-Shift Symmetry}
\label{sec:ReinterpretingSpinShiftSymm}
%--------------------------------------------------------------------------------

As discussed in \ref{sec:SpinShiftSymmetry}, it is well-known that the spinning four-point amplitude exhibits 
a so-called \textit{spin-shift symmetry} through quartic order in spin up to two-loop order. This 
means that the amplitude is invariant under the shift
\begin{equation}
    a_i^\mu \rightarrow a_i^\mu + \xi q^\mu\;,
    \label{eq:spin-shift-symmetry}
\end{equation}
for any $\xi$ with a formal counting of $\mathcal O(1/|q|^2)$. This statement may be realised at the linearised level through the action of a differential operator
\begin{equation}
    \hat{D}_{\mathrm{SS}} = q^\mu \frac{\partial}{\partial a_i^\mu}\;.
\end{equation}
In position space, the statement \eqref{eq:spin-shift-symmetry} becomes
\begin{equation}
    \hat{O}_{\mathrm{SS}} I_r(\{u_i, a_i, b\}) = 0\;, \quad \hat{O}_{\mathrm{SS}} = \Pi_{\mu \nu}\frac{\partial}{\partial b^\mu}\frac{\partial}{\partial a_i^\nu}\;,
    \label{eq:spin-shift_bspace}
\end{equation}
where we use the transverse projector
\begin{equation}
    \Pi^{\mu \nu} = \eta^{\mu \nu} + \mathcal{P}_1^{\mu\nu} + \mathcal{P}_2^{\mu\nu}\;,
    \label{eq:transverse-projector}
\end{equation}
since the Fourier transform is performed in the transverse subspace. Importantly, we have explicitly checked 
that the radial action satisfies \myeqref{eq:spin-shift_bspace} when the spin-shift symmetry 
\myeqref{eq:spin-shift-symmetry} is exhibited by the corresponding spinning four-point amplitude in momentum 
space. We have further verified that the probe limit radial action in Ref.~\cite{Gonzo:2024zxo} at 
$\mathcal{O}(G^{\infty} a_1^1  a_2^\infty)$ is invariant under such a symmetry, as the radial momentum $p_r$ 
itself is invariant.

Now the task boils down to understanding if $\hat{O}_{\mathrm{SS}}$ can be generated by any scalar function on the 
phase space. In the context of classical mechanics, physical symmetries correspond to Hamiltonian vector flows 
generated by scalar charges, say $\chi$, acting via the Poisson or Dirac bracket:
\begin{equation}
	\mathcal{V}_\chi(f) = \{f,\chi\}\;.
\end{equation}
These vector fields necessarily involve at most one derivative with respect to phase space coordinates, which are 
realised by the action of the bracket. However, any transformation involving multiple derivatives cannot live in 
the space of Hamiltonian vector fields. Thus, $\hat{O}_{\mathrm{SS}}$ cannot be interpreted as the flow generated 
by a conserved charge. Instead, this reflects a field-space redundancy of the theory, analogous to gauge 
invariances, which leave physical observables unchanged but do not correspond to physical symmetries associated with 
Noether currents. In fact, to draw a useful analogy, if we consider scattering amplitudes involving massless 
vector or graviton states of momentum $k$ and polarisation $\epsilon(k)$, then physical observables are invariant 
under a shift in the polarisation
\begin{equation}
    \epsilon^\mu(k) \rightarrow \epsilon^\mu(k) + \zeta k^\mu\;,
\end{equation}
for any $\zeta$. This reflects a gauge redundancy in the polarisation vector, rather than a symmetry of the 
Hilbert space, and thus no Noether charge is associated to this. 

It should be noted that, from the GR perspective there is no obvious motivation for such a gauge redundancy. 
In light of this, we leave a more rigorous analysis of the nature of the spin-shift symmetry to future work, 
and content ourselves with interpreting the symmetry as a possible gauge redundancy that holds to \textit{all 
orders in the PM expansion}. 

%It is perhaps unsurprising that the classical spin vector exhibits this 
%property---at least from an amplitudes perspective---since it might \textit{inherit} gauge redundancies through 
%its relation to polarisations, i.e.~$a^\mu \sim \epsilon^{\mu}_{~\nu \rho \sigma}
%p^\nu\bar{\varepsilon}^\rho \varepsilon^\sigma$. Despite this, from the GR perspective, there is no obvious 
%motivation for such a gauge redundancy. In light of this, we leave a more rigorous analysis of the nature of the 
%spin-shift symmetry to future work, and content ourselves with interpreting the symmetry as a gauge redundancy that 
%holds to \textit{all orders in the PM expansion}. 

%--------------------------------------------------------------------------------
\section{The Integrable Sector of Kerr Black Hole Scattering}
\label{sec:TheIntegrableSector}
%--------------------------------------------------------------------------------

Here we define a new asymptotic notion of classical integrability for the scattering dynamics of binary black 
holes. The dynamics of the spinning radial action $I_r$ is determined by five vectors 
$\{u_1^{\mu},u_2^{\mu},a_1^{\mu},a_2^{\mu},b^{\mu}\}$, corresponding to a $5\times 4=20$ dimensional phase space 
after using translational freedom to choose $b^{\mu} = b_2^{\mu} - b_1^{\mu}$. We then impose on-shell conditions 
$u_i^2 = 1$, the SSC $u_i \cdot a_i = 0$, and transversality $b \cdot u_i = 0$. These conditions provide $6$ 
constraints, thereby reducing the phase space to $14$ dimensions. 
This is the space in which the Dirac brackets \myeqref{eq:FundamentalDBs} are defined, but we may define 
additional conserved quantities to further constrain the phase space. 

The first thing to consider is the conservation of spin magnitudes 
$\left\langle \Delta a_i^2 \right\rangle\big|_{\mathrm{cons}} = 0$, which follows directly from the bracket 
structure because $a_i^2$ are Casimirs of the algebra. This yields two additional constraints, further reducing 
the phase space to $14 - 2 = 12$ dimensions. Translation invariance implies the conservation of centre-of-mass 
momentum
\begin{align}
    u_{\rm CM}^\mu := \frac{m_1 u_1^\mu + m_2 u_2^\mu}{M}\;, \qquad M := m_1 + m_2\;,
\end{align}
so that $\left\langle \Delta u_{\rm CM}^\mu \right\rangle\big|_{\mathrm{cons}} = 0$. It is convenient to introduce 
the \emph{relative momentum}
\begin{align}
    p_{\rm rel}^\mu := \frac{E_2 p_1^\mu - E_1 p_2^\mu}{M}\;,
\end{align}
which reduces to $(0, \boldsymbol{p}_\infty)$ in the centre-of-mass frame. Since $y$, $E_1$, $E_2$, and 
$|\boldsymbol{p}_\infty| \propto \sqrt{y^2 - 1}$ are conserved under Dirac bracket evolution, the magnitude 
of $\boldsymbol{p}_{\rm rel}$ is fixed. Then, writing $(\boldsymbol{u}_1,\boldsymbol{u}_2)$ in terms of 
$(\boldsymbol{u}_{\rm CM},|\boldsymbol{p}_{\infty}| \boldsymbol{u}_{\rm rel})$, where $\boldsymbol{u}_{\rm rel} = \boldsymbol{p}_{\rm rel} / |\boldsymbol{p}
_{\infty}|$ is a unit vector, implies four additional reductions, yielding a final phase space of dimension 
$12-4 = 8$.

Therefore we obtain the reduced phase space $\mathcal{H}^{\rm red}_{\rm 2B}$ 
\begin{align}
    \label{eq:phase-space_fin}
    &\hspace{0.8cm}\mathcal{H}^{\rm red}_{\rm 2B} = \left\{ \boldsymbol{b}_{\perp}, \boldsymbol{u}_{\rm rel}, \boldsymbol{a}_{1\perp}, \boldsymbol{a}_{2\perp} \right\}\;, \\
    &b^{\mu}_{\perp} = \Pi^{\mu \nu}b_{\nu}\;, \quad a_{i\perp}^\mu = \left( \eta^{\mu\nu} - u_i^\mu u_i^\nu \right) \frac{a_{i\nu}}{a_i^2}\;.\nonumber
\end{align}
This dynamical system, defined on a $2n$-dimensional reduced phase space endowed with a Dirac bracket structure, 
is understood to be asymptotically integrable in the Liouville sense\footnote{In our case, although there is no 
Hamiltonian in the usual sense, the radial action generates a flow on the reduced phase space that plays an 
analogous role to a Hamiltonian vector field. It would be interesting to place this concept on a rigorous 
footing.} if there exist $n$ independent phase-space functions $F_i$---said to be in \textit{involution}---that 
are all mutually commuting under the Dirac brackets
\begin{equation}
	\{F_i, F_j\}_{\mathrm{D.B}} = 0 \quad \forall \quad i,j\;.
\end{equation}
This notion is \textit{generically weaker} than bulk Liouville integrability: the existence of commuting 
asymptotic charges does not guarantee that the full equations of motion admit the same number of conserved 
quantities in involution. As seen for the spinning-probe dynamics in 
Kerr~\cite{Witzany:2019nml,Gonzo:2024zxo,Witzany:2024ttz,Skoupy:2024uan,Ramond:2022vhj,Ramond:2024sfp,
Ramond:2024ozy} and discussed earlier, the obstruction arises from curvature-dependent spin couplings that are 
absent in the asymptotic on-shell regime but are present in the local bulk dynamics, breaking the correspondence 
between asymptotic and local conservation laws.

Although the reduced phase space \myeqref{eq:phase-space_fin} is the natural one for the conservative dynamics of 
spinning black hole binary systems, we had shown earlier that there exist further conservation laws, at least for 
a spinning probe particle in Kerr. Using their asymptotic expansion \myeqref{eq:conserved-quantities-asymptotic}, we 
define 
\begin{align}
    \boldsymbol{F}^{\rm BH} = \{I_r,\tilde{L},Q,Q_Y\} \;,
\end{align}
which we use to test integrability of our spinning radial action. Within the probe limit $m_2\rightarrow\infty$ 
when both black holes are spinning, we have checked that
\begin{equation}
    \{F^{\rm BH}_i, F^{\rm BH}_j\}_{\mathrm{D.B}} = 0 \quad \forall \quad  i,j=1,2,3,4\;.
\end{equation}
The above relation holds in the probe limit at the same orders in PM and spin where conservation holds in Table~\ref{tab:core_conservation_results}.
Beyond the probe limit, when both black holes are spinning, $\tilde{L}$, $Q$, and $Q_Y$ are not in involution and 
so integrability breaks down (albeit at least at 1PM this can be restored; see \mysecref{sec:BeyondProbe&Kerr}). 
In the case of one spinning black hole, say, $a_1\rightarrow0$ ($a_2\rightarrow0$) we find that 
$Q_Y$ ($\tilde{L}$) is zero, such that the remaining bracket $\{Q,\tilde{L}\}$ ($\{Q,Q_Y\}$) 
vanishes \textit{without taking the probe limit}. We therefore conclude: 

\begin{itemize}
    \item (Conjecture) The probe dynamics of one spinning BH in the background of a Kerr BH is asymptotically integrable at all PM orders up to quartic order in spin for the probe;
    \item The scattering of one spinless BH and one spinning BH is asymptotically integrable at the 2PM order and quartic order in the BH spin, even beyond the probe limit.
\end{itemize}
This property can be made manifest by expressing the radial action in terms of the conserved charges $\tilde{L}$,
$Q$ and $Q_Y$. Again, this is done by writing an ansatz for the radial action. More precisely, at order $G^n$ we 
consider a multivariate polynomial of degree $m$ in the conserved charges and Casimirs, where $m$ is the maximum 
order in the spin multipole expansion. Denoting the multivariate polynomial as $\operatorname{Poly}$, the ansatz 
takes the form
\begin{equation}
    I^{(n)}_{r}(b) = (Q)^{-\frac{n}{2}}\operatorname{Poly}\Bigg[\frac{\tilde{L}}{Q},\frac{Q_Y}{Q},\frac{a_1^2}{Q},\frac{a_2^2}{Q}\Bigg]\;,
\end{equation}
where the prefactor dictates the overall $|b|$ scaling of the polynomial. These results are included in the 
ancillary file of Ref.~\cite{Akpinar:2025tct}.

%--------------------------------------------------------------------------------
\section{Bootstrapping the Spinning Radial Action from Symmetries}
\label{sec:BootstrappingTheRA}
%--------------------------------------------------------------------------------

An extremely intriguing consequence of our analysis is that integrability and spin-shift symmetry in the probe 
limit impose nontrivial constraints on the structure of the radial action, enabling us to bootstrap the radial 
action under the assumption that both properties hold. 

Consider the generic expression for the spinning radial action \myeqref{eq:IrGeneralAnsatz} at order $G^n$, 
describing the dynamics of a spinning probe in a Kerr background. At this order in spin, there are 70 a priori 
undetermined coefficients in the radial action. The objective is to constrain these coefficients by imposing 
conservation laws in the probe limit and enforcing spin-shift symmetry:
\begin{itemize}
    \item Conservation of the charges $Q_Y$ and $Q$ yields 40 independent constraints;
    \item Conservation of $\tilde{L}$ imposes no further constraints, because the transversality conditions $u_i \cdot b = 0$ already fix the preferred plane. Consequently, $\tilde{L}$ is automatically conserved for the radial action in the probe limit;
    \item Spin-shift symmetry imposes 15 constraints.
\end{itemize}
These constraints reduce the number of free parameters to 15, which precisely matches the number of independent 
coefficients entering the aligned-spin scattering angle. Remarkably, as a result, knowledge of the aligned-spin 
scattering angle alone is sufficient to determine all 70 coefficients of the general spinning radial action. 
Of course, this conclusion relies on the assumption that the radial action remains 
integrable and obeys spin-shift symmetry through quartic order in spin. But we have already established these 
properties at the first few PM orders and conjecture that they persist to all PM orders. It is worth noting that 
this logic is reminiscent of the tutti-frutti method developed in~\cite{Bini:2019nra,Bini:2020wpo}, although here 
symmetries play the role that, in that approach, is taken by combining different perturbative expansions.

%--------------------------------------------------------------------------------
\section{Beyond the Probe Limit and Kerr Dynamics}
\label{sec:BeyondProbe&Kerr}
%--------------------------------------------------------------------------------

Thus far, our discussion about conserved quantities has been restricted to the probe limit, i.e.~where  
$\sim 1/m_2 \rightarrow 0$. There are two natural questions that follow: 
\begin{enumerate}
	\item What happens beyond the probe limit?
    \item What happens beyond Kerr dynamics?
\end{enumerate}
Here we list additional checks of asymptotic integrability using the Dirac brackets and the radial actions 
discussed in the previous section, including terms beyond $\mathcal{O}(a_1^n a_2^m)$ for $n+m \leq 4$ and 
non-minimal couplings. 

At 2PM, the results of Ref.~\cite{Bohnenblust:2024hkw} go beyond the fourth order in the probe spin and contain 
results to $\mathcal{O}(a^{11})$. For these we find the following pattern of conservation in the probe limit 
$m_2\rightarrow\infty$:
\begin{center}
	\begin{tabular}{|c|c|c|c|c|c|c|}
		\hline
        \parbox[c][1.8cm][c]{3.9cm}{\centering 
        \small{Probe limit: 2PM~\cite{Bohnenblust:2024hkw}} \\ \footnotesize{(Minimal coupling)}
        } 
        & $\mathcal{O}(a^4)$ & $\mathcal{O}(a_1^5 a_2^0)$ & $\mathcal{O}(a_1^{11}a_2^0)$ & $\mathcal{O}(a_1^4 a_2^7)$ & $\mathcal{O}(a_1^5a_2^1)$ &$\mathcal{O}(a^{11})$\\ \hline
		$\tilde{L}$ & \tickg & n/a & n/a & \tickg& \tickg&\tickg \\ \hline
		$Q_Y$ & \tickg & \crossr & \crossr & \tickg & \crossr&\crossr\\ \hline
		$Q$ & \tickg & \tickg & \tickg & \tickg & \crossr &\crossr \\ \hline
	\end{tabular}
\end{center}
where n/a denotes cases in which the relevant conserved quantity is zero due to one of the 
black holes being spinless. The chosen powers of spin are for illustrative purposes. These results hold 
regardless of whether terms involving $|a_i|$ are included in the radial action. Beyond the probe limit the 
conservation is broken already at the orders in spin shown in \mysecref{sec:UnexpectedSymmsOfBHScattering}. 

We have also extracted the radial action from Ref.~\cite{Jakobsen:2022fcj} including non-minimal spin coupling 
terms at $\mathcal{O}(a^2)$ used to describe e.g., neutron stars. The results of our conservation checks are given 
below:
\begin{center}
	\begin{tabular}{|c|c|c|c|c|c|}
		\hline
        \parbox[c][1.8cm][c]{3.9cm}{\centering 
        \small{Probe limit: 3PM~\cite{Jakobsen:2022fcj}} \\ \footnotesize{(Non-minimal coupling)}
        } 
        & $\mathcal{O}(a_1^1a_2^0)$ & $\mathcal{O}(a_1^0a_2^1)$ & $\mathcal{O}(a_1^1a_2^1)$ & $\mathcal{O}(a_1^2a_2^0)^\ast$ & $\mathcal{O}(a_1^0a_2^2)^\ast$ \\ \hline
		$\tilde{L}$ & n/a & \tickg & \tickg & n/a & \tickg\\ \hline
		$Q_Y$& \tickg & n/a & \tickg & \crossr & n/a\\ \hline
		$Q$ & \tickg & \tickg & \tickg & \tickg& \crossr \\ \hline
	\end{tabular}
\end{center}
where an asterisk indicates the inclusion of non-minimal couplings. This table applies to the results at 1PM, 2PM, 
or 3PM since they exhibit the same pattern of conservation. These 
findings seem to be consistent with the results of Ref.~\cite{Compere:2023alp} on the (non-)existence of conserved 
quantities for generic spinning test bodies in a fixed Kerr background. As expected, we find that demanding 
conservation of the charges enforces the Wilson coefficients of the non-minimal couplings to be zero. As 
observed in \mysecref{sec:UnexpectedSymmsOfBHScattering}, the azimuthal angular momentum is always conserved in the 
probe limit. 

Furthermore, we have analysed the conservation of non-minimal terms beyond the probe limit. As before, the same 
structure persists at 1PM, 2PM, and 3PM orders:
\begin{center}
	\begin{tabular}{|c|c|c|c|c|c|}
		\hline
         \parbox[c][1.8cm][c]{4.8cm}{\centering 
        \small{Beyond probe limit: 3PM~\cite{Jakobsen:2022fcj}} \\ \footnotesize{(Non-minimal coupling)}
        } 
        & $\mathcal{O}(a_1^1a_2^0)$ & $\mathcal{O}(a_1^0a_2^1)$ & $\mathcal{O}(a_1^1a_2^1)$ & $\mathcal{O}(a_1^2a_2^0)^\ast$ & $\mathcal{O}(a_1^0a_2^2)^\ast$ \\ \hline
		$\tilde{L}$ & n/a & \tickg & \crossr & n/a & \crossr\\ \hline
		$Q_Y$& \tickg & n/a & \crossr & \crossr & n/a\\ \hline
		$Q$ & \tickg & \tickg & \crossr & \crossr& \crossr \\ \hline
	\end{tabular}
\end{center}

%-----------------------------------------------------------------------------------
\subsection{1PM Beyond Probe: A New Basis of Conserved Charges}
%-----------------------------------------------------------------------------------

The conserved quantities in \mysecref{sec:SymmetriesforScattering} are only in involution if we take the probe 
limit $m_2\rightarrow\infty$. This implies, for example, that although the tree-level radial action can be written 
in terms of the conserved quantities, integrability is not manifest outside of the probe limit. This begs the 
question: \textit{can we find a better basis of conserved quantities that are in involution beyond the probe 
limit?} At least at 1PM, the answer is yes. We find that the following three conserved quantities 
\begin{equation}
	 \mathscr{L} =\frac{m_1^2 (l \cdot a_1)}{\Sigma_{m^2}} + \frac{m_2^2 (l \cdot a_2)}{\Sigma_{m^2}}-\frac{(2 m_1 m_2 + \Sigma_{m^2} y)}{\Sigma_{m^2}}(a_1\cdot a_2)+ (u_2 \cdot a_1)(u_1 \cdot a_2)\;,
\end{equation}
\begin{equation}
	\mathscr{Q}_Y =\frac{m_2^2 (l \cdot a_1)}{\Sigma_{m^2}} + \frac{m_1^2 (l \cdot a_2)}{\Sigma_{m^2}}+\frac{(2 m_1 m_2 +\Sigma_{m^2}y)}{\Sigma_{m^2}}(a_1\cdot a_2)- (u_2 \cdot a_1)(u_1 \cdot a_2)\;,
\end{equation}
\begin{align}
	\mathscr{Q}&=-l^2+\frac{2\Delta_{m^2} y}{\Sigma_{m^2}} (l\cdot a_2)-(y^2-1)a_2^2-(u_1\cdot a_2)^2-\frac{2\Delta_{m^2}y}{\Sigma_{m^2}} (l \cdot a_1 ) \\
    &+2\left(y^2+1+\frac{4m_1 m_2 y}{\Sigma_{m^2}}\right)(a_1\cdot a_2)-2y(u_2 \cdot a_1)(u_1 \cdot a_2)-(y^2-1)a_1^2- (u_2 \cdot a_1)^2\;, \nonumber
\end{align}
where we have defined
\begin{equation}
	\Sigma_{m^2} = m_1^2 + m_2^2\;, \quad \Delta_{m^2} = m_1^2 - m_2^2\;.
\end{equation}
These quantities commute with the 1PM radial action to all orders in spin and are in involution for generic 
masses. In other words, given $F_i,F_j \in\{I_r^{(1)},\mathscr{L},\mathscr{Q}_Y,\mathscr{Q}\}$ we find 
$\{F_i,F_j\}_{\mathrm{D.B}}=0$ \textit{without} taking the probe limit. The quantities above are symmetric under 
swapping $1\leftrightarrow2$ and reduce to the usual $\tilde{L},Q_Y,Q$ when $m_2\rightarrow\infty$.
We can also rewrite the 1PM radial action in terms of these quantities as
\begin{equation}
\begin{split}
    I_{r}^{(1)}&=\frac{Gm_1m_2(1-2y^2-2y\sqrt{y^2-1})}{2\sqrt{y^2-1}}\log\left(\mathscr{C}_1\right)\\
    &+\frac{Gm_1m_2(1-2y^2+2y\sqrt{y^2-1})}{2\sqrt{y^2-1}}\log\left(\mathscr{C}_2\right)\;,
\end{split}
\end{equation}
where
\begin{align}
	\mathscr{C}_1 & = \frac{2(y+\sqrt{y^2-1})}{y^2-1}\mathscr{L}-\frac{2(y-\sqrt{y^2-1})}{y^2-1}\mathscr{Q}_Y+\frac{1}{y^2-1}\mathscr{Q}\;, \\
	\mathscr{C}_2 & =\frac{2(y-\sqrt{y^2-1})}{y^2-1}\mathscr{L}-\frac{2(y+\sqrt{y^2-1})}{y^2-1}\mathscr{Q}_Y+\frac{1}{y^2-1}\mathscr{Q}\;.
\end{align}
At 2PM or 3PM, and up to $\mathcal{O}(a^2)$ one can find three quantities that commute with the 
$\mathcal{O}(G^2)$ (or $\mathcal{O}(G^3)$) piece of the radial action beyond the probe limit. However, these 
quantities do not commute with the radial action at any other order in $G$, in particular they do not commute 
with the tree-level radial action. This then suggests that it would be possible, as in the case of the 2PN 
spinning Hamiltonian~\cite{Tanay:2020gfb}, to find some perturbative integrability for generic spinning binaries 
in the PM expansion.

%--------------------------------------------------------------------------------
\chapter{Conclusion}
\label{sec:Chapter6}
%--------------------------------------------------------------------------------
 
Scattering amplitudes have emerged as an invaluable tool for modelling the two-body problem in GR. By reformulating 
the classical scattering dynamics in the language of QFT, we gain access to a vast library of computational machinery 
developed for particle physics that enables unique insights beyond those offered by traditional GR methods. In this thesis, 
we have exploited this fact to push the frontier of precision calculations for the scattering dynamics of Kerr black holes, 
and also for investigating the symmetries and integrability of such systems. The techniques we allude to here have been 
collected and reviewed in Chapter~\ref{sec:Chapter2}.

In Chapters~\ref{sec:Chapter3} and \ref{sec:Chapter4} we showed that modelling Kerr black holes as point particles in finite-spin-$s$ representations indeed allows one to faithfully compute corrections to observables that come from the spin multipole 
expansion of Kerr black holes, at least for the first few orders in the expansion. The observables were computed by first 
extracting the radial action by means of the amplitude-action relation, followed by using the Dirac brackets proposed in 
Ref.~\cite{Gonzo:2024zxo} to obtain the linear and angular impulses. Indeed, this approach provides an extremely 
clean method for computing scattering observables at high orders in spin and perturbation theory. Importantly, these 
computations leveraged a number of techniques from modern scattering amplitudes including, but not limited to, generalised 
unitarity, the double copy, and IBP reduction. Of course, it should be pointed out that we further enjoyed various simplifications in our calculations---
especially at loop level---by virtue of taking the classical limit. 

In Chapter~\ref{sec:Chapter5} our primary goal was to initiate a novel effort towards studying conserved quantities and 
integrability from an on-shell, gauge-invariant perspective---specifically for a quadrupolar probe particle
moving in a Kerr background. This endeavour was motivated by the wealth of available data in the literature, the powerful 
computational machinery developed to date, and the observation of spin-shift symmetry through the two-loop order in perturbation 
theory. Despite the fact that our analysis is only valid at asymptotic infinity, i.e.~for scattering dynamics, it is 
nonetheless striking that the asymptotic analogues of the known bulk conserved quantities (for a quadrupolar probe in Kerr) 
remain conserved through the quartic-in-spin order of the probe---this is two orders higher than previously established 
in the bulk. By comparing the number of independent degrees of freedom with the number of conserved quantities in involution, we 
found this to be sufficient to establish Liouville integrability. Beyond this result, we have further made some progress 
in clarifying the role of spin-shift symmetry in the dynamics. In particular, this symmetry appears to persist to all orders in 
the PM expansion and to accompany integrability, although integrability itself does not explain the origin of spin-shift 
symmetry. Indeed, given the structure of the associated differential operator in position space 
(see~\mysecref{sec:ReinterpretingSpinShiftSymm}), spin-shift symmetry is perhaps more naturally interpreted as a gauge redundancy of 
the dynamics rather than as a physical constraint. A powerful consequence of this investigation becomes especially evident when 
we ask whether conservation laws and spin-shift symmetry allow us to bootstrap gauge-invariant observables, such as the radial 
action. Remarkably, the answer to this question is yes and it carries far-reaching consequences. For the scattering dynamics of 
a hexadecapolar probe in a Kerr background at arbitrary PM order, these properties show that the full non-aligned-spin dynamics 
is completely determined by the aligned-spin configuration. This is actually a profound result, since computations for 
aligned-spin configurations are substantially simpler than for generic orientations.

%--------------------------------------------------------------------------------
\section*{Future Directions}
%--------------------------------------------------------------------------------

There are a few tantalising directions that build on the results of this thesis. On the side of 
precision, pushing these computations to higher orders in perturbation theory is a particularly interesting goal. 
In fact, the state-of-the-art for spinless scattering is currently the 5PM order~\cite{Bern:2025zno, Driesse:2024xad, 
Driesse:2026qiz, Driesse:2024feo}, which has only recently been computed. In addition to this, systematically including 
dissipative effects is also important which, practically speaking, would involve computing observables via the KMOC 
formalism~\cite{Kosower:2018adc, Maybee:2019jus} or via the radiative extension to the Dirac bracket formalism~\cite{Alessio:2025flu, 
Kim:2025hpn}. Towards the direction of higher spin, probing higher orders of the Kerr multipole expansion from an amplitudes 
perspective is very much an open question. As briefly mentioned in Chapter~\ref{sec:Chapter2}, it is not clear whether amplitudes faithfully model Kerr black holes beyond the quartic order in spin because the distinction between spin-induced (minimal) and finite-size 
(non-minimal) effects is no longer straightforward. Indeed, from the amplitudes point-of-view, this follows from the fact that we cannot 
consistently write down a theory that minimally couples gravity to a massive particle of spin greater than $2$. In fact, the theory 
can only be made consistent once we include non-minimal coupling terms, which necessarily encode the finite size of the massive particle. A central 
challenge then becomes fixing the Wilson coefficients that parametrise these non-minimal terms. Perhaps our spin interpolation method will 
provide a natural notion of what we mean by \textit{minimal} coupling beyond the quartic order in spin, especially because it relies 
on spin universality, whereby lower spin amplitudes reappear in higher spin amplitudes at both the classical and quantum 
levels. This, however, is simply a hopeful statement and requires rigorous investigation. 

There are also a few open questions regarding conserved quantities and spin-shift symmetry. In particular, understanding the exact
origin of spin-shift symmetry would be (and very much is) an interesting endeavour. Despite the fact that spin-shift symmetry might be interpreted as a 
gauge redundancy in position space---perhaps inherited from spin vectors being constructed from polarisation vectors---its exact role from 
a classical GR perspective is definitely unclear. Furthermore, we could also investigate whether conserved quantities allow us to 
constrain other gauge-invariant objects, such as the Compton amplitude. Or perhaps if we can construct conserved quantities that
are conserved beyond the probe limit and/or at higher orders in probe spin. 

\appendix
\addtocontents{toc}{\protect\tocappendix}
%--------------------------------------------------------------------------------
\chapter{Yang-Mills Feynman Rules}
\label{sec:YMFeynmanRules}
%--------------------------------------------------------------------------------
Consider the Yang-Mills (YM) theory
\begin{equation}
    \mathcal{L}_{\mathrm{YM}} = -\frac{1}{4}G_{\mu \nu}^a G^{a\mu \nu} - \frac{1}{2\xi}(\partial_\mu A^{a\mu})^2\;,
\end{equation}
where $G^a_{\mu \nu} = \partial_\mu A^a_\nu - \partial_\nu A^a_\mu + g f^{abc}A^b_\mu A^c_\nu$ is the 
non-Abelian field strength tensor. Here $a = 1,\ldots, N_c^2 -1$ are adjoint colour indices 
related to the $SU(N_c)$ gauge symmetry, $A^a_\mu$ are the gluon fields, $g$ is the YM 
coupling, and $f^{abc}$ are the structure constants. We choose the normalisation of 
the generators to be $\operatorname{Tr}\{T^a T^b\} = \delta^{ab}$, such that 
\begin{equation}
    [T^a, T^b] = i\sqrt{2}f^{abc}T^c\;,
\end{equation}
where $(T^a)^{bc} = -i\sqrt{2}f^{abc}$ are the generators of the adjoint 
representation. We further define
\begin{equation}
    \tilde{f}^{abc} = i\sqrt{2}f^{abc} = \operatorname{Tr}\{[T^a,T^b]T^c\}\;,
\end{equation} 
for convenience. The gluon propagator is
\begin{align}
    & \begin{tikzpicture}
          \begin{feynman}
            % Define vertices, their positions, and their labels
            \vertex (l1) at (-1,0);
            \vertex (r1) at (1,0);
            % Draw Feynman diagram
            \diagram*{
            (l1) -- [gluon, momentum = {[arrow shorten=0.25] $k$}] (r1),
            };
          \end{feynman}
    \end{tikzpicture}
    \ = \ \frac{-i}{k^2 + i\epsilon}\left(\eta^{\mu \nu} - (1-\xi)\frac{k^\mu k^\nu}{k^2}\right)\;,
\label{eq:GluonPropagator}
\end{align}
where we choose the Feynman gauge, $\xi = 1$. The vertex rules are
\allowdisplaybreaks{
\begin{align}
\hspace{-1.5cm}
    & \begin{tikzpicture}[baseline=(current bounding box.center)]
          \begin{feynman}
            % Define vertices, their positions, and their labels
            \vertex (v1) at (0,0);
            \vertex (l1) at (2,0) {$\nu, b$};
            \vertex (r1) at (-1,1.5) {$\mu, a$};
            \vertex (r2) at (-1,-1.5) {$\rho, c$};
            % Draw Feynman diagram
            \diagram*{
            (l1) -- [gluon, momentum={[arrow shorten=0.75] $p$}] (v1),
            (r1) -- [gluon, momentum={[arrow shorten=0.75] $k$}] (v1),
            (r2) -- [gluon, momentum={[arrow shorten=0.75] $q$}] (v1),
            };
          \end{feynman}
    \end{tikzpicture}
    \ = \ \frac{ig}{\sqrt{2}} \tilde{f}^{abc} \left[\eta^{\mu \nu}(k-p)^\rho +\eta^{\nu \rho}(p-q)^\mu - \eta^{\rho \mu}(k-q)^\nu\right]\;, \\
    & \begin{tikzpicture}[baseline=(current bounding box.center)]
      \begin{feynman}
        % Define vertices, their positions, and their labels
        \vertex (v1) at (0,0);
        \vertex (l1) at (-1.5,1.2){$\mu, a$};
        \vertex (l2) at (-1.5,-1.2){$\rho, c$};
        \vertex (r1) at (1.5,1.2) {$\nu, b$};
        \vertex (r2) at (1.5,-1.2) {$\sigma, d$};
        % Draw Feynman diagram
        \diagram*{
        (l1) -- [gluon] (v1),
        (l2) -- [gluon] (v1),
        (r1) -- [gluon] (v1),
        (r2) -- [gluon] (v1),
        };
      \end{feynman}
    \end{tikzpicture}
    \ = \ \frac{ig^2}{2}\Big[\tilde{f}^{abe}\tilde{f}^{cde}\left(\eta^{\mu \rho}\eta^{\nu \sigma} - \eta^{\mu \sigma}\eta^{\nu \rho}\right) + \tilde{f}^{ace}\tilde{f}^{bde}\left(\eta^{\mu \nu}\eta^{\rho \sigma} - \eta^{\mu \sigma}\eta^{\nu \rho}\right) \nonumber \\[-1.1cm]
    & \hspace{5.5cm} + \tilde{f}^{ade}\tilde{f}^{bce}\left(\eta^{\mu \nu}\eta^{\rho \sigma} - \eta^{\mu \rho}\eta^{\nu \sigma}\right)\Big]\;.
\end{align}
}
%--------------------------------------------------------------------------------
\chapter{Relation Between Scattering Angle and Impulse}
\label{sec:ScatteringAngleAndImpulse}
%--------------------------------------------------------------------------------

\begin{figure}[h!]
	\centering
		\includegraphics[width=0.7\linewidth]{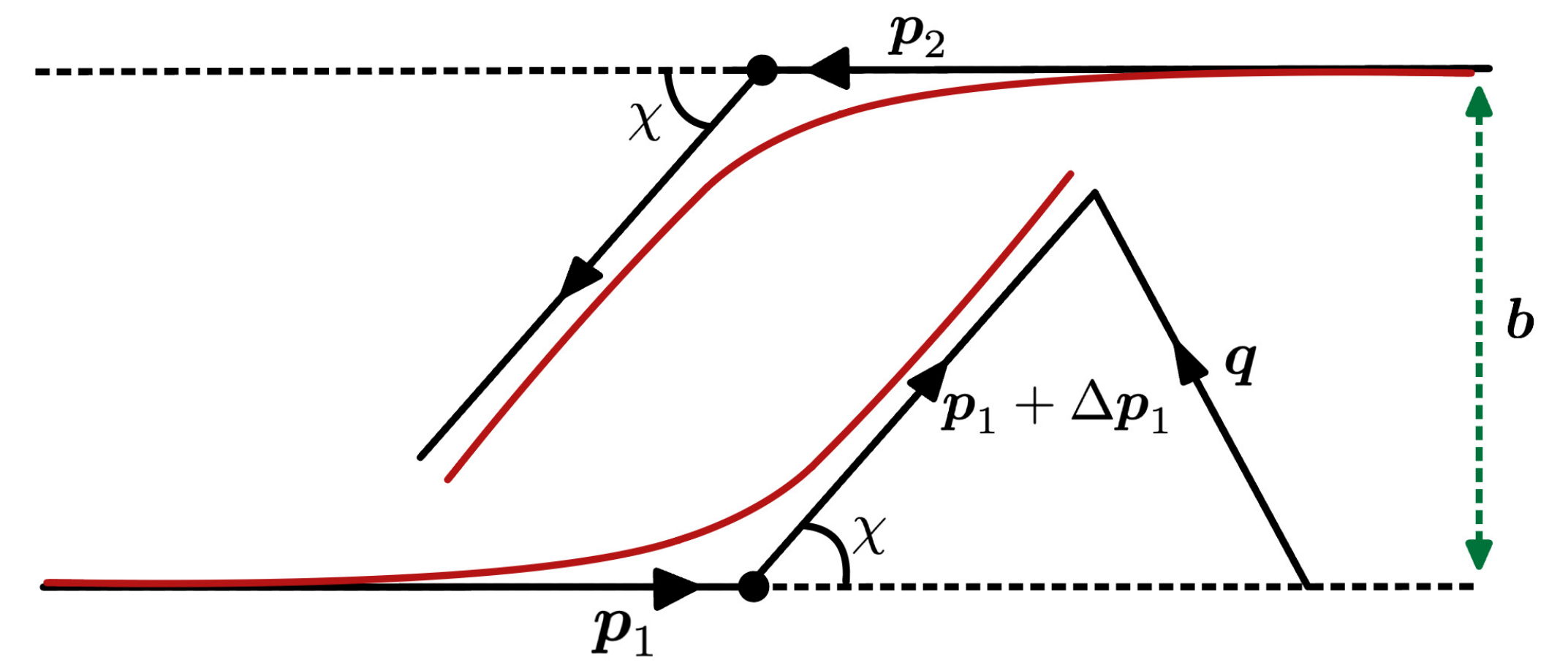}
	\caption{Black hole scattering in the centre-of-mass frame.}
	\label{fig:scatteringCOM}
\end{figure}

For the conservative scattering process in the absence of spin (or for aligned-spin configurations), 
the impulse $\Delta p_{i}^\mu$ and scattering angle $\chi$ contain exactly the same 
information~\cite{Herrmann:2021tct}. Here $i$ denotes the $i$-th particle. More specifically, we can relate 
the two by considering the components of the incoming and outgoing momenta in the centre-of-mass frame
\begin{equation}
    \begin{split}
        & p_1^\mu = (E_1, \boldsymbol{p})\;, \quad \ \ p_1^\mu + \Delta p_{1}^\mu = (E_1, \boldsymbol{p}')\;, \\
        & p_2^\mu = (E_2, -\boldsymbol{p})\;, \quad p_2^\mu + \Delta p_{2}^\mu = (E_2, -\boldsymbol{p}')\;,
    \end{split}
    \label{eq:MomentaComponents}
\end{equation}
\noindent with $|\boldsymbol{p}| = |\boldsymbol{p}'|$ such that 
\begin{equation}
    \Delta p_{1}^\mu = -\Delta p_{2}^\mu = (0, \boldsymbol{p}' - \boldsymbol{p})\;.
\end{equation}
From Fig.~\ref{fig:scatteringCOM} we find $|\boldsymbol{q}| = 2|\boldsymbol{p}| 
\operatorname{sin}{\left(\frac{\chi}{2}\right)}$ as well as $|\boldsymbol{q}| = |\boldsymbol{p}' - \boldsymbol{p}|$. 
Thus, 
\begin{equation}
    -(\Delta p_{i})^2 = (\boldsymbol{p} - \boldsymbol{p}')^2 = 4 |\boldsymbol{p}|^2 \text{sin}^2 \left( \frac{\chi}{2} \right)\;,
\end{equation}
or equivalently
\begin{equation}
	\text{sin} \left( \frac{\chi}{2} \right) = \frac{\sqrt{-(\Delta p_{i})^2}}{2|\boldsymbol{p}|}\;.
\end{equation}
This relation may be inverted to write the impulse in terms of scattering 
angle. This can conveniently be done using a decomposition
\begin{equation}
    \Delta p_{1}^\mu = \mathrm{a}_1 \frac{b^\mu}{|b|} +  \mathrm{a}_2 \check{u}_1^\mu + \mathrm{a}_3 \check{u}_2^\mu,
\end{equation}
where we define
\begin{equation}
    \check{u}_1^\mu = \frac{y u_2^\mu - u_1^\mu}{y^2 - 1}, \ \check{u}_2^\mu = \frac{y u_1^\mu - u_2^\mu}{y^2 - 1}\;.
\end{equation}
Solving for the coefficients $\mathrm{a}_i$ using the on-shell conditions for the final-state particles
\begin{equation}
	(p_i + \Delta p_i)^2 = m_i^2\;,
\end{equation}
together with momentum conservation $\Delta p_{1}^\mu = -\Delta p_{2}^\mu$ and \eqref{eq:MomentaComponents} yields 
\begin{equation}
    \Delta p_{1}^\mu = |\boldsymbol{p}|\text{sin} \left( \chi\right) \frac{b^\mu}{|b|} + |\boldsymbol{p}|^2(1-\text{cos} \left( \chi\right) )\left[ \frac{\check{u}_1^\mu}{m_1} - \frac{\check{u}_2^\mu}{m_2} \right],
\end{equation}
which can be expanded perturbatively in the coupling
\begin{equation}
\begin{split}
    & \Delta p_{1}^{\mu,(0)} = |\boldsymbol{p}| \chi^{(0)} \frac{b^\mu}{|b|}\;, \\
    & \Delta p_{1}^{\mu,(1)} = |\boldsymbol{p}| \chi^{(1)} \frac{b^\mu}{|b|} + \frac{|\boldsymbol{p}|^2}{2}(\chi^{(0)})^2\left[ \frac{\check{u}_1^\mu}{m_1} - \frac{\check{u}_2^\mu}{m_2} \right]\;, \\
    & \Delta p_{1}^{\mu,(2)} = |\boldsymbol{p}|\left( \chi^{(2)} - \frac{1}{6}(\chi^{(0)})^3 \right) \frac{b^\mu}{|b|} + |\boldsymbol{p}|^2 \chi^{(0)} \chi^{(1)}\left[ \frac{\check{u}_1^\mu}{m_1} - \frac{\check{u}_2^\mu}{m_2} \right],
\end{split}
\label{eq:ImpulseInTermsOfAngle}
\end{equation}
where $\chi^{(n)}\sim \mathcal{O}(G^{n+1})$, and the centre-of-mass three-momentum 
takes the form 
\begin{equation}
	|\boldsymbol{p}| = \frac{m_1 m_2 \sqrt{y^2-1}}{\sqrt{2m_1 m_2 y + m_1^2 + m_2^2}}\;.
\end{equation}

% --------------------------------------------------------------------------------
\chapter{Expectation Value of the Pauli-Lubanski Vector}
\label{sec:ExpectationValuePauliLubanskiVector}
% --------------------------------------------------------------------------------

In this appendix we discuss the expectation value of the Pauli-Lubanski vector following 
Ref.~\cite{Maybee:2019jus}, and derive the spin vector associated to a massive 
spin-$1$ and spin-$2$ particle. 

% --------------------------------------------------------------------------------
\section{Deriving the Spin Vector Operator}
\label{sec:DerivingTheSpinVectorOperator}
% --------------------------------------------------------------------------------
The derivation of the spin vector operator follows from QFT considerations~\cite{Maybee:2019jus}. 
The quantum mechanical interpretation of spin (up to a factor of mass) is the Pauli-Lubanski vector 
\begin{equation}
   \mathds{W}^\mu = \frac{1}{2}\epsilon^{\mu \nu \rho \sigma} \mathds{P}_\nu \mathds{J}_{\rho \sigma}\;,
   \label{eq:PauliLubanski}
\end{equation}
dependent on the translation and Lorentz generators. Within the context of QFT, the Noether charges 
associated with the respective symmetries of these generators are used to construct explicit 
field operators. For example, translation symmetry of some Lagrangian $\mathcal{L}_{\phi_s}$ 
leads to the existence of a conserved current known as the canonical stress-energy tensor 
$\Theta^{\mu \nu}$. Here the subscript ``$s$" denotes a spin-$s$ field. Then, one can determine the 
translation generator from the Noether charge
\begin{equation}
    \mathds{P}^\mu = \int \mathrm{d}^3 x \Theta^{0 \mu } = \int \mathrm{d}^3 x (\pi_s \partial^\mu \phi_s - \eta^{0\mu}\mathcal{L}_{\phi_s})\;,
    \label{eq:TranslationOperator}
\end{equation}
where $\pi_s = \partial \mathcal{L}_{\phi_s}/\partial \dot{\phi}_s$ defines the canonical momentum 
with $\dot{\phi}_s \equiv \partial_0 \phi_s$. The eigenvalues of the normal ordered field operator 
promotion of $\mathds{P}^\mu$ then defines the momentum of the particle.
In the same way, the angular momentum emerges from the conserved charge associated to Lorentz 
symmetry. The conserved current is now
\begin{equation}
    M^{\alpha \mu \nu} = x^\mu T^{\alpha \nu} - x^\nu T^{\alpha \mu}\;,
    \label{eq:LorentzTransformationCurrent}
\end{equation}
where $T^{\mu \nu}$ is the Belinfante-Rosenfeld tensor, i.e., the symmetric generalisation of 
$\Theta^{\mu \nu}$ (see Appendix~\ref{appendix:BelinfanteRosenfeldTensor} for details of its derivation). Then, 
the generator of Lorentz transformations takes the form
\begin{equation}
    \mathds{J}^{\mu \nu} = \int \mathrm{d}^3 x M^{0 \mu \nu} = \int \mathrm{d}^3 x \left( x^\mu \Theta^{0 \nu} - x^\nu \Theta^{0 \mu} + i \pi_s \mathcal{S}^{\mu \nu} \phi_s\right) \equiv \mathds{L}^{\mu \nu} + \mathds{S}^{\mu \nu}\;.
\end{equation}
The terms on the right-hand side are the \textit{orbital} and \textit{intrinsic} angular momenta, 
respectively, which cannot be distinguished in GR. The eigenvalues of the operator 
promotion define the \textit{total} angular momentum, which is the only well-defined measure of 
angular momentum. 

To isolate the spin, we need the Pauli-Lubanski vector, since orbital contributions drop 
out. To see this in action, let us consider the mode expansions of the fields and canonical 
momenta~\cite{Maybee:2019jus}
\begin{align}
    & \phi_s(x) = \sum_{\alpha} \int \mathrm{d}\Phi(k)\left(a_\alpha(k) U_\alpha(k) e^{-ik\cdot x} + b^\dagger_\alpha(k) V_\alpha(k) e^{ik\cdot x}\right), \\
    & \pi_s(x) = \sum_{\alpha} \int \mathrm{d}\Phi(k)\left(b_\alpha(k) Y_\alpha(k) e^{-ik\cdot x} + a^\dagger_\alpha(k) X_\alpha(k) e^{ik\cdot x}\right)\;, 
\end{align}
where $\alpha$ denotes the little-group index, and $a^\dagger$ and $b^\dagger$ are the particle and 
antiparticle creation operators. Then expanding $\mathds{J}^{\mu \nu}$ and calculating the inner 
product with external states yields
\begin{equation}
    \begin{split}
        \langle p',i|\mathds{J}^{\mu \nu}|p,j\rangle = & \ i \langle 0| a_i(p')\boldsymbol{:} \sum_{\alpha, \beta} \int \frac{\mathrm{d}\Phi(k)}{2E_k}\Big( X_\alpha(k) \mathcal{S}^{\mu \nu} U_\beta(k) a_\alpha^\dagger(k) a_\beta(k) \\
        & - 2x^{[\mu} k^{\nu]} X_\alpha(k) U_\beta(k) a_\alpha^\dagger(k) a_\beta(k) + \ldots \boldsymbol{:} a_j^\dagger(p)|0\rangle\;,
    \end{split}
\end{equation}
where we have used
\begin{equation}
    \int \mathrm{d}\Phi(k) = \int \frac{\hat{\mathrm{d}}^3k}{2E_k}.
\end{equation}
Now we wish to restrict ourselves to the term on the second line, which is the orbital angular 
momentum piece. Since $\mathds{P}^\mu|p,i\rangle = p^\mu|p,i\rangle$, we find
\begin{align}
    \epsilon^{\mu \nu \rho \sigma}\langle p',i|\mathds{P}_\nu \mathds{J}_{\rho \sigma}|p,j\rangle & = i \sum_{\alpha, \beta} \int \frac{\mathrm{d}\Phi(k)}{2E_k} \epsilon^{\mu \nu \rho \sigma} p^\prime_\nu k_\rho x_\sigma X_\alpha U_\beta \langle 0 | a_i(p') a^\dagger_\alpha(k) a_\beta(k) a_j^\dagger(p)|0\rangle \nonumber\\
    & = \frac{i}{E_{p^\prime}} \epsilon^{\mu \nu \rho \sigma} p^\prime_\nu p^\prime_\rho x_\sigma X_i(p') U_j(p') \hat{\delta}^{(3)} (\boldsymbol{p}- \boldsymbol{p}^\prime) \\
    & = 0\;\nonumber.
\end{align}
Therefore, the Pauli-Lubanski vector only receives contributions from the intrinsic spin part of 
the total angular momentum. Thus, we define
\begin{equation}
    \langle p',i|\mathds{W}^\mu|p,j\rangle := m \mathds{S}^\mu_{ij}(p) \hat{\delta}_{\Phi}(p-p')\;,
\end{equation}
where $\mathds{S}^\mu_{ij}$ is the spin vector operator defined in spin space, and 
\begin{equation}
    \hat{\delta}_{\Phi}(p) = 2E_p \hat{\delta}^{(3)}(\boldsymbol{p})\;.
\end{equation}
The explicit form of $\mathds{S}^\mu_{ij}$ is then obtained by computing the constant tensors 
$U_\alpha$ and $X_\alpha$ appearing in the fields and canonical momentum 
of a spin-$s$ particle. We will now do this for the case of spin-$1$ and spin-$2$.

% --------------------------------------------------------------------------------
\subsection{The Spin-1 Case}
\label{sec:SpinVector_Spin1}
% --------------------------------------------------------------------------------

Considering the Proca Lagrangian \myeqref{eq:FreeComplexProca} we find that the conjugate momentum 
is 
\begin{equation}
	\pi^\mu = \frac{\partial \mathcal{L}_{\mathrm{P}}}{\partial (\partial_0 V_\mu)} = - V^{0\mu} = - \partial_0 V^\mu + \partial^\mu V^0\;,
\end{equation}
such that the constant tensor in mode expansion is 
$-i k^0 \varepsilon^\mu_\alpha (k)$, since the second term gives a zero contribution in the expectation value. 
For the particle matrix element considered here, only the $a^\dagger a$ sector contributes. 
Indeed, for a real Proca field one may identify $b_\alpha=a_\alpha$; this identification is not 
needed for the extraction of the particle spin vector. Plugging this into the 
expectation value yields\footnote{The minus sign here comes from the fact that the constant vector 
$X_\alpha(p)$ is the conjugate of $-i k^0 \varepsilon^\mu_\alpha (p)$.}
\begin{equation}
    \langle p',i|\mathds{W}^\mu|p,j\rangle = -\frac{1}{2}\epsilon^{\mu \nu \rho \sigma}p_\nu \varepsilon^\star_{i\alpha}(p)(\Sigma_{(1) \rho \sigma})^{\alpha \beta} \varepsilon_{j\beta}(p)\hat{\delta}_{\Phi}(p-p')\;,
\end{equation}
where
\begin{equation}
    \big(\Sigma_{(1)}^{\rho \sigma}\big)^{\mu \nu} = i(\eta^{\rho \mu }\eta^{\sigma \nu} - \eta^{\sigma \mu }\eta^{\rho \nu})\;,
    \label{eq:Spin1LorentzGen}
\end{equation}
is the spin-$1$ representation of the infinitesimal Lorentz generator. Thus, equating the two 
expressions gives
\begin{equation}
    \mathds{S}^\mu_{ij}(p) = -\frac{i}{m}\epsilon^{\mu \nu \rho \sigma} p_\nu \varepsilon^\star_{i\rho}(p) \varepsilon_{j\sigma}(p),
\end{equation}
which is the form used in \myeqref{eq:SpinVectorOperator}.

% --------------------------------------------------------------------------------
\subsection{The Spin-2 Case}
\label{sec:SpinVector_Spin2}
% --------------------------------------------------------------------------------

To do the same for massive spin-2 particles, we need the canonical momentum in a 
covariant form. To derive this, we will follow Ref.~\cite{Koenigstein:2015asa} and, for 
fun, acquire the covariant Hamiltonian.

If a field $\phi_I$ is a Lorentz tensor of rank-$n$, then the conjugate momentum, which 
is a Lorentz tensor of rank-$(n+1)$, is defined as
\begin{equation}
    \pi^{\mu}_I = \frac{\partial \mathcal{L}_{\phi_I}}{\partial(\partial_\mu \phi_I)}\;.
    \label{eq:CovariantConjugateMomentum}
\end{equation}
The covariant Hamiltonian density follows as 
\begin{equation}
    \mathcal{H}(\phi_I, \pi^\mu_I) = \pi^\mu_I \partial_\mu \phi_I - \mathcal{L}_{\phi_I}\;,
    \label{eq:CovariantHamiltonian}
\end{equation}
and the EoM are acquired from Hamilton's equations
\begin{equation}
    \frac{\partial \mathcal{H}}{\partial \pi^\mu_I} = \partial_\mu \phi_I\;, \qquad \frac{\partial \mathcal{H}}{\partial \phi_I} = -\partial_\mu \pi^\mu_I\;.
\end{equation}
For a massive symmetric spin-2 field, the explicit canonical momentum reads
\begin{align}
    \pi^{\alpha \beta \gamma} & = \frac{\partial \mathcal{L}_{\text{FP}}}{\partial(\partial_\gamma \phi_{\alpha \beta})} \label{eq:Spin2CovariantConjugateMomentum} \\
    & = \frac{\partial}{\partial(\partial_\gamma \phi_{\alpha \beta})}\Bigg(\frac{1}{2}\partial_\rho \phi_{\mu \nu}\partial^\rho \phi^{\mu \nu} - \partial^\nu \phi_{\nu \rho} \partial_\mu \phi^{\mu \rho} + \partial_\mu \phi^{\mu \nu} \partial_\nu \phi - \frac{1}{2}\partial_\mu \phi \partial^\mu \phi\Bigg) \nonumber \\
    & = \partial^\gamma \phi^{\alpha \beta} - \eta^{\gamma \alpha}\partial_\mu \phi^{\mu \beta} - \eta^{\gamma \beta}\partial_\mu \phi^{\mu \alpha} + \eta^{\gamma (\alpha}\partial^{\beta)} \phi + \eta^{\alpha \beta}\partial_\mu \phi^{\mu \gamma} - \eta^{\alpha \beta} \partial^\gamma \phi \nonumber\;,
\end{align}
where in the second term of $\mathcal{L}_{\text{FP}}$ we have performed, at the level of 
the action, an integration by parts twice to move the derivatives. Notice that the 
canonical momentum is symmetric in $(\alpha,\beta)$, which should be expected since the field 
itself is symmetric. To derive the covariant Hamiltonian, it is useful to notice that 
\begin{equation}
    \mathcal{L}_{\text{FP}} = \frac{1}{2}\pi_{\alpha \beta \gamma}\partial^\gamma \phi^{\alpha \beta} - \frac{m^2}{2}(\phi_{\mu \nu}\phi^{\mu \nu} - \phi^2)\;,
\end{equation}
such that
\begin{equation}
    \begin{split}
        \mathcal{H}_{\text{FP}} & = \pi_{\alpha \beta \gamma}\partial^\gamma \phi^{\alpha \beta} - \mathcal{L}_{\text{FP}} \\
        & = \frac{1}{2}\pi_{\alpha \beta \gamma}\partial^\gamma \phi^{\alpha \beta}  + \frac{m^2}{2}(\phi_{\mu \nu}\phi^{\mu \nu} - \phi^2)\;.
    \end{split}
\end{equation}
To rewrite the kinetic piece completely in terms of canonical momenta, we make use 
of the following
\begin{align}
    & \pi_{\alpha}^{~ \alpha\gamma} = -(D-2)\partial^\gamma \phi + (D-2) \partial_\alpha \phi^{\alpha \beta} \stackrel{D = 4}{=} -2\partial^\gamma \phi + 2 \partial_\alpha \phi^{\alpha \beta}\;, \\
    & \pi^{\alpha ~ \beta}_{~\beta} = \frac{1}{2}(D-1)\partial^\alpha \phi - (D-1) \partial_\beta \phi^{\beta \alpha} \stackrel{D = 4}{=} \frac{3}{2}\partial^\alpha \phi - 3\partial_\beta \phi^{\beta \alpha}\;, \\
    & \pi_{\alpha}^{~\beta \alpha} = \frac{1}{2}(D-1)\partial^\beta \phi - (D-1) \partial_\alpha \phi^{\alpha \beta} \stackrel{D = 4}{=} \frac{3}{2}\partial^\beta \phi - 3\partial_\alpha \phi^{\alpha \beta}\;.
    \label{eq:CanonicalMomentumIdentities}
\end{align}
Thus, inverting Eq.~\eqref{eq:Spin2CovariantConjugateMomentum} for 
$\partial^\gamma \phi^{\alpha \beta}$ and using 
Eq.~\eqref{eq:CanonicalMomentumIdentities} yields
\begin{equation}
    \mathcal{H}_{\text{FP}} = \frac{1}{2}\pi_{\alpha \beta \gamma}\pi^{\alpha \beta \gamma} - \frac{1}{6}\pi_{\alpha \beta}^{~~~\alpha}\pi_{\rho}^{~\beta \rho} - \frac{1}{6}\pi_{\alpha \beta}^{~~~\beta}\pi^{\alpha ~ \rho}_{~\rho} - \frac{1}{4}\pi_{\alpha ~~ \gamma}^{~\alpha}\pi_{\rho}^{~\rho \gamma} + \frac{m^2}{2}(\phi_{\mu \nu}\phi^{\mu \nu} - \phi^2)\;.
\end{equation}

Now that we have completed our side quest of computing the Hamiltonian density, 
we focus on the conjugate momentum. Evaluating 
\myeqref{eq:Spin2CovariantConjugateMomentum} on-shell, i.e. enforcing the EoM in 
\myeqref{eq:Spin2_EOM_Set}, with $\gamma = 0$ gives
\begin{equation}
    \pi^{\alpha \beta 0} = \frac{\partial \mathcal{L}_{\text{FP}}}{\partial(\partial_0 \phi_{\alpha \beta})} \Big \rvert_{\text{on-shell}}= \partial^0 \phi^{\alpha \beta}\;.
\end{equation}
Thus, the constant tensor in the conjugate momentum is simply 
$ik^0 \varepsilon_i^{\mu \nu}(k)$, such that we find the spin-$2$ spin vector operator 
to be
\begin{equation}
    \mathds{S}^\mu_{ij} = \frac{1}{2m}\epsilon^{\mu \lambda \alpha \beta} p_\lambda \varepsilon^\star_{i \gamma \kappa}(p)\big(\Sigma_{(2)\alpha \beta}\big)^{\gamma \kappa}_{~~\rho \sigma}\varepsilon^{\rho \sigma}_j(p)\;.
\end{equation}
Note that we use the infinitesimal Lorentz generator in the spin-$2$ representation:
\begin{equation}
    \big(\Sigma_{(2)}^{\alpha \beta}\big)^{\mu \nu}_{~~\rho \sigma} = 4 i \delta^{(\mu}_{~~[\alpha} \eta^{\vphantom{\rho}}_{\beta] (\rho}\delta^{\nu)}_{~~\sigma)}\;.
    \label{eq:Spin2LorentzGen}
\end{equation}

% --------------------------------------------------------------------------------
\subsection{Lorentz Generator in the Spin-2 Representation}
\label{sec:LorentzGeneratorSpin2}
% --------------------------------------------------------------------------------

One can derive the spin-$2$ representation of the Lorentz generator by considering  
some general symmetric rank-$2$ tensor field $\gamma^{\mu \nu}(x)$ and transforming it 
under some Lorentz transformation $\Lambda^{\mu}_{~\nu}$
\begin{align}
    \bar{\gamma}^{\mu \nu}(x) & = \frac{1}{2}\Big(\Lambda^{\mu}_{~\rho} \Lambda^{\nu}_{~\sigma} + \Lambda^{\nu}_{~\rho} \Lambda^{\mu}_{~\sigma} \Big) \gamma^{\rho \sigma}(\Lambda^{-1}x) \nonumber \\ 
    & \begin{aligned}
        \ = \gamma^{\mu \nu}(x) - \frac{i}{2}\omega_{\alpha \beta}\Big[-\frac{1}{2}\Big(\big(\Sigma_{(1)}^{\alpha \beta}\big)^\gamma_{~\lambda} & x^\lambda \partial_\gamma \delta^\mu_{~\rho} \delta^\nu_{~\sigma} - \big(\Sigma_{(1)}^{\alpha \beta}\big)^\mu_{~\rho}\delta^\nu_{~\sigma} - \big(\Sigma_{(1)}^{\alpha \beta}\big)^\nu_{~\rho}\delta^\mu_{~\sigma}\\
        & + (\mu \leftrightarrow \nu)\Big) \Big]\gamma^{\rho \sigma}(x) + \mathcal{O}(\omega^2) 
    \end{aligned} \nonumber \\
    & \stackrel{!}{=} \gamma^{\mu \nu}(x) - \frac{i}{2}\omega_{\alpha \beta} \big(J^{\alpha \beta}_{(2)}\big)^{\mu \nu}_{~~\rho \sigma}\gamma^{\rho \sigma}(x) + \mathcal{O}(\omega^2) \;.
\end{align}
Then, using \myeqref{eq:Spin1LorentzGen}, we identify the Lorentz generator in the spin-$2$ 
generator by splitting the generator into the orbital and intrinsic spin pieces
\begin{equation}
    \big(J^{\alpha \beta}_{(2)}\big)^{\mu \nu}_{~~\rho \sigma} = (x^\beta P^\alpha - x^\alpha P^\beta)\delta^\mu_{~\rho} \delta^\nu_{~\sigma} + \big(\Sigma^{\alpha \beta}_{(2)}\big)^{\mu \nu}_{~~\rho \sigma}\;,
\end{equation}
where we define $P_\mu = -i\partial_\mu$ as the generator of translations\footnote{Note that 
here we are not using operator notation, since this is classical.}. Thus, we identify the 
infinitesimal Lorentz generator in the spin-$2$ representation as
\begin{equation}
    \big(\Sigma^{\alpha \beta}_{(2)}\big)^{\mu \nu}_{~~\rho \sigma} = 4 i \delta^{(\mu}_{~~[\alpha} \eta^{\vphantom{\rho}}_{\beta] (\rho}\delta^{\nu)}_{~~\sigma)},
    \label{eq:LorentzGenerator_Spin2}
\end{equation}
which agrees with literature, such as Ref.~\cite{Cangemi:2022abk}.

% --------------------------------------------------------------------------------
\section{The Belinfante-Rosenfeld Tensor}
\label{appendix:BelinfanteRosenfeldTensor}
% --------------------------------------------------------------------------------
The conserved current associated with Lorentz symmetry can be written as
\begin{equation}
    M^{\alpha \mu \nu} = x^\mu \Theta^{\alpha \nu} - x^\nu \Theta^{\alpha \mu} + S^{\alpha \mu \nu}\;,
    \label{eq:LorentzTransformationCurrent_Symm}
\end{equation}
where $\Theta^{\mu \nu}$ is the canonical Noether stress-energy tensor and 
$S^{\alpha \mu \nu}$ is the intrinsic angular momentum. The current must be conserved
\begin{equation}
    \begin{split}
        0 = \partial_\alpha M^{\alpha \mu \nu} & = \partial_\alpha(x^\mu \Theta^{\alpha \nu}) - \partial_\alpha(x^\nu \Theta^{\alpha \mu}) + \partial_\alpha S^{\alpha \mu \nu} \\
        & = \Theta^{\mu \nu} + x^\mu \partial_\alpha \Theta^{\alpha \nu} - \Theta^{\nu \mu} - x^\nu \partial_\alpha \Theta^{\alpha \mu} + \partial_\alpha S^{\alpha \mu \nu} \\
        & = \Theta^{\mu \nu} - \Theta^{\nu \mu} + \partial_\alpha S^{\alpha \mu \nu}\;,
    \end{split}
\end{equation}
since $\partial_\mu \Theta^{\mu \nu} = 0$. Thus, this enforces
\begin{equation}
    \partial_\alpha S^{\alpha \mu \nu} = - (\Theta^{\mu \nu} - \Theta^{\nu \mu})\;,
\end{equation}
or, in other words, a source of spin-current implies a non-symmetrical canonical 
stress-energy tensor. Now our goal is to construct a symmetric stress-energy 
tensor which still allows $\partial_\alpha M^{\alpha \mu \nu} = 0$. The starting point 
is to notice that the stress-energy tensor is non-unique. This means that we can 
construct a new one
\begin{equation}
    T^{\mu \nu} = \Theta^{\mu \nu} + \partial_\alpha K^{\alpha \mu \nu}\;,
    \label{eq:SymmStressEnergy}
\end{equation}
for some arbitrary tensor $K^{\alpha \mu \nu}$, which is antisymmetric in the first 
pair of indices
\begin{equation}
    K^{\alpha \mu \nu} = -K^{\mu \alpha \nu}\;.
\end{equation}
It is then easy to see that 
\begin{equation}
    \partial_\mu T^{\mu \nu} = \partial_\mu\Theta^{\mu \nu} + \partial_\mu \partial_\alpha K^{\alpha \mu \nu} = 0.
\end{equation}
In addition to this, we require $T^{\mu \nu}$ to be symmetric, such that we can write
\begin{equation}
    \Theta^{\mu \nu} - \Theta^{\nu \mu} = - \partial_\alpha ( K^{\alpha \mu \nu} - K^{\alpha \nu \mu})\;,
\end{equation}
where a simple solution is 
\begin{equation}
    \partial_\alpha ( K^{\alpha \mu \nu} - K^{\alpha \nu \mu}) = - S^{\alpha \mu \nu}\;.
    \label{eq:SimpleSolutionForIntrinsicSpin}
\end{equation}
Taking two cyclic permutations
\begin{equation}
    K^{\mu \nu \alpha} - K^{\nu \mu \alpha} = -S^{\mu \nu \alpha }\;, \quad K^{ \nu \alpha \mu} - K^{\mu \alpha \nu } = -S^{\nu \alpha \mu }\;,
\end{equation}
followed by adding \eqref{eq:SimpleSolutionForIntrinsicSpin} with the first equation and 
subtracting the second equation, yields
\begin{equation}
    K^{\alpha \mu \nu} = \frac{1}{2}(S^{\mu \alpha \nu} + S^{\nu \alpha \mu} - S^{\alpha \mu \nu})\;.
\end{equation}
Thus, the Belinfante-Rosenfeld tensor is 
\begin{equation}
    T^{\mu \nu} = \Theta^{\mu \nu} + \frac{1}{2}\partial_\alpha(S^{\mu \alpha \nu} + S^{\nu \alpha \mu} - S^{\alpha \mu \nu})\;,
\end{equation}
which is symmetric and traceless. In general, the conserved current reads
\begin{equation}
    M^{\alpha \mu \nu} = x^\mu \Theta^{\alpha \nu} - x^\nu \Theta^{\alpha \mu} - i \frac{\mathcal{L}}{\partial(\partial_\alpha F_I)} \mathcal{S}^{\mu \nu} F^I\;,
\end{equation}
where $F^I$ is some field with its respective indices, e.g. for vectors $F^I = A^\mu$, 
and the spin tensor, $\mathcal{S}^{\mu \nu} $, is anti-symmetric, i.e. 
$\mathcal{S}^{\mu \nu} = -\mathcal{S}^{\nu \mu}$. Then, we find that
\begin{equation}
    S^{\alpha \mu \nu} =  -i \frac{\partial \mathcal{L}}{\partial(\partial_\alpha F_I)} \mathcal{S}^{\mu \nu} F^I\;,
\end{equation}
such that 
\begin{equation}
    T^{\mu \nu} = \Theta^{\mu \nu} - \frac{i}{2}\partial_\alpha \left(\frac{\partial\mathcal{L}}{\partial(\partial_\alpha F_I)} \mathcal{S}^{\mu \nu} F^I- \frac{\partial\mathcal{L}}{\partial(\partial_\nu F_I)} \mathcal{S}^{\alpha \mu} F^I  - \frac{\partial\mathcal{L}}{\partial(\partial_\mu F_I)} \mathcal{S}^{\alpha \nu} F^I \right)\;.
\end{equation}
%--------------------------------------------------------------------------------
\chapter{Covariant Lorentz Transformation for Polarisation Vectors}
\label{sec:CovariantBoostedpolarisations}
%--------------------------------------------------------------------------------
Here we discuss the derivation of the Lorentz transformations used in 
\myeqref{eq:polarisationBoost}. In order to do this in a covariant way, we want to 
start by determining some general Lorentz transformation involving the appropriate 
variables, and then enforce the following
\begin{equation}
\begin{split}
    p_1 = -\bar{p}_1 + \frac{q}{2}  \longrightarrow \bar{p}_1^{\,\prime} = \frac{m_1}{\bar{m}_1}\bar{p}_1\;, \quad p_4 = \bar{p}_1 + \frac{q}{2} \longrightarrow \bar{p}_1^{\,\prime} = \frac{m_1}{\bar{m}_1}\bar{p}_1\;,
\end{split}
\end{equation}
where the normalisation $(m_1/\bar{m}_1)$ ensures that the magnitude of momentum 
is unchanged. It is then natural to define our infinitesimal generator as
\begin{equation}
  \omega^{\mu\nu} = \beta \left(\bar{p}_1^\mu q^\nu - q^\mu \bar{p}_1^{\nu}\right)\;,
\end{equation}
where $\beta$ is a free parameter still to be defined. We then write the
exact transformation via exponentiation
\begin{equation}
  \Lambda^{\mu}_{~\nu} = \left(e^{-\frac{i}{2}\omega_{\rho \sigma}\Sigma^{\rho \sigma}}\right)^\mu_{~\nu} 
  = \left(e^{\beta X}\right)^\mu_{~\nu} = \sum_{n = 0}^{\infty}{\frac{(\beta^n X^n)^\mu_{~\nu}}{n!}}\;,
\end{equation}
where the infinitesimal Lorentz generator is
\begin{equation}
	(\Sigma^{\rho \sigma})^{\mu}_{~\nu} = i(\eta^{\rho \mu}\delta^{\sigma}_{~\nu}-\eta^{\sigma \mu}\delta^{\rho}_{~\nu})\;,
\end{equation}
and generally we have
\begin{equation}
   (X^{2n+1})^\mu_{~\nu} = (-q^2 \bar{m}_1^2)^n X^\mu_{~\nu}\;, \qquad 
   (X^{2n})^\mu_{~\nu} = (-q^2 \bar{m}_1^2)^n(X^2)^\mu_{~\nu}\;,
\end{equation}
together with
\begin{equation}
  X^\mu_{~\nu} = \bar{p}_1^\mu q_\nu - q^\mu \bar{p}_{1\nu}\;, \qquad 
  (X^2)^\mu_{~\nu} = -q^2\bar{p}_1^\mu \bar{p}_{1\nu} - \bar{m}_1^2q^\mu q_\nu\;.
\end{equation}
Splitting the summation into even and odd powers, one can write the transformation
as
\begin{equation}
\begin{aligned}
  \Lambda(\psi)^{\mu}_{~\nu} & = \delta^\mu_{~\nu} + Y^\mu_{~\nu}\sum_{n = 0}^{\infty} \frac{(\psi)^{2n+1}}{(2n+1)!} 
  + (Y^2)^\mu_{~\nu}\sum_{n = 0}^{\infty} \frac{(\psi)^{2n}}{(2n)!} - (Y^2)^\mu_{~\nu} \\
   & \begin{aligned}
     & = \delta^\mu_{~\nu} + \sinh(\psi)Y^\mu_{~\nu} + (\cosh(\psi)-1)(Y^2)^\mu_{~\nu}\;,
    \end{aligned}
\end{aligned}
\end{equation}
where $\psi = \beta\sqrt{-q^2 \bar{m}^2_1}$ and $Y^\mu_{~\nu} =
\frac{1}{\sqrt{-q^2 \bar{m}^2_1}} X^\mu_{~\nu}$. Finally, to find the
appropriate angles for our transformations, we enforce the following equalities
\begin{equation}
   \Lambda(\psi)^{\mu}_{~\nu} \bar{p}_1^{\,\prime\nu} \stackrel{!}{=} -\bar{p}^\mu_1 + \frac{q^\mu}{2}\;, \qquad 
   \Lambda(\psi')^{\mu}_{~\nu}  \bar{p}_1^{\,\prime\nu} \stackrel{!}{=} \bar{p}^\mu_1 + \frac{q^\mu}{2}\;,
\end{equation}
where $\psi'$ is the angle needed for the transformation of $p_4$.
To illustrate this explicitly, let us consider solving one of the equations
above. For example, let us consider 
\begin{equation}
\begin{split}
    -\bar{p}^\mu_1 + \frac{q^\mu}{2} & \stackrel{!}{=} \Lambda(\psi)^{\mu}_{~\nu}  \bar{p}_1^{\,\prime\nu} \\
    & = \delta^\mu_{~\nu}\bar{p}_1^{\,\prime\nu} + \sinh(\psi)Y^\mu_{~\nu}\bar{p}_1^{\,\prime\nu} + (\cosh(\psi)-1)(Y^2)^\mu_{~\nu}\bar{p}_1^{\,\prime\nu} \\
    & = -\frac{m_1}{\sqrt{-q^2}}\sinh(\psi)q^\mu + \cosh(\psi)\bar{p}_1^{\,\prime\mu}\;,
\end{split}
\end{equation}
which gives two equations
\begin{equation}
    -\frac{m_1}{\sqrt{-q^2}}\sinh(\psi) - \frac{1}{2} = 0\;, \qquad \frac{m_1}{\bar{m}_1}\cosh(\psi) + 1 = 0\;, 
\end{equation}
whose solutions are
\begin{equation}
    \sinh(\psi) = -\frac{\sqrt{-q^2}}{2 m_1}\;, \qquad \cosh(\psi) = - \frac{\bar{m}_1}{m_1}\;.
\end{equation}
Then, putting these solutions together gives us
\begin{equation}
    \psi = \arctanh\left(\frac{\sqrt{-q^2}}{2\bar{m}_1}\right),
\end{equation}
which is the angle needed for the transformation of $p_1$. The angle $\psi'$ for 
the transformation of $p_4$ to $\bar{p}_1^{\,\prime}$ is simply given by $\psi' = -\psi$.
%--------------------------------------------------------------------------------
\chapter{Evaluating Tensor Integrals in Fourier Transforms}
\label{sec:EvaluatingRankedIntegrals}
%--------------------------------------------------------------------------------
Here we discuss the computation of the tensor integrals relevant for the
calculation of observables. A crucial point is that each integral is
a function of $b^2$, meaning that we can determine higher-rank integrals from
the lower-rank integrals via differentiation. As such, the first step is to
compute the scalar integral
\begin{equation}
    f_{\beta} (b^2) = \int \hat{\mathrm{d}}^D q \ \hat{\delta}(2 m_1 u_1 \cdot q) \hat{\delta}(2m_2 u_2 \cdot q)e^{i b \cdot q} (-q^2)^{-\beta}\;.
\end{equation}
which can be done using a Sudakov decomposition of the $D$-dimensional exchange momentum
\begin{equation}
    q^\mu = x_1 u_1^\mu + x_2 u_2^\mu + \boldsymbol{q}_\perp^\mu\;,
\end{equation}
where $\boldsymbol{q}_\perp^\mu$ is in the $(D-2)$-dimensional subspace
transverse to $u_1$ and $u_2$, e.g., in the decomposition
of Ref.~\cite{Herrmann:2021tct}. Then, this parametrisation yields
\begin{equation}
\begin{split}
    f_{\beta} (b^2) & = \frac{1}{N} \int \hat{\mathrm{d}}^{D-2} \boldsymbol{q}_\perp\hat{\mathrm{d}}x_1\hat{\mathrm{d}}x_2\ \hat{\delta}(x_1) \hat{\delta}(x_2) e^{i b \cdot q} (-q^2)^{-\beta}\\
    & = \frac{1}{N} \int \hat{\mathrm{d}}^{D-2} \boldsymbol{q}_\perp \ e^{-i \boldsymbol{b} \cdot \boldsymbol{q}_\perp} (\boldsymbol{q}_\perp^2)^{-\beta} \\
    & = \frac{4}{N} \frac{\Gamma(D/2 - 1 - \beta)}{2^{2\beta + 2}\pi^{(D - 2)/2}\Gamma(\beta)} \frac{1}{|b|^{D-2-2\beta}}\;,
\end{split}
\end{equation}
where $N = 4 m_1 m_2\sqrt{y^2 - 1}$. Here, the delta functions localise the two integration variables and 
forces the momentum transfer to span the $(D-2)$-dimensional subspace. Indeed, the impact
parameter is purely transverse with $\boldsymbol{b}^2 = -b^2 \equiv |b|^2$.
Then, the final integration is done by orienting the vectors appropriately and
using spherical polar coordinates. 

The higher-rank integrals follow by differentiation. To reduce clutter, we 
define
\begin{equation}
    \xi = \frac{4}{N} \frac{\Gamma(D/2 - \beta)}{2^{2\beta + 1}\pi^{(D - 2)/2}\Gamma(\beta)}\;.
\end{equation} 
For rank-$1$ we find
\begin{equation}
f^\mu_{\beta} (b^2) = \frac{1}{i} \frac{\partial f_{\beta} (b^2)}{\partial b_\mu} = -i\xi \frac{b^\mu}{|b|^{D-2\beta}}\;,
\end{equation}
for rank-$2$ tensor integrals, we obtain
\begin{align}
f^{\mu \nu}_{\beta} (b^2) & = \frac{1}{i} \frac{\partial f^\mu_{\beta} (b^2)}{\partial b_\nu} \\
& = - \frac{\xi}{|b|^{D+2-2\beta}}
\Bigg(|b|^2 \Pi^{\mu \nu} + (D-2\beta) b^\mu b^\nu \Bigg)\;, \nonumber
\end{align}
for rank-$3$ integrals, we have
\begin{align}
\hspace{-0.25cm}
f^{\mu \nu \gamma}_{\beta} (b^2) &  = \frac{1}{i} \frac{\partial f^{\mu \nu}_{\beta} (b^2)}{\partial b_\gamma} \\
& = i\xi \frac{(D-2\beta)}{|b|^{D+4-2\beta}}
\Bigg(|b|^2( b^\mu \Pi^{\nu \gamma} + b^\nu \Pi^{\gamma \mu} + b^\gamma \Pi^{\mu \nu}) + (D+2-2\beta) b^\mu b^\nu b^\gamma\Bigg)\;. \nonumber
\end{align}
and, finally, the rank-$4$ integral is
\begin{align}
f^{\mu \nu \gamma \rho}_{\beta} (b^2) =  &  \frac{1}{i} \frac{\partial f^{\mu \nu \gamma}_{\beta} (b^2)}{\partial b_\rho} \\
	& = \xi \frac{(D-2\beta)}{|b|^{D+6-2\beta}}\Bigg(|b|^4 \Big( \Pi^{\mu \nu}\Pi^{\gamma \rho}+\Pi^{\mu \gamma}\Pi^{\nu \rho}+\Pi^{\mu \rho}\Pi^{\nu \gamma}\Big) \nonumber\\
    &  \hspace{0.25cm} + (D+2-2\beta)|b|^2(b^\mu b^\rho \Pi^{\nu \gamma} + b^\nu b^\rho \Pi^{\gamma \mu} + b^\gamma b^\rho \Pi^{\mu \nu} + b^\mu b^\gamma \Pi^{\nu \rho} \nonumber \\
    & \hspace{0.25cm} + b^\nu b^\gamma \Pi^{\mu \rho} + b^\mu b^\nu \Pi^{\gamma \rho}) + (D+2-2\beta)(D+4-2\beta) b^\mu b^\nu b^\gamma b^\rho \Bigg)\nonumber \;.
\end{align}
Importantly, in all of these expressions we have replaced all instances of the metric with the
transverse projector
\begin{equation}
\begin{split}
 \Pi^{\mu}_{~\nu} & = \frac{1}{y^2 - 1}\epsilon^{\mu \rho \alpha \beta} \epsilon_{\nu \rho \gamma \delta} u_{1\alpha }u_{2\beta}u_{1}^{\gamma} u_{2}^{\delta} \\
    & =  \delta^{\mu}_{~\nu} + \frac{1}{y^2 - 1}\Big(u_1^\mu (u_{1\nu} - y u_{2\nu}) + u_2^\mu (u_{2\nu} - y u_{1\nu})\Big)\;,
\end{split}
\end{equation}
which takes us to a plane orthogonal to the velocities, i.e.~the transverse plane.
%--------------------------------------------------------------------------------
\chapter{The Carter Constant from Hamilton-Jacobi Equations}
\label{sec:CarterConstFromHJEquations}
%--------------------------------------------------------------------------------

In this appendix we wish to derive the Carter constant by solving the Hamilton-Jacobi equations. To start, the 
relativistic Hamiltonian for the geodesic motion of a spinless probe particle of mass $m$ and 
four-momentum $p_\mu$ is
\begin{equation}
	H(x,p) = \frac{1}{2} g^{\mu \nu} p_\mu p_\nu\;.
\end{equation}
This guarantees the validity of the geodesic equations
\begin{equation}
	p^\mu \nabla_\mu p^\nu = 0\;, \quad g^{\mu \nu} p_\mu p_\nu = m^2\;.
\end{equation}
Introducing Hamilton's principal function $S(x)$ via the relation $p_\mu = \partial_\mu S$, we can 
recast the mass-shell condition as
\begin{equation}
	g^{\mu \nu} \partial_\mu S \partial_\nu S = m^2\;,
	\label{eq:HJEquation}
\end{equation} 
which is the Hamilton-Jacobi equation for a massive particle. The goal is to solve this equation for $S(x)$.

Since Kerr is stationary and axisymmetric, we will search for solutions that are completely 
separable\footnote{This is consistent provided that the spacetime of interest adheres to certain properties, see 
Ref.~\cite{Chervonyi:2015ima}.}
\begin{equation}
	S(x) = -Et + L \phi + S_r(r) + S_\theta (\theta)\;,
\end{equation}
such that 
\begin{align}
	\partial_t S = - E\;, \quad \partial_\phi S = L\;, \quad \partial_r S = \frac{d S_r}{dr}\;, \quad  \partial_\theta S = \frac{d S_\theta}{d \theta}\;.
\end{align}
Plugging this into \myeqref{eq:HJEquation}, together with \myeqref{eq:InvKerrMetric}, gives
\begin{align}
	0 & = g^{tt}(\partial_t S)^2 + 2 g^{t\phi}(\partial_t S)(\partial_\phi S) + g^{rr}(\partial_r S)^2 + g^{\theta \theta}(\partial_\theta S)^2 + g^{\phi \phi}(\partial_\phi S)^2 - m^2 \nonumber \\ 
	& = \frac{1}{\Sigma \Delta}\left[r^2 + a^2 - \Delta a^2 \operatorname{sin}^2(\theta) \right] E^2 - \frac{4Mar}{\Sigma \Delta} EL - \frac{\Delta}{\Sigma}\left( \frac{d S_r}{dr} \right)^2 \\ 
	& \hspace{4cm} - \frac{1}{\Sigma}\left( \frac{d S_\theta}{d\theta} \right)^2 - \frac{\Delta - a^2 \operatorname{sin}^2(\theta)}{\Sigma \Delta \operatorname{sin}^2(\theta)}L^2 - m^2\;. \nonumber
\end{align}
Multiplying by $\Sigma$ leads to 
\begin{align}
	0 & = m^2 \Sigma + \Delta \left( \frac{d S_r}{dr} \right)^2 + \left( \frac{d S_\theta}{d\theta} \right)^2 - \left[\frac{(r^2 + a^2)^2}{\Delta} - a^2 \operatorname{sin}^2(\theta)\right] E^2 \\ 
	& \hspace{4cm} + \frac{4Mar}{\Delta} EL + \left(\frac{1}{\operatorname{sin}^2(\theta)} - \frac{a^2}{\Delta}\right) L^2 \nonumber \\
	& = m^2 r^2 + \Delta \left( \frac{d S_r}{dr} \right)^2 - \frac{(r^2 + a^2)^2}{\Delta} E^2 - \frac{a^2 L^2}{\Delta} + \frac{2(r^2 + a^2)}{\Delta}aEL - 2aEL \nonumber \\
	& \hspace{2cm} + m^2 a^2 \operatorname{cos}^2(\theta) + \left( \frac{d S_\theta}{d\theta} \right)^2 + a^2 E^2  \operatorname{sin}^2(\theta) + \frac{L^2}{\operatorname{sin}^2(\theta)} \nonumber \\
	& = m^2 r^2 + \Delta \left( \frac{d S_r}{dr} \right)^2 - \frac{1}{\Delta}\left[(r^2 + a^2)E - aL\right]^2 + \left( \frac{d S_\theta}{d\theta} \right)^2 \nonumber \\
	& \hspace{2cm} + \operatorname{cos}^2(\theta)\left[ a^2 (m^2 - E^2) + \frac{L^2}{\operatorname{sin}^2(\theta)} \right] + (aE - L)^2\;,
\end{align}
where, for the second equality, we have used the fact that 
$4Mr = 2(r^2 + a^2) - 2\Delta$, while in the third equality we added a zero 
$L - L = 0$ and used $\operatorname{sin}^2(\theta) + 
\operatorname{cos}^2(\theta) = 1$.

Rearranging all the $r$ and $\theta$ dependences onto either side of the equality, we 
quickly realise that the equality only holds if both expressions are equal to some 
constant, say $\mathcal{K}$. This constant is precisely the Carter constant in 
\mysecref{sec:SymmetriesOfTheKerrMetric}. More specifically, we find
\begin{align}
	& \mathcal{K} \equiv -\Delta p_r^2 - m^2 r^2 + \frac{1}{\Delta}\left[(r^2 + a^2)E - aL\right]^2 \;, \\
	& \mathcal{K} \equiv p_\theta^2 + \operatorname{cos}^2(\theta)\left[ a^2 (m^2 - p_t^2) + \frac{p_\phi^2}{\operatorname{sin}^2(\theta)} \right] + (ap_t + p_\phi)^2\;,
\end{align}
where we have used 
\begin{equation}
	p_t = -E\;, \quad p_\phi = L\;, \quad p_r = \frac{d S_r}{dr} \;, \quad p_\theta =  \frac{d S_\theta}{d\theta}\;.
\end{equation}
As $(ap_t + p_\phi)^2$ is conserved, it is useful to define the so-called 
\textit{Carter integral}~\cite{Gonzo:2023goe, Gonzo:2024zxo}
\begin{equation}
	\mathcal{K} - (ap_t + p_\phi)^2 \equiv p_\theta^2 + \operatorname{cos}^2(\theta)\left[ a^2 (m^2 - p_t^2) + \frac{p_\phi^2}{\operatorname{sin}^2(\theta)} \right]\;.
\end{equation}

%--------------------------------------------------------------------------------
\section*{The Hamilton-Jacobi Equation}
%--------------------------------------------------------------------------------

The equation~\myeqref{eq:HJEquation} can also be derived from the general form of the  Hamilton-Jacobi equation
\begin{equation}
	H(x, \partial_\mu S) + \frac{\partial S}{\partial \lambda} = 0\;,
\end{equation}
where $\lambda$ is an arbitrary affine parameter $x^\mu(\lambda)$. The dynamics of a massive point 
particle is conveniently encoded in a Polyakov form
\begin{equation}
	S = - \frac{1}{2} \int \mathrm{d}\lambda \left(e^{-1}g_{\mu \nu} \dot{x}^\mu \dot{x}^\nu + em^2\right) = \int \mathrm{d}\lambda \ L\;,
\end{equation} 
where $e(\lambda)$ is an einbein. The conjugate momentum is
\begin{equation}
	p_\mu = \frac{\partial L}{\partial \dot{x}^\mu} = -e^{-1} g_{\mu \nu}\dot{x}^\nu\;,
\end{equation}
which is inverted to obtain $\dot{x}^\mu = -e g^{\mu \nu} p_\nu$. The (off-shell) Hamiltonian follows as
\begin{equation}
	H = p_\mu \dot{x}^\mu - L = -\frac{e}{2}\left(g^{\mu \nu} p_\mu p_\nu - m^2\right)\;,
\end{equation}
where $e$ is completely arbitrary. To proceed, we introduce Hamilton's principal function 
$S(x, \lambda)$, which is simply the action evaluated on a classical trajectory---in other words, it satisfies the 
Euler-Lagrange equations---connecting a fixed initial point to a variable endpoint 
$x = x_{\mathrm{cl}}(\lambda)$:
\begin{equation}
	S(x, \lambda) = \int_{\lambda_0}^{\lambda} \mathrm{d} \lambda' \ L(x_{\mathrm{cl}}, \dot{x}_{\mathrm{cl}},e)\;.
\end{equation}
Varying the trajectory $x_{\mathrm{cl}}^\mu \rightarrow x_{\mathrm{cl}}^\mu + \delta x_{\mathrm{cl}}^\mu$, 
together with $\delta x_{\mathrm{cl}}^\mu (\lambda_0) = 0$ and $\delta x_{\mathrm{cl}}^\mu (\lambda) = 
\delta x^\mu \neq 0$, 
we find
\begin{align}
	\delta S(x, \lambda) & = \int_{\lambda_0}^{\lambda} \mathrm{d} \lambda' \left[\frac{\partial L}{\partial x_{\mathrm{cl}}^\mu} - \frac{d}{d\lambda'}\left( \frac{\partial L}{\partial \dot{x}_{\mathrm{cl}}^\mu}\right)\right] \delta x_{\mathrm{cl}}^\mu + \left[ \frac{\partial L}{\partial \dot{x}_{\mathrm{cl}}^\mu} \delta x_{\mathrm{cl}}^\mu\right]_{\lambda_0}^{\lambda}\;.
\end{align}
As we are evaluating the action on the classical trajectory, the EoMs are satisfied, such that only the boundary 
term survives
\begin{equation}
	\delta S(x, \lambda) = \frac{\partial L}{\partial \dot{x}^\mu_{\mathrm{cl}}(\lambda)} \delta x^\mu_{\mathrm{cl}}(\lambda) - \frac{\partial L}{\partial \dot{x}_{\mathrm{cl}}^\mu(\lambda_0)} \delta x_{\mathrm{cl}}^\mu(\lambda_0) = p_\mu \delta x^\mu\;,
\end{equation}
where we identify the conjugate momentum in the usual way
\begin{equation}
	p_\mu = \frac{\partial L}{\partial \dot{x}^\mu}\;.
\end{equation}
Finally, by identifying 
\begin{equation}
	\delta S = \frac{\partial S}{\partial x^\mu} \delta x^\mu\;,
\end{equation}
we arrive at the expression
\begin{equation}
	p_\mu = \frac{\partial }{\partial x^\mu} S(x,\lambda)\;.
\end{equation}
Substituting the Hamiltonian into the Hamilton-Jacobi equation, we find
\begin{align}
	-\frac{e}{2}\left(g^{\mu \nu} \partial_\mu S \partial_\nu S - m^2\right) + \partial_\lambda S = 0\;,
\end{align}
which reduces to 
\begin{equation}
	g^{\mu \nu} \partial_\mu S \partial_\nu S - m^2 = 0\;.
\end{equation}
This simplification follows from the action of a relativistic point particle being reparametrisation 
invariant, i.e.~$\partial_\lambda S = 0$ such that $S(x,\lambda) = S(x)$.

\backmatter

\singlespace

\phantomsection
\addcontentsline{toc}{chapter}{\bibname}
% Choose a bibliography style to suit your taste here
% This one was downloaded from http://web.reed.edu/cis/help/latex/bibtexstyles.html (June 2012)
\bibliographystyle{JHEP}
\bibliography{thesis}

\end{document}